\documentclass[twocolumn,amsmath,amssymb,10pt,a4paper]{revtex4}

\pdfoutput=1

\usepackage{hyperref}

\hypersetup{
        colorlinks=true,
        linktoc=all,
        linkcolor=violet,
        urlcolor=blue
}

\usepackage{geometry}
\usepackage{amsmath, amssymb}
\usepackage{graphicx}
\usepackage{bbold}
\usepackage{amsthm}
\usepackage{mathtools}
\usepackage{MnSymbol}
\usepackage{multirow}
\usepackage{marvosym}

\theoremstyle{definition}

\theoremstyle{remark}

\newcommand{\hobs}{\ensuremath{\hbar_\text{\Yinyang}}}

\newcommand{\be}{\begin{equation}}
\newcommand{\ee}{\end{equation}}
\newcommand{\bd}{\begin{displaymath}}
\newcommand{\ed}{\end{displaymath}}
\newcommand{\BE}{\begin{eqnarray}}
\newcommand{\EE}{\end{eqnarray}}

\newcommand{\bra}{\left\langle}
\newcommand{\ket}{\right\rangle}

\newcommand{\id}{{\rm 1\!\!I}}

\newcommand{\bs}{\ensuremath{\mathbf{s}}}
\newcommand{\bu}{\ensuremath{\mathbf{u}}}

\newcommand{\bx}{\ensuremath{\mathbf{x}}}
\newcommand{\by}{\ensuremath{\mathbf{y}}}

\newcommand{\tr}{\mathrm{Tr}}

\newcommand{\mcN}{\mathcal{N}}
\newcommand{\mcK}{\mathcal{K}}
\newcommand{\mcH}{\mathcal{H}}
\newcommand{\mcP}{\mathcal{P}}
\newcommand{\mcR}{\mathcal{R}}

\graphicspath{{../../figs/}}

\usepackage{epigraph,varwidth}
\usepackage{etoolbox}

\usepackage{tcolorbox}
\newtcolorbox{mybox}{colback=green!5!white,colframe=green!75!black}

\def\c#1{\textcolor{blue}{#1}}

\newenvironment{myquotation}{\setlength{\leftmargini}{1em}\setlength{\rightmargin}{1em}\quotation}{\endquotation}

\usepackage[toc,page]{appendix}

\newcommand{\nocontentsline}[3]{}
\newcommand{\tocless}[2]{\bgroup\let\addcontentsline=\nocontentsline#1{#2}\egroup}

\begin{document}

\title{Embodied observations from an intrinsic perspective can entail quantum dynamics}

\author{John Realpe-G\'omez}\email{john.realpe@lrc.systems}\email{john.realpe@gmail.com}
\affiliation{Laboratory for Research in Complex Systems, San Francisco, USA}
\affiliation{Complex Systems and Statistical Physics Group, School of Physics and Astronomy, The University of Manchester\footnote{Part of this work was developed while being Research Associate at The University of Manchester.}, Manchester M13 9PL, United Kingdom}
\affiliation{Instituto de Matem\'aticas Aplicadas, Universidad de Cartagena, Bol\'ivar 130001, Colombia}

\date{\today}

\begin{abstract}

After centuries of research, how subjective experience relates to physical phenomena remains unclear. Recent strategies attempt to identify the physical correlates of experience. Less studied is how scientists eliminate the ``spurious'' aspects of their subjective experience to establish an ``objective'' science. Here we model scientists doing science. This entails a dynamics formally analogous to quantum dynamics. The analogue of Planck's constant is related to the process of observation. This reverse-engineering of science suggests that some ``non-spurious'' aspects of experience remain: embodiment and the mere capacity to observe from an intrinsic perspective. A relational view emerges: every experience has a physical correlate and every physical phenomenon is an experience for ``someone''. This may help bridge the explanatory gap and hints at non-dual modes of experience.

\end{abstract}

\maketitle

\epigraph{\em To study your own self is to forget yourself. To forget yourself is to have the ``objective'' world prevail in you. To have the ``objective'' world prevail in you is to let go of your ``own'' body and mind as well as the body and mind of ``others.''}{Dogen}


\tocless\section{Introduction}\label{sm:intro}

There is mounting evidence that human experience is related to patterns of neural activity, the neural correlates.  A variety of neural correlates of concepts previously considered taboo, such as consciousness~\cite{dehaene2017consciousness,koch2016neural,mashour2018controversial,van2018threshold,joglekar2017inter} and the self~\cite{schaefer2017conscious,northoff2011self,northoff2004cortical,gusnard2001medial,legrand2009self,lenggenhager2007video,ehrsson2007experimental,blanke2012multisensory,blanke2015behavioral,blanke2009full}, have been identified. Brain stimulation techniques~\cite{selimbeyoglu2010electrical,lenggenhager2007video,ehrsson2007experimental} can induce a variety of subjective experiences, such as seeing faces, movement, and colors, having heath sensations, or feeling out of the body. Brain imagining techniques~\cite{lebedev2017brain} can directly ``read out'' thoughts. Such techniques have allowed humans to navigate virtual worlds without any traditional sensory input~\cite{losey2016navigating}, control avatars and robots by thought~\cite{cohen2014controlling,cohen2012fmri}, and extend their capabilities via direct brain-machine~\cite{lebedev2017brain,penaloza2018bmi} and brain-brain communication~\cite{stocco2015playing,jiang2018brainnet}. 

Despite such an outstanding progress, a question persists: how does subjective experience relate to physical phenomena? This is the so-called {\em hard problem of consciousness}~\cite{chalmers1995facing}. ``The problem is that physical accounts explain the structure and function as characterized from the {\em outside}, but a conscious state is defined by its subjective character as experienced from {\em inside}''~\cite{thompson2010mind} (p. 222). At the root of the problem is that the intrinsic perspective is {\em by definition} the {\em opposite} of the extrinsic perspective. It concerns how ``I'' observe or experience something from ``my'' own perspective, not how an external observer considers ``I'' should experience it---if left implicit, such external observers can become homunculi (see Appendix~\ref{s:on_internal} and Fig.~\ref{f:homunculus} therein). 

Universal principles have often been discovered through the study of specific instances of the phenomena of interest---think, e.g., of gravity and falling apples. Similarly, we could in principle learn about the intrinsic perspective by turning our attention to the best and perhaps only instance we have access to: our own.  If ``I'' manage to capture the universal features of how ``I'' experience something, ``I'' might capture how any other system with an intrinsic perspective, biological or artificial, would intrinsically experience it. 

However, scientists often study the intrinsic perspective through the analysis of other generic subjects---usually non-scientists. It is said that subjects report their (subjective) experiences, while scientists report their (objective) observations. But what if instead of generic subjects, scientists investigate other scientists with access to the same experimental resources? In this case the distinction between experiences and observations effectively dissolves. Consider, for instance, an experiment where a subject experiences an optical illusion, while the scientist realizes that it is an illusion thanks to his experimental devices. However, if the subject has access to the same experimental resources, he can also realize that it is an illusion  (cf.~Ref.~\cite{velmans2009understanding}, p. 212, and Fig.~9.2 therein; see Appendix~\ref{s:thompson-science} herein).

Furthermore, usually we take observations of measuring devices as ``public'' and experiences reported by subjects in psychological experiments as ``private''. However, observations of measuring devices could also be considered as ``private'' experiences scientists have. From this perspective, since scientists cannot directly access the experiences of each other, but only their reports, there is no way for them to be certain that their experiences of the measuring device are actually similar (the problem of other minds). Strictly speaking, we cannot rule out the possibility that they {\em might} be lying to each other. It is that we take it for granted that if scientists report the same ``private'' experience of the measuring device, then the corresponding observations are the same~\cite{velmans2009understanding} (pp. 221-222; see Appendix~\ref{s:thompson-science} herein). 

In generic subjects the intrinsic perspective is often associated with additional complexities, like cognitive, perceptual and motivational biases, that may obscure its most basic aspects. Scientists, instead, have successfully bracketed out such additional complexities through the constitution of an ``objective'' science out of their multiple intrinsic perspectives. Importantly, this does not necessarily imply that scientists have eliminated the intrinsic perspective and cognition altogether, but only those ``spurious'' aspects that prevent them from achieving ``objectivity'' (see below). Scientists investigating scientists could therefore help identify the most basic aspects that remain.

Accordingly, the fundamental regularities we call ``laws of nature'' might well in part reflect the most fundamental, ineliminable aspects of the intrinsic perspective. In principle, such aspects could be revealed by inferring from the ``laws of nature'' a model of scientists doing science that is consistent with such laws. If so, the very same equations we use to reach agreement about physical phenomena could therefore be used to reach agreement on the intrinsic perspective. This might be the closest scientists could get to learning about their own intrinsic perspective through the indirect (``third-person'') methods of science.  

Here we build on recent insights from cognitive science to develop a model of scientists doing experiments. From a materialist perspective, we ask for self-consistency: we should not neglect {\em a priori} the physics or embodiment of scientists, as if they were immaterial, but let a scientific analysis tells us {\em a posteriori} whether we can do so. Our approach does not contradict ``objectivity'' if by this we pragmatically mean that: (O1) procedures are standardized and explicit, (O2) observations are intersubjective and repeatable, and (O3) observers are dispassionate, accurate and truthful~\cite{velmans2009understanding} (p. 219; see also Refs.~\cite{varela2017embodied,thompson2014waking,bitbol2008consciousness} and Appendix~\ref{s:thompson-science} herein). In this view, the scientific method can be formulated in short as ``if you carry out these procedures you should observe or experience these results,'' which can in principle apply to the investigation of both subjective and physical phenomena. This does not imply {\em a priori} that science is observer-free as ``objectivity'' is often understood---after all, science is actively performed by scientists even if assisted by technology. It does not deny {\em a priori} either that effectively observer-free {\em theories} could be established in certain situations. 

Indeed, the dynamics of our model is formally analogous to quantum dynamics (see Appendices~\ref{s:quantum_nutshell} and \ref{s:diff-quantum}). So, our approach is consistent with current scientific knowledge and aligns with some prominent views that advocate for the central role of observers in quantum physics~\cite{mermin2014physics,fuchs2013quantum} (see below). It could also be seen as a potential reconstruction of quantum theory. Unlike current reconstructions~\cite{d2017quantum}, though, instead of working with an {\em abstract} notion of observer, our approach builds on general insights gained from the investigation of {\em actual} observers---however, it is not necessarily restricted to a specific kind of them. By working at the interface of cognitive science, artificial intelligence (AI), and physics, our approach has the potential to suggest new experiments, perspectives, and synthesis. Furthermore, it may suggest alternative roads to quantum gravity.

\

\tocless\section{Extrinsic perspective: imaginary-time quantum dynamics}\label{sm:third}
\begin{figure*}
\includegraphics[width = 0.90\textwidth]{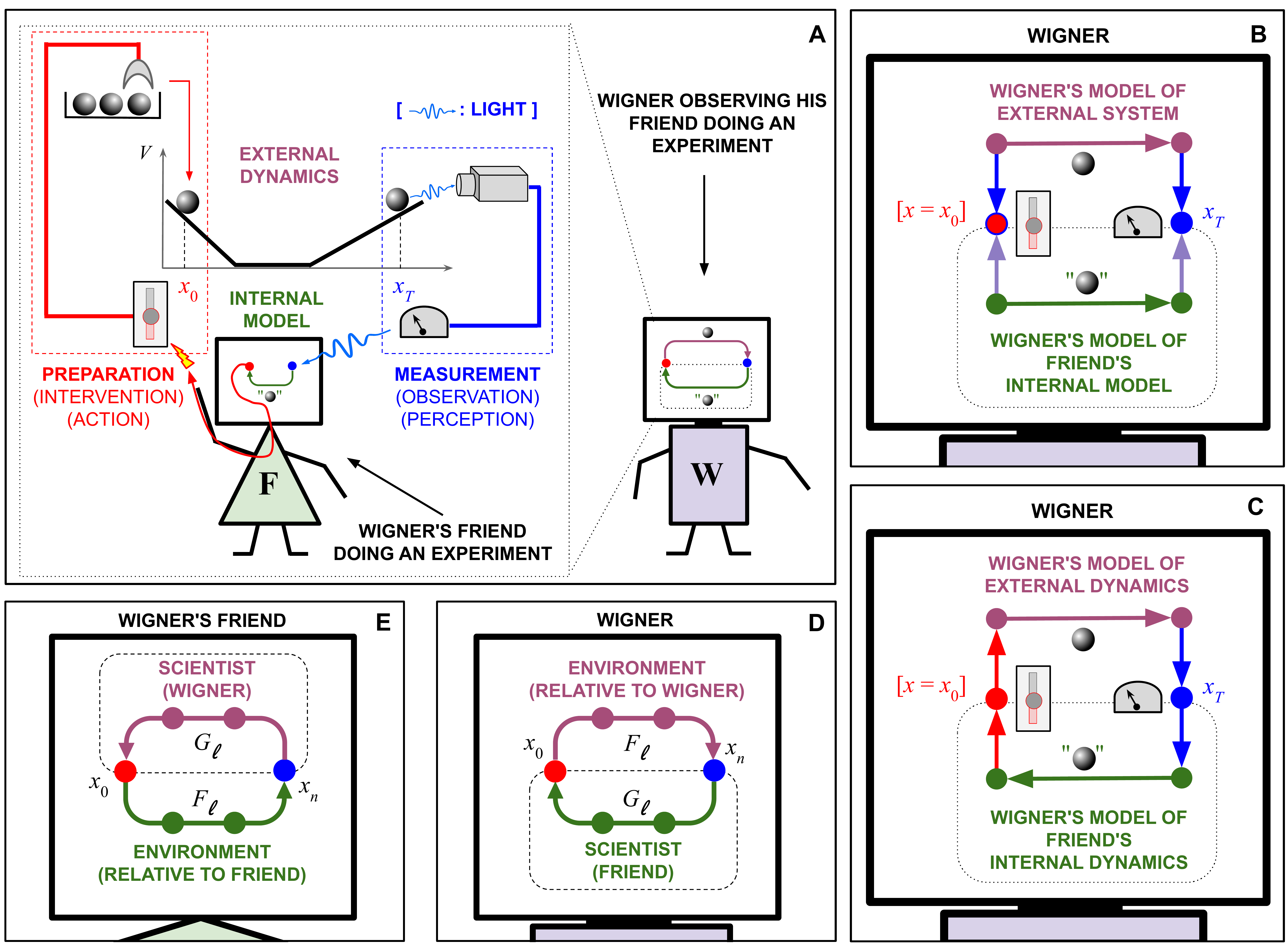}
\caption{{\em Experiments as circular processes:} (A) Wigner observes his friend doing an experiment. 
(B) Wigner's model of his friend doing an experiment in the spirit of active inference. 
The friend's actions can always prepare the same initial state, $x = x_0$, effectively implementing an intervention (denoted here as $[x=x_0]$; see Appendix~\ref{s:active} and Fig.~\ref{f:active} therein). 
(C) Enactive model of Wigner's friend and experimental system as two physical systems involved in a circular interaction (see Appendix~\ref{s:enactive} and Fig.~\ref{f:enactive} therein).  
(D) Modeling approach taken here. Arrows indicate the direction of the circular interaction, not conditional probabilities as in Bayesian networks. Factors $F_\ell$ and $G_\ell = F_{2n-1-\ell}$ correspond to the same time step for Wigner and his friend.
 (E) Wigner and his friend can exchange perspectives~\cite{velmans2009understanding}  (p. 212; see Appendix~\ref{s:relational_cognition}). 
}\label{f:circular}
\end{figure*}

\tocless\subsection{Embodied scientists doing experiments}

Here we build on enactivism whose task is ``to determine the common principles or lawful linkages between sensory and motor systems that explain how action can be perceptually guided in a perceiver-dependent world''~\cite{varela2017embodied} (p. 173; see Fig.~\ref{f:circular}A and Appendix~\ref{s:model_scientists} herein). However, this and related approaches describe embodied cognitive systems from the perspective of an {\em implicit}, unacknowledged external observer (see Appendix~\ref{s:on_internal}). Here we make such external observers {\em explicit}, which will be key to tackle the intrinsic perspective below. So, in line with the relational interpretation~\cite{Rovelli-1996} of quantum mechanics (RQM), statements like ``a scientist performs an experiment'' are considered relative to another scientist who witnesses that (like Wigner in Fig.~\ref{f:circular}A; see Appendix~\ref{s:wigner} and Fig.~\ref{f:wigner} therein). 

Figure~\ref{f:circular}A illustrates the dynamical coupling between an embodied scientist and an experimental system. This can be divided into four stages: (i) scientist's interventions on the experimental system, e.g. via moving some knobs, for preparing the desired initial state---this requires the physical interaction between the knobs and the observer's actuators; (ii) experimental system's dynamics---this is the main process traditionally analyzed in physics; (iii) scientist's measurement of the experimental system---this requires the physical interaction between the experimental system and the observer's sensors via the measuring device; (iv) scientist's internal dynamics which correlate with her experience of the experimental system. 

In the related approach of active inference~\cite{friston2010free,schwobel2018active}, experimental systems would be considered as generative {\em processes} which scientists can only access indirectly via the data generated in their sensorium (see Appendix~\ref{s:active}). Scientists can perturb such generative processes via their actions and have a generative {\em model} of their dynamics, including the effect of their own actions, which they can make as accurate as possible via learning. This is reflected in that, in Fig.~\ref{f:circular}B, the topology of the Bayesian network representing the scientist mirrors the topology of the Bayesian network representing the experimental system. In particular, both internal and external dynamics flow in the same direction (horizontal arrows in Fig.~\ref{f:circular}B; see Appendix~\ref{s:active} and Fig.~\ref{f:active} therein). 

Following enactivism~\cite{varela2017embodied, thompson2010mind,di2017sensorimotor}, instead, we give more relevance to the dynamical coupling between scientists and experimental systems. Learning scientific lawful regularities is not so much about extracting pre-existent properties of the world as about stabilizing this circular coupling and achieving ``objectivity'' (conditions (O1)-(O3) above). This may include the development of new technologies, protocols and concepts. The lawful regularities achieved in the post-learning stage are our focus here. So, our approach is independent of a specific theory of learning (see Fig.~\ref{f:circular}C; see Appendices~\ref{s:enactive}, \ref{s:relational_cognition} and Fig.~\ref{f:enactive} therein).

Of course, the scientific process generally involves many scientists and technologies. However, much as the theory of relativity can be developed without modeling all types of realistic clocks, our approach aims at capturing some general underlying principles valid beyond the particular model investigated. Indeed, enactivism does not treat the observer and the observed as two separate entities that are brought together. Rather, it treats them as {\em co-emerging} out of the circular process of observation, which is considered more fundamental~\cite{di2017sensorimotor} (p., 139). In a sense, the processes of observation can transcend individual scientists and their limitations. For simplicity, we focus here on a single scientist. However, experiments generally comprise the four stages above. So, ours can be considered as a model of a generic process of ``objective'' {\em observations}---though ignoring relativistic considerations. This process is embodied because all scientists and technologies involved are so (see Appendix~\ref{s:patterns}). 

\

\tocless\subsection{Experiments as circular processes}

We consider only the post-learning stage, when ``objectivity'' has been achieved and the scientist is just repeating the experiment a statistically significant number of times. We model this as the stationary state, $\widetilde{\mcP}(\widetilde{\boldsymbol{x}})$, of a stochastic process on a cycle, which includes deterministic systems as a particular case---this allows us to establish {\em a posteriori} which is the case, rather than forcing the model to fit {\em a priori} our (possibly deeply ingrained) assumptions. Here $\widetilde{\bx} =(x_0,\dotsc , x_{k -1})$ denotes a closed path $x_0\to x_1\to\cdots\to x_{k-1}\to x_0$ which returns to $x_{k} = x_0$ due to the scientist's causal interventions---experiments are not mere passive observations (see Appendix~\ref{s:model_scientists}). This path could be divided into two open paths $x_0\to\cdots\to x_{n}$ and $x_n\to\cdots\to x_{k}$, with $x_k = x_0$, corresponding to the environment and the scientist, respectively. Furthermore, $\widetilde{\mcP}(\widetilde{\boldsymbol{x}})$ denotes the probability to observe a path $\widetilde{\bx}$.

Since energy plays a key role in physics, we assume that the stationary state is characterized by an ``energy'' function $\mcH_\ell (x_{\ell + 1}, x_\ell)$, where $0\leq \ell\leq k$ denotes the time step. For the case of a particle in a non-relativistic potential $V$ we have 
\be\label{em:non-relativisticH}
\mathcal{H}_\ell(x_{\ell +1} ,x_\ell) = \frac{m}{2}\left(\frac{ x_{\ell +1} - x_\ell}{\epsilon}\right)^2 + \frac{1}{2}\left[V(x_\ell)+V(x_{\ell+1})\right].
\ee
for the ``external'' path ($\ell = 0,\dotsc , n-1$)---in principle, the ``internal'' path ($\ell=n,\dotsc , k$) can have a different functional form (but see below). Unlike the traditional Hamiltonian function, $\mcH_\ell$ is written in terms of consecutive position variables, $x_\ell$ and $x_{\ell +1}$, rather than instantaneous position and momentum. The potential $V$ in Eq.~\eqref{em:non-relativisticH} is symmetrized for convenience, but this does not affect the results (see Appendices~\ref{s:path}, \ref{s:App_path}). In Appendix~\ref{s:diff-quantum} we describe in detail the free-particle case ($V=0$). In Appendix~\ref{s:EM} we describe the case of a particle in an electromagnetic field, which is an example of a non-trivial complex-valued, and so ``non-stoquastic'' Hamiltonian.

We derive $\widetilde{\mathcal{P}}$ using the principle of maximum path entropy~\cite{presse2013principles}, a general variational principle analogous to the free energy principle from which a wide variety of well-known stochastic models at, near, and far from equilibrium has been derived~\cite{presse2013principles} (see Appendix~\ref{s:MaxCal}). To do so, we use the assumption, common in statistical physics, that we only know the average energy on the cycle $E_{\rm av}=\bra\frac{\epsilon}{T}\sum_\ell\mathcal{H}_\ell\ket_{\widetilde{\mathcal{P}}}$ (see below). Here $\epsilon\to 0$ is the time step size and $T = (k + 1) \epsilon $ is the total duration of a cycle. This is known~\cite{presse2013principles} to yield $\widetilde{\mcP}\propto \exp\{-\epsilon\sum_\ell\mcH_\ell/T\}$, with $\hobs = T/\lambda$, where $\lambda$ is a Lagrange multiplier fixing the average energy $E_{\rm av}$ on the cycle---the use of the subindex {\Yinyang}  here has a precise meaning, which will become clear later (see, e.g., Fig.~\ref{f:first} and discussion section).

After marginalizing over the ``internal'' path, this yields (see Appendix~\ref{s:MaxCal})
\begin{equation}\label{em:circular}
\begin{split}
\mathcal{P}(\bx) & = \sum_{x_{n+1},\dotsc , x_{k-1}}\widetilde{\mathcal{P}}(\widetilde{\bx})\\
&= \frac{1}{Z} \widetilde{F}_n(x_{0}^\prime,x_n)\cdots F_1(x_2, x_1) F_0(x_{1},x_0),
\end{split}
\end{equation}
where $Z$ is the normalization constant and we have written $x_0^\prime= x_0$ for future convenience---here we use sums to indicate either sums or integrals depending on the context. The expression $\bx =(x_0,\dotsc , x_n)$ denotes a path $x_0\to x_1\to\cdots\to x_n\to x_0$ which returns to $x_0$, but where we disregard how it does so---following physics tradition, here we focus on the environment and ignore the observer, but the same results can be obtained if, following cognitive science, we focus on the observer and considered the environment as hidden to her. Furthermore, 
\be\label{em:Ftilde}
\widetilde{F}_n(x_{0}^\prime,x_n)  = \sum_{x_{n+1},\dotsc , x_{k-1}}F_{k-1}(x_0^\prime, x_{k-1})\cdots  F_n(x_{n+1},x_n),
\ee
summarizes the dynamics internal to the observer, and
\be\label{em:F}
F_\ell(x^\prime ,x) = e^{-\epsilon \mathcal{H}_\ell(x^\prime ,x)/\hobs}/Z_\epsilon,
\ee
for $\ell=0,\dotsc , k$, where $Z_\epsilon = \sqrt{2\pi\eta\epsilon/m}$ is introduced for convenience.  As we are focusing on the ``external'' dynamics, $\hobs$ effectively characterizes the neglected environment (i.e., processes external to the experimental system), much like a temperature or a diffusion coefficient. The neglected environment causes energy fluctuations, which is why we only know the average energy---the deterministic case is included in the limit $\hobs\to 0$. We could equally focus on the ``internal'' dynamics, in which case $\hobs$ would characterize neglected processes within the observer. Either way, when we shift to the internal perspective, $\hobs$ would actually characterize the process of observation. 

\

\tocless\subsection{Reciprocal causality and imaginary-time quantum dynamics}

If we neglect the embodied observer, $\widetilde{F}_n(x_0, x_n)$ becomes a constant. In this case the cycle in Fig.~\ref{f:circular}D turns effectively into a chain and we recover the most parsimonious non-trivial case where the probability distribution in Eq.~\eqref{em:circular} is Markov with respect to a chain on variables $x_\ell$~\cite{pearl2009causality} (p. 16; see Appendices~\ref{s:QBP} and \ref{s:EQM} herein). In particular, by knowing only the initial marginal $p_0$ and the forward transition probabilities $\mcP_{\ell}^+$ from time step $\ell$ to $\ell+1$, for all $\ell$, we can readily obtain the probability for a path $\mcP_{\rm ch}(\bx) = p_0(x_0)\mcP^+_{ 0}(x_1|x_0)\cdots\mcP^+_{n-1}(x_n|x_{n-1})$. This implies in particular that we can obtain the marginal $p_{\ell +1}$ from the previous marginal $p_\ell$ via a Markovian update $p_{\ell+1}(x_{\ell+1}) = \sum_{x_\ell} \mathcal{P}_{\ell}^+(x_{\ell +1}|x_\ell) p_{\ell}(x_\ell)$. That is, via a linear transformation specified by kernels $\mathcal{P}_{\ell}^+(x_{\ell +1}|x_\ell)$ satisfying the Chapman-Kolmogorov equation---i.e., where the transition probability from $\ell$ to $\ell + 2$, for instance, can be written as $\mcP_{\ell+2 | \ell}^+(x_{\ell+2}| x_\ell) = \sum_{x_{\ell+1}}\mcP_{\ell + 1}^+(x_{\ell+2}| x_{\ell+1})\mcP_{\ell}^+(x_{\ell+1}| x_\ell)$. 

In general, when we cannot neglect the observer, we cannot write the probability of a path in the same way due to the loopy correlations. This implies in particular that we cannot obtain the marginal $p_{\ell+1}$ from the previous one $p_\ell$ via a Markovian update as above (see Appendix~\ref{s:EQM}). Indeed, it is possible to show that $\mcP(\bx) = p(x_0, x_n)\prod_{\ell = 0}^{n-2}\mcP_\ell(x_{\ell+1}|x_\ell, x_0)$ in Eq.~\eqref{em:circular}, which yields a Bernstein process where initial {\em and } final states must be specified~\cite{Zambrini-1987} (here the two-variable marginal $p$ and the transition probability $\mcP_\ell$, respectively, plays the role of $m$ and $h$ in Eq.~(2.7) therein; see Appendix~\ref{s:EQM} herein). Interestingly, the need to specify initial and final states, so common in physics, arises naturally here as this effectively turns a cycle into two chains. 

We can recover an effective Markovian-like update if, instead of marginals, we consider (real) probability matrices. For instance, if we relax the condition $x_0^\prime = x_0$ in Eq.~\eqref{em:circular} and marginalize all other variables we obtain a probability matrix $P_0(x_0^\prime , x_0) = \sum_{x_1,\dotsc , x_n}\mathcal{P}(\bx)$ with $P_0(x_0 ,x_0)= p_0(x_0)$. Interpreting factors as matrix elements, Eq.~\eqref{em:circular} yields $P_0 = \widetilde{F}_n\cdots F_1 F_0/Z$. Similarly, for $\ell=1$, $P_1=F_0\widetilde{F}_n\cdots F_1/Z$ and $P_1(x_1 , x_1)=p_1(x_1)$. This is obtained by removing the prime from $x_0$ in Eq.~\eqref{em:circular}, adding a prime to $x_1$ in $F_0$, moving $F_0(x_1^\prime, x_0)$ to the beginning of Eq.~\eqref{em:circular}, and doing the marginalization over all other variables, $P_1(x_1^\prime , x_1)=\sum_{x_0 , x_2,\dotsc , x_n}\mathcal{P}(\bx)$. 

The probability matrix $P_1=F_0\widetilde{F}_n\cdots F_1/Z$ is obtained from $P_0 = \widetilde{F}_n\cdots F_1 F_0/Z$ via the cyclic permutation of matrix $F_0$. Iterating $\ell$ times yields 
\be\label{em:matrix}
P_\ell = \frac{1}{Z} F_{\ell-1}\cdots F_1 F_0\widetilde{F}_n \cdots F_{\ell+1}F_\ell ,
\ee
where $P_\ell(x,x) = p_\ell(x)$. If $F_\ell$ is invertible we can write (see Appendix~\ref{s:EQM} for the case of pure states which, following Eq.~\eqref{em:matrix}, would be associated to non-invertible matrices $F_\ell$)
\be\label{em:P_l+1}
P_{\ell +1} = F_{\ell} P_{\ell} F_{\ell}^{-1},
\ee
for $\ell=0,\dotsc , n-1$, which is an effective Markovian-like update in that it yields $P_{\ell +1}$ via a linear transformation of $P_\ell$ alone, where the kernels $F_\ell$ satisfy the analogue of Chapman-Kolmogorov equation---i.e., the factor between time steps $\ell$ and $\ell +2$, for instance, can be written as $F_{\ell + 2 | \ell}\equiv F_{\ell +1} F_{\ell}$. So, we can effectively sidestep the circularity and keep the traditional description in terms of a causal chain, at the expense of working with probability matrices---the off-diagonal elements of such matrices contain relevant dynamical information since, if we neglect them, we cannot build $P_{\ell+1}$ from $P_\ell$ and $F_\ell$ alone. We could obtain an equation from $\ell=n\dotsc , k-1$, similar to Eq.~\eqref{em:P_l+1}, if we focus on the observer rather than the environment.

When $\epsilon\to 0$, we can assume variables $x_\ell$ and $x_{\ell +1}$ to be typically close to each other, or $F_\ell = \id + \epsilon J_\ell + O(\epsilon^2)$, where $\id$ is the identity.  For discrete variables, the {\em dynamical matrix} $J_\ell$ has non-negative off-diagonal elements.  For continuous variables $J_\ell$ is actually an operator. For instance, for $\mcH_\ell$ in Eq.~\eqref{em:non-relativisticH} we have $J_\ell \to - H/\hobs$, when $\epsilon\to 0$, where
\be\label{em:H}
H = -\frac{\hobs^2}{2 m}\frac{\partial^2}{\partial x^2} +V(x) ,
\ee
is equivalent to the quantum Hamiltonian of a non-relativistic particle in a potential $V$, and $\hobs$ plays the role of Planck's constant. This can be seen by expanding the non-Gaussian factors in the integral $[F_\ell g](x) = \int F_\ell(x, x^\prime) g(x^\prime)\mathrm{d} x^\prime$, where $g$ is a generic smooth enough function, until the integral yields terms $O(\epsilon^2)$ (see Appendices~\ref{s:QBPdyn}). 

Either way, Eq.~\eqref{em:P_l+1} becomes
\be\label{em:real_vN}
\Delta {P}_{\ell} = \epsilon [J_\ell, P_\ell] + O(\epsilon^2),
\ee
where $\Delta P_{\ell} = P_{\ell + 1} - P_\ell$ and  $[J_\ell, P_\ell] =  J_\ell P_\ell - P_\ell J_\ell $. After dividing by $\epsilon$ and taking the continuous-time limit ($\epsilon\to 0$), Eq.~\ref{em:real_vN} yields $\partial P/\partial t = [J, P]$, which is  von Neumann equation in imaginary time---i.e. with time $t=\ell\epsilon$ replaced by $i t$, where $i$ is the imaginary unit (see Appendix~\ref{s:quantum_nutshell}). Imaginary-time quantum dynamics already displays some quantum-like features~\cite{Zambrini-1987} such as, e.g., constructive interference (see Appendix~\ref{s:slits} and Fig.~\ref{f:slits} therein for an example of this in the context of the two-slit experiment).

Using belief propagation (BP)~\cite{Mezard-book-2009} (ch. 14) we can obtain the imaginary-time version of wave functions and their conjugate, given by the forward and backward BP messages $\mu_{\to\ell}$ and $\mu_{\ell\leftarrow}$, respectively, as well as of the Schr\"odinger equation and its conjugate, given by the continuous-time limit of the forward and backward BP iterations, $\mu_{\to\ell +1} = F_{\ell}\mu_{\to\ell}$ and $\mu_{\ell\leftarrow} = \mu_{\ell +1\leftarrow}F_{\ell}$, respectively (see Appendices~\ref{s:QBP} and \ref{s:EQM}, as well as Figs.~\ref{f:cavity} and \ref{f:cavity_cycle} herein). The analogue of the Born rule is given by the standard BP rule $p_\ell(x)=\mu_{\to\ell}(x)\mu_{\ell\leftarrow}(x)$. Although BP is not exact on cycles, we can in principle fix initial and final states, $p_0$ and $p_n$, to turn the cycle into two chains and search for BP messages that are consistent with these and the BP iterations~\cite{zambrini1986stochastic} (see Eqs.~(2.6), (2.16), (2.22), and (2.23) therein; $\mu_{\to\ell}$, $\mu_{\ell\leftarrow}$, and $F_\ell$ here correspond, respectively, to $\theta^\ast$, $\theta$, and $h$ therein).  However, we here focus on probability matrices because these directly yield probabilistic information, unlike BP messages that must be multiplied by another object---its ``conjugate''---to do so.

The rest of this work will build on the formalism we have established up to here. Let us just notice, though, that the Markovian and Markovian-like updates above could be considered as examples of linear and reciprocal causality, respectively.  We can in principle change the initial marginal $p_0\to p_0^\prime$ of a Markovian update without changing the ``mechanisms'' $\mcP^+_\ell$~\cite{pearl2009causality} (Sec.~1.3.1). Similarly, we can in principle change the initial probability matrix $P_0\to P_0^\prime$ for our Markovian-like update without changing the ``mechanisms''  $F_\ell$. In particular, we could choose $P_0(x, x^\prime) = \sum_\alpha \lambda_\alpha\mu^\alpha_{\to 0} (x)\mu^\alpha_{0\leftarrow}(x^\prime)$, where the BP messages $\mu_{\to 0}^\alpha$ and $\mu_{0\leftarrow}^\alpha$ correspond to a pure state and its imaginary-time conjugate, respectively, and $\lambda_\alpha\geq 0$ with $\sum_\alpha\lambda_\alpha = 1$ yields the probability associated to pure state $\alpha$ (see Appendix~\ref{s:EQM}).

\

\tocless\subsection{Constraints from ``objectivity''}

The experimental system is observed by both Wigner as an external observer and his friend as the scientist doing the experiment. However, ``objectivity'' requires that both perspectives coincide (see Appendix~\ref{s:obj_constraints}). From Wigner's perspective, factors $F_\ell$, with $\ell=0,\dotsc , n-1$, associated to paths traversed from $x_0$ to $x_n$, characterize his friend's environment (see Fig.~\ref{f:circular}D). Now, Wigner and his friend can exchange roles~\cite{velmans2009understanding} (p. 212), so they become the scientist doing the experiment and the external observer, respectively (see Fig.~\ref{f:circular}E). Standardization---condition (O1)---requires that the models, and so the factors $F_\ell$, for $\ell=0,\dotsc , k$, in both cases are the same. In particular, Wigner's internal physical correlates of his experience of the environment is characterized by factors $G_{2n - 1-\ell} \equiv F_\ell$, with $\ell = n,\dotsc , k-1$, associated to paths traversed from $x_n$ to $x_k = x_0$, i.e., effectively in the reverse direction. Intersubjectivity---condition (O2)---requires that, at each time step, what Wigner the external observer considers as his friend's environment is the same environment that Wigner the scientist observes. 

To compare each time step, the ``internal'' and ``external'' paths have to be divided into an equal number of bins, i.e. $k =2 n$. Under an exchange of roles, the initial factor $F_0(x_1, x_0)$, for instance, characterizing the environment observed by Wigner as an external observer (see Fig.~\ref{f:circular}D), corresponds to the final factor $G_0 = F_{2n -1}(x_0 , x_{2n -1})$ characterizing the environment observed by Wigner as the scientist performing the experiment (see Fig.~\ref{f:circular}E). However, the latter has to be transposed because it corresponds to paths traversed in the reversed direction. More generally, 
\be\label{em:G=FT}
G_\ell = F_{2n - \ell -1} = F_\ell^T,
\ee
with $\ell=0,\dotsc , n-1$. This implies that the two dynamics are related via the transpose operation since $P_{2n - \ell} = F_{2n-\ell-1}\cdots F_n\widetilde{F}_n^T F_{2n -1}\cdots F_{2n -\ell}/Z = P_\ell^T$---using $\widetilde{F}_n = F_{2n-1}\cdots F_n = F_0^T\cdots F_{n-1}^T$ in Eq.~\eqref{em:matrix} (see Appendix~\ref{s:obj_constraints}).

Intersubjective agreement therefore enforces the ``internal'' and ``external'' dynamics to be described by the same factors. This seems consistent with the idea that there is a pre-given ``external'' dynamics that the observer internally ``mirrors''. In particular, this seems to imply that both ``internal'' and ``external'' dynamics flow in the same direction rather than in a circular fashion (cf. Fig.~\ref{f:circular}B). However, this is not exactly so because, following enactivism, the {\em functional form} of the ``internal'' and ``external'' factors {\em co-emerge} as a lawful regularity out of the circular interaction between scientist and world. Indeed, the direction of the circular dynamics and its reversal under an exchange of roles are key in our approach to the intrinsic perspective (see Appendix~\ref{s:obj_constraints}). Finally, faithfulness---condition (O3)---requires that if the ``internal'' and ``external'' dynamics coincide, no observer reports to the contrary---something we are implicitly assuming here.

\

\tocless\section{Intrinsic perspective: real-time quantum dynamics}\label{sm:first}

\begin{figure*}
\includegraphics[width=0.80\textwidth]{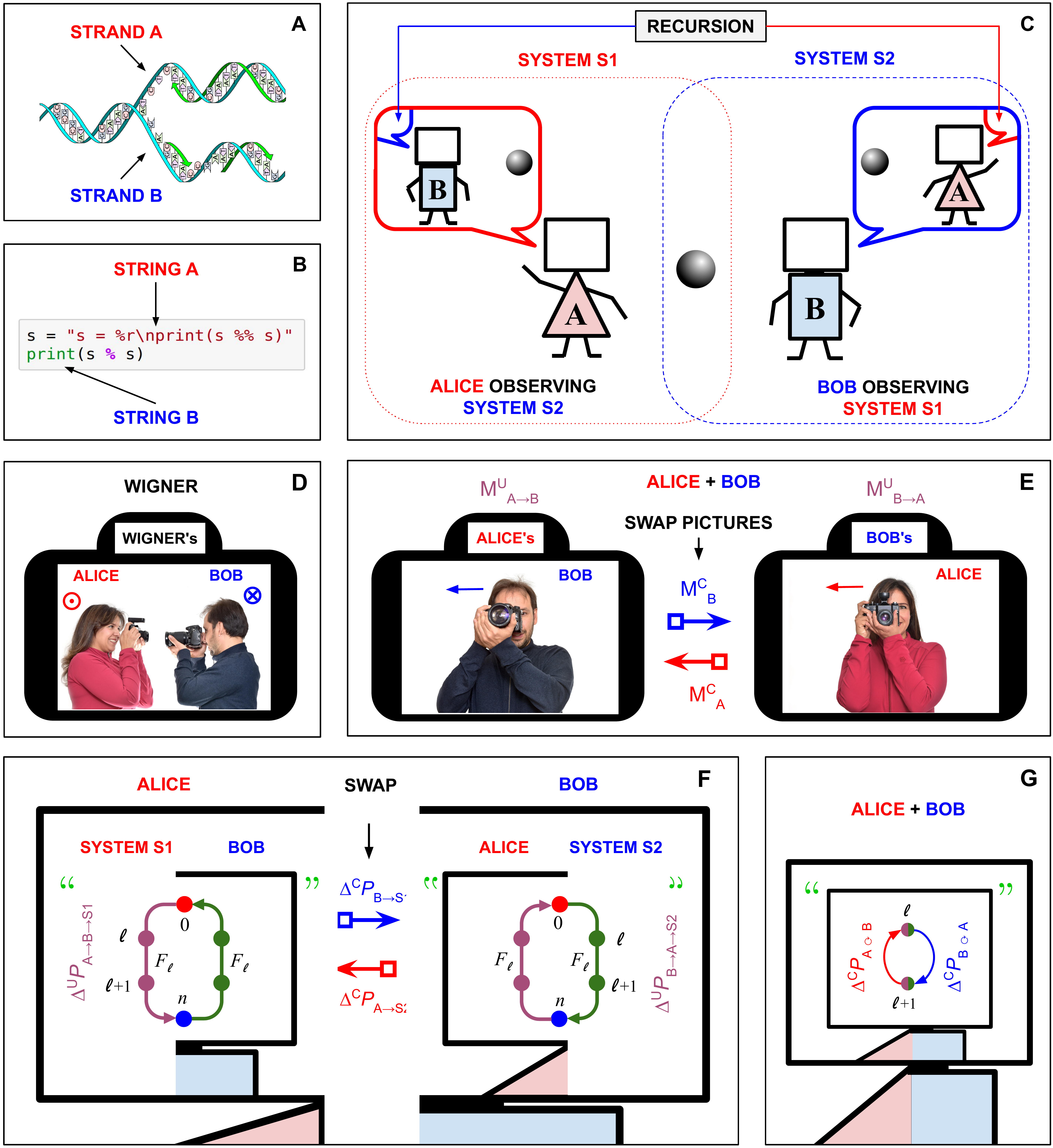}
\caption{{\it C-observers as a self-referential coupling of complementary ``sub-observers''}. Self-referential systems---like DNA molecules (A) and self-printing programs (B)---are often composed of two subsystems that mutually refer to each other (see Appendix~\ref{s:self-reference}). (C) Similarly, C-observers are composed of two ``sub-observers'' that recursively refer to each other. (D) Toy example of a C-observer (Alice+Bob) described by an E-observer (Wigner). (E) Alice and Bob take and exchange pictures of each other. They cannot detect any extrinsic motion, only their relative motion in opposite directions: outwards ($\odot$ in (D)) and inwards ($\otimes$ in (D)). However, in their pictures they appear as moving in the same direction (E). (F, G) Formal analogue of (E). Alice and Bob play complementary roles, acting both as E-observers (i.e., like Wigner in Fig.~\ref{f:circular}A; big half-heads) and as U-observers (i.e., like Wigner's friend in Fig.~\ref{f:circular}A; small half-heads). However, under a shift of perspective the circular processes effectively reverse direction (see Fig.~\ref{f:circular}D,E). From an external perspective, these two circular processes are one and the same, so this reversal has to be corrected. (G) C-observer. Figures (C, E, F) are structurally similar to the symbol \Yinyang---for instance, if the black and white colors refer to Alice and Bob, respectively. This symbol is sometimes used to refer to the intrinsic perspective in the sense intended here---i.e., as the mere capacity to observe or experience (see discussion section). Image in (A) is derivative work from \href{https://commons.wikimedia.org/wiki/File:DNA\_replication\_split.svg}{Wikimedia Commons: Madeleine Price Ball (Madprime)} [\href{https://creativecommons.org/licenses/by-sa/3.0}{CC BY-SA 3.0}]. Photos in (D,E) taken by Edwin Lemus.}
\label{f:first}
\end{figure*}

\tocless\subsection{The mere capacity to observe ``from within''}

Equation~\eqref{em:real_vN} describes the dynamics of a scientist ($F$) doing an experiment ($E$), from the perspective of another {\em external} scientist ($W$; see Fig.~\ref{f:circular}A). We can highlight this by writing $P_\ell$ in Eq.~\eqref{em:real_vN} as $P_\ell^{W\to F\to E}$. But scientific knowledge is not given by some ``god-like'' observer external to the universe. Rather, it stems from sub-systems of it: scientists, who observe the experimental systems from their own intrinsic perspectives. To do so, scientists do not necessarily require subjective experience in full, but just the mere capacity to observe from an intrinsic perspective. By ``intrinsic perspective'' we here refer to this kind of minimal phenomenal experience, which is attracting growing interest from the scientific community~\cite{thompson2014waking,metzinger2018minimal}. Importantly, like a model of gravity that is not gravity itself, our model of the intrinsic perspective is not the intrinsic perspective itself.

The problem of describing how phenomena looks from the perspective of the observer herself (or from ``my'' perspective), rather than how an external observer considers such phenomena should look to the former observer (or to ``me''), is self-referential and can easily lead to the ``homunculus fallacy'' (see Appendix~\ref{s:on_internal} and Fig.~\ref{f:homunculus} therein; see also Appendix~\ref{s:self-reference}). DNA molecules are examples of self-referential systems (see Fig.~\ref{f:first}A). These are composed of two strands that mutually refer to each other by playing complementary roles, i.e., active and passive when participating in the replication of and when being replicated by the other strand, respectively. Like DNA molecules, self-printing Turing machines (TMs) are composed of two sub-machines that mutually refer to each other, similarly playing complementary roles when printing and being printed by the other sub-machine (see Fig.~\ref{f:first}B). Such an architecture is formalized for general TMs in Kleene's recursion theorem~\cite{sipser2006introduction} (ch.~6.1; see Appendix~\ref{s:recursion} and Figs.~\ref{f:q}-\ref{f:recursion}). 

\

\tocless\subsection{Self-referential or ``conscious'' observers}

We now propose to model the intrinsic perspective in a similar way by coupling two ``sub-observers'' that, in a sense, mutually observe each other (see Fig.~\ref{f:first}C). In our relational approach, observers like Wigner's friend in Fig.~\ref{f:circular}A are ``unconscious'', or U-observers, in the sense that, like philosophical zombies, they ``exist'' only relative to an external, or E-observer (like Wigner in Fig.~\ref{f:circular}A). In contrast, observers with an intrinsic perspective, or C-observers, are (phenomenally) ``conscious'' in the sense that they ``exist'' also relative to themselves. 

Figures~\ref{f:first}D,E show a simple analogy for a C-observer, which highlights four key properties: 
(C1) It is composed of two ``sub-observers'', Alice and Bob, which play complementary roles, acting both as E- and U-observers that photograph and are photographed by each other, respectively. 
(C2) Alice and Bob exchange the pictures of each other ($M^{\rm U}_{A\to B}$ and $M_{B\to A}^{\rm U}$ in Fig.~\ref{f:first}E), to obtain a picture of themselves ($M_A^{\rm C}$ and $M_B^{\rm C}$ in Fig.~\ref{f:first}E). In this sense the combined system Alice+Bob observes itself. 
(C3) Alice and Bob can determine their relative velocities, $v_{A|B}$ and $v_{B|A}$, but not the extrinsic velocity, $v_{C|W}$, of their centroid relative to an external observer. More precisely, if Alice and Bob have velocities $v_{A|W}$ and $v_{B|W}$, respectively, relative to an external observer (Wigner in Fig.~\ref{f:first}D), the average or centroid velocity is $v_{C|W} = (v_{A|W} + v_{B|W})/2$. The corresponding velocities relative to the centroid are $v_{A|C} = v_{A|W} - v_{C|W}$ and $v_{B|C} = v_{B|W} - v_{C|W}$. For Wigner, the relative velocities between Alice and Bob are $v_{A|B} = 2 v_{A|C}$ and $v_{B|A} = 2 v_{B|C} $, respectively. Alice and Bob can determine the intrinsic velocities $v_{A|C} = -v_{B|C}$ by observing each other, but not the extrinsic velocity $v_{C|W}$ which only makes sense for Wigner.
(C4) While Alice and Bob have opposite relative velocities, in their pictures they seem to be moving in the same direction (see Fig.~\ref{f:first}E)---so, there is a minus sign that needs to be corrected. This happens because, to observe each other, they have to face each other.  This is similar to the left-right inversion that occurs when we look in a mirror. 
 
Following property (C1), we assume a C-observer is composed of two complementary sub-observers, Alice and Bob, that observe generic systems $S_2$ and $S_1$, respectively  (see Fig.~\ref{f:first}F). The systems $S_1$ and $S_2$ will later on refer to Alice and Bob, respectively, effectively implementing the {\em self-referential coupling} (SRC). Consider first symmetric dynamical matrices, i.e., $J_\ell= J^T_\ell = J_{s,\ell}$, so $J_{s, 2n-\ell - 1} = J_{s, \ell}$. Following property (C2), Alice and Bob need to swap their descriptions of each other (see Fig.~\ref{f:first}C). Similar to property (C4), there is a minus sign that needs to be corrected. Indeed, for the system $S_1$ to consistently represent Alice, i.e. for the purple arrow corresponding to $S_1$ (Fig.~\ref{f:first}F, left), which points in a counterclockwise direction, to consistently represent the purple arrow corresponding to Alice in Fig.~\ref{f:first}F (right), which points in a clockwise direction, they should actually point in the same direction. So, the orientation of one of the circles in Fig.~\ref{f:first}F has to be reversed, leading to an apparent time reversal. Since scientists are all the time observing the world from their own intrinsic perspective, we here assume that the intrinsic perspective has to be constructed at each time step. 

Let us first provide some intuition. The change in perspective and the corresponding apparent time-reversal are associated to the transpose operation (see previous section). In other words, if $\Delta^{\rm U} P_\ell = [J_{s , \ell}, P_\ell]$ is Alice's dynamics from Bob's perspective---as U- and E-observers, respectively---then $\Delta^U P_\ell^T = -[J_{s, \ell} , P_\ell^T]$ is Bob's dynamics from Alice's perspective---as U- and E-observers, respectively (cf. Fig.~\ref{f:circular}D,E). Similarly, $\Delta^C P_\ell$ and $\Delta^C P_\ell^T$, respectively, denote the dynamics of Alice and Bob as the two halves of a C-observer. So, after Alice and Bob exchange their descriptions of each other, we obtain $\Delta^C P_\ell = \Delta^U P_\ell^T$ and $\Delta^C P_\ell^T = \Delta^U P_\ell$ for Alice and Bob, respectively. We will see below that these equations describe a quantum dynamics.

Let us now be more formal. Let $P^{A\to B\to S_1}_\ell$ be the probability matrix that Alice, acting as an E-observer, assigns at time step $\ell$ to U-observer Bob observing system $S_1$ (see Fig.~\ref{f:first}F, left). Let $P^{B\to A\to S_2}_\ell$ be defined similarly (see Fig.~\ref{f:first}F, right). In our relational framework, the focus is on the corresponding changes $\Delta^U P^{A\to B\to S_1}_\ell$ and $\Delta^U P^{B\to A\to S_2}_\ell$, rather than on absolute quantities. The swapping of descriptions and the correction of the minus sign, associated to properties (C2) and (C4), respectively, lead to
\BE
\Delta^{\rm C} P^{A\to S_2}_\ell = -\Delta^{\rm U} P^{B\to A\to S_2}_\ell = -\epsilon\left[J_{s, \ell}, P^{B\to A\to S_2}_\ell\right], \label{em:1PPa}\\
\Delta^{\rm C} P^{B\to S_1}_\ell = \Delta^{\rm U} P^{A\to B\to S_1}_\ell  = \epsilon\left[J_{s, \ell}, P^{A\to B\to S_1}_\ell\right].\label{em:1PPb}
\EE
That is, Bob takes the change from Alice as is, while Alice add a minus sign to Bob's. Here the superscripts ``C'' and ``U'', respectively, indicate that these changes refer to a C-observer and a U-observer; the latter are given by Eq.~\eqref{em:real_vN}. If $\Delta^{\rm U} P^{A\to B\to S_1}$ and $\Delta^{\rm U} P^{B\to A\to S_2}$ are the analogues of $M^{\rm U}_{A\to B}$ and $M^{\rm U}_{B\to A}$, then $\Delta^{\rm C} P^{A\to S_2}$ and $\Delta^{\rm C} P^{B\to S_1}$ are the analogues of $M^{\rm C}_A$ and $M^{\rm C}_B$ (see Fig.~\ref{f:first}E,F).  

In enactivism observers always imply observed objects since they are relational to each other (see Appendix~\ref{s:patterns}). Accordingly, to implement the self-referential coupling, we now let system $S_1$ refer to observer Alice observing system $S_2$, i.e., $S_1 = A\to S_2$ (see Fig.~\ref{f:first}C). Similarly, $S_2 = B\to S_1$. Iterating, $S_1$ is Alice who is observing Bob ($A\to B$), who is observing Alice ($A\to B\to A$), who is observing Bob ($A\to B\to A\to B$), and so on {\em ad infinitum} (cf. Ref.~\cite{kauffman2009reflexivity}, p. 126). Similarly for $S_2$. Thus,
\BE
S_1 &=& A\lcirclearrowleft B \equiv A\to B\to A\to B\to\cdots,\label{em:AoB}\\
S_2 &=& B\lcirclearrowleft A \equiv B\to A\to B\to A\to\cdots,\label{em:BoA}
\EE
which aslo defines the {\em mirroring operator} $\lcirclearrowleft$. Using Eqs.~\eqref{em:1PPa} and \eqref{em:1PPb} and the relationships ${A\to B\lcirclearrowleft A = A\lcirclearrowleft B}$ and ${B\to A\lcirclearrowleft B = B\lcirclearrowleft A}$ implied by Eqs.~\eqref{em:AoB} and \eqref{em:BoA} yields 
\BE
\Delta^{\rm C} P^{A\lcirclearrowleft B}_\ell &=& -\epsilon\left[J_{s, \ell}, P^{B\lcirclearrowleft A}_\ell\right],\label{em:1PP_PAoB}\\
\Delta^{\rm C} P^{B\lcirclearrowleft A}_\ell &=& \epsilon\left[J_{s, \ell}, P^{A\lcirclearrowleft B}_\ell\right].\label{em:1PP_PBoA}
\EE

A quantum experiment starts with classical information scientists have direct access to and can agree on (see below). So, Following Eq.~\eqref{em:matrix}, the initial probability matrices can be chosen as $P^{A\lcirclearrowleft B}_0 = P^{B\lcirclearrowleft A}_0 = P_0$, where $P_0 =\widetilde{F}_n\widetilde{F}_n^T$ (see Eqs.~\eqref{em:Ftilde} and \eqref{em:G=FT}) is symmetric and its diagonal contains the probabilities characterizing the initial state. Introducing $P_0$ in the right hand side of Eqs.~\eqref{em:1PP_PAoB} and \eqref{em:1PP_PBoA} we can see that these terms are the transpose of each other. So, iterating Eqs.~\eqref{em:1PP_PAoB} and \eqref{em:1PP_PBoA} yields $P^{A\lcirclearrowleft B}_\ell \equiv P_\ell$ and $P^{B\lcirclearrowleft A}_\ell \equiv P_\ell^T$ for all $\ell$. This is reasonable since the transpose operation is related to time-reversal and swapping $A$ and $B$ reverses the time direction of the circular process (see Fig.~\ref{f:first}G and Appendix~\c{\ref{s:obj_constraints}}). 

Adding and subtracting Eqs.~\eqref{em:1PP_PAoB} and \eqref{em:1PP_PBoA} we get an equivalent pair of equations in terms of $P_{s, \ell} = (P_\ell+P^T_\ell)/2$ and $P_{a, \ell} = (P_\ell-P^T_\ell)/2$, i.e., the symmetric and anti-symmetric parts of $P_\ell$. This pair of real matrix equations can be written as (see Appendix~\ref{s:pair})
\be\label{em:vN}
i \hobs \Delta \rho_\ell = \epsilon[H_{s,\ell}, \rho_\ell] 
\ee
where  $\rho_\ell \equiv P_{s, \ell} + P_{a, \ell}/i = \rho^\dag_\ell$ and $H_{s, \ell} \equiv -\hobs J_{s, \ell} = H^\dag_{s,\ell}$ are the analogues of the Hermitian quantum density matrix and Hamiltonian operator, respectively, and we have removed the superscript ``C''---here $\dag$ denotes the conjugate transpose. After dividing by $\epsilon$ and taking the continuous-time limit, Eq.~\eqref{em:vN} yields $i\hobs\partial\rho/\partial t =[H_s, \rho]$, which is formally equivalent to von Neumann equation for real Hamiltonians. 

\

\tocless\subsection{Quantum dynamics with generic Hamiltonians}

The dynamical matrices considered here do not have negative off-diagonal entries, and so have a clear probabilistic interpretation. This is not necessarily a restriction, though, since effective off-diagonal negative entries can arise from approximations of these kinds of dynamical matrices. Consider, e.g., an electron in a coherent radiation field with potential energy proportional to $\cos(\omega t)$, described by a Hamiltonian like that in Eq.~\eqref{em:H}. Under certain conditions, the original Hamiltonian $H_{\rm rad}$ can be approximated by an effective two-level Hamiltonian $H_{\rm eff}$ whose off-diagonal entries are proportional to $\cos(\omega t)$ and so can be positive or negative~\cite{haken2005physics} (Sec.~15.3; see Appendix~\ref{s:two_atom} herein).

Furthermore, negative numbers in probabilistic expressions can sometimes have operational meaning. Consider, e.g., a coin toss where we get heads and tails with probabilities $1-p$ and $p$, respectively. We can write the probability vector as $(1-p, p) = \mathcal{R} + \mathcal{N}$ 
We can generate a sample in two stages: first generate a sample according to $\mathcal{R} = (1/2 , 1/2)$. The second vector, $\mathcal{N} = [(1-2p)/2](1, -1)$, can be interpreted as encoding transitions from states associated to the negative entry, into states associated with the positive entry. Take $1-2p>0$. In this case, if the sample generated in the first stage is in state tails, we flip it with probability $1 - 2p$. If it is in state heads, instead, we do nothing. So, the total probability for a sample to be heads is the probability for a sample to be heads in the first stage (i.e., $1/2$) plus the probability that it was tails in the first stage (i.e., $1/2$) and was flipped in the second stage (i.e., $1 - 2p$). Therefore, the probability for the sample to be in the heads state is $1/2 + (1 - 2p)/2 = 1 - p $ as required. Similarly, for tails. In a sense, $\mathcal{R}$ and $\mathcal{N}$ play a role analogous to $H_{\rm rad}-H_{\rm eff}$ and $H_{\rm eff}$ above.

Now consider dynamical matrices with an antisymmetric part, $J_{a, \ell}$---which is related to phenomena with an extrinsic directionality as it encodes the difference between time-reversed trajectories (see above and Appendix~\ref{s:obj_constraints}). These sometimes can arise from effective descriptions of systems with symmetric dynamical matrices, i.e., real Hamiltonians~\cite{vinci2017non} (see Eqs.~ (1) and (11) therein; see Appendix~\ref{s:non-stoquastic}). 

Moreover, consider the real Hamiltonian in Eq.~\eqref{em:H} with $V = 0$, related to diffusion, which acquires an imaginary part, related to drift, if we change to a reference frame moving with velocity $v$ (see Appendix~\ref{s:gaussian-quantum} and Fig.~\ref{f:diff_drift} therein). The corresponding complex Hamiltonian $H_\ell = H_{s, \ell} + H_{a, \ell}/i$ can be obtained from Eq.~\eqref{em:vN} via the corresponding change in the time derivative $\partial/\partial t\to\partial /\partial t + v\partial/\partial x$. 
Equation~\eqref{em:vN} then yields the pair of equations (see Appendix~\ref{s:pair})
\BE
\Delta P_\ell - \epsilon [J_{a, \ell} , P_\ell] &=& -\epsilon[J_{s, \ell}, P^T_\ell], \label{em:free1PPa}\\
\Delta P^T_\ell - \epsilon [J_{a, \ell}, P^T_\ell] &=& \epsilon [J_{s, \ell}, P_\ell],\label{em:free1PPb}
\EE
where $J_{s, \ell}=-H_{s,\ell}/\hobs = (\hobs/2m)\partial^2/\partial x^2$, $J_{a, \ell} = -H_{a,\ell}/\hobs = - v \partial / \partial x$, and $\rho_\ell = (P_\ell + P_\ell^T)/2 + (P_\ell - P_\ell^T)/2i$.

Since the anti-symmetric term here arises from a change of reference frame, it is analogous to the extrinsic term $v_{C|W}$ described in property (C3). This example therefore suggests that, like in property (C3), Alice and Bob cannot observe the part with extrinsic directionality when they observe each other since this only exists relative to an observer external to them. So, we finally assume that when performing the SRC associated to a general dynamical matrix $J_\ell= J_{s, \ell} + J_{a, \ell}$, we first have to subtract the anti-symmetric part, precisely as in Eqs.~\eqref{em:free1PPa} and \eqref{em:free1PPb}. This leads to the analogue of von Neumann equation with generic complex Hamiltonians, i.e., Eq.~\eqref{em:vN} with $H_{s, \ell}$ replaced by $H_\ell=H_{s, \ell} +H_{a, \ell}/i$.

\

\tocless\subsection{Generic initial quantum states and observables}

Let us first check that an initial density matrix given by $\rho_0 = P_0= \widetilde{F}_n\widetilde{F}_n^T$, which is real and so symmetric, is consistent with a generic real quantum density matrix. Indeed, a symmetric pure density matrix has the form $\rho_0^\alpha(x, x^\prime) =\sqrt{p_0^\alpha(x) p_0^\alpha(x^\prime)}$, where $p_0^\alpha(x)$ is the probability assigned to $x$. A more general symmetric density matrix is given by a mixture, i.e., $\rho_0(x,x^\prime) = \sum_\alpha\lambda_\alpha\sqrt{p_0^\alpha(x) p_0^\alpha(x^\prime)}$, where $p_0^\alpha(x)$ is the probability assigned to $x$ in the pure state $\alpha$, and $\lambda_\alpha\geq 0$, with $\sum_\alpha\lambda_\alpha = 1$, is the probability assigned to $\alpha$. Since the function $f(y, z) = \sqrt{y z}$ is concave, we have $\rho_0(x, x^\prime)\leq \sqrt{p_0(x)p_0(x^\prime)}$, where $p_0(x) = \sum_\alpha\lambda_\alpha p_0^\alpha(x)$ is the total probability assigned to $x$ in the mixture. The Cauchy-Schwarz inequality guarantees that our initial density matrix given by $\rho_0=P_0= \widetilde{F}_n\widetilde{F}_n^T$ is consistent with this. Indeed, $P_0(x, x^\prime) = \sum_y \widetilde{F}_n(x, y) \widetilde{F}_n^T(y, x^\prime)\leq \sqrt{p_0(x)p_0(x^\prime)}$, where $p_0(x) = P_0(x,x) =\sum_y \left[\widetilde{F}_n(x,y)\right]^2$ is the total probability assigned to $x$. We could in principle extend our results to asymmetric initial probability matrices, but these would not lead to consistent density matrices (see Appendix~\ref{s:EQM}). 

Being real, $\rho_0$ in principle has a standard probabilistic interpretation (see Appendix~\ref{s:squareroot}). This is not necessarily a restriction, though, since actual quantum experiments also start with classical information scientists have direct access to. A generic ``initial'' quantum state $\rho_{\rm prep}= U_{\rm prep}\rho_0 U_{\rm prep}^\dag$ is prepared only after applying a suitable quantum operation $U_{\rm prep}$ to such classical information. In principle, we can write $U_{\rm prep} = U_m\cdots U_0$, for a suitable number of time steps $m$, where each $U_\ell = \id - i\epsilon H_\ell/\hobs$ is obtained from a factor $F_\ell = \id + \epsilon J_\ell$ via $H_\ell = -\hobs (J_{s, \ell} + J_{a,\ell}/i)$---here $\ell=0,\dotsc , m$. 


We have mostly focused on one observable, i.e., position. However, ``all measurements of quantum-mechanical systems could be made to reduce eventually to position and time measurements (e.g., the position of the needle on a meter or the time of flight of a particle). Because of this possibility a theory formulated in terms of position measurements is complete enough in principle to describe all phenomena''~\cite{feynman2010quantum} (p. 96; see Appendix~\ref{s:measurements}). For instance, the momentum $p=-i\hbar\partial/\partial x$ of a particle could be measured via the position $X$ of a probe particle, initially in state $\psi_{\rm dev}(X)\propto e^{-X^2/4\sigma^2}$ with $\sigma\ll 1$, that interacts with it. Since $e^{i k x}$ are eigenfunctions of the momentum operator, with eigenvalues $\hbar k$, it is convenient to write the initial state of the system as $\psi_{\rm sys}(x)\propto \int c_k e^{i k x}\mathrm{d} k$, where $c_k$ are suitable coefficients. After a suitable interaction between the two particles, we can obtain the joint state $\Psi(x,X)\propto \int c_k e^{i k x} e^{-(X- a\hbar k)^2/4\sigma^2}\mathrm{d}x$, where $a$ is a constant, and the corresponding joint probability $\mcP(x, X) = |\Psi(x,X)|^2$~\cite{aharonov2008quantum} (ch. 7-9). However, we are {\em not} interested in observing $x$, but on {\em inferring} the momentum of the system by observing {\em only} the probe. Marginalizing $x$ yields $\mcP_{\rm dev}(X) \approx\int {|c_k|^2} \delta(X- a \hbar k )\mathrm{d}k$ for $\sigma\ll 1$, where $\delta$ is the Dirac delta function. Thus with probability $\propto\, |c_k|^2$ the position of the probe, $X= a\hbar k$, is proportional to the momentum eigenstate $\hbar k$. This is the (projective) measurement postulate of quantum theory (see Appendix~\ref{s:quantum_nutshell}).

\

\tocless\section{Discussion}\label{sm:discussion}
%


\tocless\subsection{Summary of main technical points}\label{sm:summary_main}

We have introduced a model of embodied scientists doing experiments whose dynamics is formally analogous to imaginary-time quantum dynamics---which involves only real numbers. However, the model describes such scientists from the perspective of another external scientist. Describing a scientist doing experiments from the perspective of the very same scientist is a self-referential problem. Building on insights from the mathematics of self-reference, we have proposed that an observer with an intrinsic perspective should be composed of two subsystems or ``sub-observers'' that, in a sense, mutually observe each other. This is analogous to the two strands of a ``self-replicating'' DNA molecule that help replicate each other. Roughly, the two ``sub-observers'' mutually observing each other are described by two coupled imaginary-time quantum dynamics, which could be encoded in the real and imaginary parts of a complex equation formally analogous to Schr\"odinger equation.

\

\tocless\subsection{Potential physical implications and quantum gravity}\label{sm:physical_discussion}

Our approach may provide a fresh look at the fundamental laws of nature. When $\hobs\to 0$, the most probable path in Eq.~\eqref{em:circular} satisfies Newton equation in imaginary-time (see Appendix~\ref{s:classical}). So, embodiment might help explain why the fundamental equations of physics are typically second-order in terms of a kind of variable, e.g., position only, instead of more parsimonious first-order equations. The SRC effectively implements a Wick rotation that turns imaginary-time into real-time. So, the intrinsic perspective may help explain the actual (sympletic) structure of such equations. 

The Wick rotation also plays a role in relativity. Our approach focuses on modeling how scientists do science, and scientists effectively measure ``time'' by measuring the position of a pointer on a measuring device~\cite{rovelli2011forget}. Consider two free particles with masses $m$ and $M$, positions $\bx = (x_1, x_2)$ and $\by = (y_1 , y_2)$, and (kinetic) ``energy'' function $\mcH = m(\Delta x_1^2 + \Delta x_2^2)/2\epsilon^2 + M(\Delta y_1^2 + \Delta y_2^2)/2\epsilon^2 = E$, where $\epsilon\to 0$ and $E$ are constants. Assume  $\Delta y = \sqrt{\Delta y_1^2 + \Delta y_2^2}$ is used to measure ``time'' intervals. Multiplying $\mcH$ by $2\epsilon^2/m$ we can write $\Delta x_1^2 + \Delta x_2^2+ C^{2}\Delta y^2 = \Delta  s^2$ with $\Delta s^2 = 2\epsilon^2 E / m$ and $C^2 = M/m$. This expression has the same structure of an imaginary-time spacetime interval. After a ``Wick rotation'', $y\to iy$, which can in principle be implemented via a suitable SRC, this acquires the structure of an actual (real-time) spacetime interval. A more realistic situation is where the clock particle is not free, which could in principle lead to the analogue of curved spacetime intervals. So, our approach may suggest alternative roads to quantum gravity.

Science is perhaps the best example of ``intelligence'' we know of. So, our focus on how the general structure of the ``laws of nature'' may emerge from modeling scientists doing science could provide a fresh perspective on cognitive science---for instance, how to interpret the equations obtained if we relax the ``objectivity'' constraints? In particular, it supports the idea that cognition is embodied and that this may be manifested via quantum-like features. It may also suggest routes to build ``machines that learn and think like people'' complementary to those of directly incorporating known physical laws into AI algorithms~\cite{lake2017building} or developing scientist-like AI algorithms that can learn laws {\em already} encoded in a dataset~\cite{wu2019toward}. 

In particular, our results suggest that quantum models may be relevant for ``embodied intelligence'', which hints at novel demonstrations of ``quantum supremacy.'' As a side remark, since recurrent neural networks can both encode TMs~\cite{siegelmann1995computation,siegelmann1995computational,hyotyniemi1996turing} and model biological neural networks, Kleene's recursion theorem suggests that the double-hemisphere architecture of the brain may be related to self-reference---e.g., self-modeling or self-awareness (see Appendix~\ref{s:recursion}). 

There is a close relationship between our potential reconstruction of quantum theory and QBism due to the central role both give to the observer. Because this is the key aspect in most of the resolutions proposed by QBism to the conceptual difficulties of quantum theory, such resolutions would be valid in our framework too. There are some significant differences, though. In Appendix~\ref{s:comparison}, we compare our approach to QBism and two other related approaches~\cite{d2017quantum,mueller2017law} to quantum theory. 

It is in principle possible to test whether our potential reconstruction coincides with genuine quantum theory. For instance, although our approach can naturally accommodate (non-relativistic) Hamiltonians with positive off-diagonal entries, it suggests that these arise as approximations or truncations of Hamiltonians with negative off-diagonal entries, which can be directly interpreted in probabilistic terms (see Appendix~\ref{s:negative-general}). Furthermore, what would happen in those cases where the Hermitian conjugate $\dag$ may not coincide with the conjugate transpose? There are also some subtleties in our approach that might contrast with the full quantum formalism. That is, the possibility of more general density matrices than those considered in physics (see Appendix~\ref{s:EQM}) and that interactions with measuring devices may not be too large (see Appendix~\ref{s:measurements}). 

Finally, most prominent attempts to reconstruct a quantum formalism usually stay mute about the physical origin of Planck's constant. This seems unsatisfactory because, while most of quantum theory might be read as an abstract theory of information, Planck's constant directly connects to the physics---e.g., by setting the relevant energy scales. In contrast, the key role played by embodied observers in our approach suggests $\hobs$ somehow characterizes the physical process of observation (see below). 

\

\tocless\subsection{Potential phenomenological implications and the hard problem of consciousness}\label{sm:phenomenal_dis}

Mathematics historically tried to avoid self-reference and its puzzles by holding a strict hierarchy where functions can operate on numbers but not the other way around. In computation this is reflected in the distinction between data and programs. However, at least in arithmetics and computation, this does not work. Incompleteness and undecidability, as well as the power of formal systems and Turing machines, are intimately related to self-reference~\cite{moore2011nature} (ch.~7.2). Indeed, $\lambda$-Calculus is a manifestly self-referential formulation of computation, which dissolves the data-program distinction: there are only strings operating on strings~\cite{moore2011nature} (ch.~7.4). Strings are both data and programs and can operate on themselves. No string is assumed to play a special role (see Appendix~\ref{s:self-reference} and Fig.~\ref{f:duality} therein). 

In science self-reference has been scrupulously avoided by strictly relying on third-person methods. Although scientists are considered physical systems, they are implicitly assumed to play the special role of observing other physical systems from a disembodied perspective. Generally, scientists investigate human subjects different from themselves (see Appendix~\ref{s:nonduality}). 

Why not rather embrace self-reference? Much as strings operating on strings dissolves the data-program distinction, scientists investigating scientists can effectively dissolve the distinction between experiences and observations. This symmetric situation allows the subjects under investigation to achieve ``objectivity'' and the role of scientists in the doing of science to be acknowledged. Our approach indicates that we can take experience as the starting point of ``objective'' science and still be consistent with current scientific knowledge. This is in line with neurophenomenology~\cite{varela2017embodied,velmans2009understanding,bitbol2008consciousness,thompson2014waking} (see Appendix~\ref{s:physical-phenomenal}). 

Our approach aligns with RQM's view that every physical phenomenon is relative to an observer~\cite{Rovelli-1996}---an embodied observer with an intrinsic perspective, though. Since observers are embodied, every experience has a physical correlate. Since observations are made from an intrinsic perspective, every physical phenomenon is an experience for ``someone''. This is not to suggest that, e.g., rocks have phenomenal experiences, but that rocks are rocks for ``someone'' who experience them as such. However, here ``someone'' does not necessarily refer to an individual or ``self'' but to a kind of ``awareness'' (see below). 

If the quantum formalism---the foundation on which the skyscraper of science stands---already integrates physical phenomena with a minimal form of phenomenal experience, the question of how subjective experience ``emerges'' out of physics would become more tractable. What would emerge is {\em not} experience {\em as such} but increasingly complex {\em contents} of experience---some of which could refer to the experience of being a self. This would parallel the emergence of increasingly complex physical phenomena---some of which could correlate with the experience of being a self---from the ``basic constituents of matter'' (cf. Ref.~\cite{thompson2014waking}, pp. xxxii, 63; see Appendices~\ref{s:thompson-consciousness} and \ref{s:summary} herein). If so, new quantum-inspired metrics might be devised to help assess whether a system is (phenomenally) conscious. For instance, if there are regimes where it behaves {\em as if} it knew quantum theory, perhaps it is conscious.

Scientists establish an ``objective'' description of the physical world via refined third-person methods. Such methods reveal that our ordinary conceptions are only valid in a narrow ``normal'' range of observation, beyond which more fundamental, sometimes highly counter-intuitive phenomena is observed. In principle, it is also possible to establish an ``objective'' description of the phenomenal world via refined first-person methods. Indeed, like $\lambda$-Calculus that allows strings to operate on themselves, a manifestly self-referential science should allow scientists to investigate themselves, i.e., their own phenomenal experience~\cite{wallace2007hidden,velmans2009understanding,blackmore2018consciousness}. 

Neurophenomenology combines insights from refined first- and third-person methods. Recent research~\cite{thompson2014waking,thompson2020not} suggests that our ordinary conceptions of the phenomenal world are only valid in a narrow ``normal'' range, beyond which our current assumptions seem to lose validity (see Appendices~\ref{s:physical-phenomenal} and \ref{s:nonduality} and Table~\ref{t:comparison}). In particular, it suggests that, instead of the usual taxonomy based on the unconscious, access consciousness, and self-awareness~\cite{dehaene2017consciousness}, a more appropriate taxonomy for consciousness is based on ``pure awareness'', contents of awareness, and self-awareness. While the former assumes that (access) consciousness is the outcome of a process of increasing complexity~\cite{dehaene2017consciousness}, the latter assumes that (phenomenal) consciousness is fundamentally devoid of cognitive complexity~\cite{thompson2014waking,metzinger2018minimal}. Indeed, here ``pure awareness'' is just the mere potential to become aware of something and the ability to apprehend whatever appears~\cite{thompson2014waking,metzinger2018minimal,ramm2019pure,josipovic2019nondual,josipovic2014neural,blackmore2018consciousness}. This could be considered a more precise definition of what we are referring to as the intrinsic perspective. 
 
The experience of pure awareness is considered to be non-dual~\cite{thompson2014waking,metzinger2018minimal,ramm2019pure,josipovic2014neural,loy2012nonduality}---i.e. such that perceived dualities, like the distinction between subject and object, are absent~\cite{tang2015neuroscience}---and can be contentless~\cite{thompson2014waking,metzinger2018minimal}---in this sense, it seems analogous to the physical ``vacuum'' (see below). Being the very precondition for any phenomenal experience to manifest, pure or non-dual awareness is said to be all-pervasive and irreducible to lower level experiences (see Appendices~\ref{s:thompson-consciousness}-\ref{s:summary}). Curiously, non-dual awareness is sometimes represented by the symbol \Yinyang, wich has the same structure of Fig.~\ref{f:first}.

An analogy with lucid dreams is sometimes made~\cite{thompson2014waking} (p. 143, 164, 174; Appendix~\ref{s:phenomenal} and Fig.~\ref{f:lucid} herein). In normal dreams, dreamers usually identify with their dream body. This frames the dreaming experience as ``I am here and the world is there''. However, this I-world distinction is somehow inaccurate as the dream state is composed of both. In lucid dreams, dreamers can in principle realize that they are dreaming and that they are not the dream body---no matter how real this appeared before reaching lucidity. Their sense of self encompasses the whole dream state, in a rather ``non-dual'' way.  
There is evidence that the state of lucid dreaming can be paralleled by a form of lucidity in deep and dreamless sleep~\cite{thompson2014waking,windt2016does, windt2015just,metzinger2018minimal}, a state which is not only devoid of the self-other distinction required for experiencing oneself as separate from the world, but also from any kind of content altogether. So, it is in principle possible to experience a form of non-dual awareness during lucid deep and dreamless sleep. This is said to be possible also in deep meditative states~\cite{thompson2014waking,metzinger2018minimal}.

From a mainstream neuroscience perspective, the experience of both self and world in the waking state are associated to neural correlates ``within'' the same individual. Furthermore, the self-other distinction apparent at gross macroscopic scales dissolves ``into a frenzied, self-organizing dance of smaller components''~\cite{theise2005now} at subtler microscopic scales. So, the possibility of non-dual modes of experience might not be so far-fetched.

The physical correlates of the intrinsic perspective, or non-dual awareness, are likely the physical manifestations of $\hobs$ (see Appendices~~\ref{s:thompson-correlates}-\ref{s:parallels}). We understand this comparison may rise a substantial amount of healthy skepticism, an attitude that may be reinforced by the many unfounded claims made in the past on this regard. However, in contrast to previous claims, our discussion is based on a concrete and precise conceptual and mathematical framework. In this sense, we think, it is not just a shot in the dark. We do this because we are convinced that science advances by going through controversy, not by avoiding it, and that further research could potentially unlock social impact on a grand scale (see below). 

Indeed, consider a thought experiment where the ``particles'' of a generic system described by factors $F\propto e^{\epsilon\mcH/\hobs}$ are progressively removed. Notice that $\hobs$ remains the same during the whole process, as do the SRC which implements the intrinsic perspective (see Fig.~\ref{f:first}F). Moreover, it seems natural to assume that $\hobs$ remains the same even after all particles are gone, in which case only the SRC remains---like two empty cameras facing each other in Fig.~\ref{f:first}E. Being the very precondition for observing every physical phenomenon, the SRC is all-pervasive and irreducible to lower level physical phenomena---this has the flavor of incompleteness and undecidability (see Appendix~\ref{s:planck-self}). All this seems consistent with the idea that Planck's constant characterizes irreducible and all-pervasive fluctuations that persists even in the ``vacuum'', a relevant concept in quantum theory. Finally, the SRC is composed of two subsystems, Alice and Bob in Fig.~\ref{f:first}C,F, that play alternative roles as subject and object, in a seemingly non-dual way. That is, U- and E-observers, respectively, play the role of an object (or ``body'') and a subject (or ``mind'') that ``cognizes'' that object. A C-observer integrates both roles.

Interestingly, RQM's view that even the process of observation is relative to another external observer has some similarities with the notion of ``conceptual dependence'' in neurophenomenology~\cite{thompson2014waking} (pp. 331-332; see Appendices~\ref{s:thompson-science}, \ref{s:summary} and \ref{s:parallels} and Table~\ref{t:comparison}). Indeed, the interaction being modeled between U-observer and experimental system might be considered as the ``basis of designation'' for the process of observation, the E-observer might be associated to the ``designating cognition'', and the model itself might be considered as the ``term'' use to designate it.

New kinds of experiments combining first- and third-person methods might thus be envisioned, where scientists can explore potential quantum-like features of their own experience. Meditation techniques and biofeedback might facilitate the observation of quantum-like effects in ``quantum cognition'' experiments~\cite{bruza2015quantum,wang2014context}, for instance, or in experiments of photon detection by humans~\cite{tinsley2016direct,dodel2017proposal,holmes2018testing,vivoli2016does}. Similarly, psychophysics experiments might be devised to measure $\hobs$ and test whether $\hobs = \hbar$. If so, much like Brownian fluctuations provide indirect evidence about the ``existence'' of atoms, quantum fluctuations might provide indirect evidence about the ``existence'' of non-dual awareness. Admittedly these kinds of experiments may be hard to realize for now~\cite{wallace2007hidden,contemplative}. However, the history of science is full of examples of our astonishing human capacity to transform our reality whenever the stakes are high. 

By grounding our worldview, scientific paradigms can hugely impact society. While neglecting phenomenal experience, materialism has extraordinarily improved our lives through the systematic transformation of the physical world. However, such transformations by themselves are neither good nor bad; this entirely depends on our mindset. If phenomenal experience is as fundamental as physical phenomena, perhaps an equally extraordinary, systematic transformation of our minds is possible. This may empower us to balance our mastery of nature with a similar mastery of ourselves. Some meditation techniques are claimed to help develop an ``objective'' mind, which seems very relevant in this age of massive misinformation enabled in part by our mastery of AI and information technologies.
Non-dual modes of experience are claimed to be the most radical methods to develop a peaceful and virtuous mind, less fearful and biased, more harmoniously connected to others and the world. If so, the stakes are high: further research could help us tackle some of the biggest challenges we face today---e.g., strong social divisions, extreme inequality, and a lack of agreement to fight climate change~\cite{mind}.

\

\begin{widetext}
\tocless\section{Epilogue}

\begin{center}

{\em May all beings have happiness and the causes of happiness.

May all beings be free from suffering and the causes of suffering.

May all beings not be separated from sorrowless bliss.

May all beings abide in equanimity, free from bias, greed and hatred.}

\

{The Four Immeasurables}
\end{center}
\end{widetext}

\

\clearpage

\newpage

\tocless\acknowledgements

I thank Marcela Certuche, Harold Certuche and Blanca Dominguez for their support, financial and otherwise, during a substantial part of this project. I thank Tobias Galla, Alan J. McKane, and the University of Manchester for their support at the beginning of this project. I thank Guen Kelsang Sangton for insightful discussions on Buddhist philosophy. I thank Marcin Dziubi\'nski for his brief but useful lessons on recursion and self-reference. I thank Shailesh Date for comments and support. I thank Diana Chapman Walsh, Alejandro Perdomo-Ortiz, Addishiwot Woldesenbet Girma, Delfina Garc\'ia Pintos, Marcello Benedetti, Kenneth Augustyn, John Myers, Marcus Appleby, Nathan Killoran, Markus M\"uller, Michael R. Sheehy, Nathan Berkovitz, Eduardo Pont\'on, Roberto Kraenkel, Camila Sardeto Deolindo, Cerys Tramontini, Hernan Ocampo, Oscar Bedoya, Gonzalo Ordo\~nez, Maria Schuld and Robinson F. Alvarez for comments and constructive criticism. I thank Mariela G\'omez Ram\'irez and Nelson Jaramillo G\'omez for bringing my attention to these ideas. This research is funded in part by the Gordon and Betty Moore Foundation (Grant GBMF7617) and by the John Templeton Foundation as part of the Boundaries of Life Initiative (Grant 60973). I thank FAPESP grant 2016/01343-7 for funding my visit to ICTP-SAIFR from 20-27 January 2019 where part of this work was done.

\



\appendix
\noappendicestocpagenum
\tableofcontents

\appendix

\section{Overview}\label{s:overview}

At the core of this multi-year work is the integration of ideas from different disciplines, such as cognitive science, neurophenomenology, quantum physics, philosophy, mathematics and computer science. We have not yet found a way to sidestep this without undermining the central message of the work. Some readers may feel that this work involves ideas and methodologies from a range of fields somewhat wider than usual in interdisciplinary works. What some readers may find rather obvious, others may find a bit obscure. For instance, some well-versed readers in phenomenology may not be very familiar with the kinds of mathematical calculations involved in physics and vice versa. Furthermore, to some readers some of the concepts involved, such as enactivism, self-reference, ``quantumness'' and non-dual awareness, may seem a bit subtle or counter-intuitive. Moreover, some readers may feel that some of the ideas on which this work builds are scattered throughout a wide variety of literature and that it may take a bit longer than usual to go over it and put all the pieces together. 

With this in mind we try to provide in these appendices as much detail as possible, with the hope that this may facilitate the reading of the work. Unfortunately, the price to pay for this is a rather long document. There is no free lunch after all. We expect that, depending on their background, some readers can obviate some appendices and focus on others. We now provide an overview to help navigate this document. 

In Appendix~\ref{s:on_internal} we describe how standard approaches to the intrinsic perspective can easily lead to an infinite regress. In Appendix~\ref{s:physical-phenomenal} we review some insights from neurophenomenology~\cite{thompson2014waking,velmans2009understanding,bitbol2008consciousness,blackmore2018consciousness} regarding the nature of both science and consciousness that are relevant to this work. This phenomenological reflection suggests, in particular, that the scientific method could be formulated in a pragmatic way that applies to both the physical and the phenomenal worlds. We also review a neurophenomenological framework recently proposed~\cite{thompson2014waking} for a potentially more fundamental understanding of consciousness and highlight the potential parallels with our approach discussed in the main text. As this appendix may seem a bit long and subtle, we include a summary in Appendix~\ref{s:summary}. Some readers may prefer to read the summary first.

In Appendix~\ref{s:quantum_aspects} we describe some aspects of quantum theory that are relevant to this work, and illustrate the type of conceptual problems associated to the theory. In particular, we mention Rovelli's relational interpretation of quantum mechanics, which aligns with the relational approach taken in the main text. 
Furthermore, we show how to write the von Neumann equation as a pair of equations in terms of real matrices. These are interpreted in the main text as describing two ``sub-observers'' mutually observing each other to implement the intrinsic perspective. We also show how some well-known and non-trivial examples of quantum dynamics can be written in terms of real kernels with non-negative entries, which can be interpreted in probabilistic terms. Appendix~\ref{s:diff-quantum} complements Appendix~\ref{s:quantum_aspects} by providing a comprehensive discussion of one of the simplest quantum systems, i.e., a non-relativistic free particle, which is closely related to classical diffusion. We use this example to highlight some aspects that are relevant to our approach, in general, and to take insight to deal with complex Hamiltonians, in particular. As these two appendices are necessarily technical, we try to provide as much mathematical detail as possible.

In Appendix~\ref{s:model_scientists} we first briefly review the frameworks of active inference and enactivism. Afterwards we discuss the more relational approach we take in this work to model scientists doing science and the constraints imposed by objectivity. This allows us to implement the self-referential coupling in the main text. We also discuss the famous two-slits experiment from the perspective offered by our framework. This provides some insights on the phenomenon of quantum interference. Finally, we discuss how enactivism focuses on the process of observation itself, rather than on the observer and observed poles of the subject/object polarity, and how this suggests our approach may transcend single observers and their cognitive limitations.

In Appendix~\ref{s:MP_recasted} we discuss the principle of maximum path entropy~\cite{presse2013principles} and the associated graphical models.
We also show that the belief propagation (BP) algorithm~\cite{Mezard-book-2009} on chain-like graphs can be written in a way formally analogous to imaginary-time quantum dynamics~\cite{Zambrini-1987}. 
However, in this case, the BP messages on the leaves of the chain, which initiate the BP iteration, can always be chosen constant. 
We then discuss the more interesting case of cycles. Although the BP algorithm is not guaranteed to be exact anymore~\cite{Weiss-2000}, we can choose initial and final conditions for the probability marginals to effectively turn the cycle into two chains. This yields the formal analogue of the general imaginary-time quantum dynamics considered in Ref.~\cite{Zambrini-1987}. We also discuss how real and so symmetric density matrices may be interpreted in classical terms and how the ``classical'' limit, $\hobs\to 0$, of the extrinsic-perspective dynamics can lead to imaginary-time classical dynamics.

Since the concept of self-reference~\cite{moore2011nature} does not seem to be very common in the literature, in Appendix~\ref{s:self-reference} we review some of its main aspects and highlight the role it might play in science, as discussed in the main text. In particular, we use the examples of Turing machines and $\lambda$-Calculus to briefly discuss the concepts of duality and non-duality. 
We also briefly discuss some potential relationships between quantum theory and some aspects of self-reference.

The analysis in the main text was restricted to factors with non-negative entries, which can be interpreted in probabilistic terms. Genuine quantum dynamics does not seem to have this restriction, though. 
In Appendix~\ref{s:negative-general} we argue that our approach is not necessarily restricted to factors with non-negative entries either. We discuss further the case of Hamiltonians with complex entries too. 
 
In this work we mostly focus on one observable: position. However, genuine quantum theory deals with different kinds of observables, e.g., momentum or energy. However, strictly speaking, abstract notions like momentum or energy eigenstates have to be implemented in the laboratory in terms of things we are familiar with, e.g., the position of a pointer in a measuring device.
In Appendix~\ref{s:measurements} we argue, along with Feynman~\cite{feynman2010quantum}, that our focus on position variables only should in principle allow for the description of all (non-relativistic) phenomena.
 We briefly discuss some aspects of quantum measurement theory and how the concept of (momentum) eigenstates can naturally arise from measurements of position of a pointer in a measuring device suitably interacting with the system. Finally, we discuss how the Hamiltonian operators required to implement the quantum measurement naturally arise from non-negative real kernels. Appendices~\ref{s:negative-general} and \ref{s:measurements} may be useful to compare our approach to the genuine (non-relativistic) quantum formalism.

In Appendix~\ref{s:comparison} we briefly discuss two prominent approaches to quantum theory---QBism~\cite{debrota2018faqbism,fuchs2013quantum} and a recent derivation from information principles~\cite{d2017quantum}. We also briefly discuss a different approach~\cite{mueller2017law} to model the observer based on algorithmic information theory. Although still at a toy-model level, this recent algorithmic approach also seems to point to some quantum-like phenomenology. We focus on how we currently understand these alternative approaches relate to ours.

These appendices are not intended to form a single coherent document. Rather, they are intended to provide the context, background, details, or proofs for the claims made in the main text. As such, the main text is the thread that connects these appendices. 

\section{On modeling the intrinsic perspective}\label{s:on_internal}
\begin{figure}
\includegraphics[width = \columnwidth]{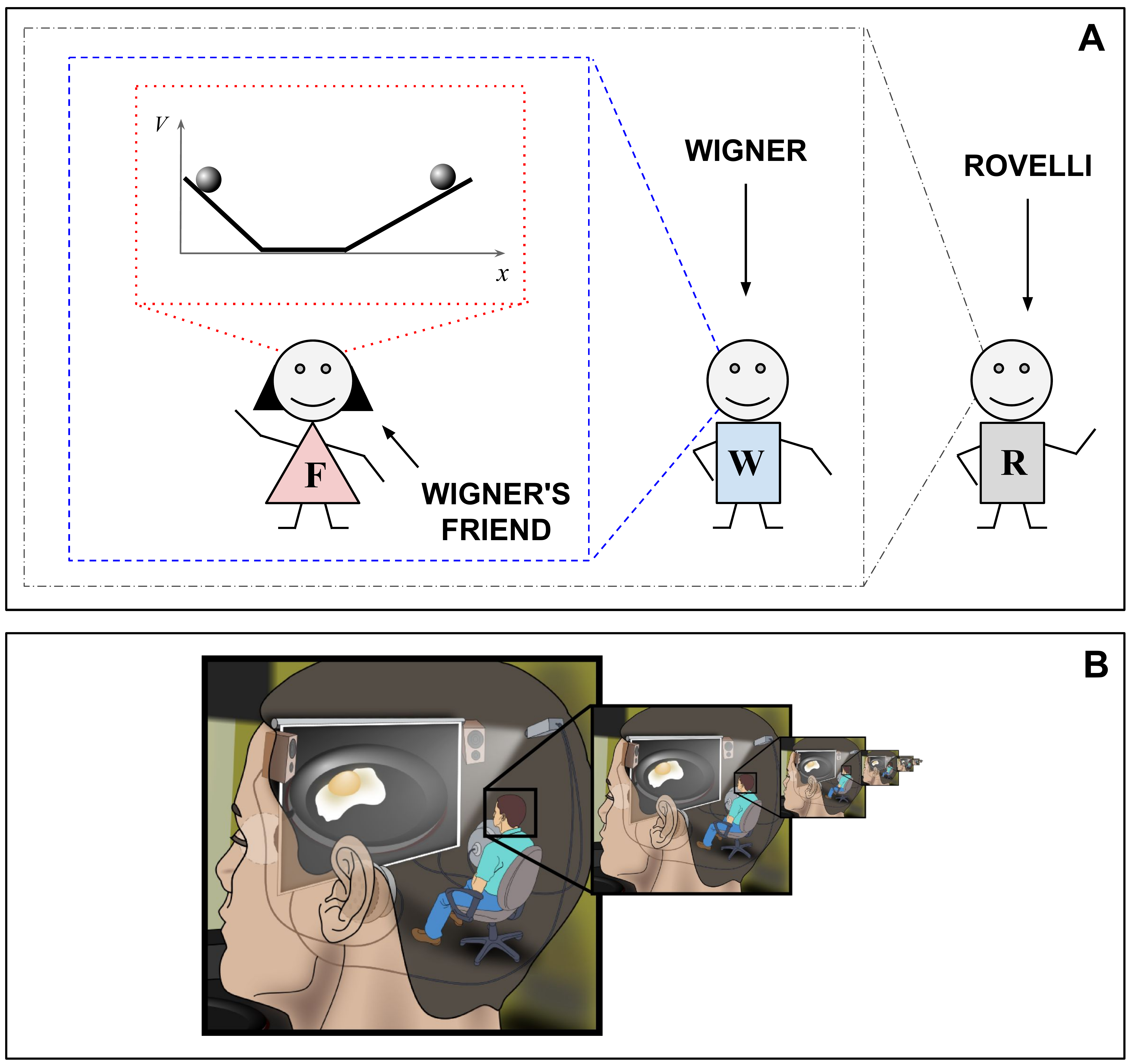}
\caption{{\em The intrinsic perspective and the homunculus fallacy:} (A) Traditionally, physics analyzes the experimental system alone (inner red dotted box) without explicitly taking into account the observer. Traditionally, cognitive science analyzes the observer (Wigner's friend) interacting with an environment or experimental system (middle blue dashed box) without explicitly taking into account the external observer (Wigner) that observes the former. We do take explicit account of Wigner (external black dashed-dotted box; cf. Fig.~\ref{f:circular}A in the main text). However, this more relational perspective, hinted at by Rovelli~\cite{Rovelli-1996} (see Appendix~\ref{s:wigner}), would not take explicit account of the observer who is looking at Wigner. By adding another observer (Rovelli) we are head to an infinite regress (cf. Fig. 5.1 in Ref.~\cite{rovelli2007quantum}). (B) The homunculus fallacy, which is similar to the situation in (A) except that it leads to an infinte regress of internal, rather than external, observers. Image in (B) is taken from \href{https://commons.wikimedia.org/wiki/File:Cartesian\_Theater.svg}{Wikimedia Commons: derivative work: Pbroks13Original: Jennifer Garcia (Reverie)} [\href{https://creativecommons.org/licenses/by-sa/2.5}{CC BY-SA 2.5}] }
\label{f:homunculus}
\end{figure}

Here we discuss in a bit more detail our approach to the intrinsic perspective and how standard approaches can easily lead to an infinite regress. 
With good reason there has historically been an insistence on defining scientific expressions from a privileged extrinsic or third-person perspective (see Appendices~\ref{s:nonduality}, \ref{s:physical-phenomenal}, \ref{s:Muller}). However, the intrinsic or first-person perspective is, {\em by definition}, the {\em opposite} of the extrinsic perspective we have traditionally used in science. So, we should try to avoid falling into the understandable temptation to describe the intrinsic perspective in terms of its opposite, the extrinsic perspective, in a straightforward way. We better first turn our attention to the intrinsic perspective itself. To do so, it may be useful to remember that science has often advanced by investigating specific instances of a phenomenon, say an apple falling from a tree, to find general features that transcend the specific example investigated, e.g., the law of gravity. 

Now, the only instance of intrinsic perspective that we have direct access to is our own. So, one way to investigate the intrinsic perspective is by turning our attention from the usual external objects of observation, including other observers, to ourselves, the subjects who observe. This does {\em not} imply that a specific scientist that is analyzing his own intrinsic perspective is special in any way, {\em nor} that we are introducing any kind of subjective bias. It {\em neither} implies that we humans are special in some sense. Instead, this is only a strategy to investigate which features of the intrinsic perspective we find universal when we compare our own experience of it with the experiences reported by others (see Appendices~\ref{s:physical-phenomenal} and \ref{s:nonduality}). 

As in the example above, where the law of gravity transcends the special case of a falling apple, such universal features can potentially transcend the specific example of our own intrinsic perspective. Indeed, one of the main theories of consciousness~\cite{tononi2016integrated} follow a qualitatively similar approach. It starts by trying to identify the ``essential properties of phenomenal experience'' and this is effectively done by comparing our own phenomenal experience with that reported by others. 

Our modeling of scientists doing science attempts to capture the idea that the intrinsic perspective concerns how ``I'' experience something, not to how {\em other} considers  ``I'' experience it. If ``I'' manage to model the universal features of how ``I'' experience something, ``I'' would be modeling the universal features of how everyone else intrinsically experiences that something for themselves. The problem of describing how phenomena looks from the perspective of the observer herself, rather than how an external observer considers such phenomena should look to the former observer, is self-referential. A na\"ive approach to it can easily lead to an infinite regress (see Fig.~\ref{f:homunculus}A). 

Consider, for instance, a model of an observer observing a certain phenomenon (like Wigner's friend in Fig.~\ref{f:homunculus}A). We might be tempted to say that we are modeling the ``first-person perspective'' of Wigner's friend (see Appendix~\ref{s:Muller}). However, in our view this effectively describes the perspective that an external observer (like Wigner in Fig.~\ref{f:homunculus}A) assigns to the observer under investigation. So, strictly speaking, this is not the intrinsic perspective of Wigner's friend. In this example {\em Wigner observes his friend who in turn observes an experimental system}. Acknowledging the presence of Wigner was hinted at by Rovelli in his relational interpretation of quantum mechanics~\cite{Rovelli-1996}, and this is a key step in our modeling of the intrinsic perspective. However, in our view, this is not enough because we can now ask: who observes Wigner? If we were to add another external observer (named Rovelli in Fig.~\ref{f:homunculus}A) we would be headed to an infinite regress, as we would need to keep on adding observers. Indeed, this would be a variation of the so-called ``homunculus fallacy.''

The homunculus fallacy~\cite{deacon2011incomplete} is the failed attempt to explain a person's analysis of sensory inputs and decision on appropriate responses in terms of another little person (an homunculus) inside the former, who is responsible for doing so (see Fig.~\ref{f:homunculus}B). How does this homunculus analyze his sensory inputs and decide on his appropriate responses? If we add another homunculus within the previous one, we are headed to an infinite regress. 

These kind of problems are common when dealing with self-referential systems. Consider, for instance, the task of finding a program in Python 3 that prints itself. An homuncular attempt would be to print a program, say 
\be\label{e:hello1}
\texttt{print(``Hello world!'')},
\ee
by simply adding a print operator. This yields
\be\label{e:hello2}
\texttt{print(`print(``Hello world!'')')},
\ee
which prints the program in Eq.~\eqref{e:hello1}, but does not actually print itself, i.e., it does not print the program in Eq.~\eqref{e:hello2}. If we try to add another print  operator we are headed to an infinite regress. In this example, the print operator plays the role of the observer and the homunculus in Figs.~\ref{f:homunculus} A and B, respectively.

A program in Python 3 that does print itself is shown in Fig.~\ref{f:first}B in the main text. It is composed of two strings that mutually refer to each other. Kleene's recursion theorem shows that it is possible in general to build self-referential Turin machines by composing two sub-machines that mutually refer to each other (see Appendix~\ref{s:recursion}).  
In this work we propose that we can similarly escape the homuncular infinite regress associated to the intrinsic perspective by allowing Wigner's friend in Fig.~\ref{f:homunculus}A to also observe Wigner. In this way, Wigner and his friend can mutually observe each other, effectively observing themselves. Like the two sub-machines of a self-referential Turing machine, here Wigner and his friend do not refer anymore to two individuals, but to two halves of a single individual. In other words, the architecture of a self-referential observer should be composed of two sub-systems that refer to each other (see Fig.~\ref{f:first} in the main text).

\section{The rigorous investigation of the physical and phenomenal worlds}\label{s:physical-phenomenal}

Here we review and discuss in more detail some insights from phenomenology regarding the nature of both science and consciousness we have relied on in the main text. This phenomenological reflection suggests that the scientific method could be formulated in a pragmatic way that applies to both the physical and the phenomenal worlds. We also discuss a neurophenomenological framework recently proposed for a potentially more fundamental understanding of consciousness. Since this appendix may be a bit long, subtle and technical, we then present a summary of the main points. The author may prefer to read this summary first and then decide what aspects to focus on. Here we rely heavily on quoting experts in the field. Our goal up to here is to facilitate the reading of the main aspects of the literature relevant for our work.

Finally, in Appendix~\ref{s:parallels} we discuss in more detail some of the potential parallels with our approach that might help put the phenomenological framework summarized here into more quantitative terms. Since this framework is rather recent and challenges some assumptions of the mainstream approach, this discussion is necessarily speculative and a healthy skepticism is implied. We have included this discussion here because we think it holds the potential to suggest new lines of inquiry that may help expand our current understanding of consciousness and science more generally.

Strictly speaking, consciousness can mean either phenomenal consciousness or access consciousness~\cite{block1995confusion,fazekas2018perceptual}. Phenomenal consciousness is the subjective experience itself or ``what it feels like'' to be in a particular state. It essentially means {\em ``experience in all its forms across waking, dreaming, deep sleep, and meditative states of awareness''}~\cite{thompson2014waking} (pp. 15-16). 
Access consciousness, on the other hand, is characterized by the availability of perceptual information for use in a wide-range of cognitive tasks such as reporting its content, or reasoning or acting on the bases of it~\cite{block1995confusion,phillips2018methodological}. The mainstream perspective in neuroscience is largely based on access consciousness, as this is associated with the verbal reports or other behavioral responses that can be measured in the lab---we will briefly discuss this below. On the other hand, recent developments in neurophenomenology have placed a focus on phenomenal consciousness, leading to an alternative characterization of consciousness that we will discuss further below. Except when specified otherwise, {\em here the term consciousness refers to phenomenal consciousness} since this is our focus here.

This neurophenomenological framework tries to parallel the way we have reached our current fundamental understanding of the physical world, which has been possible in part because physics does not constraint itself {\em a priori} to what is believed to be ``normal''. Rather, there is a constant push for more and more refined techniques of exploration of the phenomena to be studied. Furthermore, qualitatively different phenomena usually requires qualitatively different exploration techniques---e.g., microscopes for the very small and telescopes for the very far. This approach has revealed that our ordinary concepts of, e.g., space, time, and matter are only approximately valid in a narrow ``normal'' range of observation, beyond which things become much subtler and counter-intuitive. Furthermore, such subtler and counter-intuitive phenomena are considered more fundamental than the ``normal'' phenomena, not the other way around, because they are based on more refined exploration techniques that allow for a wider range of observation. 

The main phenomena investigated in the science of consciousness is subjective experience. So, refined techniques for the exploration of the phenomenal world, beyond the narrow range considered ``normal'' today, may be required to push towards a more fundamental understanding of consciousness. Considering what we have learned from the physical world, we should be prepared to find counter-intuitive phenomena which may defy our current assumptions and perhaps be difficult to grasp. One such refined technique for exploring the phenomenal world is meditation. Advance meditative techniques can lead to absorptions into so-called ``altered'' states of consciousness, as opposed to what we consider ``normal'' today~\cite{bitbol2015altered,thompson2014waking,metzinger2018minimal,blackmore2018consciousness}, which indeed suggest some counter-intuitive phenomena. Interestingly, there appears to be some common core concepts underlying some counter-intuitive aspects of both the physical and the phenomenal world, e.g., relationalism (see below and Appendix~\ref{s:nonduality}). 

Although some of the most refined first-person methods known to date, e.g., meditation, has been typically used by contemplative traditions like Buddhism, this is not to be read as an attempt to bring ``mysticism'' into science. Quite the contrary, it can be read as an attempt to bring the rigor of science to the utilization of these methods, if possible, which may be of great benefit to society. Paraphrasing Neil deGrasse Tyson, perhaps we may know enough to think we are right about everything regarding these traditions, but not enough to realize that we may have missed or misunderstood some scientifically useful and relevant points. 

The 14th Dalai Lama, who has been actively participating in regular conversations between leading scientists and contemplative practitioners, has manifested a rather scientific attitude in this regard (emphasis ours): 

\begin{myquotation}My confidence in venturing into science lies in my basic belief that as in science so in Buddhism, understanding the nature of reality is pursued by means of critical investigation: {\em if scientific analysis were conclusively to demonstrate certain claims in Buddhism to be false, then we must accept the findings of science and abandon those claims.} (Ref.~\cite{lama2005universe}, pp. 2-3)
\end{myquotation}

Since the language used by these traditions may seem confusing, it is useful to focus on the actual meaning of the words used and the context in which they are used. It is also useful not to disregard the whole tradition because we find something we completely disagree with. Fortunately, some philosophers and scientists have made a great effort to cut through the language and extract those aspects that could be relevant for science. We discuss some of them here. 

We start by reviewing some of the main reasons provided by Thompson~\cite{thompson2014waking,varela2017embodied}, Velmans~\cite{velmans2009understanding}, and Bitbol~\cite{bitbol2008consciousness} to argue that there is both a primacy of the physical and a primacy of the phenomenal. We also discuss why this suggests that a more fundamental scientific description of nature should put both the physical and the phenomenal in an equal, though relational or non-dual footing. This more fundamental description should avoid the extremes of solipsism, panpsychism, dualism, and bare materialism. We then move on to discuss Thompson's neurophenomenological framework for (phenomenal) consciousness. Afterwards, we review some preliminary ideas on the potential physical correlates of ``consciousness-as-such'', ``pure awareness'' or ``non-dual awareness''. We finally provide a brief summary of the main points relevant for our final discussion on the potential parallels with our approach. 

\subsection{Interdependence of the physical and the phenomenal}\label{s:thompson-science}

Here we present some insights from the field of phenomenology related to how the scientific method is actually performed in practice. To proceed, it may be useful to temporarily put aside all our assumptions, no matter how accurate they may seem, and try to ``describe the real factual situation'', as advised by Einstein. To begin, Bitbol~\cite{bitbol2008consciousness} argues that experience plays a more fundamental role in science than usually acknowledged:

\begin{myquotation} An analysis of ordinary and scientific knowledge shows us that objective domains of knowledge are elaborated in two steps, with conscious experience as an implicit departure point. Firstly, one progressively pushes aside any feature of experience on which conscious subjects cannot agree, such as individual tastes, community values, or the emotional tinge which is associated by individuals and communities with particular situations. Secondly, one only retains a sort of structural residue of conscious experience that can be the object of a consensus, and of a collectively efficient use as a predictive tool. At its most abstract, this structural residue is mathematical; but it can also consist of general propositions stating various types of relations between entities and predicates, such as ``brains are bioelectrical organs made of neurons and glial cells''. (Ref.~\cite{bitbol2008consciousness})
\end{myquotation}

As an illustration of this process of ``objectification'', Bitbol discusses how thermodynamic variables such as temperature and pressure have been extracted from their experiential basis:

\begin{myquotation}
In the beginning, there were bodily ``sensations'', ordinary practices, and an overabundance of qualitative observations about color of metals, fusion or ebullition of materials, expansion of liquids according to whether they are cold or hot etc. Heat and temperature were hardly distinguished from one another, and from the feeling of hotness. As for pressure, it was little more than a name for felt strain on the skin. But, progressively, a new network of quantitative valuations emerged from this messy experiential background, together with the laws that connect them (such as the ideal gas law). Even though sensations of hotness and strain still acted as a root and as a last resort for these valuations, they slipped farther and farther away from attention, being the deeper but less reliable stratum in a growingly organized series of criteria for assessing thermodynamic variables. At a certain point, the sensation of hotness no longer played the role of an implicit standard at all; it was replaced by phase transitions of water taken as references for a scale of variable dilatations in liquid thermometers. This scale, which posits a strict order relation of temperatures, replaced the mixture of non-relational statements of hot or cold and partial order relation of hotter and colder which tactile experience together with qualitative observation of materials afford. Accordingly, the visual experience of graduation readings, or rather the invariant of many such visual perceptions, was given priority over the tactile experience of hotness. Later on, when the function ``Heat'' was clearly distinguished from the variable ``temperature'', and its variation defined as the product of the ``heat capacity'' times the variation of temperature, tactile experience was submitted to systematic criticism: the feeling of hotness was now considered as a complex and confused outcome of heat transfer between materials of unequal heat capacities and the skin, and also of the physiological state of the subject. From then on, declarations about tactile experience, which had acted initially as the tacit basis of any appraisal of thermic phenomena, were pushed aside and locked up in the restrictive category of so-called ``subjective'' statements~\cite{peschard2008heat}. (Ref.~\cite{bitbol2008consciousness})
\end{myquotation}

Furthermore, Bitbol argues that we tend to forget the long scientific process and to take its ``end result'' as the starting point, the ``objective truth'' out of which the actual starting point of the process, i.e., subjective experience, has to be derived (emphasis his):

\begin{myquotation}The creators of objective knowledge become so impressed by its efficacy that they tend to {\em forget} or to minimize that conscious experience is its starting point and its permanent requirement. They tend to forget or to minimize the long historical process by which contents of experience have been carefully selected, differentiated, and impoverished, so as to discard their personal or parochial components and to distillate their universal fraction as a structure. They finally turn the whole procedure upside down, by claiming that experience can be explained by one of its structural residues. Husserl severely criticized this forgetfulness and this inversion of priorities, that he saw as the major cause of what he called the ``crisis'' of modern science~\cite{husserl1970crisis}. According to him, it is {\em in principle} absurd to think that one can account for subjective conscious experience by way of certain objects of science, since objectivity has sprung precisely from what he calls the ``life-world'' of conscious experience. (Ref.~\cite{bitbol2008consciousness})
\end{myquotation}

Bitbol finally provides a series of suggestions to develop an alternative framework or stance. In particular, he invites us to reflect on how ``objectivity'' arises from intersubjective debate, the role of intersubjectivity as a common ground for the investigation of both the phenomenal and the physical worlds, and how a systematic training of experience---e.g., via meditation or bio-feedback---could potentially expand the basis of possible intersubjective agreement (see original for Bitbol's references to other works):

\begin{myquotation}
[First, s]how how objectivity arises from a universally accepted procedure of intersubjective debate. Do not construe it as a transcendent resource of which intersubjective consensus is only an indirect symptom. Draw inspiration from a careful reflection about physics: either from the process of emergence of objective temperature valuations from an experiential underpinning... or from the model of quantum mechanics construed as a science of inter-situational predictive invariants rather than a science of ``objects'' in the ordinary sense of the word.... Then, recognize that intersubjectivity should be endowed with the status of a common ground for both phenomenological reports and objective science. Start from this common ground in order to elaborate the amplified variety of knowledge that results from embedding phenomenological reports and objective findings within a unique structure...

[Second, d]o not rely on a minimal and most elementary form of intersubjective consent, but try to amplify the criteria of intersubjective understanding by refining the stability and sharpness of subjective experience. After all, the reason why numerical values and ratios are privileged as objects of intersubjective agreement is likely to be the fact that they are not too difficult to be agreed upon, even among subjects with a poorly cultivated experience. But if experience is systematically trained and educated, either in the first person by meditation, or in the second person by making explicit unsuspected features of experience in dialogue, or in a combination of first- and third-person modes by bio-feedback, the basis of possible intersubjective consensus is likely to expand beyond recognition. (Ref.~\cite{bitbol2008consciousness})
\end{myquotation}

Velmans~\cite{velmans2009understanding} reaches similar conclusions by analyzing the roles played by human beings in psychophysics experiments, wherein an experimentalist ``objectively'' observes a subject ``subjectively'' experiencing, e.g., images. He asks what makes a human being a ``subject'' ($S$) and another an ``experimenter'' ($E$) and finds no fundamental distinction (see Appendix~\ref{s:velmans}). Like Bitbol, Velmans notices that without first-person ``experiences'' there cannot be third-person ``observations'' and therefore no science (emphasis his): 

\begin{myquotation}Although reductionists pretend otherwise,``external observers'' are also ``experiencing subjects'' and ``experiencing subjects'' are also ``external observers''. In a typical psychophysical experiment they simply play different {\em roles}. External observers are normally interested in events external to themselves (for example the mental states of other people) and consequently focus on what their observations (of other people) {\em represent}. Subjects are typically asked to focus on the nature of the experiences themselves. However, in terms of {\em phenomenology} there is no difference between a given individual's ``observations'' and ``experiences''... Your visual observations and visual experiences of this [article], for example, are one and the same. One cannot reduce first-person experiences to third-person observations for the simple reason that, {\em without first-person experiences one cannot have third-person observations}. Empirical science {\em relies} on the ``evidence of the senses''. Eliminate experiences and you eliminate science! (Ref.~\cite{velmans2009understanding}, p. 316)
\end{myquotation}

Usually we take observations of measuring devices as ``public'' and those reported by subjects in psychological experiments as ``private''. However, Velmans points out that observations of measuring devices are, strictly speaking, ``private'' experiences scientists have. It is just that we take it for granted that if scientists report the same ``private'' experience of the measuring device, then the corresponding observations are the same. However, since scientists cannot directly access the experiences of each other, but only their reports, there is no way for them to be certain that their experiences of the measuring device are actually similar. Strictly speaking, we cannot rule out the possibility that they {\em might} be lying to each other. We may argue against Velmans that natural selection might in principle eliminate the liars in the long-term, or that liars could perhaps be forced to tell the truth by asking them to place bets---like in QBism. But the point remains: scientific observations are strictly speaking experiences (emphasis his):

\begin{myquotation}[The experimenter ($E$) and the subject ($S$)] know what it is like to have their own experiences, but they can only access the experiences of others indirectly via their verbal descriptions or nonverbal behaviour. This applies to {\em all} observed phenomena [including physical phenomena, e.g., a pointer in a measuring device] ... As $E$ does not have direct access to $S$'s experience of the [pointer] and vice versa, there is no way for $E$ and $S$ to be {\em certain} that they have a similar experience (whatever they might claim). $E$ might nevertheless {\em infer} that $S$'s experience is similar to his own on the assumption that $S$ has similar perceptual apparatus, operating under similar observation arrangements, and on the basis of $S$'s similar observation reports. $S$ normally makes similar assumptions about $E$. It is important to note that this has not impeded the development of physics and other natural sciences, which simply ignore the problem of ``other minds'' (uncertainty about what other observers actually experience). They just take it for granted that if {\em observation reports} are the same, then the corresponding {\em observations} are the same. The success of natural science testifies to the pragmatic value of this approach... 

Ironically, psychologists have often agonised over the merits of observation reports {\em when produced by subjects}, although, like other scientists, they take them for granted {\em when produced by experimenters}, on the grounds that the observations of subjects are ``private and subjective'', while those of experimenters are ``public and objective''. As experimenters do not have access to each other's experiences any more than they have access to the experiences of subjects, this is a fallacy, as we have seen. Provided that the observation conditions are sufficiently standardised, the observations reported by subjects can be made public, intersubjective and repeatable amongst a community of subjects in much the same way that observations can be made public, intersubjective and repeatable amongst a community of experimenters. This provides an epistemic basis for a science of consciousness that includes its phenomenology. (Ref.~\cite{velmans2009understanding}, pp. 222-223)
\end{myquotation}

Velmans proposes to formulate the scientific method in a way that more accurately reflects the pragmatic features described above. He argues that in so doing the scientific method could equally apply to both the investigation of the physical and the phenomenal worlds (emphasis his):

\begin{myquotation}[O]bservations can be ``objective'' in the sense of {\em intersubjective}, and the observers can ``be objective'' in the sense of being dispassionate, accurate and truthful. Procedures can also ``be objectified'' in the sense of being standardised and explicit. No observations, however, can be objective in the sense of being {\em observer-free}. Looked at in this way, there is no unbridgeable, epistemic gap that separates physical phenomena from psychological phenomena. 

In short, once the {\em empirical method} is stripped of its dualist trappings, it applies as much to the science of consciousness as it does to the science of physics in that it adheres to the following principle:

\

If observers $E_{1 \textrm{ to } n}$  (or subjects $S_{1 \textrm{ to } n}$) carry out procedures $P_{1 \textrm{ to } n}$ under observation conditions $C_{1 \textrm{ to } n}$ they should observe (or experience) result $R$

\

(assuming that $E_{1 \textrm{ to } n}$ and $S_{1 \textrm{ to } n}$ have similar perceptual and cognitive systems, that $P_{1 \textrm{ to } n}$ are the procedures which constitute the experiment or investigation, and that $C_{1 \textrm{ to } n}$ include {\em all} relevant background conditions, including those internal to the observer, such as their attentiveness, the paradigm within which they are trained to make observations and so on). 

Or, to put it more simply:

{\em If you carry out these procedures you should observe or experience these results.}~\cite{velmans2009understanding} (pag. 219).
\end{myquotation}

Notice that this notion of ``objective''---which is the one we use in this work---allows for the possibility to achive an ``objective'' description of the phenomenal world. Like in traditional science---Velmans argues---to achieve such an ``objective'' view may require effort and ingenuity, it may not be obvious at first sight (emphasis his):

\begin{myquotation}[T]he {\em phenomena} of consciousness provide data that are potentially public, intersubjective and repeatable. Consequently, the need to use and develop methodologies appropriate to the study of such phenomena does not place them beyond science. Rather, it is part of science. (Ref.~\cite{velmans2009understanding}, p. 221)
\end{myquotation}

Velmans refers to his approach as ``critical phenomenology'' to emphasize---in the same spirit of Varela's neurophenomenology---the relevance of both first- and third-person methods:

\begin{myquotation}[C]ritical phenomenology [CP] adopts a form of ``psychological complementarity principle'' in which first-person descriptions of experience and third-person descriptions of correlated brain states provide accounts of what is going on in the mind that are complementary and mutually irreducible. A complete account of mind requires both ... 
[W]hile CP takes subjective experiences to be real, it remains cautious about the veridical nature of phenomenal reports in that it assumes neither first- nor third-person reports of phenomena to be incorrigible, complete, or unrevisable---and it remains open about how such reports should be interpreted within any given body of theory. 

CP is also open to the possibility that first-person investigations can be improved by the development of more refined first-person investigative methods, just as third-person investigations can be improved by the development of more refined third-person methods. CP also takes it as read that first- and third-person investigations of the mind can be used conjointly, either providing triangulating evidence for each other, or in other instances to inform each other.

Finally, CP is {\em reflexive}, taking it for granted that {\em experimenters} have first-person experiences and can describe those experiences much as their subjects do. And crucially, experimenters' {\em third-person reports of others} are based, in the first instance, on their {\em own first-person experiences}... 

If this analysis is correct, the ``phenomena'' observed by experimenters are as much a part of the world that they experience as are the ``subjective experiences'' of subjects. If so, the {\em whole} of science may be thought of as an attempt to make sense of the phenomena that we observe or experience. (Ref.~\cite{velmans2009understanding}, pp. 228-229)
\end{myquotation}

In our view, grounding the whole of science on experience is somehow analogous to the way $\lambda$-Calculus grounds the whole of computation on strings (see Appendix~\ref{s:nonduality}). $\lambda$-Calculus indicates that the distinction between data and code on which Turing machines rely is only apparent and unnecessary (see Fig.~\ref{f:duality}). Such a distinction is similar to the distinction between ``objective'' observations and ``subjective'' experiences. It may facilitate analysis and communication but perhaps it is unnecessary if we expand the domain of science to incorporate the whole of experience, including ``objective'' observations as a particular case---as Bitbol suggested above. By explicitly incorporating both scientists and the world they investigate into a single, relational framework, such a reflexive or manifestly self-referential science should allow a form of {\em self-referential experiments}, where scientists can rigorously investigate themselves via, e.g., meditation or bio-feedback. This would be similar to the way string can operate on themselves in $\lambda$-Calculus. We expect our approach could serve as a step in this direction. 

In principle, scientists could achieve ``objectivity''---in Velmans' sense---about their own experiences much like mathematicians agree on the proof of a theorem~\cite{wallace2007hidden}. Mathematicians can do this even though theorems are mental constructs that, strictly speaking, cannot be shown to others in the same way we show a pointer in an experimental device. Indeed, some meditative traditions have been exploring for centuries such a kind of rigorous communication protocols about personal conscious experiences~\cite{wallace2007hidden,thompson2014waking,varela2017embodied}. Moreover, such math-like intersubjective agreements can be compared to third-person measurements of the corresponding physical correlates, for instance, as an alternative way to cross-check them. On this regard, the 14th Dalai Lama says:

\begin{myquotation}A comprehensive scientific study of consciousness must therefore embrace both third-person and first-person methods: it cannot ignore the phenomenological reality of subjective experience but must observe all the rules of scientific rigor: So the critical question is this: Can we envision a scientific methodology for the study of consciousness whereby a robust first-person method, which does full justice to the phenomenology of experience, can be combined with objectivist perspective of the study of the brain?

Here I feel a close collaboration between modern science and the contemplative traditions, such as Buddhism, could prove beneficial. Buddhism has a long history of investigation into the nature of the mind and its various aspects---this is effectively what Buddhist meditation and its critical analysis constitute. Unlike that of modern science, Buddhism’s approach has been primarily from first-person experience. The contemplative method, as developed by Buddhism, is an empirical use of introspection, sustained by rigorous training in technique and robust testing of the reliability of experience. All meditatively valid subjective experiences must be verifiable both through repetition by the same practitioner and through other individuals being able to attain the same state by the same practice. If they are thus verified, such states may be taken to be universal, at any rate for human beings. (Ref.~\cite{lama2005universe}, p. 134)
\end{myquotation}

Such self-referential experiments might allow scientists to have direct contact with deeper forms of phenomenal consciousness, which may help them collectiveley identify its universal qualitative features to guide the search of its physical correlates. Interestingly, in line with Velmans, some traditions like Buddhism maintain that once meditators achieve an ``objective'' view of the world, they realize it is not ``observer-free''---as suggested also by our approach. 

In the next subsection we discuss a potentially more fundamental, though admittedly counter-intuitive, framework for phenomenal consciousness proposed by Thompson~\cite{thompson2014waking} based on reports from advanced meditators (see also Ref.~\cite{thompson2020not}). Like Velmans, Thompson also proposes a non-dual scientific framework, building partly on Bitbol's ideas. While Velmans elaborates on the scientific method in a way we find rather systematic, Thompson elaborates on phenomenal consciousness in a way we find quite profound. 

Thompson notices that scientific evidence strongly suggests that consciousness is contingent upon physical phenomena, i.e., its neural correlates. In this sense there is a {\em primacy of the physical}. However, he argues that, strictly speaking, there is also an epistemological and methodological {\em primacy of the phenomenal}~\cite{varela2017embodied,thompson2014waking}: without consciousness there is no observation and without observation there is no data; everything perceived, believed, theorized, researched, intersubjectively agreed upon, and known is done so by conscious observers~\cite{varela2017embodied}. In Thompson's words:

\begin{myquotation}[T]he implicit departure point and always-present background condition for science is our concrete, sensuous experience of the life-world. In creating classical science, we set aside features of this kind of experience that vary individually and cannot be made the object of a stable consensus. Using logic and mathematics, we create an abstract and formal representation of certain invariant and structural features of what we experience under rigorously controlled conditions that we impose, and this formal model becomes an object of consensus and the basis for an objective description. [... S]cientific models are distillations of our embodied experience as observers, modelers, and interveners. [... S]cientific knowledge is not the exhibition of the nature of reality as it is in itself; it is an expression of the relation between our embodied cognition and the world that it purports to know. (Ref.~\cite{varela2017embodied}, p. xxvii).
\end{myquotation}

We can implement the so-obtained models via technological devices to intervene, manipulate, measure, and control phenomena {\em as registered in conscious experience}. In this sense, ``lived experience is the point of departure and return for the production of objective knowledge''. For instance, to develop a quantitative approach regarding lengths, we select an arbitrary conscious experience as a unit of measurement, e.g., the experience of a platinum bar, and compare all other related conscious experiences, e.g., a bar of wood, to that standard. The comparison itself is a conscious experience too.  

Now, precisely because of its third-person, objective approach, it seems quantitative science can only deal with the contents of consciousness, which can be compared among them, not with consciousness as such---according to Thompson, consciousness cannot be objectified, because it is that by which any object shows up for us at all. We can never step outside consciousness to see how it measures up to something else. In this sense, consciousness is like space: we can develop quantitative accounts of the objects ``in'' space by comparing them among themselves, but we cannot step outside space to compare it to something else. However, much as we have indirectly learned about space by studying the objects ``in'' space, we can in principle learn about subjective experience by investigating how scientists can establish an ``objective'' science out of their multiple subjective perspectives, as we argue in this work. 

Thompson also challenges the common view that consciousness is nothing other than a brain process, i.e., that every conscious experience is identical to some pattern of brain activity:

\begin{myquotation}Neuroscience itself doesn't demonstrate this identity; rather, the identity is a metaphysical interpretation of what neuroscience does show, namely, the contingency or dependence of certain kinds of mental events on certain kinds of neuronal events. In every neuroscience experiment on consciousness, the evidence is always of the co-occurrence of mental events and neuronal events, and that isn't sufficient to establish their identity. Even the causal manipulations, strictly speaking, go both ways, from neuronal events to mental events and from mental events to neural ones. We can alter a person's mental states by acting on her brain (through direct electrical brain stimulation, drugs, surgery, and so on), and we can alter a person's brain activity by acting on her mental states (by asking her to imagine something, by having her direct her attention in a certain way, and so on). (Ref.~\cite{thompson2014waking}, pp. 101-102).
\end{myquotation}

However, Thompson argues that we {\em cannot} infer from the epistemological primacy of consciousness that it has {\em ontological} primacy in the sense of being the primary reality out of which everything is composed or the ground from which everything is generated. That the world we know is always a world for consciousness does not entail that the world is made out of consciousness. 

In summary, for Thompson we can never step outside consciousness when investigating and knowing the physical world, and consciousness never shows up independently of some physical basis. In this sense, there is both a primacy of the physical and a primacy of the phenomenal. From this perspective, and in line with relationalism, the physical and the phenomenal could be said to co-emerge in relationship to each other, i.e., in a non-dual way, like two sheaves of reeds propping each other up~\cite{thompson2014waking}.  Hence, Thompson invites us to explore non-dual frameworks that can hold these two points of view, the physical and the phenomenal, without privileging one over the other (emphasis ours):

\begin{myquotation}Since consciousness by nature is experiential, and experience is primary and ineliminable, consciousness cannot be reductively explained in terms of what is fundamentally or essentially nonexperiential, like classical science. Yet classical science, as well as much (though not all) of modern science, has conceived of physical phenomena as being fundamentally, in themselves, essentially devoid of any relation to anything experiential. Bridging from nature so conceived to consciousness is an impossible task, for the two concepts mutually exclude each other...

[T]he scientific concept of matter may need to be modified in order to account for the ultimately nondual relationship between consciousness and matter, that is, for the way that ``subtle consciousness'' [i.e. non-dual awareness] and ``subtle energy'' [i.e. the hypothesized physical basis of non-dual awareness] are contingent upon each other. In my view, this proposal is best understood not as pointing back to something like nineteenth-century ``vitalism'', according to which living things posses a special nonphysical element or substance, but instead as pointing toward the need to {\em rethink what we mean by ``physical'' so that physical being is understood as naturally including, at its most fundamental level, the potential for consciousness or experiential being.} (Ref.~\cite{thompson2014waking}, pp. 103-104)
\end{myquotation}

Our approach suggests the mere potential for experience is already incorporated at the most fundamental levels of the physical world. So, we may not need to rethink what we mean by ``physical'', but to more carefully consider the implications of the quantum nature of the universe. We will discuss the expressions ``non-dual awareness'' and ``subtle energy'' in the quote above in the next two subsections. Importantly, Thompson makes clear he is advocating for neither dualism nor panpsychism, but for a non-dual framework (emphasis ours):

\begin{myquotation}Dualism says that matter and consciousness have totally different natures; panpsychism says that every physical phenomenon possesses some measure of experience as part of its intrinsic nature. Neither position is attractive to me [since they don't sit well with scientific evidence...]  

{\em [W]e need to work our way to a new understanding [that] would replace our present dualistic concepts of consciousness and physical being, which exclude each other from the start,  with a nondualistic framework in which physical being and experiential being imply each other or derive from something that is neutral between them.} (Ref.~\cite{thompson2014waking}, pp. 104-105)
\end{myquotation}

Up to now we have discussed the relationalism between the physical and the phenomenal world. However, according to the Madhyamaka school of Buddhism, relationalism in principle characterizes all phenomena (see {\em Appendix}~\ref{s:self-relationalism}). This is usually referred to as {\em ``dependent arising''}. Thompson~\cite{thompson2014waking} describes that, according to the Madhyamaka school of Buddhism, dependent arising happens at three levels. The first one is {\em causal dependence}. For instance, ``[i]n the case of a cell, many causes and conditions---environmental, genetic, metabolic, and so forth---contribute to its coming into existence, its continuing to exist for a time, and its ceasing to exist''~\cite{thompson2014waking} (p. 330). The second level is {\em whole/part dependence}. For example, a cell depends on its parts: its membrane, organelles, molecular constituents, etc. But such parts also depend on the whole:  ``the specific metabolic pathways and molecules they synthesize depend on the functioning of the entire cell for their existence, for these pathways and molecules can't persist for long outside the special environment the cell maintains within its membrane. Thus part and whole co-arise and mutually specify each other''~\cite{thompson2014waking} (p. 330).

The third level of dependent arising is {\em conceptual dependence}. Thompson points out that, ``[a]ccording to the subschool of Madhyamaka called Pr\=asa\.ngika Madhyamaka, the identity of something as a single whole depends on how we conceptualize it and refer to it with a term. We can also add that its identity depends on a scale of observation. Thus the cell has no intrinsic identity independent of a conceptual scheme and scale of observation that individuate it as a unit''~\cite{thompson2014waking} (p. 330). For instance, ``[o]n one level, cells are indivisible things; on another, they dissolve into a frenzied, self-organizing dance of smaller components''~\cite{theise2005now}. More precisely (emphasis ours): 

\begin{myquotation}What we mark off as a system depends on our cognitive frame of reference and the concepts we have available. 

Such conceptual dependence doesn't mean that nothing exists apart from our words and concepts, or that we make up the world with our minds. On the contrary, in order to designate something as a cell or as a fluid, or more generally as a system of whatever kind, there must be some basis for that designation in what we observe. The subtle point, however, is that what shows up for us as a system of whatever kind depends not just on a basis of designation but also on how we conceptualize that basis and use words to talk about it. For this reason, the full statement of dependent arising as conceptual dependence, according to Pr\=asa\.ngika Madhyamaka, is that {\em whatever is dependently arisen depends for its existence on a basis of designation, a designating cognition, and a term use to designate it}. (Ref.~\cite{thompson2014waking}, pp. 331-332)
\end{myquotation}

The level of causal dependence is in line with the focus of science in general, and physics in particular, on causality. Emergence and complex systems theory, according to Thompson, is conceptually related to the level of part/whole co-dependence. The level of conceptual dependence is subtler. Our approach suggests that it might be related to quantum theory, as we discuss further in Appendix~\ref{s:parallels}. 

\subsection{A potential road to a more fundamental understanding of consciousness}\label{s:thompson-consciousness}

The discussion in the previous subsection suggests it may be possible to investigate the phenomenal world with the same rigor that the physical world has been investigated. This requires the development of refined first-person methods and rigorous, standardized protocols to reach intersubjective agreement and establish ``objectivity''. Here we discuss Thompson's recent neurophenomenological framework for (phenomenal) consciousness, which attempts to do this. As this framework contrasts with that of mainstream access consciousness neuroscience, we first briefly discuss the latter.

\subsubsection{Mainstream perspective: unconscious, conscious, and self-related processes}\label{s:access}

Access consciousness aligns with the mainstream view that consciousness is the outcome of a process of increasing cognitive complexity:  ``raw data'' coming, e.g., from the external world are processed unconsciously in order to ``extract'' high-level features that might afterwards become consciously accessible. From this perspective,  consciousness ``emerges'' at the end of such a process.

Indeed, some mainstream perspectives in access consciousness neuroscience~\cite{dehaene2017consciousness} posit three levels of information processing. First, it is possible for a human subject to react to subliminal information even though she reports being unaware of it 
---this is referred to as unconscious processing and is sometimes associated to ``feature extraction''. Second, conscious access takes place when information presented to a human subject enters her awareness and becomes reportable to others. 
Finally, some processes may refer to self-related tasks---e.g., self-monitoring.

\subsubsection{Neurophenomenological perspective: non-dual awareness, contents of awareness, and self-awareness}\label{s:phenomenal}

\begin{figure*}
\includegraphics[width=0.8\textwidth]{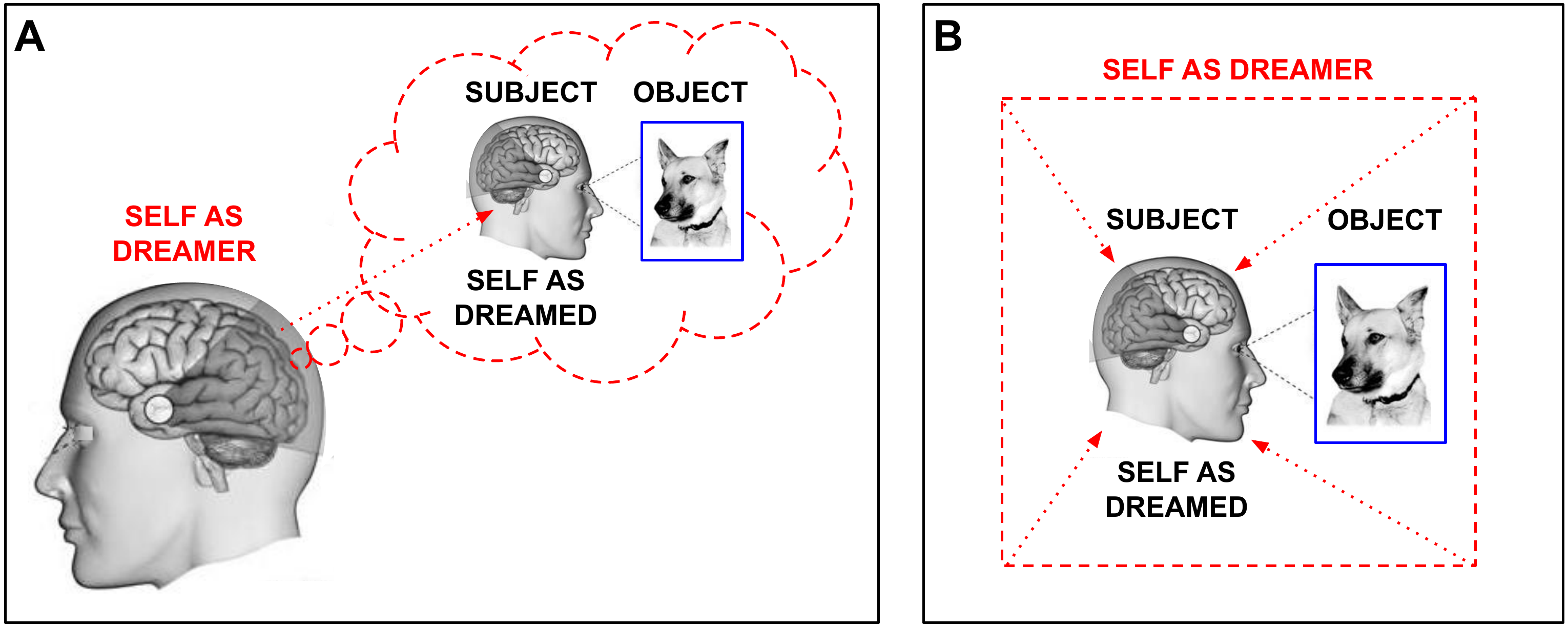}
\caption{{\em Lucid dreaming and non-dual awareness}: In normal (non-lucid) dreams we, the self-as-dreamer, usually identify with the dream body, or self-as-dreamed. Here we illustrate this both from an external (A) and an internal (B) perspective. The dashed box in (B) represents the internal perspective of the self-as-dreamer. The identification of the self-as-dreamer with the self-as-dreamed is represented with dotted arrows. However, both the self-as-dreamed and the world (here a dog) are manifestations of our dreaming mind with associated neural correlates. In a lucid dream we can in principle become aware that we are dreaming. Furthermore, we can become aware that the self-as-dreamed is not who we really are, no matter how convincing, or `real', this seems to be before we reach lucidity. Instead, our sense of self can encompass the awareness of the whole dream state---dashed box in (B). The whole dream state is composed of both the self-as-dreamed and the world (here a dog), which contrasts with the usual dual perspective that frames our dreaming experience as `I am here and the world is there'. In this sense, this is a form of non-duality. Something similar can in principle happen in the waking state. We might say that we can become aware that both the sense of self and the world we perceived in the waking state are both produced by the corresponding neural correlates taking place ``inside us''. Of course, we may not experience the neural correlates themselves, but since these are associated to mind states, we might equally say that we become aware that the sense of self and the world are both manifestations of the mind. Images of head and dog taken from \href{https://commons.wikimedia.org/wiki/File:Neural\_Correlates\_Of\_Consciousness.jpg}{Wikimedia Commons: Christof Koch---an exact copy of Figure 1.1 in Ref.~\cite{koch2004quest}} [\href{https://creativecommons.org/licenses/by-sa/3.0}{CC BY-SA 3.0}].}\label{f:lucid}
\end{figure*}

Instead of going from the unconscious to the conscious via increasing cognitive complexity, as in mainstream access consciousness neuroscience, Thompson's framework in a sense goes in the reverse direction. Based on insights from the rigorous investigation of the phenomenal world enabled by meditative techniques, Thompson suggests that (phenomenal) consciousness, at its most fundamental level, is totally devoid of cognitive complexity. We now discuss Thompson's framework in detail.

In everyday life, at ordinary scales of observation, the physical world appears to be composed of separated objects with well-defined properties which interact among them, space appears to be independent of time, and space and time appear to be independent of matter. In particular, the intuition that the physical world is made up of separated objects with well defined properties is implicit in the search for the ``fundamental particles'' out of which all other objects are composed. 

However, after we learned to develop more refined techniques of observation to transcend our senses, we also learned that the physical world usually defies common sense. Space and time are not independent of each other; according to the special theory of relativity they form a single entity---spacetime. Not even spacetime is a separate entity; according to the general theory of relativity it depends on matter which in turn is influenced by spacetime. The ``fundamental particles'' described by quantum physics do not look like ordinary objects, e.g., they depend on the way we observe them---so, strictly speaking, we cannot say these are (outer) observations of an observer-independent physical world~\cite{proietti2019experimental,frauchiger2018quantum,brukner2018no}. So, contrary to everyday intuition, the physical world at its most fundamental level seems to be relational~\cite{Rovelli-1996,rovelli2007quantum}. This includes, in particular, the interdependence or relationalism between the observer and the observed seemingly implied by quantum physics.

Similarly, the world of phenomenal experience, or the phenomenal world, in everyday life appears to be composed of separated ``entities''. For instance, there seems to be a pre-existing world ``out there'' and a separated pre-existing subject ``in here'' that perceives it. But, given the lessons we have learned from science, should we uncritically accept this as the way the phenomenal world is and reject {\em a priori} any attempt to explore this in a more rigorous way? Or should we first look for more refined exploration techniques to confirm whether this is actually the case?

Indeed, the field of neurophenomenology combines the tools of neuroscience with refined techniques of observation of the phenomenal world developed by some millenarian traditions, i.e., meditation techniques---although, strictly speaking, meditation cannot be considered either as the (inner) observation of an observer-independent phenomenal world~\cite{varela2017embodied}. Interestingly, some advanced meditation techniques has also led to the conclusion that, at its most fundamental level, the phenomenal world too is relational (see Appendix~\ref{s:self-relationalism}). This includes, in particular, the interdependence or relationalism between subject and object, also referred to as subject/object non-duality, implicit in the ``altered state'' of consciousness called ``non-dual awareness''~\cite{thompson2014waking,metzinger2018minimal,ramm2019pure,josipovic2019nondual,dahl2015reconstructing,loy2012nonduality} (see Appendix~\ref{s:self-non}). 

Non-dual awareness has been recently defined in a {\em Nature Reviews Neuroscience} article~\cite{tang2015neuroscience} as ``a state of awareness in which perceived dualities, such as the distinction between subject and object, are absent''. Recently, scientists like Christof Koch~\cite{koch2019feeling} (see ch. 10 therein) and Thomas Metzinger~\cite{metzinger2018minimal} have joined the efforts of Evan Thompson and the late Francisco Varela, among others, to tackle the concept of non-dual awareness. Due to its relatively low cognitive complexity, non-dual awareness has been suggested~\cite{forman1998does,metzinger2018minimal} to be to consciousness studies as the {\em E. coli} is to biology, i.e., it may be among the simplest versions of consciousness and so can be a source of insight that is simpler to investigate. In a {\em Trends in Cognitive Science} article~\cite{dahl2015reconstructing}, Richard J. Davidson and collaborators identify ``non-dual oriented practices'', which can in principle lead to the experience of non-dual awareness, as ``designed to elicit an experiential shift into a mode of experiencing in which the cognitive structures of self/other and subject/object are no longer the dominant mode of experience''. Non-dual awareness is also referred to as ``pure awareness'', ``pure consciousness'', ``pure experience'', ``luminous awareness'', ``subtle consciousness'', ``consciousness-as-such'', ``minimal phenomenal experience'', among others~\cite{thompson2014waking,metzinger2018minimal,ramm2019pure,josipovic2019nondual,dahl2015reconstructing,loy2012nonduality}. Here we will mostly stick to the expressions non-dual awareness, pure awareness or consciousness-as-such. 

Before discussing further the admittedly subtle concept of non-dual awareness, let us briefly mention the experience of lucid dreaming which may serve as a metaphor for it (see Fig.~\ref{f:lucid}); we will closely follow Thompson~\cite{thompson2014waking} (see chapters 4-6 and references therein for further details; see also Ref.~\cite{windt2016does}). In normal---i.e., non-lucid---dreams, dreamers tend to identify with the dream ego, or self as dreamed, which often appears in the form of the dream body, and to take what we experience in the dream as real (see Fig.~\ref{f:lucid}). In a lucid dream the dreamer can in principle become aware that she is dreaming, influence the dream ego and remember waking life. There is evidence that the lucid dreamer can communicate---while lucid dreaming---with scientists in a lab via movements of the eyes or other parts of the body. This in principle allows for communication using the Morse code or to address questions such as ``do dream actions take the same amount of time as waking actions?'' There is also evidence that communication can go the other way around, i.e., from scientists in the lab to lucid dreamers by, e.g., stimulation via high and low tones. 

The key point is that in a lucid dream we can become aware that our dream ego, our self as dreamed, is not who we really are as we usually believe during a normal, non-lucid dream (see Fig.~\ref{f:lucid}). Rather, we can recognize that we encompass the whole dream state. In other worlds, we can realize that both our dream ego and the rest of the dream world are actually manifestations of our own mind. In Thompson's words:

\begin{myquotation}When we dream, we see the dreamscape from the perspective of the dream ego. Although the entire world of the dream exists only as the content of our awareness, we identify ourselves with only a portion of that content---the dream ego that centers our experience of the dream world and presents itself as the locus of our awareness [see Fig.~\ref{f:lucid}]. 

In a lucid dream, however, we experience  another kind of awareness with a different locus. This awareness witnesses the dream state, but without being immersed in the dream world the way the dream ego is. No matter what contents come and go, including the forms taken by the dream ego, we can tell they are not the same as the awareness witnessing the dream state. From the vantage point this provides, we can observe dream images precisely as dream images, i.e., as manifestations of the mind---not the mind of the dream ego but the mind of the dreamer who imagines the dream ego. In this way, we no longer identify only with the ego within the dream; our sense of self now encompasses the witness awareness of the whole dream state. (Ref.~\cite{thompson2014waking}, p. 143)
\end{myquotation}

From a neuroscientific perspective, both the dream ego and the dream world have associated neural correlates. Strictly speaking, according to mainstream neuroscience, the same is true in the waking state: both the sense of self and the world that we perceive have associated neural correlates. Were those neural correlates to disappear, the sense of self and the world we perceive would also disappear {\em for us}. 

In principle, something analogous to the disidentification with the dream ego that can happen in a lucid dream can also happen in the waking state. Although this might sound strange at first, one way we might look at this from the perspective of mainstream neuroscience is that we can become aware that the sense of self and the world {\em we perceive} are both due to associated physical phenomena taking place ``inside us'', i.e., their neural correlates. Even though we may not actually experience the neural correlates themselves, these are associated to mental phenomena. So, we could also say that we can become aware that our sense of self and the world we perceive in the waking state are both manifestations of our mind (see Fig.~\ref{f:lucid}). Importantly, this is {\em not} to deny the conventional reality of the sense of self and the world, but to point out the way such a sense of self and world can exist {\em for us}. From Thompson's neurophenomenological perspective (emphasis ours):

\begin{myquotation}In an ordinary dream, we identify with our dream ego and take what we experience to be real... Whatever we see or feel seems to exist apart from us with its own being or intrinsic nature. This confused state of mind serves {\em as a model} for our waking ignorance of the nature of reality. We think our waking ego exists with its own separate and essential nature, but this belief is delusional, for our waking ego  is no less an {\em imaginative construction} than our dream ego, formed by imaginatively projecting ourselves into the past in memory and into the future in anticipation. The ``I-Me-Mine'' standing over against the world as a separate thing or entity can function as a distorted mental construct in the waking state, not just in dreams. The dream world and the waking world both seem real and solid, yet in neither case do we realize that whatever we take to be real and solid is always a mode of appearance---something that appears real in one way or another---and that modes of appearance by their very nature can't be separated from the mind. 

By contrast, full and complete lucidity---where we wake up in the dream state and directly experience it as luminous appearance, empty of substantiality---offers a traditional metaphor for liberation and enlightenment [see below]... This metaphor {\em isn't mean to deny the conventional reality of the waking world} . It aims rather to effect a fundamental shift in our understanding of what it means for something to be real. {\em ``Real'' is the name we give to certain stable ways that things appear and continue to appear when we test them, not the name for some essence hidden behind or within appearances}. (Ref.~\cite{thompson2014waking}, p. 174)
\end{myquotation}

The insights obtained from the rigorous investigation of the phenomenal world suggests a shift in perspective which does not necessarily contradict the scientific knowledge accumulated to date. Indeed, our derivation of a quantum formalism by explicitly modeling scientists doing science provides evidence that such a shift in perspective can be completely compatible with our current scientific knowledge. 

Thompson discusses a technique called ``dream yoga'' which is said to help have lucid dreams in order to practice meditation in the dream state. This kind of meditation is considered as a powerful method for learning to recognize the state of non-dual awareness. He suggests lucid dreaming and dream yoga offers a way to investigate consciousness disentangled from current sensory input, and envisions a new science of dreaming~\cite{thompson2014waking,windt2016does}. In Thompson's words (emphasis ours):

\begin{myquotation}Meditative lucid dreaming or ``dream yoga'' brings contemplative insight into the dream state. For centuries Tibetan Buddhists have cultivated dream yoga as a practice of mental transformation. Combining their ancient practices with modern methods of lucid dream communication, we can envision a new kind of dream science that integrates dream psychology, neuroscience, and dream yoga.  

But dream yoga doesn't just offer new tools for dream science. It strikes deeper by challenging the assumption that reality is independent of the mind. Dream yoga asks us to view waking experience as a dream while also teaching us how to wake up within the dream state. In this way... {\em [D]ream yoga tries to show us how the waking world  is not outside and separate from our minds; it is brought forth and enacted through our imaginative perception of it.} (Ref.~\cite{thompson2014waking}, pp 164-165)
\end{myquotation}

We now return to the discussion of non-dual awareness. There is evidence that the state of lucid dreaming can be paralleled by a form of lucidity in deep and dreamless sleep~\cite{windt2016does,thompson2014waking}. This state of deep and dreamless sleep is not only devoid of the self-other distinction required for experiencing oneself as separate from the world, but also from any kind of content altogether. So, it is in principle possible to experience a form of non-dual awareness during lucid deep and dreamless sleep~\cite{thompson2014waking,windt2016does,windt2015just}. This is said to be possible also in deep meditative states~\cite{thompson2014waking}. 

It is understandable that some of us may consider the concept of non-dual awareness perhaps senseless, counter-intuitive, or to belong to the domain of philosophy or religion rather than science. However, this does not need to be so. Indeed, similar things could have been said in the past about the concept of the origin of the universe. We cannot travel back to that moment and directly measure what happened. Moreover, not even some physical concepts such as space and time may have existed at the very moment the universe originated. However, guided by a theory, we can deduce certain consequences that we can observe today like, e.g., cosmic microwave background radiation. 

Similarly, although strictly speaking non-dual awareness is said not to be experienced with a subject/object structure, it may still be possible, perhaps guided by a theory, to experience or measure certain consequences before and after the experience of non-dual awareness takes place.  Indeed, there has been some recent attempts to characterize the concept of non-dual awareness in more scientific terms and to identify its potential physical correlates~\cite{thompson2014waking,koch2019feeling, windt2015just, josipovic2014neural, metzinger2018minimal,josipovic2019nondual,loy2012nonduality}.

Thompson argues that the existence of a kind of awareness in deep and dreamless sleep would challenge the conscious/unconscious taxonomy used in mainstream consciousness neuroscience, which tends to classify deep dreamless sleep as unconscious. In particular, he emphasizes the distinction between phenomenal and access consciousness and points out that what mainstream neuroscience usually considers as unconscious might actually qualify as phenomenally conscious. In Thompson words (emphasis his):

\begin{myquotation}Consciousness can mean awareness in the sense of {\em subjective experience} [i.e., phenomenal consciousness] or awareness in the sense of {\em cognitive access} [i.e., access consciousness] [...] 

[Y]ou could be phenomenally aware of something while lacking cognitive access to that awareness. Perhaps you experience the image on the screen, but it went so fast you weren't able to form the kind of memory needed for verbal report of exactly what it was [...] 

On way to think about the Indian yogic idea of subtle consciousness [i.e., non-dual awareness] is to see it as pointing to deeper levels of phenomenal consciousness to which we don't ordinarily have cognitive access, especially if our minds are restless and untrained in meditation. According to this way of thinking, much of what Western science and philosophy would describe as unconscious might qualify as conscious, in the sense of involving subtle levels of phenomenal awareness that could be accessible through meditative mental training. (Ref.~\cite{thompson2014waking}, p. 7-8)
\end{myquotation}

Furthermore, Thompson emphasizes that, strictly speaking, typical approaches in mainstream neuroscience do not actually address consciousness as such:

\begin{myquotation}[B]inocular rivalry doesn't give us a contrast between the presence and absence of consciousness; it give us a contrast between the presence of a particular visual content within consciousness and the absence of that content from consciousness. When you're in a binocular rivalry set up, you're awake with a coherent field of awareness, and you report the coming and going of a particular content within that field. Your consciousness as such, however, never disappears; on the contrary, the experimental setup depends precisely on your being conscious the entire time and being able to report the changing contents of your awareness [...] 

Given that specific patterns of brain activity correlate with states deemed to be conscious because individuals have cognitive access to their contents, and thus can report those contents, we infer that these patterns of brain activity are reliable neural correlates of consciousness. So, strictly speaking, we're not correlating consciousness itself with brain activity; rather, we're correlating something that we already take to be a reliable indication or expression of consciousness---verbal reports or some other cognitive performance---with brain activity. (Ref.~\cite{thompson2014waking}, pp. 64, 98)
\end{myquotation}

After carefully investigating the detailed descriptions of the phenomenal world obtained via refined meditative techniques, Thompson proposes a three-fold taxonomy of consciousness~\cite{thompson2014waking}: (non-dual) awareness, contents of awareness, and self-awareness (or self-experience). This taxonomy differs from the conscious/unconscious one, traditionally assumed in mainstream access consciousness neuroscience. Thompson describes awareness as essentially {\em contentless} and as that which is the crucial precondition for any content or appearance to manifest (i.e., it is {\em ``luminous''}), is able to apprehend those appearances in one way or another (i.e., it is {\em ``knowing''}), and in so doing is self-appearing and prereflectively self-aware` (i.e., it is {\em ``reflexive''}). Furthermore, awareness is {\em non-dual}, i.e., it does not poses a subject-object structure.

\subsection{Non-dual awareness and its potential physical correlates}\label{s:thompson-correlates}

With the expression ``subtle energy'' in Appendix~\ref{s:thompson-science}, Thompson is referring to a conjecture made by the 14th Dalai Lama at the {\em Massachusetts Institue of Technology} as a concession to neuroscience, which has not found convincing evidence that consciousness can be independent of a physical substrate, as Buddhists tend to believe. In this, the Dalai Lama sticks to his statement that ``if scientific analysis were conclusively to demonstrate certain claims in Buddhism to be false, then we must accept the findings of science and abandon those claims''~\cite{lama2005universe} (p. 3). Here we discuss in more detail this potential physical correlate of non-dual awareness. In Appendix~\ref{s:parallels}, we discuss them from the fresh perspective provided by our approach. 

More precisely, the meditative insights reported by the Indian and Tibetan Buddhist traditions suggest there are different levels of consciousness. On one extreme, there is the everyday ``gross'' consciousness, which the Dalai Lama readily concedes can certainly be contingent on the brain. However, at the other extreme, there is ``subtle consciousness'', or non-dual awareness, which he believes cannot be contingent on the brain but on more subtle physical processes (emphasis ours): 

\begin{myquotation}Just as there are many different levels of consciousness [the Dalai Lama says], there are many different levels of energy. Broadly speaking, from the Vajrayana perspective [considered to be the highest standpoint, the most precise and reliable for the phenomenal investigation of the mind/matter relationship], so long as an event or phenomenon is a conscious one, it is necessarily contingent upon a physical event or physical phenomenon. In general, we can make that broad statement. But it will have to be followed immediately with a ``however'' clause. The ``however'' or caveat is that the physical basis---the energy---for subtle consciousness is also of a very subtle kind. {\em It carries all movement or excitation}, even at the level of brain cells. Mental movement is also due to that energy. So the scientific concept of matter needs to be modified in order to appreciate this subtle energy.

One possible avenue [to scientifically investigate this kind of consiousness], he says, is to experiment on meditators who are in what the Tibetan Buddhist tradition calls the ``clear light state''. (Ref.~\cite{thompson2014waking}, pp. 84-86)
\end{myquotation}

The properties ascribed to such a ``subtle energy'' seem to parallel the properties of quantum fluctuations in several respects, as we discuss in Appendix~\ref{s:parallels}. Admittedly, this seems far removed from the traditional perspectives in both mainstream neuroscience and contemplative traditions. Thompson mentions Francisco Varela's position on the matter: 

\begin{myquotation}Shortly before his death, Francisco Varela talked about the Tibetan Buddhist notion of ``subtle consciousness''... [S]ubtle consciousness isn't an individual consciousness; it's not  an ordinary ``me'' or ``I'' consciousness. It's sheer luminuous and knowing awareness beyond any sensory or mental content. It's rarely seen by the ordinary mind, except occasionally in special dreams, intense meditation, and at the very moment of death, when one's ordinary ``I'' or ``me'' consciousness falls apart. It's the foundation for every other type of consciousness, and it's believed to be independent of the brain. Neuroscience can't conceive of this possibility, while for Tibetan Buddhists it's unthinkable to dismiss their accumulated experience testifying to the reality of this primary consciousness...

Varela's position is to suspend judgement. Don't neglect the Buddhist observations and don't dismiss what we know from science. Instead of trying to seek a resolution or an answer, contemplate the question and let it sit there. Have the patience and forbearance to stay with the open question... 

[S]taying with the open question means turning it around and examining it from all sides, without trying to force any particular answer or conclusion. But it also means not being afraid to follow wherever the argument goes. (Ref.~\cite{thompson2014waking}, pp. xxiv-xxv)
\end{myquotation}

Staying with the open question, Thompson allows himself to conjecture that this subtle energy might be associated to the electromagnetic fields produced by living cells, specially the neuroelectrical fields produced by the brain and the bioelectrical fields produced by the heart (emphasis ours):

\begin{myquotation}From [a] bioelectrical perspective, evolution occurs not just in structures, organs, and bodies but also in dynamic electrical fields. Whereas life in general comprises the emergence of self-organizing bioelectrical fields, animal evolution comprises the emergence of self-organizing neuroelectrical fields. 

The Dalai Lama [said] that the physical basis for pure awareness is a subtle energy whose presence can be felt in the body. This energy [...] is said to carry all excitation and movement, including at the level of cells. Although the Dalai Lama suggested that the scientific concept of matter may need to be modified in order to appreciate this energy, I'm inclined to think that this energy is already known to science as  {\em the electromagnetic fields produced by living cells, specially the neuroelectrical fields produced by the brain and the bioelectrical fields produced by the heart}. What has barely begun to be investigated, however, is how meditative practices sensitize one to subtle experiential aspects of these bioelectrical fields, as well as how these practices enable one to alter bioelectromagnetic processes in the brain, the heart, and the rest of the body. (Ref.~\cite{thompson2014waking}, pp. 342-344)
\end{myquotation}

Staying with the open question, here we also allow ourselves to explore how our approach may potentially embodied several of the main points of Thompson neurophenomenological framework (see main text and Appendix~\ref{s:parallels}). In so doing, we try not to be afraid to follow wherever the argument goes. To facilitate the comparison, we now summarize the main points discussed up to here.

\subsection{Summary of main points}\label{s:summary}

Here we summarize some of the main ideas we have discussed at length in Appendices~\ref{s:thompson-science}-\ref{s:thompson-correlates}. Based on this summary, we will discuss in Appendix~\ref{s:parallels} some potential parallels between these ideas and our approach. Here by (phenomenal) consciousness we refer just to subjective experience.

First, as pointed out by Dehaene~\cite{dehaene2014consciousness}, one of the main ingredients that has turned (access) consciousness from a ``philosophical mystery'' into a laboratory phenomenon is taking subjective or first-person reports seriously. Tononi's integrated information theory~\cite{tononi2016integrated}, one of the main theories of consciousness, takes subjectivity so seriously that it ``starts from the essential properties of phenomenal experience, from which it derives the requirements for the physical substrate of consciousness''. How to determine the ``essential properties of phenomenal experience''? One way is to analyze our own everyday phenomenal experience and compare to what others report from their own analysis---this is essentially what Tononi has done (see below). 

This focus on everyday experience is analogous to the focus of early Newtonian physics on everyday phenomena---e.g., an apple falling from a tree---which led to the discovery of some general principles---e.g., the law of gravity. However, more refined (third-person) exploration techniques taught us that a more fundamental understanding of the physical world defies common sense---think of quantum physics and Einstein's theory of relativity. Similarly, it seems natural to expect that a more fundamental understanding of consciousness can be reached if we look for more refined (first-person) techniques for exploring the phenomenal world, rather than focusing only on everyday experience. We should be prepared to find results that defy common sense as well.

The field of {\em neurophenomenology}~\cite{varela2017embodied,thompson2014waking,blackmore2018consciousness} aims at integrating refined first- and third-person methods to have a more comprehensive understanding of consciousness. Although some of the most refined first-person methods known to date, e.g., meditation, has been typically used by contemplative traditions like Buddhism, this is not to be read as an attempt to bring ``mysticism'' into science. Quite the contrary, it can be read as an attempt to bring the rigor of science to the utilization of these methods which may be of great benefit to society. Interestingly, these more refined first-person methods suggest that, like the physical world, the phenomenal world appears to be relational.

Second, there is both a {\em primacy of the physical}, i.e., consciousness never shows up apart from a physical substrate, and a {\em primacy of the phenomenal}, i.e., without consciousness there is no observation and without observation there is no data---furthermore, the scientific method presupposes consciousness in that everything perceived, believed, theorized, researched, intersubjectively agreed upon, and known is done so by conscious observers. In line with this two-fold primacy, and with the relationalism suggested by the critical investigation of both the physical and the phenomenal worlds, Thompson~\cite{thompson2014waking}---as well as Velmans~\cite{velmans2009understanding} and Bitbol~\cite{bitbol2008consciousness}---invites us to search for a non-dual framework that gives equal weight to the physical and the phenomenal worlds. 

In particular, we should avoid falling into the extreme of solipsism, i.e. assuming that the world is nothing but a projection of the mind, and the extreme of bare materialism, i.e., assuming that there is an observer-independent material world---an extreme that the ultimate quantum nature of the physical world also suggests should be avoided~\cite{proietti2019experimental,frauchiger2018quantum,brukner2018no}. Moreover, this search should fall neither into the extreme of dualism, i.e., assuming matter and consciousness are two intrinsically existing substances with totally different natures, nor the extreme of panpsychism, assuming {\em ad hoc} that physical particles have ``micro-experiences''. Rather, the sought for {\em non-dual framework} should be based on an interdependence of the physical and the phenomenal that allows these two to imply each other, like two sheaves of reeds propping each other up. 

Such a framework should not only rely on refined third-person methods, traditionally used in science, but also in refined first-person methods, like meditation, that can potentially enable a more rigorous and systematic exploration of the phenomenal world. Velmans~\cite{velmans2009understanding} argues this could be achieved if by ``objective'' we pragmatically mean that procedures are (O1) standardized and explicit, (O2) that observations are intersubjective and repeatable, and (O3) that observers are dispassionate, accurate and truthful---instead of {\em a priori} equating ``objective'' to ``observer-free''. He suggests to formulate the scientific method essentially as ``if you carry out these procedures you should observe or experience these results''---the act of carrying out the procedures can itself be seen as an experience. This might ground the whole of science on experience, with ``objective'' observation as a particular case---much like $\lambda$-Calculus grounds the whole of computation on strings only, effectively eliminating the distinction between data and code. In such a manifestly self-referential formulation of science, explorations of the phenomenal world via refined first-person methods could be considered as self-referential experiments where scientists investigate themselves---much like strings in $\lambda$-Calculus operate on themselves.

Third, relationalism is sometimes referred to as {\em dependent arising} in the Madhyamaka school of Buddhism. According to Thompson~\cite{thompson2014waking}, there are three levels of dependent arising: (i) {\em causal dependence}, (ii) {\em part/whole dependence}, and (iii) {\em conceptual dependence}. In short, whatever is dependently arising (i) depends on causes and conditions to exist, (ii) has its parts and whole co-arising and mutually specifying each other, and (iii) depends for its existence on a {\em basis of designation}, a {\em designating cognition}, and a {\em term to designate it}. Causal dependence is in line with science in general due to its focus on understanding causality. Thompson suggests that parts/whole dependence is related to complex systems theory and emergence. Our approach suggest that conceptual dependence might be related to quantum theory, as discussed in Appendix~\ref{s:parallels}. 

Fourth, based on a careful review of some insights obtained from such refined first-person exploration techniques, Thompson~\cite{thompson2014waking} proposes a {\em three-fold taxonomy of (phenomenal) consciousness}: consciousness-as-such or non-dual awareness, contents of awareness, and self-awareness. This differs from the unconscious/access-conscious/self-conscious taxonomy traditionally assumed in mainstream access consciousness neuroscience. However, in contrast to Thompson's, neither Dehaene's nor Tononi's frameworks, which are two of the most prominent theories of access consciousness, are based on these more refined first-person techniques. In this sense, Dehaene's and Tononi's frameworks, while extremely relevant, might turn out to be to Thompson's framework as Newtonian physics is to quantum physics. However, Thompson's framework is much more recent and, unlike Dehaene's and Tononi's, it is not yet formalized into a mathematical theory---although we hope our approach may be a step in this direction (see Appendix~\ref{s:parallels}).

Non-dual awareness has been defined in a {\em Nature Reviews Neuroscience} article~\cite{tang2015neuroscience} simply as ``a state of awareness in which perceived dualities, such as the distinction between subject and object, are absent'' (see Fig.~\ref{f:lucid}). However, according to Thompson's more thorough characterization~\cite{thompson2014waking}, {\em non-dual awareness} is: {\em ``luminous''}, i.e., it is the crucial precondition for any content or appearance to manifest, the mere potential to become aware of something; {\em ``knowing''}, i.e., it is able to apprehend whatever appears; {\em ``reflexive''}, i.e., it is self-appearing and prereflectively self-aware; {\em contentless}, i.e., it is prior to any content or appearance; {\em non-dual}, i.e., it is not experienced in a subject/object structure; {\em irreducible}, i.e., it is the most fundamental level of consciousness on which all other forms of consciousness depend, it cannot be explained in terms of lower-level concepts, it cannot be objectified or known as an object of observation because it is the very precondition for any object to be observed.  Finally, the {\em contents of awareness} are whatever we happen to be aware of from moment to moment, and {\em self-awareness} is the experience of some of the contents of awareness as referring to a self. 

Finally, in line with scientific evidence and contrary to widespread Buddhist belief, the 14th Dalai Lama points out that the system they consider as the most precise and reliable for investigating the relationship between mind and matter suggests there is indeed a physical basis for non-dual awareness, though it should be ``of a very subtle kind''---a {\em ``subtle energy''}. How to interpret this {\em ``subtleness''} of the physical correlates of non-dual awareness?  The subtle energy referred to by the Dalai Lama could be said to be {\em all-pervading} in that it is said to carry all excitation and movement, including at the level of cells---in particular, it should not be constraint to specific neural systems or the brain only. In a sense, this parallels the ``luminous'' quality of non-dual awareness, i.e., its being always present as a precondition for any phenomenal content to manifest. 
 
{\em Thompson's conjecture} is that such a subtle energy might be associated to the electromagnetic fields produced by living cells, which are physical yet, strictly speaking, not material. Thompson suggests this hypothesis could be tested by measuring how meditation can sensitize one to subtle aspects of these fields and allow one to alter them. Besides Thompson's, there are other recent proposals for the physical correlates of non-dual awareness. For instance, Metzinger~\cite{metzinger2018minimal} suggests that it is about the ascending reticular arousal system, which essentially turns on the brain. Koch~\cite{koch2019feeling} suggests it is about the causal power of interacting neurons, captured in terms of integrated information theory~\cite{tononi2016integrated}, that could fire to generate contents of consciousness but do not. Josipovic~\cite{josipovic2014neural,josipovic2019nondual} suggests it is about the central precuneous network, which appears to play a key role in the global organization of the brain. However, here we focus on Thompson's proposal as it is the only one that appears to be consistent with our approach. We leave a more thorough comparison for future work.

There are some additional potential aspects of the physical correlates of non-dual awareness, paralleling some of their phenomenal counterparts, which we find relevant.  Indeed, in the same way that non-dual awareness cannot be objectified, i.e., made a content of awareness, nor reduce to lower-level experiences, its corresponding physical correlates, though of course physical, should not be reducible to objective, lower-level physical phenomena. Paralleling the reflexivity and non-duality of awareness, their physical correlates should be associated to a self-referential process and integrate subject and object in a relational, interdependent way.

\subsection{Potential parallels with the quantum formalism}\label{s:parallels}

\begin{table*}\centering
        \begin{tabular}{l | p{0.42\textwidth} | p{0.43\textwidth} }
		\cline{2-3}
                 & Critical investigation of the physical world &  Critical investigation of the phenomenal world \\ 
                \hline
Methodology & Enhancing experimental techniques & Enhancing meditative techniques \\ 
& Logical reasoning &  Logical reasoning \\
& Intersubjective agreement & Math-like intersubjective agreement \\
                \hline
		Similarities & Relationalism, interdependence 		    	    & Relationalism, interdependence \\ 
			 & Physical causality & Causal dependence \\
		 	 & Emergence, complex systems & Part/whole dependence \\
			 & Observations relative to E-observer:         & Conceptual dependence: \\
	 		 & $\bullet$ U-observer/observed interaction                  & $\bullet$ Basis of designation\\
			 & $\bullet$ ``Cognition'' by E-observer         & $\bullet$ Designating cognition\\
			 & $\bullet$ Model of U-observer-observed interaction         & $\bullet$ Designating term \\
			 & ``Quantumness'': 					    & Non-dual awareness: \\
			 & $\bullet$ Involves mere capacity to observe ``from within'' 		    & $\bullet$ ``Luminous'' and ``knowing''\\
			 & $\bullet$ Observer/observed interdependence 		    & $\bullet$ Subject/object non-duality\\
			 & $\bullet$ All-pervasive fluctuations 		    & $\bullet$ All-pervasive awareness\\
			 & $\bullet$ Irreducible to lower-level physical phenomena  & $\bullet$ Irreducible to lower-level phenomenal experiences\\
			 & $\bullet$ Involves self-reference 			    	    & $\bullet$ Reflexive\\
			 & $\bullet$ Can refer to ``the vacuum'' 		            & $\bullet$ Can be contentless\\
		\hline
		Differences & Primacy of the physical---but observer-dependent & Primacy of the phenomenal---but ``subtle energy'' associated \\
			    & Public---though in principle inferred from invariants of private experiences. & Private---though in principle inferable from publicly available physical correlates \\
		\hline
        \end{tabular}
\caption{\small {\em Critical investigation of the physical and phenomenal worlds}: 
Here we summarize some potential parallels and discrepancies suggested by the critical investigation of the physical and phenomenal worlds, as well as our approach (see Appendices~\ref{s:summary} and~\ref{s:parallels}). Here ``luminous'' refers to being the mere potential to become aware of something, and ``knowing'' refers to being able to apprehend whatever appears.
} \label{t:comparison}
\end{table*}


Here we discuss in more detail the potential relationship between the ideas summarized in Appendix~\ref{s:summary} and our approach (see Table~\ref{t:comparison}). In doing so, we try to follow Varela's advice to stay with the open question and avoid being afraid to follow wherever the argument goes. We are aware that the type of comparison attempted here may understandably rise a substantial amount of healthy skepticism, an attitude that may be reinforced by the many unfounded claims made in the past on this regard. However, in contrast to previous claims, our discussion is based on a concrete and precise conceptual and mathematical framework and, in this sense, we think, is not just a shot in the dark. So, our invitation is to have an attitude like the one that allowed the origin of the universe to become a subject of scientific discourse. 

Moreover, we see a few reasons why we should not {\em a priori} shy away from this discussion. First, there is recent evidence that humans can be sensitive to single quanta of light, or photons~\cite{tinsley2016direct}. There is also some evidence that certain aspects of human behavior appear to display some quantum-like features~\cite{bruza2015quantum,wang2014context}---which is usually referred to as ``quantum cognition''. This suggests that quantum-like phenomena may not be that far removed from human experience as previously thought. 

Second, the critical investigation of both the physical and phenomenal worlds lead to strikingly similar conclusions, like relationalism, as we have discussed above. Should we uncritically take this just as a mere coincidence or should we investigate whether there is an underlying reason for this to be so---especially, taking into account that the process of observation appears to play a key role in both cases? 

Finally, the widespread assumption that consciousness emerges from ``matter'' rests on whatever definition of ``matter'' physics may provide. However, the fundamentally quantum nature of the physical world, strictly speaking, suggests a concept of ``matter'' far removed from the ordinary concept we may find ``normal'' in everyday life. Indeed, there is a kind of circularity in the statement that consciousness emerges from ``matter'' because, according to quantum theory, at the most fundamental level, whatever we call ``matter'' depends on how we observe it~\cite{bitbol2008consciousness}. 

Our approach indicates that we can take experience as the starting point of science---as suggested by Velmans~\cite{velmans2009understanding}, Bitbol~\cite{bitbol2008consciousness} and Thompson~\cite{thompson2014waking}---and still be consistent with current scientific knowledge. In this view, what is inconsistent with science is not experience itself, but experience that is not ``objective'' in Velmans' sense. Scientific observations effectively amount at ``objective'' experiences. Indeed, we use Velmans' notion of ``objectivity'' to obtain a quantum formalism, as we have restricted our approach to those experiences that are ``objective''. In other words, we have focused on scientists investigating scientists. 

Additionally, our approach seems to be consistent with the notion of {\em conceptual dependence}, the third level of {\em dependent arising}. More precisely, it suggests that scientifically observed phenomena may depend for its existence on a basis of designation, a designating cognition, and a term use to designate it. Figure~\ref{f:circular}A in the main text shows the interaction between an embodied scientist (Wigner's friend, an U-observer) and an experimental system---this could be considered as the {\em basis of designation} for the phenomenon observed. However, this interaction is ``modelled'' or ``cognized'' by another observer (Wigner, an E-observer)---this could be considered as the {\em designating cognition}. Wigner's ``model'' itself might be consider the symbol or ``term'' use to designate the phenomenon. However, both Wigner and his friend turn into ``sub-observers'' and become integrated after shifting to the intrinsic perspective, so the three levels of conceptual dependence become integrated too.

We have found that two ``non-spurious'' aspects of experience that would be absent in an ``observer-free'' scientific theory remain: embodiment and the mere capacity to experience or observe from an intrinsic perspective---which we are referring to simply as the ``intrinsic perspective''. That observers are embodied is consistent with embodied cognition, enactivism, and the idea that every experience has a physical correlate. That observations are made from an intrinsic perspective suggests that every physical phenomenon is an experience for someone. This is in line with Rovelli's RQM, which posits that every physical phenomenon is relative to an observer. Except that observers in RQM are considered just a physical system more---e.g., a rock could count as an observer.

This mere capacity to experience could be defined more precisely, following Thompson's neurophenomenological analysis, as the mere potential to become aware of something and the ability to apprehend whatever appears, which are the ``luminous'' and ``knowing'' qualities of non-dual awareness. This potential connection between quantum theory and non-dual awareness does not have to be as strange as it may sound. After all, in the relational scientific framework discussed above, the investigation of both the physical and the phenomenal worlds rely on experience, and the intrinsic perspective is by definition the very precondition that makes experience possible. Importantly, neither the intrinsic perspective, nor embodiment {\em per se}, involve any additional subjective complexities, like cognitive, perceptual or motivational biases. So, by themselves they do not hinder ``objectivity''---in Velmans' sense. 

Our approach also suggests a potential physical correlate for non-dual awareness---what the Dalai Lama referred to as a ``subtle energy''. The features of such a physical correlate should parallel those that characterize the concept of non-dual awareness. In particular, being the mere capacity to observe something, non-dual awareness is said to be independent of its contents. Indeed, at its most fundamental level, it can be contentless. The self-referential coupling (SRC) implements the intrinsic perspective, so it should have similar features. Indeed, no matter how many ``particles'' there are in the system investigated, the implementation of the SRC is always the same; in this sense it is independent of its contents. If we removed all content from Fig.~\ref{f:first}F only the SRC between Alice and Bob would remain. That is, only the half-heads completely empty of any content---or two empty cameras facing each other in the toy example presented in Fig.~\ref{f:first}E. Being self-referential, the SRC has a reflexive flavor. Furthermore, in our approach the SRC is the crucial precondition for a scientist to observe any physical phenomenon from her intrinsic perspective; as such it cannot be reduced to any of those contents---this has the flavor of incompleteness and undecidability results (see Appendix~\ref{s:planck-self}). Finally, the SRC is composed of two subsystems, Alice and Bob in Fig~\ref{f:first}C, that play interdependent roles as subject and object, in a seemingly non-dual way. 

Now, consider a generic system of $N$ particles described by a transition kernel proportional to the factor $e^{\epsilon\mcH/\hobs}$ (cf. Eq.~\eqref{em:F} in the main text), where $\mcH $ is the corresponding Hamiltonian. Imagine that we remove the $N$ particles one by one until they are all gone---this would be a sort of ``vacuum'', a relevant concept in quantum theory. Notice that $\hobs$, being constant, remains the same during the whole process. It makes sense to imagine that it remains the same once we have removed all particles---after all, the ``vacuum'' in quantum theory is not ``classical'' but ``quantum''. So, from this perspective, $\hobs$ seems to characterize the physical processes that support the SRC between Alice and Bob. 

This suggests the physical correlate of non-dual awareness would be a form of quantum fluctuations characterized by $\hobs$. Interestingly, quantum fluctuations are irreducible to lower level physical phenomena as well as all-pervading since they affect every physical phenomenon. This situation would be in line with science, as it does not negate the physical basis of awareness, and with the meditative insights from the Tibetan Buddhist tradition, because, even though there would be a physical basis for the subtlest level of consciousness, this basis cannot be reduce to lower-level physical phenomena. It is physical but indeterminate.  

As we have discussed in previous subsections, in the same way that strings can operate on themselves in $\lambda$-Calculus, an approach to scientists investigating scientists should allow for scientists to investigate themselves. Humans can {\em learn} to explore ``objectively'' the physical world, e.g., by doing a career in science, even if they are not born being ``objective''. In principle, humans can also {\em learn} to explore the phenomenal world in an ``objective'' way---again, ``objective'' in Velmans' sense, not in the sense of ``observer-free''. Furthermore, scientists can carefully design an experimental system to effectively neutralize unwanted influences, simplify its investigation, and isolate a phenomenon of interest. Similarly, it is in principle possible---e.g., through meditation techniques---to calm down the mind and concentrate single-pointedly on a specific phenomenal experience of interest---e.g., the breath or an image.

This suggests that new kinds of experiments could be envisioned at the interface of neurophenomenology, consciousness neuroscience, quantum physics, and quantum cognition, where highly trained meditators collaborate with scientists to explore potential quantum-like features of the phenomenal world. For instance, we could explore whether meditation might enhance the quantum-like effects reported in quantum cognition experiments, or facilitate the observation of quantum-like features in experiments of single photon detection by humans. We could also search for novel, psychophysics experiments to measure $\hobs$ and test whether indeed $\hobs = \hbar$. Consistent with Thompson's conjecture, we could also investigate whether, through meditation, humans could sensitize to and alter subtler and subtler aspects of the electromagnetic fields that underlie our cognitive processes, down to the quantum level. Further theoretical and experimental developments are required to understand which possibilities are available, if any. We hope our mathematical framework can help guide the kind of questions and predictions that could be addressed.  

\section{Quantum mechanics recast}\label{s:quantum_aspects}
Here we first describe in Appendix~\ref{s:quantum_nutshell} some aspects of quantum theory relevant to this work for the reader that may not be familiar with them. Afterwards, in Appendix~\ref{s:wigner} we illustrate the type of conceptual problems associated to the theory. We do this via the so called Wigner's friend thought experiment wherein an external scientist (Wigner) observers his friend observing a physical system. We also mention Rovelli's relational interpretation of quantum mechanics (RQM), which aligns with the relational approach taken in the main text (see Fig.~\ref{f:circular}A). Rovelli suggests his relational interpretation can solve the conceptual puzzle associated to Wigner's friend thought experiment. 

In Appendix~\ref{s:pair} we then discuss how the von Neumann equation, which characterizes the dynamics of closed quantum systems, can be understood as a pair of equations in terms of real matrices. This pair of equations are interpreted in the main text in the spirit of Kleene's recursion theorem (see Appendix~\ref{s:recursion}) as describing two ``sub-observers'' mutually observing each other to implement the intrinsic perspective. In Appendix~\ref{s:examples} we show how, following this decomposition of the von Neumann equation in terms of a pair of real equations, some well-known and non-trivial examples of quantum dynamics can be written in terms of real kernels with non-negative entries. To keep the discussion as simple as possible, the technical details associated to such examples are presented separately in Appendix~\ref{s:details}.

\subsection{Quantum theory in a nutshell}\label{s:quantum_nutshell}
In Appendix~\ref{s:diff-quantum} we describe in detail a non-relativistic quantum free particle, one of the simplest quantum systems, in close parallel with its classical analogue, a diffusive particle. Here we complement Appendix~\ref{s:diff-quantum} by discussing some more general aspects of quantum theory for the reader that may not be familiar with these. 

A quantum system differs from a classical stochastic system in that~\cite{realpe2019can}: (i) its state is necessarily described in general by a {\em density matrix}, $\rho$, whose diagonal contains all probabilistic information; (ii) the off-diagonal entries of $\rho$ can in general be complex and $\rho$ is Hermitian, i.e. $\rho^\dag = \rho$ where $\rho^\dag$ represents the adjoint of $\rho$---often $\rho^\dag = [\rho^\ast]^T$ coincides with the conjugate transpose of $\rho$---this is the case, e.g., of finite dimensional systems; (iii) the dynamics of the system is given by 
\be\label{e:UrhoU}
\rho^\prime = U\rho U^\dag,
\ee
where $U$ has in general complex entries and is unitary, i.e. $U U^\dag = U^\dag U =\id$. When the time-step size $\epsilon$ tends to zero we can write $U=\id - i\epsilon H/\hbar$, where $H = H^\dag$ is a suitable Hermitian matrix called the Hamiltonian, $\hbar$ is Planck's constant, and $i$ is the imaginary unit. Equation~\eqref{e:UrhoU} then becomes
\be\label{e:vNApp}
i\hbar\frac{\partial\rho}{\partial t} = [H, \rho],
\ee
Equation~\eqref{e:vNApp} is called von Neumann equation. 

Some relevant questions to understand quantum theory are~\cite{realpe2019can}: Why does $\rho$ need to be a matrix? Why does only the diagonal of $\rho$ contain the probabilistic information---which is known as the Born rule? Why is $\rho$ complex and Hermitian? Why does $\rho$ satisfy von Neumann equation, Eq.~\eqref{e:vNApp}? 

When $\rho$ has only one eigenvector $\left|\psi\ket$ with non-zero eigenvalue, it is said to describe a {\em pure state}. Since the diagonal of $\rho$ represents probabilities, such a non-zero eigenvalue is actually equal to one; so we can write $\rho = \left|\psi\ket\bra\psi\right|$, where $\bra\psi\right| = [\left|\psi\ket]^\dag$ is the Hermitian conjugate of $\left|\psi\ket$. In this case, if we write $\rho^\prime = \left|\psi^\prime\ket\bra\psi^\prime\right|$, Eq.~\eqref{e:UrhoU}  becomes 
\be\label{e:psi'}
\left|\psi^\prime\ket = U\left|\psi\ket .
\ee

Similarly, for pure states Eq.~\eqref{e:vNApp} becomes equivalent to Schr\"odinger equation
\be\label{e:SchApp}
i\hbar\frac{\partial\left|\psi\ket}{\partial t} = H\left|\psi\ket .
\ee

An important point for the implementation of the SRC (i.e., the self-referential coupling) in the main text is that, without loss of generality, the initial state of a quantum experiment can be considered as a symmetric density matrix, $\rho_{\rm sym}$, whose diagonal elements yield the classical probabilities for the system to be in the corresponding state. Indeed, a general density matrix, $\rho$, can be written as $\rho = U_{\rm prep}\rho_{\rm sym}U_{\rm prep}^\dag$ via a suitable unitary preparation operator $U_{\rm prep}$. Consider a quantum experiment with a general initial density matrix $\rho$ which evolves to a final state $\rho^\prime = U\rho U^\dag$. This is equivalent to an experiment with initial density matrix $\rho_{\rm sym}$, which evolves to the same final state $\rho^\prime = U_{\rm eff} \rho_{\rm sym} U_{\rm eff}^\dag$, via the effective unitary operator $U_{\rm eff} = U U_{\rm prep}$.

Consider the measurement of a general observable characterized by the Hermitian operator
\be
O = \sum_k \lambda_k \left|k\ket\bra k\right|,
\ee
where $\lambda_k\in\mathbb{R}$ and $\left|k\ket$ are the corresponding eigenvalues and eigenvectors, respectively---the index $k$ can also be continuous in which case the sum is changed by an integral. According to the (projective) measurement postulate of quantum theory, if the normalized quantum state is given by
\be
\left|\psi\ket = \sum_k c_k \left| k\ket ,
\ee
then a measurement of $O$ would yield outcome $\lambda_k$ with probability $|c_k|^2$ and the quantum state would update or ``collapse'' to $\left|\psi\ket\to\left| k\ket$. So, the expected value of $O$ is
\be\label{e:<O>}
\bra O\ket_\psi \equiv \bra\psi\right| O \left|\psi\ket = \tr\left[\rho_\psi O \right] = \sum_k |c_k|^2 \lambda_k ,
\ee
where $\rho_\psi = \left|\psi\ket\bra\psi\right|$ is the density matrix associated to pure state $\left|\psi\ket$. 

More generally, consider a mixed state $\rho = \sum_\alpha \lambda_\alpha\left|\psi_\alpha\ket\bra\psi_\alpha\right|$, which could be read as ``with probability $\lambda_\alpha\geq 0 $ the system is in pure state $\left|\psi_\alpha\ket$''. Equation~\eqref{e:<O>} can be straightforwardly generalized to $\bra O\ket_\rho = \tr\left[\rho O\right]$.

Although this may give the impression that quantum states such as $\left| k\ket$ exists in the abstract, this is not necessarily so. As discussed in Appendix~\ref{s:measurements}, states such as these can be more pragmatically considered as shorthand notations for the kind of manipulations we have to do in the laboratory to actually implement such measurements. In particular, measurements are generally implemented via suitable interactions between the system of interest and a measurement device. Such interactions can only be characterized in terms of things we are familiar with, such as the position of a pointer or the angle between two mirrors.

Under a {\em Wick rotation}, i.e. under the change $t = i \tau$ Eqs.~\eqref{e:vNApp} and \eqref{e:SchApp} become 
\BE
\hbar\frac{\partial\widetilde{\rho}}{\partial \tau} &=& [H, \widetilde{\rho}],\label{e:vN_Wick}\\
\hbar\frac{\partial\left|\widetilde{\psi}\ket}{\partial \tau} &=& H\left|\widetilde{\psi}\ket\label{e:Sch_Wick} ,
\EE
which are now real equations, so all quantities involved become real too. This is termed {\em imaginary-time} or {\em Euclidean} quantum dynamics~\cite{Zambrini-1987,zambrini1986stochastic}. 

The entries of the vector $\left|\psi\ket$ can be written as $\psi(x) = \bra x|\psi\ket = \sqrt{p(x)} e^{i \varphi}$, which is called the wave function. Under a Wick rotation the $i$ disappears, so Wick-rotated, imaginary-time, or Euclidean wave functions can be written as $\widetilde{\psi}(x) = \bra x|\widetilde{\psi}\ket = \sqrt{p(x)} e^\phi(x)$ (see Appendices~\ref{s:quantum_diff}, \ref{s:slits}, and \ref{s:MP_recasted}). When $\phi(x) = 0$ we have $\widetilde{\psi}(x) = \theta (x) \equiv \sqrt{p(x)}$ (cf. Appendix~\ref{s:squareroot}).

\subsection{Wigner's friend thought experiment and Rovelli's relational interpretation}\label{s:wigner}
\begin{figure}
\includegraphics[width=\columnwidth]{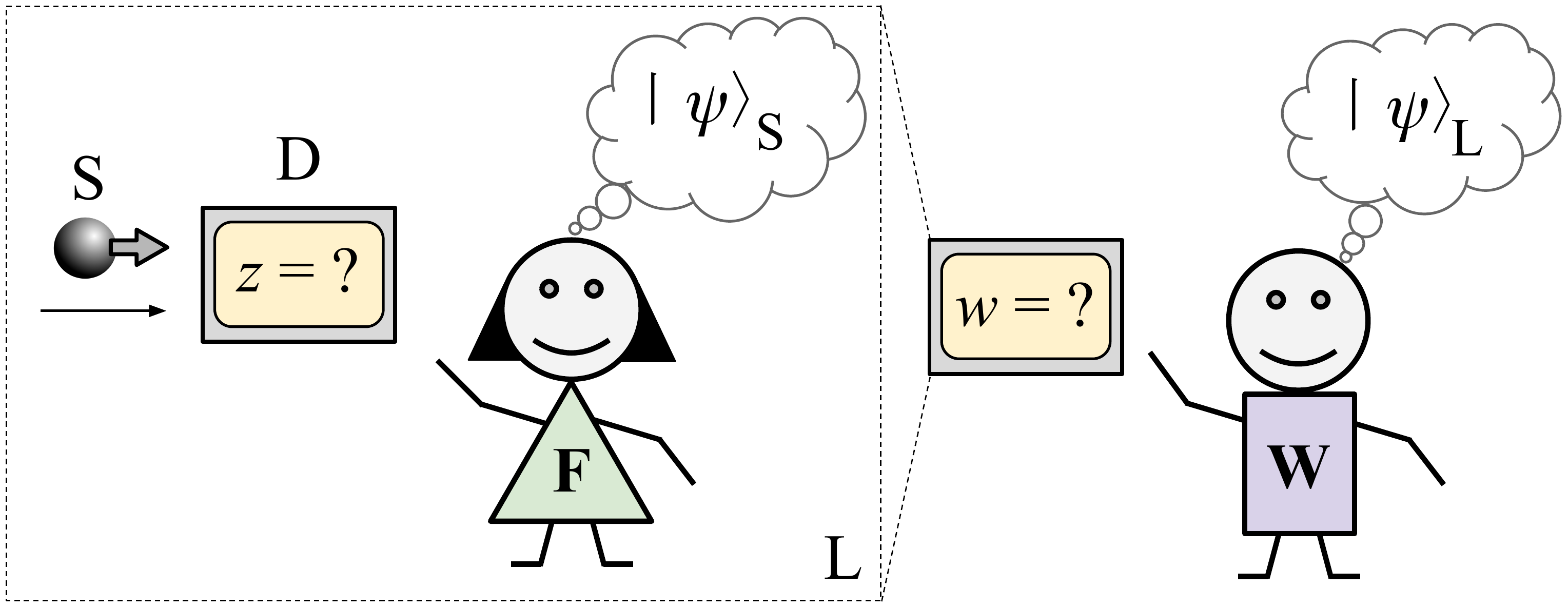}
\caption{{\em Wigner's friend thought experiment:} Wigner's friend ($F$) measures the spin ($S$) of a particle in the vertical direction and obtains outcome $z$. From Wigner's friend perspective, the spin is in one of the two pure states $\left|\psi\ket_S$ defined in Eq.~\eqref{e:psi_S}. Wigner observes his friend's lab ($L = S + F$) from the outside, considering it as a quantum system (dashed box). Since Wigner have no access to $z$, he would assign to $L$ a superposition state $\left|\Psi\ket_L$ (see Eq.~\eqref{e:psi_L}). Has event $z$ happen or not? (cf. Fig.~1 in Ref.~\cite{frauchiger2018quantum}). The way Wigner's friend is treated here is an example of how observers are usually treated in physics as any other physical system, in sharp contrast with cognitive science; in this case Wigner's friend is treated as a particle with spin. Figure~\ref{f:circular}A in the main text depicts a similar but {\em classical} situation based on the enactive approach to cognitive science. }
\label{f:wigner}
\end{figure}

Here we illustrate the type of problems associated to the conceptual interpretation of quantum theory. To do so, consider the so-called Wigner's friend thought experiment (see Fig.~\ref{f:wigner}). This consists of two observers: Wigner's friend ($F$) who measures a system ($S$) in her lab ($L$), and Wigner ($W$) who simultaneously measures the composed system $S+F$, i.e. his friend's lab $L$. Suppose Wigner's friend measures a two-state property of the system $S$ denoted by a variable $z\in\{-1, +1\}$. For instance, the system $S$ could be an electron and $z$ could denote whether its spin points up, $z=+1$, or down $z=-1$. If the outcome of the measurement is $z=+1$ or $z=-1$, then Wigner's friend would say that the system $S$ is in state (cf. Eq. (1) in Ref.~\cite{frauchiger2018quantum})
\be\label{e:psi_S}
\left|\psi\ket_S = \left|\uparrow\ket_S\equiv\begin{pmatrix} 1\\ 0\end{pmatrix}_S \hspace{0.4cm}\textrm{	or	}\hspace{0.4cm}\left|\psi\ket_S = \left|\downarrow\ket_S \equiv\begin{pmatrix} 0\\ 1\end{pmatrix}_S,
\ee
respectively. 

Wigner cannot measure the system $S$ alone but only the lab as a whole, $L = S+F$---for simplicity, we are not explicitly modeling the detector ($D$) here as this would be redundant for our purpose: the detector state determines with absolute certainty the state of Wigner's friend, i.e. what she knows about the system $S$ (see Eq.~(2) in Ref.~\cite{frauchiger2018quantum}). Mathematically, the state of the detector and of Wigner's friend are essentially replicated variables and it is enough for us to keep only one of them (c.f. Ref.~\cite{Rovelli-1996} and Sec.~5.6 of Ref.~\cite{rovelli2007quantum}). 

Assume that, from Wigner's perspective, his friend is initially in a generic pure state $\left|\textsc{init}\ket_F$ and that the lab remains isolated during his friend's measurement. Thus, according to quantum theory, Wigner can describe his friend's measurement as a quantum interaction via a unitary (see Eq.~\eqref{e:psi'})
\BE
U_{S\to L}\left|\downarrow\ket_S\otimes\left|\textsc{init}\ket_F &\equiv &\left|\downarrow\ket_S\otimes\left|z = -1\ket_F, \label{e:Udown}\\
U_{S\to L}\left|\uparrow\ket_S\otimes\left|\textsc{init}\ket_F &\equiv &\left|\uparrow\ket_S\otimes\left|z = +1\ket_F .\label{e:Uup}
\EE
These equations say that if after Wigner's friend observes outcome $z=-1$, i.e. she is in state $\left|z=-1\ket_F$, after her measurement then the state of system $S$ is $\left|\downarrow\ket_S$, and similarly for $z=+1$ (see Eq.~\eqref{e:psi_S}). This is essentially what is meant in practice by state and observation.

Now, assume that Wigner knew that the state of the system $S$, right before his friend measures it, was $\left|\to\ket_S\equiv\sqrt{1/2}(\left|\downarrow\ket_S + \left|\uparrow\ket_S)$. So, by the linearity of the unitary dynamics (see Eq.~\eqref{e:psi'}), the state that Wigner would assign to the lab after his friend's measurement has been completed is
\be\label{e:psi_L}
\left|\Psi\ket_L = \frac{\left|\downarrow\ket_S\otimes\left|z = -1\ket_F + \left|\uparrow\ket_S\otimes\left|z = +1\ket_F}{\sqrt{2}},
\ee
i.e. a linear superposition of the states defined in Eqs.~\eqref{e:Udown} and \eqref{e:Uup}. This equation expresses the fact that, according to Wigner, his friend has measured system $S$. So, Wigner knows that his friend knows the state of system $S$. However, Wigner himself does not know the state of $S$ and that is why he describes the state of the lab by a superposition, $\left|\Psi\ket_L$. In contrast, since Wigner's friend has indeed measured system $S$, for her $S$ is in a definite state: either $\left|\downarrow\ket_S$ or $\left|\uparrow\ket_S$, not a superposition of these. Assume she has actually observed $z=-1$, i.e. $S$ is in state $\left|\downarrow\ket_S$. We can ask~\cite{rovelli2007quantum} (see Sec.~5.6.1 therein): Has event $z=-1$ happened or not?

According to Rovelli's relational interpretation of quantum mechanics~\cite{Rovelli-1996} (see also Sec.~5.6 of Ref.~\cite{rovelli2007quantum}), this apparent puzzle captures the core conceptual difficulty of the interpretation of quantum theory, i.e., reconciling the possibility of quantum superpositions with the fact that the world we observe is characterized by definite values of physical quantities. However, Rovelli further argues that the puzzle dissolves if we admit that states of physical systems are not determined in an absolute sense but relative to another physical system, an ``observer''. Importantly, for Rovelli even the fact that an observer observes a certain phenomenon is itself relative to another observer (see Appendix~\ref{s:relational_cognition}). In contrast to common wisdom in cognitive science, though, for Rovelli ``observation'' is just a physical interaction and observers are just physical systems with no special properties, e.g. rocks and electrons classify as ``observers''~\cite{Rovelli-1996} (see Fig.~\ref{f:wigner} herein). 

\subsection{Von Neumann equation as a pair of real matrix equations}\label{s:pair}

Here we show how the von Neunmann equation, Eq.~\eqref{e:vNApp}, can generally be written in terms of two equations describing two real probability matrices related to the real and imaginary parts of the density matrix (cf. Appendix~\ref{s:gaussian-matrix}). Following the discussion in the main text, these pair of equations can be interpreted as implementing the intrinsic perspective. In a sense, this section goes in the reverse direction of the main text. In the next subsections we show explicit examples of quantum dynamics written in terms of non-negative real kernels that can be interpreted in probabilistic terms.

We assume that we can represent the adjoint operation $\dagger$ by the combination of transpose $T$ and complex conjugate $\ast$ operations, i.e. if $\mu$ is a generic matrix with complex entries, then $\mu^\dagger = (\mu^T)^\ast = (\mu^\ast)^T$. Notice that a generic Hermitian matrix can be written as $\mu = M_s + M_a/ i$, where $M_s = M_s^T$ is a real symmetric matrix and $M_a = -M_a^T$ is a real antisymmetric matrix; indeed ${\mu^\dagger = M_s^T -  M_a^T/i = M_s +  M_a/i = \mu}$. Furthermore, since any generic real matrix $M$ can be decomposed into symmetric and antisymmetric parts, i.e. $M = M_s + M_a$, then we can write $M_s = (M+M^T)/2$ and $M_a = (M-M^T)/2$. From this perspective, we can consider a generic Hermitian matrix $\mu$ as a convenient representation of a generic real matrix $M$ that allows us to keep explicit track of the symmetric and antisymmetric parts of the latter via the real and imaginary parts of the former, respectively. 

So, we can write $\rho $ and $H$ in Eq.~\eqref{e:vNApp} as $\rho = P_s +  P_a/ i$ and $H = -\hbar J_s +  -\hbar J_a/ i$. Here the symmetric and antisymmetric matrices corresponding to $\rho$ and $H$ can be written in terms of real matrices $P$ and $J$, respectively, as done for the generic matrix $M$ above. Since $\tr \rho = 1$ and the diagonal elements of an antisymmetric matrix are zero, we have $\tr\rho=\tr P_s = \tr P$, so $\tr P = 1$. We have written $H$ in terms of a real matrix $\hbar J$ so we do not have to worry about $\hbar$ in the equations below. We will refer to $J$ as the {\em dynamical matrix}. 

In this way, Eq.~\eqref{e:vNApp} can be written as
\begin{equation}\label{e:PsPa}
i \frac{\partial}{\partial t}(P_s + P_a/ i) = - [J_s + J_a/i , P_s +  P_a/i],
\end{equation}
where $\hbar$ has been absorved in $J = - H/\hbar$. Equating the real and imaginary parts of Eq.~\eqref{e:PsPa} we get a pair of equations
\begin{eqnarray}
\frac{\partial P_s}{\partial t} &=&   [J_s , P_a] + [J_a , P_s] ,\label{e:DtPs} \\
\frac{\partial P_a}{\partial t} &=&  - [J_s , P_s] + [J_a , P_a] .\label{e:DtPa}
\end{eqnarray} 

By adding and subtracting Eqs.~\eqref{e:DtPs} and \eqref{e:DtPa}, we obtain an equivalent pair of equations in terms of the real matrix $P$, i.e.
\BE
\frac{\partial P}{\partial t} &=&    - [J_s , P^T]  + [J_a , P]    ,\label{e:DtP} \\
\frac{\partial P^T}{\partial t} &=&   [J_s , P]    + [J_a , P^T]  .\label{e:DtPT}
\EE 
While Eq.~\eqref{e:DtPT} is the transpose of Eq.~\eqref{e:DtP}, we can also write these two equations as corresponding to two different observers $A$ and $B$ who describe the experiment with probability matrices $P_A$ and $P_B$, respectively; i.e. 
\BE
\frac{\partial P_A}{\partial t} &=&  - [J_s , P_B] + [J_a , P_A] ,\label{e:DtPA} \\
\frac{\partial P_B}{\partial t} &=&   [J_s , P_A] +  [J_a , P_B] ,\label{e:DtPB}
\EE 
under the condition that $P_B = P_A^T$ at time $t=0$, which guarantees that at all next time steps we have $P_B = P_A^T$ (cf. Eqs.~\eqref{e:DtPAfreepar} and \eqref{e:DtPBfreepar}). This condition can in principle be satisfied for any experiment as discussed in the main text and Appendices~\ref{s:EQM} and \ref{s:obj_constraints}.

As described in the main text, $A$ and $B$ can indeed be considered as two complementary ``sub-observers''.  In Appendix~\ref{s:examples} we recast some typical examples of quantum dynamics to show that these equations can be formulated in terms of real non-negative kernels that therefore can in principle be interpreted probabilistically (see also Appendix~\ref{s:MP_recasted}).

\

\noindent{\bf Remark:} Notice that since for small time steps $\epsilon$ we can write unitary evolution operators as ${U_\epsilon = \id - i\epsilon H}/\hbar$. So, we can write commutators with $H$ in terms of commutators with $U_\epsilon$ since 
\be\label{e:remark}
{[U_\epsilon,\rho] = -i\epsilon [H,\rho]/\hbar};
\ee
this is also true for other types of evolution kernels. Using Eq.~\eqref{e:remark}, we can then write the von Neumann equation, Eq.~\eqref{e:vNApp}, as 
\be\label{e:vN_Ueps}
\frac{\partial\rho}{\partial t} =\frac{1}{\epsilon}[U_\epsilon , \rho], 
\ee
where the limit $\epsilon\to 0$ is understood. 
 
This observation is useful when dealing with Gaussian kernels, for instance, because a Gaussian kernel $\mathcal{K}(x,x^\prime)$ with vanishing variance $\sigma^2$ cannot be straightforwardly expanded in a Taylor series as $\mathcal{K}(x, x^\prime)\approx \delta(x- x^\prime) + {O}(\sigma^2)$, where $\delta(x)$ is the Dirac delta. However, we can straightforwardly write $\mathcal{K}(x, x^\prime) = \delta(x- x^\prime) - \mathcal{L}(x,x^\prime)$, where $\mathcal{L}(x,x^\prime) = \delta(x-x^\prime) - \mathcal{K}(x, x^\prime) $. Although $\mathcal{L}$ is not ${O}(\sigma^2)$, its convolution with a smooth function yields a term ${O}(\sigma^2)$. In Appendices~\ref{s:path}, \ref{s:EM}, \ref{s:App_path} and \ref{s:App_EM} we obtain von Neumann-like equations similar to Eq.~\eqref{e:vN_Ueps}, where commutators are directly written in terms of real Gaussian kernels $\mathcal{K}$ with variance $\sigma^2 = O(\epsilon)$.

\

\subsection{Some examples of quantum dynamics in terms of non-negative real kernels}\label{s:examples}
Here we briefly discuss how some well-known examples of quantum systems can be described in terms of non-negative real kernels, including systems associated to complex non-stoquastic Hamiltonian operators. In Appendix~\ref{s:details} we provide the details of the derivations. In Appendix~\ref{s:negative-general} we discuss why the non-negativity of the kernels is not necessarily a restriction in our approach. Although we here focus on the position observable only, in Appendix~\ref{s:measurements} we discuss why this is not necessarily a restriction in our approach either.

\subsubsection{From non-relativistic path integrals to real convolutions}\label{s:path}

Consider the Schr\"odinger equation
\be\label{e:SchEqn}
i\hbar \frac{\partial\psi(x,t)}{\partial t} = -\frac{\hbar^2}{2m}\frac{\partial^2\psi(x,t)}{\partial x^2} + V(x)\psi(x, t)
\ee
 of a particle of mass $m$ in a one-dimensional non-relativistic potential $V$. In terms of the short-time path integral representation~\cite{feynman1948space}, with time step $\epsilon\to 0$,  can be written as (see Appendix~\ref{s:App_path} for all technical details)
\be\label{e:Main_GaussianKernel_vN}
\frac{\partial \rho}{\partial t} = \frac{i}{\epsilon}\left[(\mathcal{K}\ast\rho)-(\rho\ast\mathcal{K})\right],
\ee
where $\rho(x, x^\prime , t)$ is the density matrix and $\mathcal{K}(x,x^\prime)$ is a {\em real} kernel given by
\be\label{e:Main_K_T+V}
\mathcal{K}(x, x^{\prime}) = \frac{1}{|\mathcal{A}|}\exp{\left[-\frac{\mathcal{H}(x, x^{\prime})\epsilon}{\hbar} \right]},
\ee
with $\mathcal{A} = \sqrt{i 2\pi\hbar\epsilon /m }$ and
\be\label{e:Main_classicalH}
\mathcal{H}(x, x^\prime) = \frac{m}{2}\frac{(x - x^\prime)^2}{\epsilon^2} + V(x),
\ee
the corresponding ``energy'' function, which is a function of the position at two consecutive times, {\em not} of position and momentum. 

In contrast to Eq.~\eqref{e:remark}, Eq.~\eqref{e:Main_GaussianKernel_vN} contains a factor $i$. This is because the kernel $\mcK$ is real while $U$ is complex. We have {\em effectively} done a Wick rotation to get the real non-negative kernel $\mcK$, and then reintroduce the $i$ as a factor in Eq.~\eqref{e:Main_GaussianKernel_vN}, which essentially implements the SRC.

Equation~\eqref{e:Main_classicalH} would lead to an asymmetric kernel in Eq.~\eqref{e:Main_K_T+V} due to the potential $V$. However, since the contribution of the potential to the exponent of $\mathcal{K}$ in Eq.~\eqref{e:Main_K_T+V} is already of order $\epsilon$, we can straightforwardly symmetrize it by replacing $V(x)\to [V(x)+V(x^\prime)]/2 = V(x) + O(\epsilon)$ without affecting the relevant equations.

In Eq.~\eqref{e:Main_GaussianKernel_vN} we have introduced the convolutions 
\begin{widetext}
\BE
\left[\mathcal{K}\ast\rho\right] (x,x^\prime) &=& \frac{1}{|\mathcal{A}|}\int \exp{\left[-\frac{\mathcal{H}(x, x^{\prime\prime})\epsilon}{\hbar} \right]}\rho(x^{\prime\prime},x^\prime, t)\mathrm{d} x^{\prime\prime},\label{e:int_K*rho}\\
\left[\rho\ast\mathcal{K}\right] (x, x^\prime ) &=& \frac{1}{|\mathcal{A}|}\int \rho(x, x^{\prime\prime}, t)\exp{\left[-\frac{\mathcal{H}(x^{\prime\prime}, x^\prime )\epsilon}{\hbar} \right]}\mathrm{d} x^{\prime\prime}. \label{e:int_rho*K} 
\EE
\end{widetext}
Notice that the integration variables in $\mathcal{K}\ast\rho$ and $\rho\ast\mathcal{K}$ are, respectively, the first and second arguments of $\rho$, which yields the analogous of left and right matrix multiplication. 

Following the discussion in Appendix~\ref{s:pair}, Eq.~\eqref{e:Main_GaussianKernel_vN} can be written as a pair of real matrix equations (see more general example in Sec.~\ref{s:EM}). The point we want to make here is that the kernel appearing in such pair of equations can be real and non-negative, as we can see in Eq.~\eqref{e:Main_K_T+V}. In the next section we show this is also true in more general cases where the Hamiltonian is complex.

\

\noindent{\bf Remark:} While the kernel $\mathcal{K}$ defined in Eq.~\eqref{e:Main_K_T+V} is real and non-negative, it is {\em not} normalized. Indeed, we have (e.g. take $\psi(x^\prime, t) = 1$ in Eq.~\eqref{e:realGaussian})
\be\label{e:intK_T+V}
\int \mathcal{K}(x,x^\prime) \mathrm{d} x = 1-\epsilon V(x)/\hbar + O(\epsilon^2).
\ee
This fact has sometimes been used to argue against the viability of any probabilistic interpretation of the Euclidean, or imaginary-time, Schr\"odinger equation \cite{Zambrini-1987, zambrini1986stochastic,zambrini1986variational}. 

However, we will show in Appendix~\ref{s:MP_recasted} that the proper probabilistic analogue of $\mathcal{K}$ is not a transition probability but factors in a graphical model or something closer to the squared root of the product of forward and backward transition probabilities (cf. Eqs.~\eqref{e:K_P+*P-}). 


\subsubsection{Particle in an electromagnetic field via asymmetric real kernels}\label{s:EM}

The Schr\"odinger equation of a particle of charge $e$ interacting with an electromagnetic field can be written as 
\be\label{e:Main_SchrodingerEM}
\begin{split}
i\hbar\frac{\partial \psi(\bx,t)}{\partial t}=& -\frac{\hbar^2}{2m}\left(\nabla - i \frac{e}{\hbar c}\mathbf{A}\right)^2\psi(\bx,t)\\
 &+ e V(\bx,t)\psi(\bx,t),\\
\end{split}
\ee
%
where $\bx$ denotes the position vector in three dimensional space, while $V$ and $\mathbf{A}$ denote the scalar and vector fields respectively. Notice that the Hamiltonian associated to Eq.~\eqref{e:Main_SchrodingerEM} now contains an imaginary part given by the terms linear in $\mathbf{A}$ arising from the expansion of ${(\nabla - i e\mathbf{A} /\hbar c)^2\psi(\bx,t)}$.

As shown in full detail in Appendix~\ref{s:App_EM}, and following Appendix~\ref{s:pair}, the von Neumann equation corresponding to Eq.~\eqref{e:Main_SchrodingerEM} can be written as a pair of real matrix equations
\BE
\frac{\partial P_A}{\partial t} &=&  - \frac{1}{\epsilon}[\mathcal{K}_s, P_B]  + \frac{1}{\epsilon}[\mathcal{K}_a, P_A]  ,\\
\frac{\partial P_B}{\partial t} &=&   \frac{1}{\epsilon}[\mathcal{K}_s, P_A]   + \frac{1}{\epsilon}[\mathcal{K}_a, P_B]  ,
\EE
where the two probability matrices satisfy $P_A = P$ and $P_B = P^T$, and ${\rho = (P + P^T)/2 + (P-P^T)/2 i}$ yields the density matrix. Here, $\mathcal{K}_s$ and $\mathcal{K}_a$, are the symmetric and anti-symmetric parts of a real kernel ${\mathcal{K}=\mathcal{K}_s+\mathcal{K}_a}$ given by 
\be\label{e:Main_KrealEM}
\mathcal{K}(\bx , \bx^\prime) = \frac{1}{|\mathcal{A}_{EM}|}\exp{\left[-\frac{\epsilon}{\hbar}\mathcal{H}_{\rm EM}(\bx , \bx^\prime) \right]},
\ee
where the real electromagnetic ``energy'' function is given by 
\begin{widetext}
\be\label{e:Main_HrealEM}
\mathcal{H}_{\rm EM}(\bx , \bx^\prime) = \frac{m}{2}\left(\frac{\bx - \bx^\prime}{\epsilon}\right)^2 + V\left(\frac{\bx+\bx^\prime}{2}, t\right) + \frac{e}{c}\left(\frac{\bx-\bx^\prime}{\epsilon}\right)\cdot\mathbf{A}\left(\frac{\bx+\bx^\prime}{2}, t\right) +\frac{e^2}{m c^2} \left[\mathbf{A}\left(\frac{\bx+\bx^\prime}{2}, t\right)\right]^2 . 
\ee
\end{widetext}
So, even in the case of a charged particle in an electromagnetic field, whose Hamiltonian operator is complex (and so non-stoquastic), can be thought of as arising from a real non-negative kernel real $\mathcal{K}$.  It is no clear at this point, though, how to interpret $\mathcal{H}_{\rm EM}$ defined in Eq.~\eqref{e:Main_HrealEM}---how it relates to the actual energy of the system---nor the real kernel $\mathcal{K}$ defined in Eq.~\eqref{e:Main_KrealEM}. It seems to suggests a probabilistic interpretation of electromagnetic phenomena. We leave this for future work. 

\

\subsection{Derivation of real kernel representations in Appendix~\ref{s:examples}}\label{s:details}

\subsubsection{From non-relativistic path integrals to real convolutions}\label{s:App_path}

As originally described by Feynman, the non-relativistic Schr\"odinger equation can be derived from the path integral via (see e.g. Eq. (18) in Ref~\cite{feynman1948space})
\begin{equation}\label{e:path1D}
\psi(x_{\ell+1}, t+\epsilon) = \frac{1}{\mathcal{A}}\int\exp{\left[\frac{i}{\hbar}S(x_{\ell+1},x_\ell)\right]}\psi(x_\ell,t)\mathrm{d} x_\ell,
\end{equation}
where for the one-dimensional case in Appendix~\ref{s:path} we can set the short-time action as
\be\label{e:action}
S(x_{\ell+1}, x_\ell) = \frac{m\epsilon}{2}\left(\frac{x_{\ell + 1}- x_\ell}{\epsilon} \right)^2 - \epsilon V(x_{\ell+ 1 }),
\ee
and 
\be
\mathcal{A} = \sqrt{i 2\pi\hbar\epsilon /m };
\ee
furthermore, $x_\ell$ represents the position of the particle at time $t=\ell\epsilon$. Notice that by iterating Eq.~\eqref{e:path1D} we can obtain the path integral representation. 

It is possible to rewrite these equations in terms of {\it real} Gaussian convolutions, which can be naturally interpreted in probabilistic terms. Before we show how to do this, we will outline the main steps in the derivation of Schr\"odinger equation, Eq.~\eqref{e:SchEqn}, from Eq.~\eqref{e:path1D}, described in detailed in e.g. Ref.~\cite{feynman1948space}. This closely parallels the derivation in terms of real Gaussian convolutions to be described afterwards.  

First, expanding $\psi(x_{\ell+1}, t+\epsilon)$ to first order in $\epsilon$, we can write Eq.~\eqref{e:path} as 
\be\label{e:path_step1}
\begin{split}
\epsilon\frac{\partial\psi(x_{\ell+1}, t)}{\partial t} =& \frac{1}{\mathcal{A}}\int\exp{\left[\frac{i}{\hbar}S(x_{\ell+1},x_\ell)\right]}\psi(x_\ell,t)\mathrm{d} x_\ell  \\
& - \psi(x_{\ell + 1}, t).
\end{split}
\ee
Since $\epsilon\to 0$, the complex Gaussian factor associated to the kinetic term in Eq.~\eqref{e:action} {\it oscillates very fast} except in the region where $|{x_{\ell +1} - x_\ell| = O(\sqrt{\hbar\epsilon/m})} $. So, to estimate the integral to first order in $\epsilon$, the term $\psi(x_\ell, t)$ in the right hand side of Eq.~\eqref{e:path_step1} need be expanded around $x_{\ell + 1}$ to second order in ${x_{\ell + 1} - x_\ell}$. Consistent with this approximation to first order in $\epsilon$, we also do $\exp{[-i V(x)\epsilon / \hbar]} = 1 -i V(x)\epsilon / \hbar + O(\epsilon^2) $. This leads to Eq.~\eqref{e:SchEqn}. 

However, as we know from the derivation of the diffusion equation approximation for a random walk~\cite{van1992stochastic,risken1989fpe}, a real Gaussian has an equivalent cancelling effect, not because of fast oscillations but because of exponentially small terms (see Appendix~\ref{s:diff_nodrift}). More precisely, we will argue that if we introduce the energy function
\be\label{e:classicalH}
\mathcal{H}(x, x^\prime) = \frac{m}{2}\frac{(x - x^\prime)^2}{\epsilon^2} + V(x),
\ee
we can do the replacement, which amounts at a Wick rotation $\epsilon\to -i\epsilon$,
\be\label{e:replacement}
\frac{1}{\mathcal{A}}\exp{\left[\frac{i}{\hbar}S(x,x^\prime)\right]}\to\frac{1}{|\mathcal{A}|} \exp{\left[-\frac{\mathcal{H}(x, x^\prime)\epsilon}{\hbar} \right]};
\ee
notice that the kinetic term in the right hand side of Eq.~\eqref{e:replacement} leads to a real Gaussian with variance $\hbar\epsilon / m$ and normalization constant given precisely by $|\mathcal{A}|$.

Due to this Gaussian term, the integral

\begin{widetext}
\be\label{e:realGaussian}
\begin{split}
\frac{1}{|\mathcal{A}|}\int \exp{\left[-\frac{\mathcal{H}(x, x^\prime)\epsilon}{\hbar} \right]}\psi(x^\prime, t)\mathrm{d} x^\prime  &= \left[1-\frac{\epsilon}{\hbar} V(x)\right]\left[\psi(x, t) + \frac{\hbar \epsilon }{2 m}\frac{\partial^2 \psi(x,t)}{\partial x^2} \right]+ O (\epsilon^2) \\
& = \psi(x, t) -\frac{\epsilon}{\hbar} V(x)\psi(x,t) + \frac{\hbar \epsilon }{2 m}\frac{\partial^2 \psi(x,t)}{\partial x^2} + O (\epsilon^2), 
\end{split}
\ee
\end{widetext}
can be approximated to first order in $\epsilon$ in a way similar to that of the integral in Eq.~\eqref{e:path_step1}. Indeed, since $\epsilon\to 0$, the real Gaussian factor associated to the kinetic term in Eq.~\eqref{e:classicalH} {\it is exponentially small} except in the region where $|{x - x^\prime | = O(\sqrt{\hbar\epsilon/m})} $. This has allowed us to estimate the integral to first order in $\epsilon$ by expanding the term $\psi(x^\prime, t)$ in the left hand side of Eq.~\eqref{e:realGaussian} around $x$ up to second order in ${x - x^\prime}$. Consistent with this approximation to first order in $\epsilon$, we have also done $\exp{[- V(x)\epsilon / \hbar]} = 1 - V(x)\epsilon / \hbar + O(\epsilon^2) $. 

This implies that Eq.~\eqref{e:SchEqn} can be written in terms of real Gaussian convolutions in a way similar to Eq.~\eqref{e:path_step1} as
\begin{widetext}
\be\label{e:GaussianSchrodinger}
\epsilon\frac{\partial\psi(x_{\ell+1}, t)}{\partial t} = i \left\{\frac{1}{|\mathcal{A}|}\int \exp{\left[-\frac{\mathcal{H}(x_{\ell + 1}, x_\ell)\epsilon}{\hbar} \right]}\psi(x_\ell, t)\mathrm{d} x_\ell  - \psi(x_{\ell + 1}, t) \right\}.
\ee
\end{widetext}

Indeed, by replacing the term with the integral in the right hand side of Eq.~\eqref{e:GaussianSchrodinger} by the right hand side of Eq.~\eqref{e:realGaussian}, and multiplying both sides of Eq.~\eqref{e:GaussianSchrodinger} by $i\hbar/\epsilon$, we obtain Schrodinger equation, Eq.~\eqref{e:SchEqn}. In this way we have essentially bring the imaginary unit $i$ from the exponential in Eq.~\eqref{e:path_step1} down to turn it into a linear factor in Eq.~\eqref{e:GaussianSchrodinger}. In contrast to Eq.~\eqref{e:path}, however, it does not seem possible to obtain a real path integral by iterating Eq.~\eqref{e:GaussianSchrodinger}. 

Finally, we now show that Eq.~\eqref{e:GaussianSchrodinger} leads to an equation analogous to the von Neumann equation (Eq.~\eqref{e:vNApp}) for the density matrix ${\rho(x, x^\prime , t) = \psi(x, t)\psi^\ast(x^\prime, t)}$, in terms of {\em real} Gaussian convolutions instead of differential operators. Indeed, taking the time derivative of this density matrix yields
\be\label{e:Psi_time} 
\frac{\partial \rho(x, x^\prime , t)}{\partial t} = \frac{\partial\psi(x, t)}{\partial t}\psi^\ast(x^\prime, t) + \psi(x, t)\frac{\partial\psi^\ast(x^\prime, t)}{\partial t};
\ee 
now, replacing the time derivatives of the wave function $\psi$ and its conjugate $\psi^\ast$ in Eq.~\eqref{e:Psi_time}, respectively, by the right hand side of Eq.~\eqref{e:GaussianSchrodinger} and its conjugate we obtain
\begin{widetext}
\be\label{e:Gaussian_vN}
\begin{split}
\frac{\partial \rho(x, x^\prime , t)}{\partial t} =& \frac{i}{\epsilon}\left\{\frac{1}{|\mathcal{A}|}\int \exp{\left[-\frac{\mathcal{H}(x, x^{\prime\prime})\epsilon}{\hbar} \right]}\rho(x^{\prime\prime},x^\prime, t)\mathrm{d} x^{\prime\prime}  - \rho(x, x^\prime , t) \right\} \\
& - \frac{i}{\epsilon}\left\{\frac{1}{|\mathcal{A}|}\int \exp{\left[-\frac{\mathcal{H}(x^\prime, x^{\prime\prime})\epsilon}{\hbar} \right]}\rho(x, x^{\prime\prime}, t)\mathrm{d} x^{\prime\prime}  - \rho(x, x^\prime , t) \right\}.
\end{split}
\ee
\end{widetext}

Clearly, the terms $\rho(x, x^\prime , t)$ in the right hand side cancel out, and we can write Eq.~\eqref{e:Gaussian_vN} in compact form as
\be\label{e:GaussianKernel_vN}
\frac{\partial \rho}{\partial t} = \frac{i}{\epsilon}\left[(\mathcal{K}\ast\rho)-(\rho\ast\mathcal{K})\right],
\ee
where we have introduced the kernel 
\be\label{e:K_T+V}
\mathcal{K}(x, x^{\prime}) = \frac{1}{|\mathcal{A}|}\exp{\left[-\frac{\mathcal{H}(x, x^{\prime})\epsilon}{\hbar} \right]},
\ee
and the convolutions 
\begin{widetext}
\BE
\left[\mathcal{K}\ast\rho\right](x,x^\prime, t) &= \frac{1}{|\mathcal{A}|}\int \exp{\left[-\frac{\mathcal{H}(x, x^{\prime\prime})\epsilon}{\hbar} \right]}\rho(x^{\prime\prime},x^\prime, t)\mathrm{d} x^{\prime\prime}, \\
\left[\rho\ast\mathcal{K}\right](x^\prime , x, t)  &=\frac{1}{|\mathcal{A}|}\int \exp{\left[-\frac{\mathcal{H}(x^\prime, x^{\prime\prime})\epsilon}{\hbar} \right]}\rho(x, x^{\prime\prime}, t)\mathrm{d} x^{\prime\prime}.
\EE
\end{widetext}
to represent, respectively, the first and second integrals in the right hand side of Eq.~\eqref{e:Gaussian_vN}. Notice that the integration variables in $(\mathcal{K}\ast\rho)$ and $(\rho\ast\mathcal{K})$ are, respectively, the first and second arguments of $\rho$, which yields the analogous of left and right matrix multiplication. 

\

\subsubsection{Particle in an electromagnetic field via asymmetric real kernels}\label{s:App_EM}

\noindent{\em Prelude: Hermitian kernel via a complex ``energy'' function}: The Schr\"odinger equation of a particle of charge $e$ interacting with an electromagnetic field can be written as 
\be\label{e:SchrodingerEM}
\begin{split}
i\hbar\frac{\partial \psi(\bx,t)}{\partial t}=& -\frac{\hbar^2}{2m}\left(\nabla - i \frac{e}{\hbar c}\mathbf{A}\right)^2\psi(\bx,t)\\
 &+ e V(\bx,t)\psi(\bx,t),
\end{split}
\ee
where $\bx$ denotes the position vector in three dimensional space, while $V$ and $\mathbf{A}$ denote the scalar and vector fields respectively.  Notice that the Hamiltonian operator associated to Eq.~\eqref{e:SchrodingerEM} now contains an imaginary part given by the terms linear in $\mathbf{A}$ arising from the expansion of ${(\nabla - i e\mathbf{A} /\hbar c)^2\psi(\bx,t)}$.

Equation~\eqref{e:SchrodingerEM} can be derived via the path integral formulation from the extension of Eq.~\eqref{e:path} to three-dimensional space (i.e. by doing the substitution $x\to\bx$)
\begin{equation}\label{e:path}
\psi(\bx_{\ell+1}, t+\epsilon) = \frac{1}{\mathcal{A}_{EM}}\int\exp{\left[\frac{i}{\hbar}S_{EM}(\bx_{\ell+1},\bx_\ell)\right]}\psi(\bx_\ell,t)\mathrm{d}^3 \bx_\ell,
\end{equation}
with action
\begin{widetext}
\be\label{e:actionEM}
S_{EM}(\bx_{\ell+1}, \bx_\ell) = \frac{m\epsilon}{2}\left(\frac{\bx_{\ell + 1}- \bx_\ell}{\epsilon} \right)^2 + \frac{e \epsilon}{c}\left(\frac{\mathbf{x}_{\ell+1} - \mathbf{x}_{\ell}}{\epsilon}\right)\cdot\mathbf{A}\left(\frac{\bx_{\ell+ 1 }+\bx_\ell}{2},t\right) - \epsilon V\left(\frac{\bx_{\ell+ 1 }+\bx_\ell}{2},t\right),  
\ee
\end{widetext}
and 
\be\label{e:A_EM}
\mathcal{A} \to\mathcal{A}_{EM} \equiv (i 2\pi\hbar\epsilon / m)^{3/2};
\ee
here we are using a midpoint discretization for the action.

As in the previous section, it is possible to derive Eq.~\eqref{e:SchrodingerEM} by doing a replacement similar to that in Eq.~\eqref{e:replacement}, i.e.
\begin{widetext}
\be\label{e:replacementEM}
\frac{1}{\mathcal{A}_{EM}}\exp{\left[\frac{i}{\hbar}S_{EM}(\bx,\bx^\prime)\right]}\to\frac{1}{|\mathcal{A}_{EM}|} \exp{\left[-\frac{\epsilon}{\hbar}\widetilde{\mathcal{H}}_{EM}(\bx, \bx^\prime)\right] },
\ee
\end{widetext}
but with a complex ``energy'' function

\newpage

\begin{widetext}
\be\label{e:classicalH_EM}
\widetilde{\mathcal{H}}_{EM}(\bx, \bx^\prime) = \frac{m}{2}\left(\frac{\bx - \bx^\prime}{\epsilon}\right)^2 + V\left(\frac{\bx+\bx^\prime}{2}, t\right)
- i\frac{e}{c}\left(\frac{\bx-\bx^\prime}{\epsilon}\right)\cdot\mathbf{A}\left(\frac{\bx+\bx^\prime}{2}, t\right).
\ee
\end{widetext}
As with Eq.~\eqref{e:replacement}, this corresponds to a Wick rotation $\epsilon\to -i\epsilon$. We can recognize that the real part is the three-dimensional version of the energy function defined in Eq.~\eqref{e:classicalH}. We could think of $\mathcal{H}_{EM}$ as an ``energy'' function with a complex interaction energy whose real and imaginary parts correspond to the electric and magnetic fields, respectively. This seems similar to writing the electromagnetic field as $\mathbf{E} + i\mathbf{B}$ which allows to write Maxwell's equations in compact form (see e.g. Eq. (7.10) in Ref.~\cite{doran2003geometric}). 

As in the previous section, we can derive Eq.~\eqref{e:SchrodingerEM} from an equation analogous to Eq.~\eqref{e:GaussianSchrodinger}, i.e. 
\begin{widetext}
\be\label{e:GaussianSchrodingerEM}
\epsilon\frac{\partial\psi(\bx_{\ell+1}, t)}{\partial t} = i \left\{\frac{1}{|\mathcal{A}_{EM}|}\int \exp{\left[-\frac{\epsilon}{\hbar}\widetilde{\mathcal{H}}_{EM}(\bx_{\ell + 1}, \bx_\ell) \right]}\psi(\bx_\ell, t)\mathrm{d}^3 \bx_\ell  - \psi(\bx_{\ell + 1}, t) \right\}.
\ee
\end{widetext}
Indeed, as in the previous section, the Gaussian factor in the complex kernel (see Eqs.\eqref{e:A_EM} and~\eqref{e:classicalH_EM})
\be\label{e:C}
\mathcal{C}(\bx , \bx^\prime) = \exp{\left[-\frac{\epsilon}{\hbar}\widetilde{\mathcal{H}}_{EM}(\bx, \bx^\prime) \right]} / {|\mathcal{A}_{EM}|},
\ee
associated to the kinetic term in Eq.~\eqref{e:classicalH_EM} allows us to expand the other factors in the integral in Eq.~\eqref{e:GaussianSchrodingerEM} around $\bx_{\ell +1}$ up to second order in $|\bx_{\ell +1} - \bx_\ell|$ or to first order in $\epsilon$. More precisely, by introducing the variable ${\bu = \bx_{\ell+1}-\bx_\ell}$, so ${(\bx_{\ell+1}+\bx_\ell)/2 = \bx_{\ell + 1} - \bu/2}$ as well as ${\bx_\ell = \bx_{\ell+1} - \bu}$, and doing ${\bx = \bx_{\ell+1}}$, $\bx^\prime = \bx_\ell$ to avoid cluttering the equations with indexes, we can write

\begin{widetext}
\be\label{e:C*psi}
[\mathcal{C}\ast\psi](\bx , t) =\frac{1}{|\mathcal{A}_{EM}|}\int \exp\left(-\frac{m \bu^2}{2 \hbar\epsilon } \right) f(\bx, \bu, t)\left[\psi(\bx ,t) - \bu\cdot\nabla\psi(\bx ,t) + \frac{1}{2}\bu\cdot \mathbf{H}\psi(\bx, t)\cdot\bu \right] + O(\epsilon^2),
\ee
\end{widetext}
where the convolution $\mathcal{C} \ast\psi$ denotes the integral in the right hand side of Eq.~\eqref{e:GaussianSchrodingerEM}, $\mathbf{H}\psi$ stands for the Hessian or matrix of second derivatives of $\psi$. Furthermore, the function

\begin{widetext}
\be
f(\bx, \bu, t) = \left\{1-\frac{\epsilon}{\hbar}V(\bx, t) + i\frac{e}{\hbar c}\bu\cdot\mathbf{A}(\bx,t) - i\frac{e}{2\hbar c}\bu\cdot\nabla\mathbf{A}(\bx,t)\cdot\bu -\frac{1}{2}\left[\frac{e}{\hbar c}\bu\cdot\mathbf{A}(\bx, t)\right]^2 \right\},
\ee
\end{widetext}
is the expansion up to first order in $\epsilon$ or second order in $\bu$ of the exponential factors in the complex kernel $\mathcal{C}$, which correspond to the interaction terms in Eq.~\eqref{e:classicalH_EM}, i.e. those containing $V$ and $\mathbf{A}$.
\newpage

Taking into account that the first two moments of $\bu$ are $\bra u_j\ket_\bu = 0$ and $\bra u_j u_k\ket_\bu = \delta_{j k}\hbar\epsilon/m$, where $\langle\cdot\rangle_\bu$ refers to the average taken with the Gaussian ${\exp(-m\bu^2/2\hbar\epsilon)/|\mathcal{A}_{EM}|}$, and that terms containing $\epsilon \bu^2$ and $\bu^3$ or higher can be neglected, the integral in Eq.~\eqref{e:C*psi} yields 
\begin{widetext}
\be
[\mathcal{C}\ast\psi](\bx, t) = \left(1-\frac{\epsilon}{\hbar} V\right)\psi +\frac{\hbar\epsilon}{2 m}\nabla^2\psi - i\frac{e\epsilon}{mc}\mathbf{A}\cdot\nabla\psi - i\frac{e \epsilon}{2 m c}\nabla\cdot\mathbf{A} \psi - \frac{e^2\epsilon}{2 \hbar m c^2 }\mathbf{A}^2\psi
\ee
\end{widetext}

Furthermore, taking into account that
\be 
\begin{split}
\left(\nabla - i\frac{e}{\hbar c}\mathbf{A}\right)^2\psi =& \nabla^2\psi - \left(\frac{e}{\hbar c}\right)^2\mathbf{A}^2\psi - \\
& i \frac{e}{\hbar c}\left[2 \mathbf{A}\cdot\nabla\psi + (\nabla\cdot\mathbf{A})\psi\right],
\end{split}
\ee
we obtain
\begin{widetext}
\be\label{e:C*psi_final}
[\mathcal{C}\ast\psi](\bx , t) =\psi(\bx ,t) + \frac{\epsilon}{\hbar}\left[\frac{\hbar^2}{2 m}\left(\nabla - i\frac{e}{\hbar c}\mathbf{A}\right)^2 \psi(\bx ,t) - V(\bx, t)\psi(\bx ,t)\right] + O(\epsilon^2).
\ee
\end{widetext}

So, we can indeed write the Schr\"odinger equation Eq.~\eqref{e:SchrodingerEM} as 
\be\label{e:partial}
\epsilon\frac{\partial\psi}{\partial t} = i [\mathcal{C}\ast\psi - \psi],
\ee
which is the analogous of Eq.~\eqref{e:GaussianSchrodinger}. To see this, we can replace $\mathcal{C}\ast\psi$ in Eq.~\eqref{e:partial} by the right hand side of Eq.~\eqref{e:C*psi_final}, cancel out the $\psi$ terms, and multiply the remaining equation by $i\hbar/\epsilon$, which yields Eq.~\eqref{e:SchrodingerEM}.

By doing the same analysis that led from Eq.~\eqref{e:GaussianSchrodinger} to Eqs.~\eqref{e:Gaussian_vN}~and~\eqref{e:GaussianKernel_vN} we get
\be\label{e:C*rho}
\frac{\partial\rho}{\partial t} = \frac{i}{\epsilon} [\mathcal{C}\ast\rho - \rho\ast\mathcal{C}]\equiv \frac{i}{\epsilon}[\mathcal{C}, \rho] .
\ee
Notice, however, that contrary to the kernel $\mathcal{K}$ in Eq.~\eqref{e:GaussianKernel_vN}, which is real and symmetric, the kernel $\mathcal{C}$ in Eq.~\eqref{e:C*rho} is still complex and asymmetric. Indeed, from Eqs.\eqref{e:classicalH_EM}~and~\eqref{e:C} we can see that by exchanging the two arguments of $\mathcal{C}$ we get
\be\label{e:C_asymmetric}
\mathcal{C}(\bx^\prime ,\bx ) = [\mathcal{C}(\bx ,\bx^\prime )]^\ast,
\ee
which, considering $\mathcal{C}$ as a matrix, also shows that $\mathcal{C}$ is Hermitian. We could also see that $\mathcal{C}$ is Hermitian by writing Eq.~\eqref{e:C} as $\mathcal{C} = \mathcal{K}_s + \mathcal{K}_a / i$ with
\begin{widetext}
\BE
\mathcal{K}_s(\bx , \bx^\prime) &=& \frac{1}{|\mathcal{A}_{EM}|}\exp{\left[-\frac{m({\bx - \bx^\prime})^2}{2\hbar\epsilon} -\frac{\epsilon}{\hbar} V\left(\frac{\bx+\bx^\prime}{2}, t\right) \right]}\cos\left[\frac{e}{\hbar c}\left({\bx-\bx^\prime}\right)\cdot\mathbf{A}\left(\frac{\bx+\bx^\prime}{2}, t\right)\right],\label{e:Ks} \\
\mathcal{K}_a(\bx , \bx^\prime) &=& -\frac{1}{|\mathcal{A}_{EM}|}\exp{\left[-\frac{m({\bx - \bx^\prime})^2}{2\hbar\epsilon} -\frac{\epsilon}{\hbar} V\left(\frac{\bx+\bx^\prime}{2}, t\right) \right]}\sin\left[\frac{e}{\hbar c}\left({\bx-\bx^\prime}\right)\cdot\mathbf{A}\left(\frac{\bx+\bx^\prime}{2}, t\right)\right], \label{e:Ka}
\EE
\end{widetext}
which are clearly symmetric and antisymmetric, respectively, due to the symmetry properties of the cosinusoidal and sinusoidal functions.

\noindent{\em Postlude: real asymmetric kernel via a real ``energy'' function}: Since the kernel $\mathcal{C}$ defined in Eq.~\eqref{e:C} is Hermitian (see Eq.~\eqref{e:C_asymmetric}), Eq.~\eqref{e:C*rho} has the same structure of Eq.~\eqref{e:vNApp}. So, as described in Appendix~\ref{s:pair} (see also the remark therein), we can write Eq.~\eqref{e:C*rho} as a pair of real equations in terms of a real kernel $\mathcal{K} = \mathcal{K}_s + \mathcal{K}_a$, which in this case is given by the sum of the right hand sides of Eqs.~\eqref{e:Ks}~and~\eqref{e:Ka},
\begin{widetext}
\be
\mathcal{K}(\bx , \bx^\prime) = \frac{1}{|\mathcal{A}_{EM}|}\exp{\left[-\frac{m({\bx - \bx^\prime})^2}{2\hbar\epsilon} -\frac{\epsilon}{\hbar} V\left(\frac{\bx+\bx^\prime}{2}, t\right) \right]}[\cos z - \sin z],\label{e:Kpre}
\ee
\end{widetext}
where 
\be\label{e:z}
z = \frac{\epsilon }{\hbar }\frac{e}{ c}\left(\frac{\bx-\bx^\prime}{\epsilon}\right)\cdot\mathbf{A}\left(\frac{\bx+\bx^\prime}{2}, t\right).
\ee

Due to the very sharp Gaussian factor (since ${\epsilon\to 0}$), we can expand the sinusoidal and cosinusoidal functions up to second order in their argument since the rest gives contributions of order higher than $\epsilon$. Now, up to second order we have 
\be\label{e:cossinexp}
\cos z - \sin z = \exp(-z-z^2) + O(z^3).
\ee

Equations~\eqref{e:Ks}-\eqref{e:z} show that, even in the case of a charged particle in an electromagnetic field, whose Hamiltonian operator is complex (and so non-stoquastic), can be thought of as arising from a non-negative real kernel ${\mathcal{K}=\mathcal{K}_s+\mathcal{K}_a}$. More precisely, following Appendix~\ref{s:pair} the corresponding von Neumann equation can be understood as a pair of real matrix equations
\BE
\frac{\partial P_A}{\partial t} &=& - \frac{1}{\epsilon}[\mathcal{K}_s, P_B] +\frac{1}{\epsilon}[\mathcal{K}_a, P_A]  ,\\
\frac{\partial P_B}{\partial t} &=& \frac{1}{\epsilon}[\mathcal{K}_s, P_A]  + \frac{1}{\epsilon}[\mathcal{K}_a, P_B]  ,
\EE

\

\noindent where the two probability matrices satisfy $P_A = P$ and $P_B = P^T$, and ${\rho = (P + P^T)/2 + (P-P^T)/2i}$ yields the density matrix. Following the approach introduced in the main text, these equations can be interpreted in probabilistic terms. 

However, the term $z^2$ in Eq.~\eqref{e:cossinexp} leads via Eq.~\eqref{e:z} to a term with a quadratic factor $(\epsilon / \hbar)^2$, which prevents us from writing $\mathcal{K}\propto\exp{(-\epsilon\mathcal{H}_{\rm EM}/\hbar)}$ with a Hamiltonian-like function $\mathcal{H}_{\rm EM} $ independent of $\epsilon$ and $\hbar$ (cf. Eq.~\eqref{e:C}). It is possible to go around this issue, though, by noticing that (remember that $\bu = \bx-\bx^\prime$, so $(\bx + \bx^\prime)/2 = \bx - \bu/2$)
\begin{widetext}
\be\label{e:averages}
\bra z^2\ket_\bu = \left(\frac{e}{\hbar c}\right)^2\bra\left[\bu\cdot\mathbf{A}\left(\bx-\frac{\bu}{2}, t\right)\right]^2\ket_\bu = \frac{\epsilon}{\hbar}\frac{e^2}{ m c^2}[\mathbf{A}(\bx, t)]^2 + O(\epsilon^2),
\ee
\end{widetext}
where $\langle\cdot\rangle_\bu$ refers to the average taken with the Gaussian ${\exp(-m\bu^2/2\hbar\epsilon)/|\mathcal{A}_{EM}|}$, and we have used the results $\bra u_j\ket_\bu = 0$ and $\bra u_j u_k\ket_\bu = (\hbar\epsilon / m) \delta_{j k} $. 

Since the first term in the right hand side of Eq.~\eqref{e:averages} is already of order $\epsilon$, we can safely replace
\be
z^2 \to \frac{\epsilon}{\hbar}\frac{e^2}{m c^2}\left[\mathbf{A}\left(\frac{\bx+\bx^\prime}{2}, t\right)\right]^2 + O(\epsilon^2),
\ee
so
\be
e^{-z^2}\to\exp\left(-\frac{\epsilon}{\hbar}\frac{e^2}{ m c^2} \mathbf{A}^2\right),
\ee
in Eq.~\eqref{e:cossinexp}, since when we expand in $\bu$ only the term independent of $\bu$ remains as the other terms in the expansion lead to terms of higher order in $\epsilon$.

So, we can indeed write the asymmetric real kernel ${\mathcal{K}=\mathcal{K}_s+\mathcal{K}_a}$ in Eq.~\eqref{e:Kpre} associated to the complex kernel $\mathcal{C}$ in Eq.~\eqref{e:C} as
\be\label{e:KrealEM}
\mathcal{K}(\bx , \bx^\prime) = \frac{1}{|\mathcal{A}_{EM}|}\exp{\left[-\frac{\epsilon}{\hbar}\mathcal{H}_{\rm EM}(\bx , \bx^\prime) \right]}
\ee
where the {\it real} electromagnetic ``energy'' function (with no tilde) is given by
\begin{widetext}
\be\label{e:HrealEM}
\begin{split}
\mathcal{H}_{\rm EM}(\bx , \bx^\prime) &= \frac{m}{2}\left(\frac{\bx - \bx^\prime}{\epsilon}\right)^2 + V\left(\frac{\bx+\bx^\prime}{2}, t\right) + \frac{e}{c}\left(\frac{\bx-\bx^\prime}{\epsilon}\right)\cdot\mathbf{A}\left(\frac{\bx+\bx^\prime}{2}, t\right) +\frac{e^2}{ m c^2} \left[\mathbf{A}\left(\frac{\bx+\bx^\prime}{2}, t\right)\right]^2 
\end{split}
\ee
\end{widetext}
%

\section{Classical diffusion vs. non-relativistic quantum free particle}\label{s:diff-quantum}

In Appendix~\ref{s:quantum_aspects} we provide a review of some concepts of quantum theory that are relevant for the discussion in the main text. In particular, we show how von Neumann equation, which characterizes the quantum dynamics of the density matrix, can be written in terms of a pair of equations describing the dynamics of a pair of real matrices related to the real and imaginary parts of the density matrix. We also provide some non-trivial examples. Here we complement Appendix~\ref{s:quantum_aspects} by providing a comprehensive discussion of one of the simplest quantum systems---i.e., a non-relativistic free particle---which is closely related to a classical diffusive particle. We use this example to highlight some aspects that are relevant to the discussion in the main text. In doing so, we hope to provide some useful intuition for the reader that may not be familiar with quantum theory. 

\subsection{Classical diffusion and real Gaussian kernels}\label{s:diff} 

Here we present the example of a classical diffusive particle in a way different from the usual presentation. Although at some point this may seem a rather contrived presentation, it is purposely done this way to highlight certain aspects which are relevant for the discussion in the main text. In particular, we distinguish between symmetric and antisymmetric changes of the probability density and the corresponding operators that generate them. Furthermore, we highlight the antisymmetric changes that can be associated to relative motion.

\subsubsection{No drift: symmetric probability changes only}\label{s:diff_nodrift}

Here we discuss the case of a diffusive particle with no drift. This concerns only symmetric changes of the particle's probability density and, therefore, only symmetric operators. Consider a diffusive particle with zero drift and diffusion coefficient $D= k_B T/\gamma$, where $k_B$, $T$, and $\gamma$ are Boltzmann constant, temperature, and the coefficient of friction, respectively. The probability density to transition from position $x^\prime$ to position $x$ during a time step of size $\epsilon$ is given by a Gaussian kernel
\be
\begin{split}\label{e:diff_kernel}
\mathcal{P}_\epsilon^{0} (x|x^\prime) &= \frac{1}{\sqrt{4\pi D\epsilon}}e^{-(x-x^\prime)^2/4 D\epsilon} \\
&\equiv \frac{1}{Z_\epsilon}e^{-\tfrac{\epsilon}{h_{\rm diff}}\mathcal{H}_{\rm diff}^0(x,x^\prime)}, 
\end{split}
\ee
where $h_{\rm diff} = 2 k_B T$ and
\be\label{e:Hdiff}
\mathcal{H}_{\rm diff}^0(x, x^\prime) = \frac{\gamma}{2}\frac{(x - x^\prime)^2}{\epsilon^2} ,
\ee
play an analogous role to Planck's constant and the Hamiltonian function of a free particle of mass $\gamma$, respectively. Here $Z_\epsilon = \sqrt{2\pi h_{\rm diff}\epsilon / \gamma}$ is the normalization constant. 

If $p^0(x^\prime, t)$ is the probability density for the diffusive particle with zero drift to be at position $x^\prime$ at time $t$, the probability density for it to be at position $x$ at time $t+ \epsilon$ is given by
\be\label{e:diff_dyn}
p^0(x,{t + \epsilon}) = \int \mathcal{P}^{0}_\epsilon (x|x^\prime)p^0(x^\prime ,t)\mathrm{d}x^\prime .
\ee
For $\epsilon\to 0$ the Gaussian kernel in Eq.~\eqref{e:diff_kernel} is exponentially small except when $|x-x^\prime |= O(\sqrt{h_{\rm diff}\epsilon/ \gamma})$. So, we can expand $p^0_t(x^\prime , t)$ up to second order around $x$ to obtain the right hand side of Eq.~\eqref{e:diff_dyn} up to first order in $\epsilon$. This yields
\be\label{e:diff_eps}
p^0(x, {t + \epsilon}) = p^0(x, t) + \epsilon D\frac{\partial^2 p^0(x , t)}{\partial x^2} + O(\epsilon^2),
\ee
or, taking the limit $\epsilon\to 0$, using $D=h_{\rm diff}/2 \gamma$, and multiplying both sides by $h_{\rm diff}$,
\be\label{e:diff}
h_{\rm diff}\frac{\partial p^0(x, t)}{\partial t} = \frac{h_{\rm diff}^2}{2 \gamma}\frac{\partial^2 p^0(x, t)}{\partial x^2} .
\ee
Equation~\eqref{e:diff} is similar to the Schr\"odinger equation of a non-relativistic free particle (see Eq.~\eqref{e:free} below), except that the latter has an imaginary unit and a minus sign in the left and right hand sides, respectively. Unlike Planck constant that has units of energy and time, $h_{\rm diff}$  has units of energy only; this is related to $\gamma$ having units of mass over time.

Equation~\eqref{e:diff} can be written as (cf. Eq.~\eqref{e:Hfree_psi0} below)
\be\label{e:Jdiff_p0}
h_{\rm diff}\frac{\partial p^0}{\partial t} = \widetilde{J}^0 p^0,
\ee
where
\be\label{e:Jdiff}
\widetilde{J}^0 =  \frac{h_{\rm diff}^2}{2 \gamma}\frac{\partial^2 }{\partial x^2} ,
\ee
is analogous to the Hamiltonian operator associated to a free particle (see Eq.~\eqref{e:Hfree} below). Indeed, if we do a {\em Wick rotation}, i.e., if we do $t = i\tau$, Eq.~\eqref{e:Jdiff_p0} turns into Schr\"odinger equation, Eq.~\eqref{e:Hfree_psi0}, with Hamiltonian operator $H^0=-\widetilde{J}^0$ and time variable $\tau$.

\subsubsection{Drift: antisymmetric probability changes and moving reference frames}\label{s:diff-drift}
\begin{figure}
\includegraphics[width=\columnwidth]{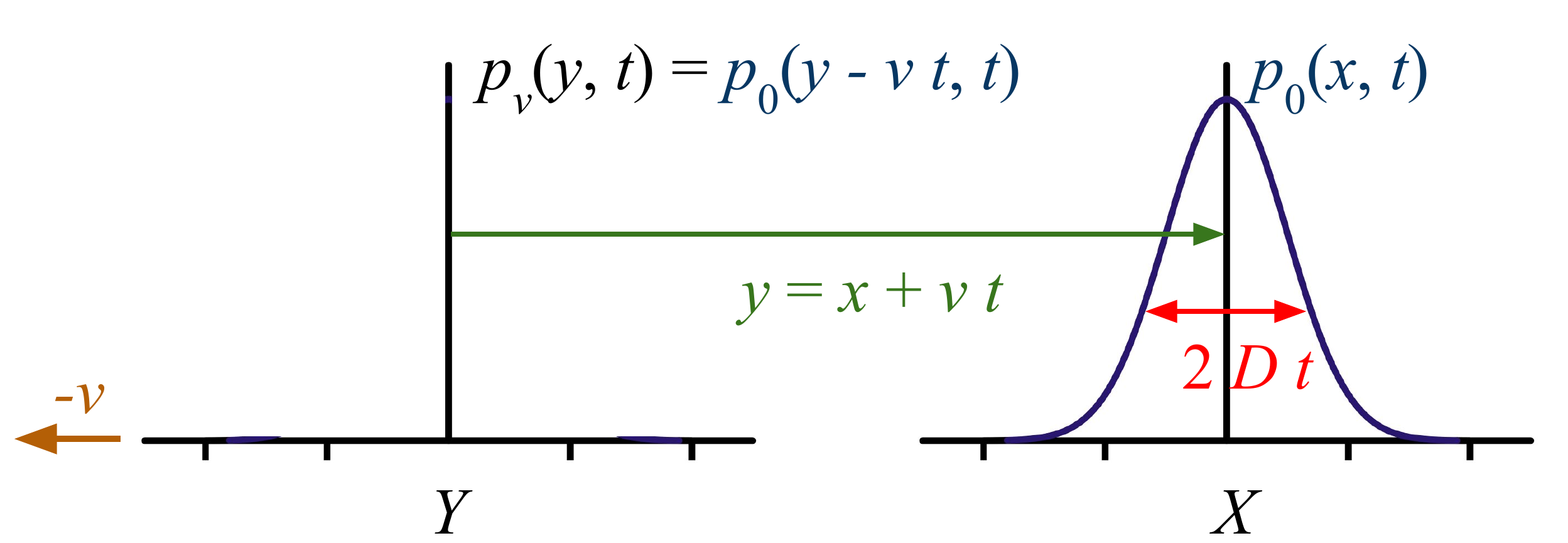}
\caption{{\em Drift induced by a change of reference frame:} A diffusive particle with no drift in a reference frame $X$, described by probability density $p^0(x,t)$, acquires a drift when observed from a reference frame $Y$ that has a velocity $-v$ with respect to $X$. In the reference frame $Y$ the diffusive particle is described by the probability density $p^v(y,t)=p^0(y-v t, t)$. }\label{f:diff_drift}
\end{figure}

Here we discuss the case of a diffusive particle with drift. The drift is associated to antisymmetric changes of the particle's probability density and, therefore, to antisymmetric operators. Furthermore, in this case the drift can be considered as due to a change of reference frame (see Fig.~\ref{f:diff_drift}). 

Indeed, consider the same diffusive particle discussed in the previous subsection, which has no drift in the reference frame $X$ utilized therein. Imagine that this particle is now described from a different reference frame $Y$, which is moving with a constant velocity $v$ along the negative $x$ direction with respect to reference frame $X$ (see Fig.~\ref{f:diff_drift}). Due to this relative motion, the same diffusive particle is observed with a drift $v$ along the positive $x$ direction. This change of reference frame amounts at a coordinate transformation 
\be\label{e:x->y}
x\to y  = x+v t.
\ee

The diffusive particle in reference frame $X$ is described by probability density $p^0(x,t)$, which satisfies Eq.~\eqref{e:diff}. So, in reference frame $Y$ the same diffusive particle is described by the probability density
\be\label{e:p_v_p0}
p^v(y,t) = p^0(y-v t, t).
\ee
The equation satisfied by $p^v$ can be obtained from Eq.~\eqref{e:diff} by observing that the change of coordinates in Eq.~\eqref{e:x->y} transforms the time and space derivatives as
\be\label{e:derivatives_v}
\left.\frac{\partial}{\partial x}\right|_{t} = \left.\frac{\partial}{\partial y}\right|_{t}\hspace{0.5cm}\text{and}\hspace{0.5cm}\left.\frac{\partial}{\partial t}\right|_{x} =\left.\frac{\partial}{\partial t}\right|_{y} +v\left.\frac{\partial}{\partial y}\right|_{t}.
\ee
So Eq.~\eqref{e:diff} transforms into 
\be\label{e:diff_drift}
\begin{split}
h_{\rm diff}\frac{\partial p^v(y, t)}{\partial t} &= \widetilde{J}^v p^v(y, t) \\
&\equiv \frac{h_{\rm diff}^2}{2 \gamma}\frac{\partial^2 p^v(y, t)}{\partial y^2} -v h_{\rm diff}\frac{\partial p^v(y ,t)}{\partial y}.
\end{split}
\ee
Equivalently, the operator $\widetilde{J}^0$ in Eq.~\eqref{e:Jdiff} transforms into (cf. Eq.~\eqref{e:Hfree_v} below)
\be\label{e:Jdiff_v}
\widetilde{J}^v = \widetilde{J}^v_s + \widetilde{J}^v_a . 
\ee
Here (cf. Eqs.~\eqref{e:Hfree_v_s} and \eqref{e:Hfree_v_a})
\BE
\widetilde{J}^v_s &=& \frac{h_{\rm diff}^2}{2 \gamma}\frac{\partial^2 }{\partial y^2},\label{e:Jdiff_v_s}\\
\widetilde{J}^v_a &=&  - v h_{\rm diff}\frac{\partial }{\partial y} ,\label{e:Jdiff_v_a}
\EE
are the symmetric and antisymmetric parts of $\widetilde{J}^v$, which describe diffusion and drift respectively (see below). 

We now describe a different way to obtain this result which serves to illustrate some ideas we use in the main text to obtain a quantum dynamics. Under the change of coordinates in Eq.~\eqref{e:x->y} the transition probability density $\mathcal{P}_\epsilon^{0}$ in Eq.~\eqref{e:diff_kernel} transforms into 
\be\label{e:mcPv}
\mcP_\epsilon^{v}(y|y^\prime) \equiv \frac{1}{Z_\epsilon}e^{-\tfrac{\epsilon}{h_{\rm diff}}\mcH_{\rm diff}^v(y,y^\prime)} ,
\ee
where (cf. Eq.~\eqref{e:p_v_p0})
\be\label{e:Hdiff_drift}
\begin{split}
\mathcal{H}_{\rm diff}^{v}(y, y^\prime) &= \mathcal{H}_{\rm diff}^{0}(y-v(t+\epsilon), y^\prime -v t) \\
&= \frac{\gamma}{2}\frac{(y - y^\prime )^2}{\epsilon^2} - \gamma v \frac{(y - y^\prime )}{\epsilon} + \frac{\gamma}{2} v^2 .
\end{split}
\ee

The second term in the second line of Eq.~\eqref{e:Hdiff_drift} is antisymmetric under permutations of $y$ and $y^\prime$, wich reflects the intrinsic directionality associated to the drift. In contrast to $\mcH_{\rm diff}^0$ in Eq.~\eqref{e:Hdiff}, which is symmetric under such permutations, $\mcH_{\rm diff}^{v}$ in Eq.~\eqref{e:Hdiff_drift} contains both terms which are antisymmetric (second term in second line) and terms which are symmetric (first and third terms in second line).  

In line with this, Eq.~\eqref{e:diff_eps} transforms into
\be\label{e:DP_diff_drift}
\Delta p^v(y, t) = \Delta_s p^v(y,t) + \Delta_a p^v(y,t),
\ee
where $\Delta p^v(y,t) = p^v(y ,t+\epsilon)-p^v(y,t)$ and
\BE
\Delta_s p^v(y,t) &\equiv& \int K_{\epsilon}^{\rm sym}(y ,y^\prime) p^v(y^\prime ,t)\mathrm{d} y^\prime\label{e:DsPdiff} \\
&=& \epsilon \frac{\epsilon h_{\rm diff}}{2\gamma}\frac{\partial^2 p^v(y,t)}{\partial y^2} + O(\epsilon^2),\nonumber\\
\Delta_a p^v(y,t) &\equiv& \int K_{\epsilon}^{\rm anti}(y ,y^\prime)p^v(y^\prime ,t)\mathrm{d} y^\prime \label{e:DaPdiff}\\
&=& - \epsilon v\frac{\partial p^v(y,t)}{\partial y} +O(\epsilon^2),\nonumber
\EE
are the symmetric and antisymmetric contributions to $\Delta p^v(y,t)$, respectively.  Here
\begin{widetext}
\BE
K_\epsilon^{\rm sym}(y,y^\prime) &\equiv& \frac{1}{2}\left[\mcP_\epsilon^{v}(y|y^\prime) +\mcP_\epsilon^{v}(y^\prime |y)\right]
= \frac{1}{Z_\epsilon}{e^{-\tfrac{\gamma (y-y^\prime)^2}{2\epsilon h_{\rm diff}} - \tfrac{\epsilon\gamma v^2}{2 h_{\rm diff}}}}\cosh\left[\frac{\gamma v (y-y^\prime)}{h_{\rm diff}} \right],\label{e:Kdiff_s}\\
K_\epsilon^{\rm anti}(y,y^\prime) &\equiv& \frac{1}{2}\left[\mcP_\epsilon^{v}(y|y^\prime) -\mcP_\epsilon^{v}(y^\prime |y)\right]
= \frac{1}{Z_\epsilon}{e^{-\tfrac{\gamma (y-y^\prime)^2}{2\epsilon h_{\rm diff}} - \tfrac{\epsilon\gamma v^2}{2 h_{\rm diff}}}}\sinh\left[\frac{\gamma v (y-y^\prime)}{h_{\rm diff}} \right],\label{e:Kdiff_a}
\EE
\end{widetext}
are the symmetric and antisymmetric parts of the corresponding transition probability density in Eq.~\eqref{e:mcPv} and we have used Eqs.~\eqref{e:mcPv} and \eqref{e:Hdiff_drift}.

To obtain the final expressions in Eqs.~\eqref{e:DsPdiff} and \eqref{e:DaPdiff} we expanded both $p^v(y^\prime , t)$ and the hyperbolic functions in Eqs.~\eqref{e:Kdiff_s} and \eqref{e:Kdiff_a} up to second order around $y$ and proceeded as in the derivation of Eq.~\eqref{e:diff_eps}.
Using Eqs.~\eqref{e:DsPdiff} and \eqref{e:DaPdiff} we can take the $\epsilon\to 0$ limit of Eq.~\eqref{e:DP_diff_drift} which, after multiplying both sides by $h_{\rm diff}$, yields Eq.~\eqref{e:diff_drift}.
A solution $p^v(x,t)$ to Eq.~\eqref{e:diff_drift} yields the probability density to find the diffusive particle in position $y$ at time $t$ in reference frame $Y$.

\subsubsection{Circular processes and imaginary-time quantum-like diffusion}\label{s:quantum_diff}

Here we briefly discuss the case of a process on a cycle 
\be\label{e:diff_factor}
\mcP_{\rm diff}(\by) = \widetilde{F}_n(y_0 , y_n) \prod_{\ell = 0}^{n-1}F_\ell(y_{\ell +1} , y_{\ell}) ,
\ee
with Gaussian factors $F_\ell$. The additional factor $\widetilde{F}_n$ turns the chain into a loop---here $y_{n+1} = y_0$. This case is associated to imaginary-time quantum dynamics and, as such, can lead to real-time quantum dynamics after implementing the SRC. In this sense, this is the proper classical analogue of the quantum dynamics of a free particle.

As discussed in Appendix~\ref{s:EQM}, it is in general not possible to write Eq.~\eqref{e:diff_factor} in a way compatible with a standard Markov chain. A way to obtain something similar to a Markov chain is to condition on the initial and final probability densities, $p^v(y_0, t)$ and $p^v(y_n , t+n\epsilon)$, so we can effectively open the circular graphical model. This kind of stochastic processes with given initial and final data were originally studied long ago by Schr\"odinger in an attempt to find the closest classical analog to Schr\"odinger equation~\cite{schrodinger1932theorie, schrodinger1931umkehrung}. So, they are often referred to as Schr\"odinger bridges---however, as far as we know, Schr\"odinger did not consider them as a way to turn a circular graphical model into one with the topology of a chain, as we do here. 

It turns out Schr\"odinger bridges can be described in terms of what we today call belief propagation algorithms (see Appendix~\ref{s:EQM}). More precisely, assume the factors $F_\ell(y,y^\prime) = e^{-\epsilon\mcH_{\rm diff}^v(y,y^\prime)}/Z_\epsilon$, where $\mcH_{\rm diff}^v$ is given in Eq.~\eqref{e:Hdiff_drift}. So, the probability density 
\be\label{e:diff_Born}
p^v(y, t) = \mu_{\to}^v(y,t)\mu_{\leftarrow}^v(y, t),
\ee
of the Schr\"odinger bridge is described in terms of messages $\mu_\to^v$ and $\mu_{\leftarrow}^v$; here $t_0\leq t\leq t_f$, where $t_0$ and $t_f$ are the initial and final times, respectively. These messages satisfy the equations
\begin{widetext}
\BE
\mu_{\to}^v(y,{t + \epsilon}) &=& \frac{1}{Z_\epsilon}\int e^{-\tfrac{\epsilon}{h_{\rm diff}}\mcH_{\rm diff}^v(y, y^\prime)}  \mu_{\to}^v(y^\prime ,t)\mathrm{d}y^\prime ,\label{e:diff_drift_dyn_mu->}\\
\mu_{\leftarrow}^v(y,{t }) &=& \frac{1}{Z_\epsilon}\int e^{-\tfrac{\epsilon}{h_{\rm diff}}\mcH_{\rm diff}^v(y^\prime, y)}  \mu_{\leftarrow}^v(y^\prime ,t +\epsilon  )\mathrm{d}y^\prime ,\label{e:diff_drift_dyn_mu<-}
\EE
\end{widetext}
as well as the initial and final conditions 
\BE
p^v(y,t_0) &=& \mu_{\to}^v(y,t_0)\mu_{\leftarrow}^v(y, t_0), \label{e:bridge_0}\\
p^v(y,t_f) &=& \mu_{\to}^v(y,t_f)\mu_{\leftarrow}^v(y, t_f).\label{e:bridge_N}
\EE
In this particular case $F_\ell(y,y^\prime) = \mcP_\epsilon^v(y|y^\prime)$ in Eq.~\eqref{e:mcPv}. In general, however, factors $F_\ell$ are not normalized. In the $\epsilon\to 0$ limit Eqs.~\eqref{e:diff_drift_dyn_mu->} and \eqref{e:diff_drift_dyn_mu<-} become (cf. Eq.~\eqref{e:diff_drift} and Appendix~\ref{s:EQM})
\BE
h_{\rm diff}\frac{\partial\mu_{\to}(y,t)}{\partial t} &=& \left[\widetilde{J}^v\mu_{\to}\right](y,t)\label{e:diff_drift_mu->}\\
&\equiv & \frac{h_{\rm diff}^2}{2 \gamma}\frac{\partial^2 \mu_{\to}(y, t)}{\partial y^2} -v h_{\rm diff}\frac{\partial \mu_\to(y ,t)}{\partial y},\nonumber\\
-h_{\rm diff}\frac{\partial\mu_{\leftarrow}(y,t)}{\partial t} &=& \left[\mu_{\leftarrow}\widetilde{J}^{v}\right](y,t)\label{e:diff_drift_mu<-}\\
&\equiv & \frac{h_{\rm diff}^2}{2 \gamma}\frac{\partial^2 \mu_{\leftarrow}(y, t)}{\partial y^2}  + v h_{\rm diff}\frac{\partial \mu_\leftarrow(y ,t)}{\partial y}.\nonumber
\EE
Notice the change in the sign of $v$ in the second equation.

Equations~\eqref{e:diff_drift_dyn_mu<-} and \eqref{e:diff_drift_mu<-} are the time reversal of Eqs.~\eqref{e:diff_drift_dyn_mu->} and \eqref{e:diff_drift_mu->}, respectively. This reflects the fact that $\mu_{\to}^v$ and $\mu_{\leftarrow}^v$ carry information about the initial and final conditions forward and backward in time, respectively. It is then convenient to write 
\BE
\mu_{\to}^v(y,t) &=& e^{R(y,t) + S(y,t)},\label{e:im-R+S}\\
\mu_{\leftarrow}^v(y,t) &=& e^{R(y,t)-S(y, t)},\label{e:im-R-S}
\EE
where $R(y,t) \to R(y,-t)$ and $S(y,t) \to -S(y,-t)$ are symmetric and antisymmetric, respectively, under time reversal (cf. Ref.~\cite{Zambrini-1987} after Eq.~(2.20); see also Eq.~(2.47) therein). Furthermore, according to Eq.~\eqref{e:diff_Born}, we have $p^v(y, t) = e^{2 R(y, t)} $, so 
\BE
\mu_{\to}^v(y , t) &=& \sqrt{p^v(y, t)}e^{S (y, t)},\label{e:im-p+S}\\
\mu_{\leftarrow}^v(y , t) &=& \sqrt{p^v(y, t)}e^{-S (y, t)}.\label{e:im-p-S}
\EE
%

\subsubsection{``Form invariant'' equations}\label{s:diff_inv}

Equations~\eqref{e:diff_drift_mu->} and \eqref{e:diff_drift_mu<-} correctly describes the Schr\"odinger bridge in reference frame $Y$. However, physicists prefer that equations describing ``fundamental laws of nature'' remain ``form invariant'' under changes of coordinates like that in Eq.~\eqref{e:x->y}. Granted, Eqs.~\eqref{e:diff_drift_mu->} and \eqref{e:diff_drift_mu<-} are not considered fundamental equations in this sense. However, the Schr\"odinger equation which we obtain based on them is indeed considered a fundamental equation. 

The main point we want to make is that we can obtain the dynamics of the probability density in reference frame $Y$ either by explicitly solving Eqs.~\eqref{e:diff_drift_mu->} and \eqref{e:diff_drift_mu<-} or by using ideas of ``form invariance''; the choice is optional. In the quantum analogue discussed below, more relevance is usually given to ``form invariance''. However, in that case too we can pragmatically obtain the correct wave function by solving the analogues of Eqs.~\eqref{e:diff_drift_mu->} and \eqref{e:diff_drift_mu<-}.

So, we now illustrate how Eqs.~\eqref{e:diff_drift_mu->} and \eqref{e:diff_drift_mu<-}, describing the system in reference frame $Y$, can be kept formally equivalent to the corresponding equations in reference frame $X$, i.e. to Eqs.~\eqref{e:diff_drift_mu->} and \eqref{e:diff_drift_mu<-} with $v=0$. This closely mirrors the situation with the Schr\"odinger equation described in Appendix~\ref{s:gaussian-complex}.

To do so, notice that
\be\label{e:Hv_H0}
\mcH_{\rm diff}^v(y, y^\prime) = \mcH_{\rm diff}^0(y,y^\prime) +  \frac{\Delta f_\gamma (y,y^\prime)}{\epsilon} ,
\ee
where $\Delta f_\gamma(y,y^\prime) = f_\gamma(y,t+\epsilon) - f_\gamma(y^\prime,t)$ with (here $c=\gamma$)
\be\label{e:f_z}
f_c(y,t) = -c v y + \frac{c}{2}v^2 t . 
\ee
Following Eqs.~\eqref{e:Hv_H0} and \eqref{e:f_z}, we can rewrite Eqs.~\eqref{e:diff_drift_dyn_mu->} and \eqref{e:diff_drift_dyn_mu<-} as (cf. Eq.~\eqref{e:free_dyn'})
\begin{widetext}
\BE
e^{f_\gamma(y, t+ \epsilon)/h_{\rm diff}}\mu_{\to}^v(y,{t + \epsilon}) &=& \frac{1}{Z_\epsilon}\int e^{-\tfrac{\epsilon}{h_{\rm diff}}\mcH_{\rm diff}^0(y, y^\prime)} \left[e^{f_\gamma(y^\prime, t)/h_{\rm diff}} \mu_{\to}^v(y^\prime ,t)\right]\mathrm{d}y^\prime ,\label{e:diff_drift_dyn_mu->f}\\
e^{-f_\gamma(y, t)/h_{\rm diff}}\mu_{\leftarrow}^v(y,{t }) &=& \frac{1}{Z_\epsilon}\int e^{-\tfrac{\epsilon}{h_{\rm diff}}\mcH_{\rm diff}^0(y^\prime, y)} \left[e^{-f_\gamma(y^\prime, t+\epsilon)/h_{\rm diff}} \mu_{\leftarrow}^v(y^\prime ,t +\epsilon  )\right]\mathrm{d}y^\prime .\label{e:diff_drift_dyn_mu<-f}
\EE
\end{widetext}
So, we can see that the modified messages  (notice the tildes)
\BE
\widetilde{\mu}_{\to}(y,{t }) &=& e^{f_\gamma(y, t)/h_{\rm diff}}\mu_{\to}^v(y,{t }),\label{e:mu->tilde}\\
\widetilde{\mu}_{\leftarrow}(y,{t }) &=& e^{-f_\gamma(y, t)/h_{\rm diff}}\mu_{\leftarrow}^v(y,{t }),\label{e:mu<-tilde}
\EE
satisfy the driftless diffusion equation and its conjugate, i.e. Eqs~\eqref{e:diff_drift_mu->} and \eqref{e:diff_drift_mu<-} with $\widetilde{J}^v$ replaced by $\widetilde{J}^0$. Importantly, this modification does not change the probabilities since $\widetilde{\mu}_\to(y,t)\widetilde{\mu}_\leftarrow(y,t) = \mu_\to^v(y,t)\mu_\leftarrow^v(y,t) = p^v(y,t)$. So, we can say that the dynamical equation remains ``form invariant'' under transformations like that in Eq.~\eqref{e:x->y}, but the messages transform according to Eqs.~\eqref{e:mu->tilde} and \eqref{e:mu<-tilde}. 

\subsubsection{Formulation in terms of real probability matrices}\label{s:prob-matrix}

Finally, it will be useful to extend this analogy with quantum mechanics by describing the free particle using probability matrices, in terms of which we implement the SRC in the main text. This is because density matrices, the quantum analogue of probability matrices, have a more direct connection to probabilities than ``wave functions'' do. Indeed, while the diagonal of a density matrix already yields the relevant probabilistic information, a wave function needs to be multiplied by ``another'' wave function, its conjugate, to obtain probabilistic information. 

introducing the real probability matrix (cf. Eq.~\eqref{e:rho_free} below)
\be
P^v(y,y^\prime; t) = \mu_{\to}^v(y,t)\mu_{\leftarrow}^v(y^\prime , t),
\ee
whose diagonal gives the probabilities $P^v(y,y;t ) = p^v(y,t)$. Taking the time derivative of $P^v$ and using Eqs.~\eqref{e:diff_drift_mu->} and \eqref{e:diff_drift_mu<-} yields
\be\label{e:realvNfree}
h_{\rm diff}\frac{\partial P^v}{\partial t} = [\widetilde{J}^v, P^v] = [\widetilde{J}^v_s, P^v] + [\widetilde{J}^v_a, P^v],
\ee
where $[A, B] = AB - BA$ is the commutator between operators, or matrices, $A$ and $B$. Similarly, Eq.~\eqref{e:DP_diff_drift} becomes 
\be\label{e:DPmatrix_diff_drift}
\Delta P^v = \Delta_s P^v +\Delta_a P^v,
\ee
where 
\be\label{e:DsP_DaP}
\Delta_s P^v = [J^v_s, P^v]\hspace{0.5cm}\text{and}\hspace{0.5cm} \Delta_a P^v = [J_a^v ,P^v],
\ee
with $J_s^v = \widetilde{J}_s^v/h_{\rm diff}$ and $J_a^v = \widetilde{J}_a^v/h_{\rm diff}$ (notice the tildes;  cf. Eqs.~\eqref{e:Drho_free_drift} and~\eqref{e:Ds_rho_Da_rho} below).

%

\subsection{Non-relativistic quantum free particle in terms of real Gaussian kernels}\label{s:gaussian-quantum}
\subsubsection{Standard formulation using complex Gaussian kernels}\label{s:gaussian-complex}
To provide some intuition to those who may not be familiar with quantum mechanics, we here discuss the case of a non-relativistic free particle, which closely parallels the case of a classical diffusive particle described in Appendix~\ref{s:diff}. We also discuss how a drift due to a change of reference frame leads to antisymmetric changes accompanied by a factor $i$. In the main text we take insight from this example to deal with complex Hamiltonians. 

The probability {\em amplitude} for a free particle of mass $m$ to transition from position $x$ at time $t$ to position $x^\prime$ at time $t+\epsilon$ is given by a complex Gaussian kernel (cf. Eq.~\eqref{e:diff_kernel})
\be
\begin{split}\label{e:free_kernel}
\mathcal{G}_\epsilon^{0} (x, x^\prime) &= \frac{1}{\sqrt{2 i \pi \hbar \epsilon/m}} e^{i m (x-x^\prime)^2/2\hbar\epsilon}\\
&\equiv \frac{1}{\mathcal{A}_\epsilon}e^{i\tfrac{\epsilon}{\hbar}\mathcal{L}_{\rm free}^0(x,x^\prime)}, 
\end{split}
\ee
where $\hbar$ is Planck constant and (cf. Eq.~\eqref{e:Hdiff})
\be\label{e:Lfree}
\mathcal{L}_{\rm free}^0(x, x^\prime) = \frac{m}{2}\frac{(x - x^\prime)^2}{\epsilon^2} ,
\ee
is the so-called Lagrangian which, for the case of a free particle, coincides with the Hamiltonian function, i.e. its (kinetic) energy. Here $\mathcal{A}_\epsilon = \sqrt{i 2\pi \hbar\epsilon / m}$ plays the role of a normalization factor. 

If $\psi_0(x^\prime, t)=\sqrt{p^0(x,t)}e^{i\varphi(x,t)}$ is the probability {\em amplitude}, or wave function, for the free particle to be at position $x^\prime$ at time $t$, the corresponding probability is given by the Born rule $p^0(x,t) = \psi_0^\ast(x,t) \psi_0^\ast(x,t)$. Here the real function $\varphi(x,t)$ is the phase of the wave function and $\psi_0^\ast(x,t) = \sqrt{p^0(x,t)}e^{-i\varphi(x,t)}$ is the complex conjugate of $\psi_0(x,t)$. The probability amplitude for the particle to be at position $x$ at time $t+ \epsilon$ is given by (cf. Eq.~\eqref{e:diff_dyn})
\be\label{e:free_dyn}
\psi_0(x,{t + \epsilon}) = \int \mathcal{G}^{0}_\epsilon (x ,x^\prime)\psi_0(x^\prime ,t)\mathrm{d}x^\prime .
\ee
For $\epsilon\to 0$ the complex Gaussian kernel in Eq.~\eqref{e:free_kernel} oscillates very fast except in the region $|x-x^\prime |= O(\sqrt{\hbar\epsilon/ m})$ which makes contributions outside this range negligible. So, as in the case of a diffusive particle, we can expand $\psi_0(x^\prime, t)$ up to second order around $x$ to obtain the right hand side of Eq.~\eqref{e:free_dyn} up to first order in $\epsilon$. This leads to similar results as those of the diffusive particle, only that in this case the ``variance'' of the Gaussian kernel is complex, i.e. $\sigma^2 = \tfrac{i\hbar\epsilon}{m}$. So (cf. Eq.~\eqref{e:diff_eps})
\be\label{e:free_eps}
\psi_0(x, {t + \epsilon}) = \psi_0(x, t) + i \epsilon \frac{\hbar}{2 m}\frac{\partial^2 \psi_0(x , t)}{\partial x^2} + O(\epsilon^2),
\ee
or, taking the limit $\epsilon\to 0$ and multiplying both sides by $i \hbar$ (cf. Eq.~\eqref{e:diff}),
\be\label{e:free}
i \hbar\frac{\partial \psi_0(x, t)}{\partial t} = -\frac{\hbar^2}{2 m}\frac{\partial^2 \psi_0(x, t)}{\partial x^2} .
\ee
Equation~\eqref{e:free} has the same form of Eq.~\eqref{e:diff}, except for the imaginary unit and the minus sign in the left and right hand sides, respectively. Planck constant, $\hbar$, has units of energy and time, unlike $h_{\rm diff}$ which has units of energy only; this is related to the coefficient of friction, $\gamma$ in Eq.~\eqref{e:diff} having units of mass over time instead of mass only.

Equation~\eqref{e:free} is usually written as (cf. Eq.~\eqref{e:Jdiff_p0})
\be\label{e:Hfree_psi0}
i \hbar\frac{\partial \psi_0}{\partial t} = H^0\psi_0,
\ee
where
\be\label{e:Hfree}
H^0 =  -\frac{\hbar^2}{2 m}\frac{\partial^2 }{\partial x^2} ,
\ee
is the Hamiltonian operator associated to a free particle. From Eq.~\eqref{e:Jdiff} we see that $H^0=-\widetilde{J}^0$.

We now illustrate how a drift due to a change of reference frame leads to antisymmetric changes accompanied by an $i$. We used this example in the main text to extend our results to complex Hamiltonians. So, consider an observer moving with a constant velocity $-v$ along the positive $x$ direction, who would therefore observe the same particle with an additional velocity $v$. This amounts at the coordinate transformation in Eq.~\eqref{e:x->y} which, using Eq.~\eqref{e:derivatives_v}, transforms Eq.~\eqref{e:free} into (cf. Eq.~\eqref{e:diff_drift})
\be\label{e:free_drift}
i \hbar\frac{\partial \psi_v(y, t)}{\partial t} = -\frac{\hbar^2}{2 m}\frac{\partial^2 \psi_v(y, t)}{\partial y^2} -i v\hbar \frac{\partial \psi_v(y ,t)}{\partial y} .
\ee
Equivalently, the Hamiltonian operator $H^0$ in Eq.~\eqref{e:Hfree} transforms into (cf. Eq.~\eqref{e:Jdiff_v})
\be\label{e:Hfree_v}
H^v = H^v_s +  H^v_a / i
\ee
where
\BE
H^v_s  &=& -\frac{\hbar^2}{2 m}\frac{\partial^2 }{\partial y^2},\label{e:Hfree_v_s} \\
H^v_a  &=&  v\hbar\frac{\partial }{\partial y} ,\label{e:Hfree_v_a}
\EE
are associated to the symmetric and antisymmetric parts of an underlying stochastic process---indeed, notice that $H^v = - (\widetilde{J}^v_s +  \widetilde{J}^v_a / i) $ (see below; cf. Eqs.~\eqref{e:Jdiff_v_s} and \eqref{e:Jdiff_v_a}). 

As in the case of the diffusive particle, Eq.~\eqref{e:free_drift} can be obtained in a different way by noticing that the function $\mathcal{L}_{\rm free}^0$ in Eq.~\eqref{e:Lfree} transforms into (cf. Eq.~\eqref{e:Hdiff_drift} and \eqref{e:Hv_H0})
\be\label{e:Lfree_drift}
\begin{split}
\mathcal{L}_{\rm free}^{v}(y, y^\prime) &= \frac{m}{2}\frac{(y - y^\prime - \epsilon v)^2}{\epsilon^2} \\
&= \frac{m}{2}\frac{(y - y^\prime )^2}{\epsilon^2} - m v \frac{(y - y^\prime )}{\epsilon} + \frac{m}{2} v^2 \\
&= \mathcal{L}_{\rm free}^0 + \frac{\Delta f_m}{\epsilon},
\end{split}
\ee
where $\Delta f_m = f_m(y,t+\epsilon) - f_m(y^\prime,t)$ and $f_m$ is obtained from Eq.~\eqref{e:f_z} by doing $c=m$. The second term in the second line of Eq.~\eqref{e:Lfree_drift} is antisymmetric under permutations of $y$ and $y^\prime$, which reflects the intrinsic directionality associated to the moving reference frame. In contrast to $\mathcal{L}_{\rm free}^0$ in Eq.~\eqref{e:Lfree}, which is symmetric under such permutations, $\mathcal{L}_{\rm free}^{v}$ in Eq.~\eqref{e:Lfree_drift} contains both terms which are antisymmetric (second term in second line) and terms which are symmetric (first and third terms in second line). 

This turns Eq.~\eqref{e:free_dyn} into
.%
\be\label{e:free_drift_dyn}
\psi_v(y,{t + \epsilon}) = \int \frac{1}{\mathcal{A}_\epsilon}e^{i\tfrac{\epsilon}{\hbar}\mathcal{L}_{\rm free}^{v}(y ,y^\prime)}\psi_v(y^\prime ,t)\mathrm{d}y^\prime ,
\ee
Equation~\eqref{e:free_drift} can be obtained from Eq.~\eqref{e:free_drift_dyn} in the same way that Eq.~\eqref{e:free} was obtained from Eq.~\eqref{e:free_dyn}. However, we will derive this result using real Gaussian kernels in Appendix~\ref{s:gaussian-real}, which more closely parallels the case of the classical diffusive particle. 
Using Eq.~\eqref{e:free_drift_dyn} and the third line of Eq.~\eqref{e:Lfree_drift} we can write (cf. Eqs.~\eqref{e:diff_drift_dyn_mu->f} and \eqref{e:diff_drift_dyn_mu<-f})
\begin{widetext}
\be\label{e:free_dyn'}
\psi_v(y,{t + \epsilon})e^{-i f_m(y, t+\epsilon)/\hbar} = \int \mathcal{G}_{\epsilon}^{0}(y ,y^\prime)\left[ \psi_v(y^\prime ,t)e^{-i f_m(y^\prime, t)/\hbar}\right]\mathrm{d}y^\prime .
\ee
\end{widetext}
Comparing Eq.~\eqref{e:free_dyn'} to Eq.~\eqref{e:free_dyn} we see that 
\be\label{e:free_trans}
\psi_0(x,t) = \psi_v(x,t) e^{-i f_m(x,t)/\hbar}.
\ee
So, we can equivalently say that Schr\"odinger equation remains ``invariant'', but the wave function between the two observers transform according to Eq.~\eqref{e:free_trans}, in such a way that if we know one of them we can compute the other through this equation. We want to emphasize, though, that from a practical point of view, we can also obtain the wave function that yields the correct probabilities by directly solving Eq.~\eqref{e:free_drift}. 

The probability density associated to a wave function is obtained via the Born rule $p^v(y,t) = \psi_v(y,t)\psi_v^\ast(y,t)$, where $\psi_v^\ast$ is the complex conjugate of $\psi_v$. A way to deal simultaneously with both the quantum dynamics and the Born rule is to work in terms of a complex probability matrix, or {\em density matrix}
\be\label{e:rho_free}
\rho_v(y,y^\prime; t) = \psi_v(y,t)\psi_v^\ast(y^\prime , t),
\ee
whose diagonal yields precisely the probability density associated to the corresponding wave function, i.e. $p^v(y, t) = \rho_v(y,y; t)$. We can obtain the equation governing the dynamics of $\rho_v$ by taking the time derivative of Eq.~\eqref{e:rho_free} and using Eq.~\eqref{e:free_drift} and its conjugate to replace the time derivatives of $\psi_v$ and $\psi_v^\ast$, respectively. This yields (see Eq.~\eqref{e:Hfree_v})
\be\label{e:vNfree}
i\hbar\frac{\partial\rho_v}{\partial t} = [H^v, \rho_v] = [H_s^v, \rho_v] + [H_a^v,\rho_v] / i ,
\ee
where $[A,B] = A B-B A$ is the commutator between operators, or matrices, $A$ and $B$. Equation~\eqref{e:vNfree} is the von Neumann equation equivalent to Schr\"odinger equation.
\subsubsection{Formulation in terms of real Gaussian kernels}\label{s:gaussian-real}

The close analogy between the derivation of Eqs.~\eqref{e:free} and \eqref{e:diff} suggests it is possible to derive the latter in terms of real Gaussian kernels too. Indeed, by following the same steps that led to Eq.~\eqref{e:diff} we can show that 
\be\label{e:eps_psi}
\epsilon\frac{\partial\psi_0(x,t)}{\partial t} = i \left\{\left[\mcK_\epsilon^0\ast\psi_0\right](x,t)  - \psi_0(x,t)\right\},
\ee
with
\BE
\mcK_\epsilon^0(x,x^\prime) &=& \frac{1}{|\mathcal{A}_\epsilon|}e^{-\tfrac{\epsilon}{\hbar}\mathcal{L}_{\rm free}^0},\label{e:mcK0} \\
\left[\mcK_\epsilon^0\ast\psi_0\right](x,t) &=& \int\mcK_\epsilon^0(x,x^\prime)\psi_0(x^\prime,t)\mathrm{d}x^\prime, 
\EE
leads precisely to Eq.~\eqref{e:Hfree_psi0}---in more general situations, the exponent would not be given by the same Lagrangian function but by a new type of ``energy'' function (see Appendix~\ref{s:EM} for the nontrivial example of a quantum charged particle in a classical electromagnetic field).

As in the case of a diffusive particle, here the kernel associated to an observer moving with velocity $v$ in the positive $x$ direction is obtained form Eq.~\eqref{e:mcK0} by simply changing $\mathcal{L}_{\rm free}^0$ for the corresponding $\mathcal{L}_{\rm free}^v$ in Eq.~\eqref{e:Lfree_drift}, i.e.
\be\label{e:mcKv}
\mcK_\epsilon^v(y,y^\prime) = \frac{1}{|\mathcal{A}_\epsilon|}e^{-\tfrac{\epsilon}{\hbar}\mathcal{L}_{\rm free}^v (y,y^\prime)},
\ee
which, except for the parameters $m$ and $\hbar$, is the same as Eq.~\eqref{e:mcPv}.

However, as already hinted by Eq.~\eqref{e:Hfree_v}, in contrast to Eq.~\eqref{e:DP_diff_drift} the antisymmetric contributions need to be divided by the imaginary unit (cf. Appendix~\ref{s:pair}). So, the analogous of Eq.~\eqref{e:DP_diff_drift} is
\be\label{e:Dpsi_free_drift}
\Delta \psi_v(y, t) = i\left[ \Delta_s \psi_v(y,t) + \Delta_a \psi_v(y,t) / i\right] ,
\ee
where $\Delta \psi_v(y,t) = \psi_v(y ,t+\epsilon)-\psi_v(y,t)$  and the $i$ outside the square brackets comes from the structure of Eq.~\eqref{e:eps_psi}, and is associated to the SRC in the main text. Furthermore (cf. Eqs.~\eqref{e:DsPdiff} and \eqref{e:DaPdiff}),
\BE
\Delta_s \psi_v(y,t) &\equiv& \int \mcK_{\epsilon}^{\rm sym}(y ,y^\prime) \psi_v(y^\prime ,t)\mathrm{d} y^\prime\label{e:Ds_psi_free} \\
&=& \frac{\epsilon\hbar}{2 m} \frac{\partial^2 \psi_v(y,t)}{\partial y^2} + O(\epsilon^2),\nonumber\\
\Delta_a \psi_v(y,t) &\equiv& \int \mcK_{\epsilon}^{\rm anti}(y ,y^\prime)\psi_v(y^\prime ,t)\mathrm{d} y^\prime \label{e:Da_psi_free}\\
&=& -\epsilon v\frac{\partial \psi_v(y,t)}{\partial y} +O(\epsilon^2),\nonumber
\EE
are the symmetric and antisymmetric contributions to $\Delta \psi_v(y,t)$, respectively. Here (cf. Eqs.~\eqref{e:Kdiff_s} and \eqref{e:Kdiff_a})
\begin{widetext}
\BE
\mathcal{K}_\epsilon^{\rm sym}(y,y^\prime) &\equiv& \frac{1}{2}\left[\mcK_\epsilon^{v}(y,y^\prime) +\mcK_\epsilon^{v}(y^\prime ,y)\right]
= \frac{1}{Z_\epsilon}{e^{-\tfrac{m (y-y^\prime)^2}{2\epsilon \hbar} - \tfrac{\epsilon m v^2}{2 \hbar}}}\cosh\left[\frac{m v (y-y^\prime)}{\hbar} \right],\label{e:Kfree_s}\\
\mathcal{K}_\epsilon^{\rm anti}(y,y^\prime) &\equiv& \frac{1}{2}\left[\mcK_\epsilon^{v}(y,y^\prime) -\mcK_\epsilon^{v}(y^\prime ,y)\right]
= \frac{1}{Z_\epsilon}{e^{-\tfrac{m (y-y^\prime)^2}{2\epsilon \hbar} - \tfrac{\epsilon m v^2}{2 \hbar}}}\sinh\left[\frac{m v (y-y^\prime)}{\hbar} \right],\label{e:Kfree_a}
\EE
\end{widetext}
are the symmetric and antisymmetric parts of the corresponding transition kernel $\mcK_\epsilon^v$ in Eq.~\eqref{e:mcKv}. To obtain the final expressions in Eqs.~\eqref{e:Ds_psi_free} and \eqref{e:Da_psi_free} we expanded both $\psi_v(y^\prime , t)$ and the hyperbolic functions in Eqs.~\eqref{e:Kfree_s} and \eqref{e:Kfree_a} up to second order around $x$ and proceed as in the case of a diffusive particle. Using Eqs.~\eqref{e:Ds_psi_free} and \eqref{e:Da_psi_free} we can take the $\epsilon\to 0$ limit of Eq.~\eqref{e:Dpsi_free_drift} which, after multiplying both sides by $\hbar$, yields Eq.~\eqref{e:free_drift}. 

\subsubsection{Formulation in terms of real probability matrices}\label{s:gaussian-matrix}
Equation~\eqref{e:Dpsi_free_drift} is better understood in terms of Eq.~\eqref{e:vNfree}, which can be written as (cf. Eq.~\eqref{e:DPmatrix_diff_drift})
\be\label{e:Drho_free_drift}
\Delta\rho_v = i\left(\Delta_s\rho_v + \Delta_a\rho_v / i\right)
\ee
where (cf. Eq.~\eqref{e:DsP_DaP})
\be\label{e:Ds_rho_Da_rho}
\Delta_s\rho_v = [{J}_s,\rho_v],\hspace{0.5cm}\text{and}\hspace{0.5cm}
\Delta_a\rho_v = [{J}_a,\rho_v],
\ee
with $J_s = -H_s/\hbar$ and ${J}_a = -H_a/\hbar$. The main differences between Eqs.~\eqref{e:Drho_free_drift} and \eqref{e:DPmatrix_diff_drift} are that: (i) $\rho_v$ is complex while $P^v$ is real; (ii) the antisymmetric contriubution $\Delta_a\rho_v$ is accompanied by an $i$ while $\Delta_a P^v$ is not; (iii) There is an external $i$ outside the brackets of the right hand side of Eq.~\eqref{e:Drho_free_drift} that is absent in Eq.~\eqref{e:DPmatrix_diff_drift}. 

Let us write $\rho_v = P_s + P_a / i$ in terms of symmetric and antisymmetric matrices, $P_s = P_s^T$ and $P_a = -P_a^T$. Separating Eq.~\eqref{e:Drho_free_drift} into its real and imaginary parts we obtain a pair of equations (cf. Appendix~\ref{s:pair})
\BE
\Delta P_s &=& [J_s, P_a] + [J_a , P_s],  \\
\Delta P_a &=& -[J_s P_s] + [J_a ,  P_a ] .  
\EE
Furthermore, by writing $P_a = (P + P^T)/2$ and $P_a = (P-P^T)/2$ as the symmetric and antisymmetric parts of a probability matrix $P$ we obtain
\BE
\Delta P^{A} &=& -[J_s , P^B] + [J_a , P^A], \label{e:DtPAfreepar} \\
\Delta P^B &=& [J_s , P^A ] + [J_a, P^B]. \label{e:DtPBfreepar}
\EE
Here $P^A = P$ and $P^B = P^T$ can be interpreted as the real probability matrices of two observers that mutually observe each other in order to be able to refer to themselves (see main text). These equations are somehow similar to Eq.~\eqref{e:DPmatrix_diff_drift}.

\

\section{Modeling scientists doing experiments}\label{s:model_scientists}

Here we discuss some details of  our approach to model scientists doing experiments based on some concepts of modern cognitive science. We first discuss two well-known modeling frameworks, i.e., active inference (Appendix~\ref{s:active}) and enactive cognition (Appendix~\ref{s:enactive}), that are relevant for our purpose. However, we take a more relational approach than traditionally done in these two modeling frameworks. Indeed, in line with the relational interpretation of quantum mechanics (RQM)~\cite{Rovelli-1996} (see Appendix~\ref{s:wigner}), the modeling of a scientist doing an experiment is done from the perspective of another scientist (Appendix~\ref{s:relational_cognition}). 
This allows us to implement the self-referential coupling discussed in the main text. We expect our approach can offer some fresh perspective on cognitive science models. 

\subsection{Active inference: world as a generative process, scientists as generative models}\label{s:active}
\begin{figure*}
\includegraphics[width=0.8\textwidth]{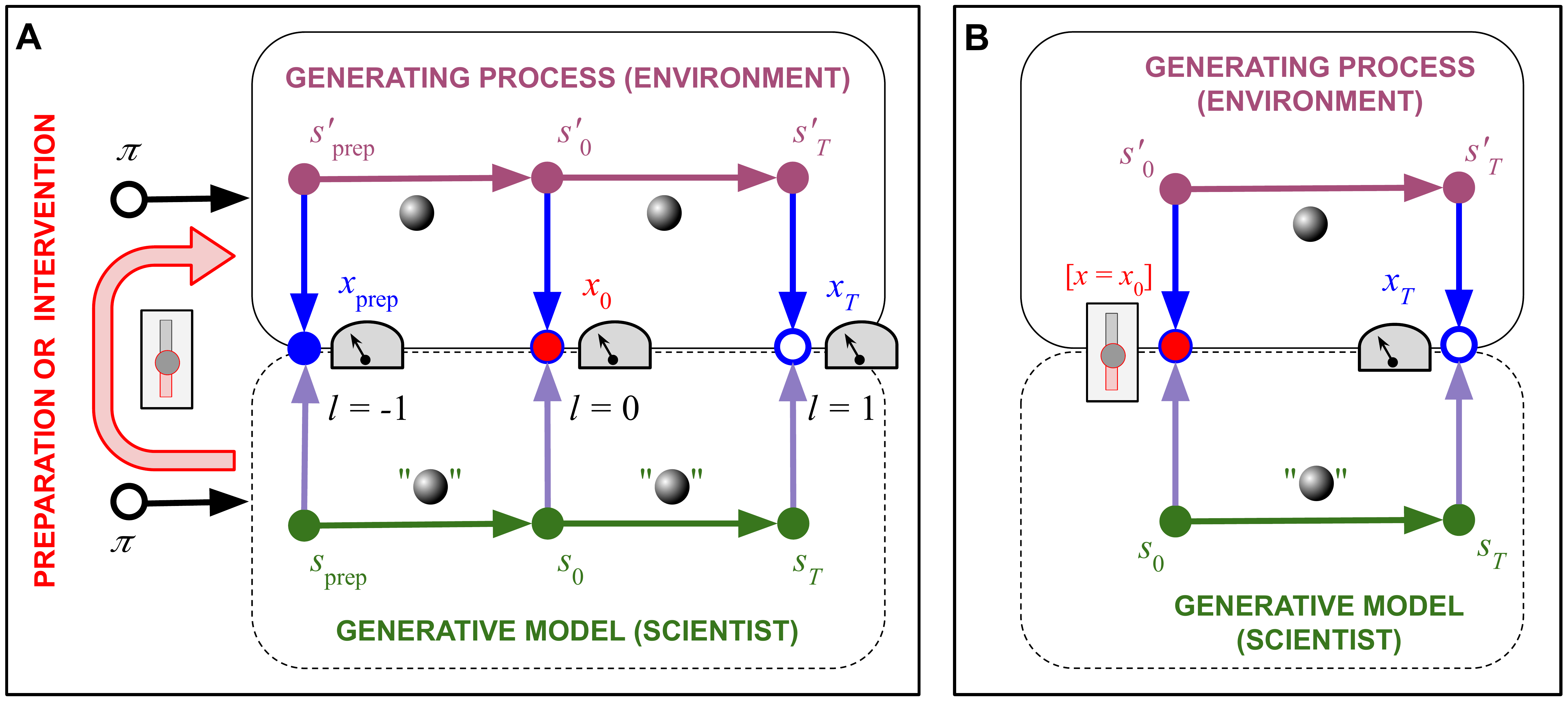}
\caption{{\em Active inference:} (A) Graphical model characterizing active inference (cf. Figs. 1 and 2 in Ref.~\cite{friston2017graphical} as well as Figs.~2 and 3 in Ref.~\cite{schwobel2018active}). The upper graphical model enclosed within a solid line is the {\em generative process} associated to the external system. The only accessible information about this generative process is the data it generates on the observer's sensors. The lower graphical model enclosed within a dashed line is the {\em generative model} the observer has about the external world. (B) When the scientist performs the required actions to consistently transform the variable position $x_{\rm prep}$ into the same initial position $x_0$, she effectively removes all causal dependencies before the start of the experiment at time step $\ell=0$. This could be interpreted as a form of causal intervention on the system. We denote this here as $[x=x_0]$ to technically differentiate it from Pearl's do-calculus, which does not necessarily model the scientist implementing the intervention~\cite{pearl2009causality}. }\label{f:active}
\end{figure*}

Here we briefly discuss some aspects of active inference in the framework of a scientist carrying out an experiment. Although we present some technical details for the reader that may not be familiar with it, our main purpose is to highlight the main underlying concepts. In active inference the external world---an experimental system in this case---is considered as a generative {\em process}, while the organism---here a scientist---perceiving, interacting with, and learning about such an external world is considered as (or to have) a generative {\em model} (see Fig.~\ref{f:active}; cf. Fig.~2 in Ref.~\cite{friston2017graphical} and Figs. 1 and 2 in Ref.~\cite{schwobel2018active}). We discuss these in the next subsections, closely following Ref.~\cite{schwobel2018active}.

\subsubsection{Experimental systems as generative processes}\label{s:gen_process}

Following active inference, the scientist's (controlled) environment, i.e. the experimental system, is considered hidden to her; she can only indirectly access it by the data it generates in her sensorium via her observations. In Fig.~\ref{f:active}A we represent the environment by a Bayesian network enclosed within a solid rounded rectangle, which depends on the actions of the organism (external arrow pointing towards the solid rounded rectangle; cf. Fig. 2 in Ref.~\cite{friston2017graphical}; see Sec.~2.1 in Ref.~\cite{schwobel2018active}). Accordingly, the state of the environment at time step $\ell$ is described by hidden variables $s_\ell^\prime$ (top dark magenta circles) which can generate an observation $x_\ell$ (center blue and red circles) with a probability $\Omega_\ell(x_\ell | s_\ell^\prime)$ (blue arrows pointing downwards).

The environment dynamics is specified by the transition probability $\Theta_\ell(s_{\ell+1}^\prime | s_\ell^\prime , a_\ell)$ that the environment is in state $s_{\ell +1}^\prime$ at time step $\ell +1$, given that at the previous time step its state was $s_\ell^\prime$ and the scientist performed action $a_\ell$, e.g., by moving some knobs. The dynamical dependency between hidden variables is represented in Fig.~\ref{f:active}A by the top horizontal dark magenta arrows. The dependency of these dynamics on the scientist's actions is represented by the black arrow external to the solid rounded rectangle and pointing towards it. This is to emphasize that the scientist can select a whole sequence of actions according to a behavioral policy~\cite{schwobel2018active,friston2017graphical}, $\pi$, as discussed in the next subsection. 

To keep the discussion at the minimal level of complexity required to illustrate the relevant concepts for our purpose, we focus here only on three time steps, $\ell = -1,0, 1$ (see Fig.~\ref{f:active}A). However, each transition from a time step $\ell$ to the next $\ell+1$ can be partitioned into as many time steps as desired~\cite{schwobel2018active}.

\subsubsection{Scientists as generative models}\label{s:gen_model}

Following active inference, the scientist is considered to be, or to have physically encoded in her neural system and perhaps body, a generative model of her (controlled) environment, i.e. of the experimental system. This generative model is represented in Fig.~\ref{f:active}A by a Bayesian network within a dashed rounded rectangle, which mirrors the Bayesian network representing the environment. The generative model is defined as a joint probability distribution over observations $x_\ell$ (middle blue and red circles), internal ``copies'' $s_\ell$ of the environment's hidden states $s_\ell^\prime$ (bottom green circles), which are encoded in the scientist's neural system or body, and behavioral policies $\pi$ (black node external to the solid rounded square). The latter could be specified, for instance, by a sequence of control states $u_\ell$ (see Sec.~2.2 in Ref.~\cite{schwobel2018active}), i.e. $\pi = (u_{-1}, u_0 , u_1)$, which denote a subjective abstraction of an action, such as a neuronal command to execute a specific action in the environment~\cite{schwobel2018active}. In Ref.~\cite{schwobel2018active} a one-to-one mapping is assumed between a selected control state $u_\ell$ and executed action $a_\ell$ in each time step $\ell$.

The generative model is represented in Fig.~\ref{f:active}A by a Bayesian network within a dashed rounded rectangle, which mirrors the Bayesian network representing the environment. It can be written as~\cite{schwobel2018active} (see Eq.~(2.4) therein)
\begin{widetext}
\be\label{e:GM}
\mathcal{P}^{\rm gen}(\bx, \bs, \pi) = p^{\rm pol}(\pi) p_{-1}(s_{-1})\prod_{\ell = 0}^1 \mathcal{P}^{\rm obs}_\ell(x_\ell | s_\ell)\mathcal{P}_\ell^{\rm dyn}(s_\ell |s_{\ell-1}, \pi ),
\ee
\end{widetext}
where $\bx = (x_{-1},x_0, x_1)$ and $\bs = (s_{-1}, s_0 , s_1)$. Here $\mcP^{\rm dyn}$ (bottom horizontal green arrows in Fig.~\ref{f:active}A) specifies the scientist's model of the environment's hidden dynamics, which can be affected by the actions the scientist performs according to the behavioral policy $\pi$. Furthermore, $\mcP_\ell^{\rm obs}$ (bottom purple arrows pointing upwards in Fig.~\ref{f:active}A) specifies the model of how hidden states of the environment generate observations. Finally, $p_{-1}$ and $p^{\rm pol}$ are priors over the initial state of the environment and the policy, respectively.

Now, when carrying out an experiment a scientist first prepares the state of the experimental system at the start of the experiment, i.e., at time step $\ell= 0$. Say the experimental system is a particle in a piece-wise linear potential (see Fig.~\ref{f:circular}A in the main text). This could be done, for instance, by performing a measurement at a previous time step, $\ell = -1$, say of the position of the particle $x_{-1} = x_{\rm prep}$ as displayed in a reading device---this would correspond to the first time step in Fig.~\ref{f:active}A. Afterwards, the scientist can act on the system to consistently obtain a desired observation, $x_0=x_0^\ast$, at time step $\ell =0 $ when the experiment starts. 

For instance, the scientist can generate some commands that would lunch a mechanism that moves the particle an amount $x_0^\ast-x_{\rm prep}$ in such a way that the scientist consistently observes a given position, $x_0=x_0^\ast$, as displayed on a reading device, modulo experimental error. Different observations $x_{\rm prep}$ at time step $\ell =-1$ would lead to different actions. The aim of those actions is precisely that an observation at time step $\ell=0$ always yields the same result, $x_0=x_0^\ast$. Since the observation at time step $\ell = 0$ yields consistently the same result, this effectively removes the dynamical dependencies before this time step, when the experiment starts. This amounts at a form of causal intervention. We denote this here as $[x_0=x_0^\ast]$ to technically differentiate it from Pearl's do-calculus, which denote causal interventions as $\textsc{do}[x_0 = x_0^\ast]$ to imply a very specific manipulation of a directed acyclic graph. Pearl's do-calculus does not necessarily model the scientists implementing the causal interventions~\cite{pearl2009causality} (see Fig.~\ref{f:active}B).

\subsection{Enactivism: dynamical coupling between scientist and world}\label{s:enactive}
\begin{figure*}
\includegraphics[width=0.8\textwidth]{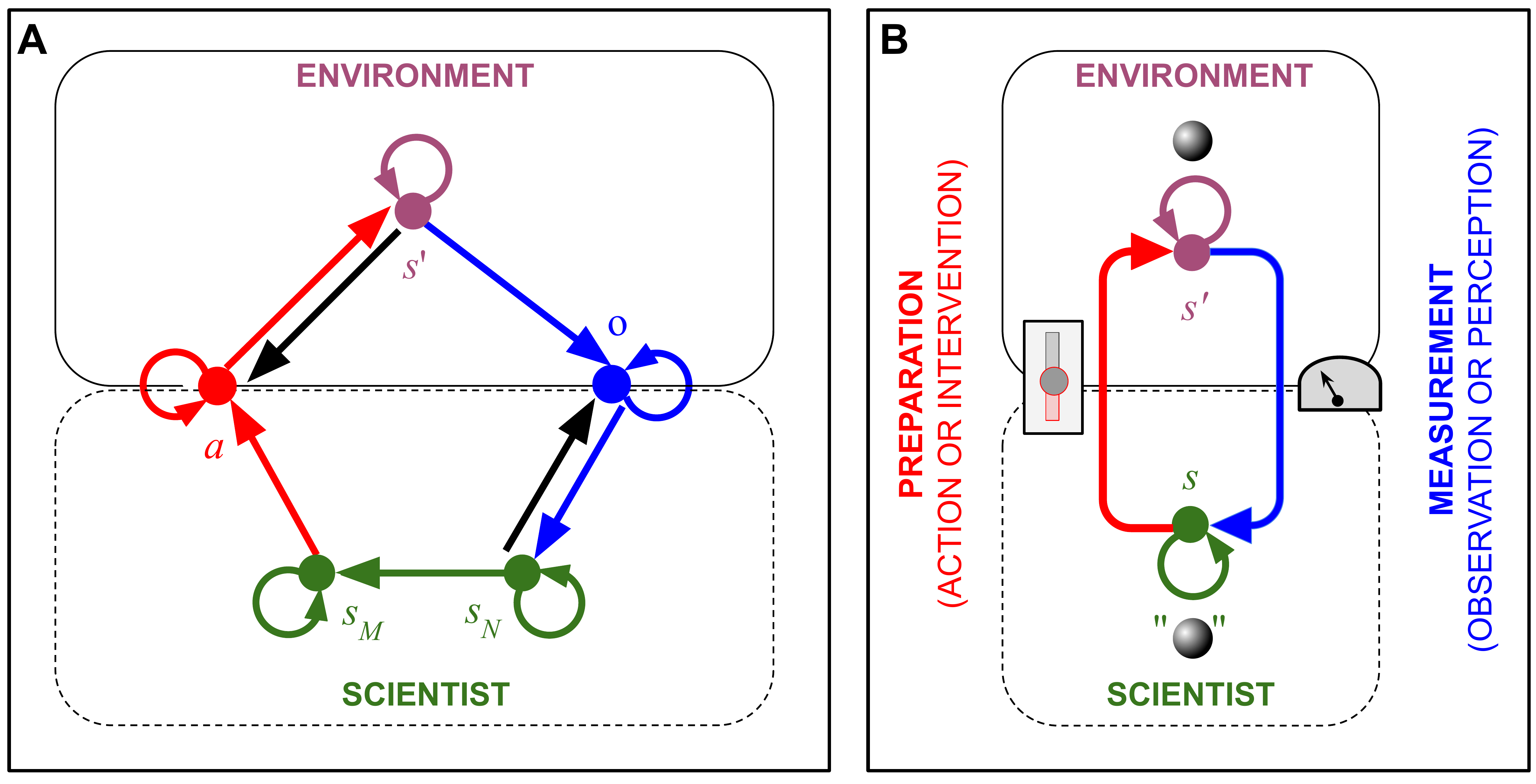}
\caption{{\em Enactivist framework:} (A) Dependency graph of the enactive cognitive model described by Eqs.~\eqref{e:env}-\eqref{e:act}, as presented in Ref.~\cite{di2017sensorimotor} (see Ch. 3 and Fig.~3.5 therein; see also Ref.~\cite{buhrmann2013dynamical}). Nodes represent variables. An arrow indicates that the variable it points to depends on the variable in its tail---in particular, circular arrows indicate recurrent dependencies. This dependency graph represents a circular interaction: scientist's actions, $a$, influence the environment's state, $s^\prime$; environment's states influence the scientist's sensor activity, $o$, via observations; sensor activity influences neural activity, $s_N$; neural activity influences motor activity, $s_M$, i.e. outflowing movement-producing signals; finally, motor activity influences scientist's actions, which closes the interaction loop. Although, internal neural activity and environment's states can influence back, respectively, sensor activity and scientist's actions---e.g. by changing body configuration---the global dynamics is clockwise. (B) Simplified dependency graph that only shows the circular dependency between environment's states, $s^\prime$, and scientist's internal states, $s= (s_M, s_N)$. An action can  prepare a desired state of the experimental system, e.g., a hand movement to turn a knob that places a particle in a desired location---in this sense it may be considered as a form of causal intervention. An observation can be mediated via a reading device, e.g. to determine the final position of the particle (see Fig.~\ref{f:circular}A in the main text).  }\label{f:enactive}
\end{figure*}

Active inference, as briefly described above, still has a representationalist flavour in that the task of the scientist is to learn a model, i.e. a representation, as accurate as possible of the environment's dynamics, including how her own actions affect it. The environment, which is described by the {\em fixed} probability distributions $\Theta_\ell$ and $\Omega_\ell$ in Sec.~\ref{s:gen_process}, is considered as something externally given. This is reflected in that the topology of the Bayesian network representing the scientist mirrors the topology of the Bayesian network representing the environment. In particular, the internal and external dynamics (horizontal arrows in Fig.~\ref{f:active}) flow in the same direction. 

In contrast, the enactive approach~\cite{varela2017embodied,di2017sensorimotor,gallagher2017enactivist} puts a stronger emphasis on the dynamical coupling between scientist and environment~\cite{buhrmann2013dynamical, di2017sensorimotor}. The focus is often on the particular sensor and motor systems of an individual like, e.g., a human or a robot. However, scientists manage to transcend their own sensorimotor limitations with the aid of technological devices that therefore enable them to couple to the world in ``more fundamental'' ways. For instance, the kind of manipulations and observations associated to light-matter interaction experiments are enabled by, e.g., lasers and electron microscopes. These kinds of couplings between scientists and world are hardly possible without such technologies. Such technologies are created by scientists themselves in their quest for lawful regularities. In this quest scientists have to learn how to build suitable experimental devices, how to stabilize the experimental system and achieve repeatability, how to obtain a decent measurement precision, etc. In general, how to achieve objectivity---conditions (O1)-(O3) in the main text.

One of the enactive approaches to learning builds on ideas from Piaget, which is said to avoid ``the extreme conceptions of empiricism (the structure of the world must be learned by extracting patterns and regularities from it) and intellectualism (the agent imposes its notions on the world to order it following some pre-existing categories)''~\cite{di2017sensorimotor} (p. 107). Our work is focused only on the post-learning stage, though. So, it does not depend on a specific theory of enactive learning. A recent description of an enactive approach to learning can be found in Ref.~\cite{di2017sensorimotor} (ch. 4).

In an attempt to clarify some of these ideas, we now briefly comment on a quote by Di Paolo {\em et al.} on the enactive approach: 

\begin{myquotation}``Action in the world is always perceptually guided. And perception is always an active engagement with the world. The situated perceiver [e.g., a scientist] does not aim at extracting properties of the world as if these were pregiven, but at understanding the engagement of her body [possibly enhanced by technological devices] with her surroundings, usually in an attempt to bring about a desired change in relation between the two. To understand perception [e.g., what scientists observe in experiments] is to understand how these sensorimotor regularities or contingencies are generated by the coupling of body and world [possibly mediated by technologies that can enhance motor and sensory capabilities] and how they are used in the constitution of perceptual and perceptually guided acts. 

``According to [Ref.~\cite{varela2017embodied}] the task of the enactive approach is therefore `to determine the common principles or lawful linkages between sensory and motor systems [a.k.a., sensorimotor contingencies] that explain how action can be perceptually guided in a perceiver-dependent world.' '' ~\cite{di2017sensorimotor} (p. 42).
\end{myquotation}

From this perspective, we could consider both the scientist and the environment as physical systems involved in a circular interaction possibly enabled by technological devices (see Fig.~\ref{f:enactive}; cf. Fig.~3.5 in Ref.~\cite{di2017sensorimotor}). Our results in the main text suggest that what we call laws of nature may be associated to such lawful linkages between sensory and motor systems, or sensorimotor contingencies. 

However, to the best of our knowledge, the mathematical formalization of enactivism is not as well developed as that of active inference. Indeed, we are aware of only a couple of rather recent works~\cite{buhrmann2013dynamical, di2017sensorimotor} that attempt to do that. Here we briefly discuss some of the main concepts underlying enactivism, closely following~\cite{di2017sensorimotor} (see Ch. 3 therein; see also Ref.~\cite{buhrmann2013dynamical}). 

For instance, the (controlled) environment or experimental system could be described by state variables $s^\prime$, e.g., the position of a particle in a piece-wise linear potential (see Fig.~\ref{f:circular}A in the main text). Similarly, we could use variables $a$ to represent actions the scientist perform on the experimental system, e.g. by moving her hand to turn a knob that puts the particle in a desired position---these kinds of actions could be considered effectively as state preparations or causal interventions. The dynamics of the environment can then be described by~\cite{di2017sensorimotor} 
\be\label{e:env}
\frac{\mathrm{d} s^\prime}{\mathrm{d} t} = \mathcal{E}(s^\prime , a),
\ee
where the function $\mathcal{E}$ captures the dependency of the environment's current state on its previous state and the scientist's previous actions. 

The scientist's sensor activity, here denoted by variables $o$, is influenced by the environment via her observations that stimulate her sensorium.  Furthermore, in Refs.~\cite{di2017sensorimotor, buhrmann2013dynamical} the scientist is assumed to have an internal neural dynamics, here described by variable $s_N$, which modulates the sensors activity. The scientist's sensors' dynamics can then be described by
\be\label{e:obs}
\frac{\mathrm{d} o}{\mathrm{d} t} = \mathcal{O}(s^\prime , s_N),
\ee
where the function $\mathcal{O}$ captures the dependency of the scientist's sensor dynamics on the state of both the environment and the scientist's internal neural dynamics.

The dynamics of neural activity is assumed to depend on sensor activity and on the neural activity itself, i.e.
\be\label{e:neu}
\frac{\mathrm{d} s_N}{\mathrm{d} t} = \mathcal{N}(s , s_N),
\ee
where the function $\mathcal{N}$ captures such dependencies. Additionally, the scientist's outflowing movement-producing signals, or motor activity, denoted here by $s_M$, is assumed to be influenced by the neural activity, $s_N$, i.e.
\be\label{e:mot}
\frac{\mathrm{d} s_M}{\mathrm{d} t} = \mathcal{M}(s_N),
\ee
where the function $\mathcal{M}$ captures such an influence. 

Finally, the interaction loop is closed by assuming the scientist's actions, which can be implemented via body configurations, depend on the current actions she performs, on her internal motor activity, and on the state of the environment. So, the scientist's actions dynamics can be described as
\be\label{e:act}
\frac{\mathrm{d} a}{\mathrm{d} t} = \mathcal{A}(a, s_M, s^\prime),
\ee
where the function $\mathcal{A}$ captures such dependencies.

\subsection{Relational enactivism: beyond active inference and enactivism}\label{s:relational_cognition}

Here we discuss in more detail how we build on insights from active inference, enactive cognition, and relationalism to develop a minimal model of scientists doing experiments. In particular, we discuss in more detail the constraints from ``objectivity'', i.e., conditions (O1)-(O3) in the main text, and the phenomenon of quantum interference that takes place in the two slits experiment. We also discuss how our approach could actually refer to generic patterns of observations that can transcend specific observers and their potential cognitive limitations.

\subsubsection{Experiments as stationary circular processes}\label{s:relational_circular}

The main feature of both active inference and enactive cognition is that they emphasize a circular interaction, or sensorimotor loop~\cite{di2017sensorimotor,baltieri2019generative,baltieri2017active}, between the scientist and the (controlled) environment, i.e. the experimental system (see Figs.~\ref{f:active} and \ref{f:enactive}). The former usually relies on probabilistic inference, while the latter usually relies on dynamical systems. These two approaches may turned out not to be as different, though~\cite{baltieri2019generative,baltieri2017active}.

Our approach incorporates this core circular dynamics, but in a way we find more parsimonious. For instance, models in active inference often incorporate not only the basic variables---e.g., position---but also multiple derivatives of them---e.g., velocity, acceleration, and other ``generalized coordinates''~\cite{friston2017graphical}. In contrast, our approach only deals with the basic variables. Moreover, the dynamics in active inference arises from (approximately) inverting the generative model the agent has of her environment. Following enactivism, we do not assume that there is an actual model that needs to be inverted. Similarly, enactive models tend to have a more complex structure, with bidirectional influences between some variables, but not others, and recurrent influences in all variables (see Fig.~\ref{f:enactive}A). In contrast, our approach has influences flowing in a single direction, with the recurrent influences explicitly unwrapped into chains of temporal transitions (see Fig.~\ref{f:enactive}B in this appendix and Fig.~\ref{f:circular}C, D in the main text). 

More parsimonious models are to be expected since we focus on some of the most fundamental regularities that can be found in nature. In contrast, active inference and enactivism usually focus on modelling specific agents, e.g., bacteria or robots. Fundamental physical laws are usually associated with the bare properties that remain after all unnecessary complexity from the phenomena of interest has been eliminated. So, it is likely that such fundamental regularities are somehow associated with the most parsimonious, non-trivial models possible. 

\subsubsection{Models of scientists doing experiments are relative to other scientists}\label{s:relational_sci}

In RQM~\cite{Rovelli-1996,rovelli2007quantum}, the process of observation itself is relative to an ``external'' observer who observes the first observer interact with or observe an experimental system---like Wigner and his friend, respectively, in Fig.~\ref{f:circular}A in the main text (see Appendix~\ref{s:wigner} and Fig.~\ref{f:wigner} therein). According to Rovelli (emphasis ours):

\begin{myquotation}The absolute state of affairs of the world is a meaningless notion; asking about the absolute relation between two descriptions is precisely asking about such an absolute state of affairs of the world [...] 

[T]he fact that a certain quantity $q$ has a value with respect to [an observer] $O$ is a physical fact; {\em as a physical fact, its being true, or not true, must be understood as relative to an observer, say $P$.} (Ref.~\cite{Rovelli-1996}, Sec. II D)
\end{myquotation}

Velmans has taken a similar view in his attempt to identify what are the differences between the internal and external perspectives in consciousness research~\cite{velmans2009understanding} (see, e.g., Fig.~9.2 and page 212 therein; cf. Fig.~5.1 in Ref.~\cite{rovelli2007quantum}; see below). In particular, Velmans has considered the possibility that Wigner and his friend can exchange roles (see Fig.~\ref{f:circular}D,E in the main text). Velmans argues that, {\em phenomenologically speaking}, there can be no actual difference in the subjective versus objective status of the phenomena ``experienced'' by Wigner's friend and that ``observed'' by Wigner (see Appendix~\ref{s:velmans} below).

In this work we take this relational view too. This contrast with traditional active inference and enactivism where, to our knowledge, the model of an agent interacting with an environment is {\em not} understood as relative to another scientist. In these approaches models are usually considered in an absolute sense. For instance, in active inference the agent's environment is considered as a generative process, which is often treated as a ``ground truth'' the agent has to learn, as illustrated in Fig.~\ref{f:active} (but see Refs.~\cite{baltieri2019generative,baltieri2017active}). Similarly, while enactivism emphasizes the circular interaction between agent and environment without assuming that the agent is learning a model of a ``ground truth'' environment, to our knowledge such a circular interaction is {\em not} considered as relative to another agent that is observing it, as illustrated in Fig.~\ref{f:enactive}. Something similar could be said of active inference too.

\subsubsection{Wigner and his friend can exchange roles}\label{s:velmans}

Velmans' analysis suggests some mathematical constraints for our relational approach. He considers a situation similar to that presented in Fig.~\ref{f:circular}A in the main text and asks what makes one human being---like Wigner's friend---a ``subject'' and another---like Wigner---an ``experimenter''~\cite{velmans2009understanding} (p. 212)?  Velmans argues that the difference has to do with the fact that Wigner's friend is required to focus only on her own experiences (of the environment or the experimental system), which she needs to respond to or report on in an appropriate way. In contrast, Wigner is interested primarily in the subject's experiences, and in how these depend on the environment stimulus or brain states that he can ``observe.''

But the roles of Wigner and his friend are interchangeable. To exchange roles they ``merely have to turn their heads'', Velmans notices, so that Wigner focuses exclusively on the environment and describes what he experiences, while his friend focuses her attention not just on the environment (which she now thinks of as a ``stimulus'') but also on events that she can observe in Wigner's brain, and on Wigner's reports of what he experiences (see Figs.~\ref{f:circular}D,E and \ref{f:first}C). So, Wigner and his friend turn now into the ``subject'' and the ``experimenter'', respectively. Velmans further argues that in this situation Wigner's friend would now be entitled to think of her observations (of the environment and Wigner's brain) as ``public and objective'' and to regard Wigner's experiences of the environment as ``private and subjective''.

Velmans finds this outcome absurd, as to him the {\em phenomenology} of the environment remains the same, viewed from the perspective of either Wigner or his friend, whether it is {\em thought of} as an ``observed stimulus'' or as an ``experience''. To Velmans nothing has changed in the character of the environment that Wigner and his friend can observe other than the focus of their interest. In other words, that regarding {\em phenomenology} there is no difference between ``observed phenomena'' and ``experiences''. 

Velmans emphasizes that his analysis concerns only the phenomenal world, as opposed to the physical world. This is relevant in situations where, for instance, the ``subject'' directly experiences an optical illusion, while the ``experimenter'' realizes that it is an illusion thanks to his experimental devices. However, if both ``subject'' and ``experimenter'' are scientists with access to the same experimental resources, they both can realize that it is an optical illusion. 

In the main text we allow both ``subject'' and ``experimenter'' to be scientists with the same access to resources. In this situation, the distinction between phenomenal and physical world effectively collapses in that the experiences of the former are effectively the same as the observations of the latter.

\subsubsection{Constraints from ``objectivity''}\label{s:obj_constraints}

Exchanging the roles of Wigner and his friend is relevant for implementing the conditions for ``objectivity.'' As discussed in the main text, this leads to some mathematical constraints on our relational model (see Figs.~\ref{f:circular}D,E). From the perspective of an external observer (Wigner), a scientist (Wigner's friend) is ``objective'' if she---equipped with the required experimental devices---accurately reflects what actually happens ``outside'' of her, as repeatedly observed by Wigner and his colleagues. For instance, if Wigner repeatedly observes a new distant start through a telescope but his friend reports that there is none, Wigner has a reason to doubt the objectivity of his friend. This impression is reinforced if a large community of scientists agree with Wigner that there is indeed a new star. 

As discussed in the main text, the conditions for ``objectivity'' imply that the chain of factors $F_\ell$ associated to the environment (see Fig.~\ref{f:circular} therein) ``mirrors'' the chain of factors $G_\ell = F_{2n - 1- \ell}$ associated to the scientist, i.e., $G_\ell = F_\ell^T$ or 
\be\label{e:F=FT}
F_{2n - \ell -1} = F_\ell^T.
\ee
Here $\ell= 0,\dotsc , n-1$. More precisely, 
\begin{widetext}
\be\label{e:obj_constraints}
\widetilde{\mcP}(\widetilde{\bx}) = \frac{1}{Z} F^T_0 (x_0, x_{2 n-1})\cdots F^T_{n-1} (x_{n+1}, x_n) F_{n-1} (x_n ,x_{n-1})\cdots F_0 (x_1 ,x_0).
\ee
\end{widetext}
This implies that $P_{2n - \ell}  = P_\ell^T$, for $\ell = 0, \dotsc , n-1$, i.e., that the dynamics from time $n$ to time $2n$ is given by the reverse transposed of the dynamics from time $0$ to time $n$. Indeed, using Eq.~\eqref{e:F=FT}, Eq.~\eqref{em:matrix} in the main text can be written for $\ell=0,\dotsc , n-1$ as (with $\widetilde{F}_n = F_{k-1}\cdots F_n $, $k=2n$)
\be
\begin{split}
P_\ell &= \frac{1}{Z} F_{\ell-1} \cdots F_0  F_{2n-1}\cdots F_n F_{n-1} \cdots F_\ell\\
&= \frac{1}{Z} F_{\ell-1} \cdots F_0  F_{0}^T\cdots F_{n-1}^T F_{n-1} \cdots F_\ell .
\end{split}
\ee
Further permuting factors we can obtain the dynamics from time $n$ on, which yields for time step $2n - \ell \geq 0$ (with $\ell= 0, \dotsc , n-1$)
\be\label{e:P_2n-l}
\begin{split}
P_{2n - \ell } &= \frac{1}{Z} F_{2n - \ell-1} \cdots F_n F_{n-1} \cdots F_{0} F_{2n-1}\cdots  F_{2n - \ell}\\
&= \frac{1}{Z} F_{\ell}^T \cdots F_{n-1}^T  F_{n-1} \cdots F_{0} F_{0}^T \cdots F_{\ell-1}^T = P_\ell^T .
\end{split}
\ee

Assuming $F_\ell^T$ is invertible, Eq.~\eqref{e:P_2n-l} can also be written as (with $\ell = 0,\dotsc n-1$) 
\be\label{e:PT_eq}
P_{2n - \ell } = F_{\ell}^T P_{2n - \ell - 1} \left[F_{\ell}^T\right]^{-1}.
\ee
Equivalently, multiplying Eq.~\eqref{e:PT_eq} by $F_\ell^T$ and its inverse, respectively from the right and the left, and doing $P_{2n-\ell} = P_\ell^T $ we can write the reverse dynamics from time $2n$ to $n$ as
\be
P_{\ell + 1}^T = \left[F_{\ell}^T\right]^{-1} P_{\ell}^T \left[F_{\ell}^T\right],
\ee
which is the transpose of Eq.~\eqref{em:P_l+1} in the main text.  

The initial and final probability matrices in this case are symmetric, i.e., $P_0 = P_0^T$ and $P_n = P_n^T$. However, as discussed in Appendix~\ref{s:EQM}, we can in principle fix arbitrary initial and final probabilities which turns the cycle into two effective chains where belief propagation can be run. In this case of more general initial and final probabilities, the ``external'' and ``internal'' dynamics are still the transpose of each other, but $P_{2n}^{\rm int} = \left[P_0^{\rm ext}\right]^T$ and $P_{n}^{\rm int} = \left[P_{n}^{\rm ext}\right]^T$. This property still allows for the self-referential coupling (SRC) to lead to a quantum dynamics characterized by von Neumann equation, Eq.~\eqref{e:vNApp}, as discussed in the main text and Appendix~\ref{s:pair}. However, as also discussed in Appendix~\ref{s:EQM}, for $P_0$ to be consistent with a standard density matrix, it may have to be symmetric.

Anyways, this seems to bring us back to the formulation of active inference presented in Appendix~\ref{s:active}, where there is a pregiven environment dynamics, specified by a generative process, that the scientist's generative model tries to mirror (see Fig.~\ref{f:active}). 
However, this is not exactly so because the functional form of both $F_\ell$ and $G_\ell$ co-emerge as a lawful regularity out of the many interactions scientists have with the environment during the learning stage, when they are familiarizing with the system under investigation. This learning stage may involve the invention and fine-tuning of new protocols, devices, and even concepts (e.g., spacetime curvature) that enable scientists to couple to the environment in ways that were not possible before, and thus to enact new kinds of lawful regularities. 

For instance, the kind of regularities associated to quantum and relativity theories, which are invisible to the naked eye, are enabled by sophisticated experimental protocols and devices, as well as conceptual frameworks, all developed by scientists themselves. The factors $F_\ell$ and $G_\ell$ characterize the end result of this process. That is, the situation when the necessary protocols, devices, and concepts are already invented and standardized, the experiment is so well designed that repeatability is achieved, and scientists need only run the experiment a statistically significant number of times. In other words, when ``objectivity'' has been achieved.

In contrast, in the case of active inference as presented in Appendix~\ref{s:active}, although the generative process $\Theta_\ell$ depends on the scientist's actions, its {\em functional} form is usually considered fixed. The agent can transform the functional form of its generative model via learning, but it cannot transform the functional form of the generative process. In the enactive view, instead, it is {\em as if} the agent could transform the generative process as well---e.g., via the development of new technologies---leading potentially to new kinds of environment dynamics (e.g., lasers) and observation mechanisms (e.g., Geiger counters). In other words, ``the world plays a role in learning that is different from that of providing inputs to internal processing. Nothing [...] prevents aspects of the dynamics of the world forming constitutive parts of the learnt sensorimotor schemes''~\cite{di2017sensorimotor} (p. 105). Of course, we could enlarge the notion of action in active inference to include changes in the functional form of the generative process $\Theta_\ell$, but to our knowledge this is not how it is usually interpreted.

\subsubsection{Example: Two-slit experiment}\label{s:slits}
\begin{figure*}
\includegraphics[width=0.8\textwidth]{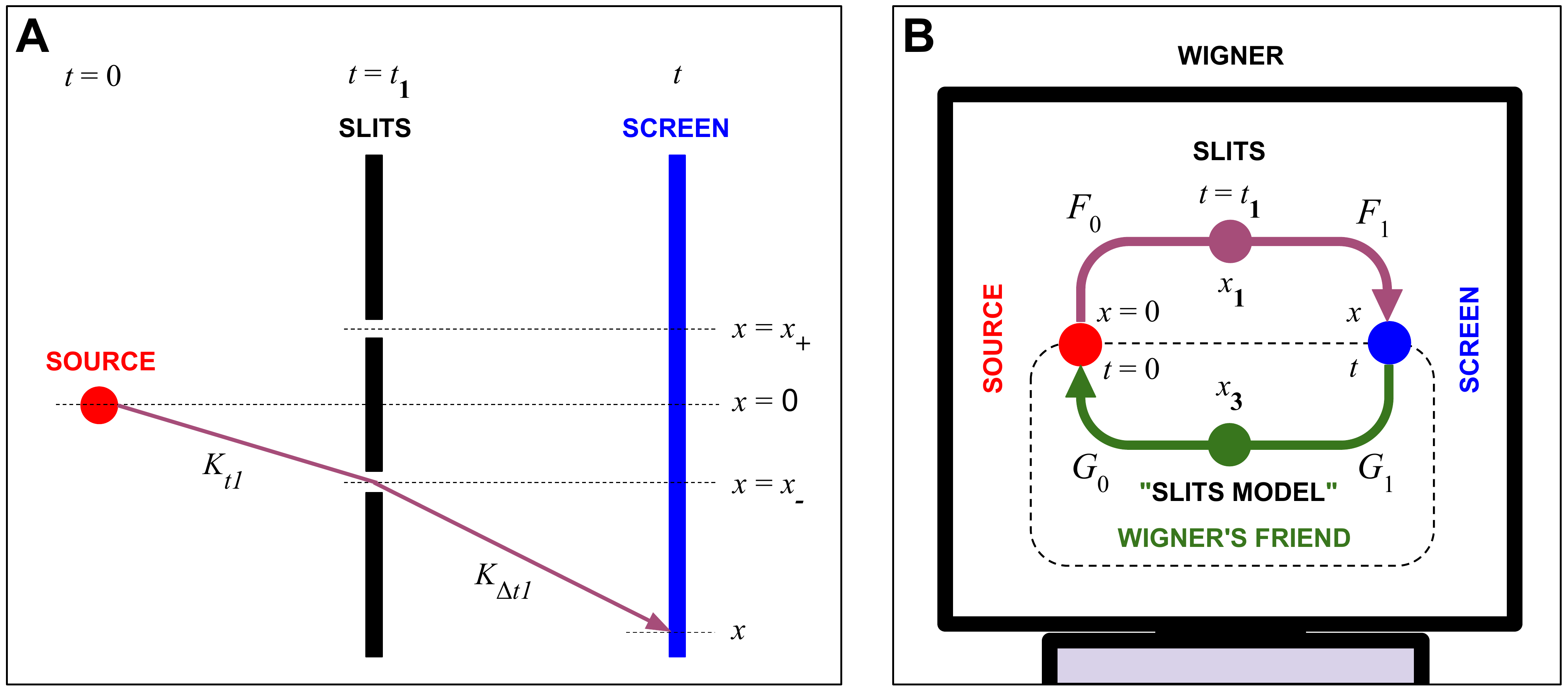}
\caption{{\em Two-slit experiment}: (A) Standard (real-time) version of the two-slit experiment. A particle initially at (vertical) position $x=0$ at time $t=0$ goes at time $t = t_1$ through a barrier with two slits, located at positions $x_\pm = \pm x_{\rm slit}$, and hits a screen at a generic position $x$ at a generic time $t$. Slits can be open or closed. 
(B) Factor graph associated to a scientist (Wigner's friend) performing a two-slit experiment, as modelled by an external scientist (Wigner). The real non-negative factors $F_0$ and $F_1$ capture the ``external'' dynamics between the source and the slits, and between the slits and the screen, respectively. Factors $G_0$ and $G_1$ captures the dynamics ``internal'' to Wigner's friend. Here $x_3$ refers to the ``internal'' physical correlate of the position of the slits.  When only one slit is open, say the slit located at $x_+$, consistency requires that $x_1 = x_3 = x_+$. However, when both slits are open, $x_1$ does not have to equal $x_3$ anymore, which yields the imaginary-time version of quantum interference.  } \label{f:slits}
\end{figure*}

Here we will first briefly review the two-slit experiment to illustrate the phenomenon of {\em quantum interference}, which is sometimes considered one of the hallmarks of quantum physics (see Fig.~\ref{f:slits}A). Afterwards, we discuss how the imaginary-time version of this phenomenon naturally arises in our framework using factor graph models like those described by Eq.~\eqref{em:circular} in the main text (see Fig.~\ref{f:slits}B). So, the formal analogue of standard quantum interference arises after implementing the SRC associated to this model to shift to the intrinsic perspective, as discussed in general in the main text. Interference phenomena has also been obtained using other approaches (see, e.g., Ref.~\cite{spekkens2007evidence}). However, to our knowledge, none actually yields the full mathematical form of this phenomenon. Indeed, if we ignore the SRC we can also obtain (imaginary-time) quantum interference which, though seemingly non-classical, does not completely coincides with standard quantum interference.

\

\noindent{\em Two-slit experiment:} Consider a particle that travels from a source to a screen through a barrier with two slits (see Fig.~\ref{f:slits}A). Let the probability amplitude to go from $x^\prime$ at time $t^\prime$ to $x$ at time $t$ be given by $K_{\Delta t}(x, x^\prime) $, with $\Delta t = t- t^\prime$. Assume the particle is initially localized in the source at (vertical) position $x=0$ (at time $t=0$)---i.e., its initial state is $\psi(x, 0) = \delta(x)$, where $\delta$ stands for the Dirac delta function. Then the state at time $t$ is given by
\be
\psi(x, t) = \int K_t(x, x^\prime)\psi(x^\prime , 0)\mathrm{d} x^\prime = K_t(x, 0).
\ee

Now imagine that one of the two slits is closed, so the particle can only go through one of them at time $t_1$. That is, $x_1 = x_\pm$ where the upper and lower sign denotes the situation where the particle has gone through the upper and lower slit, respectively. In this case, the state of the particle at time $t$ is
\be
\psi_\pm (x, t ) = K_{\Delta t}  (x , x_\pm) K_{t_1} (x_\pm , 0),
\ee
where $\Delta t = t - t_1$. The probability to find the particle at position $x$ in the screen at time $t$ is 
\be
\mcP_\pm(x,t) = |\psi_\pm(x, t)|^2 .
\ee
So we can write
\be\label{e:wave+-}
\psi_\pm(x,t) = \sqrt{\mcP_\pm(x, t)} e^{i\varphi_\pm(x,t)},
\ee
which defines the phases $\varphi_\pm (x, t)$.

If the two slits are open, instead, then the state of the particle at time $t$ is given by the superposition
\be
\psi_{\rm both} (x, t ) = \frac{1}{\sqrt{2}}\left[\psi_+(x, t) + \psi_-(x, t)\right].
\ee
Here the factor $1/\sqrt{2}$ is to guarantee that $\psi_{\rm both}$ is normalized. So, the probability to find the particle at position $x$ at time $t$ is
\begin{widetext}
\be\label{e:q-interf}
\mcP_{\rm both}(x, t) = |\psi_{\rm both}(x,t)|^2 = \frac{1}{2}\left[\mcP_+(x, t) + \mcP_-(x,t)\right] +  \sqrt{\mcP_+(x, t)\mcP_-(x,t)}\cos\left[\Delta \varphi(x,t)\right].
\ee
\end{widetext}
where $\Delta\varphi (x,t) = \varphi_+(x,t) - \varphi_-(x,t)$. 

Equation~\eqref{e:q-interf} shows that the probabilities associated to quantum states do not satisfy the sum rule of classical probability theory, i.e., 
\be\label{e:P!=P+P}
\mcP_{\rm both}(x, t)\neq \frac{1}{2}\left[\mcP_+(x, t) +\mcP_-(x, t) \right].
\ee
which is the phenomenon of {\em quantum interference}. The factor $1/2$ in the right hand side of Eq.~\eqref{e:P!=P+P} indicates that, due to symmetry between slits, the particle can go through either slit with probability $1/2$. The last term in Eq.~\eqref{e:q-interf}, which is proportional to $\cos\left[\Delta\varphi(x,t)\right]$, is the interference term. 

\

\noindent{\em Factor graph model associated to the two slit-experiment:} We now discuss a factor graph model associated to the two-slit experiment (see Fig.~\ref{f:slits}B). We will see that it coincides with the imaginary-time version of the standard quantum two-slit experiment described above---i.e., with a hyperbolic cosine term instead of the cosine term in Eq.~\eqref{e:q-interf}, as $\cosh(z) = \cos(i z)$. Consider a path $\widetilde{\bx} = (0, x_1, x, x_3)$ that starts as before at the source located at $x=0$ at $t=0$ and goes through $x_1\in\{x_+ , x_-\}$, $x$, and $x_3$ at times $t_1$, $t$, and $t_3$, respectively, to return to $x=0$. Here $x_3$ is associated to the physical correlates of the slits ``internal'' to Wigner's friend in Fig.~\ref{f:slits}B. Following Eq.~\eqref{em:circular} in the main text, the probability associated to this path is
\be\label{e:Pfactor-slits}
\widetilde{\mcP}(0, x_1, x, x_3 ) = F_0 (x_1 , 0) F_1(x, x_1) G_1(x_3 , x) G_0(0, x_3 ).
\ee
In general, factors $F_\ell$ and $G_\ell$ above do not have to be related, but ``objectivity'' requires $G_\ell = F_\ell^T$ (see Eq.~\eqref{e:F=FT}).

If one of the two slits is closed, the particle can only go through one of them at time $t_1$. That is, $x_1 = x_\pm$ where the upper and lower sign denotes the situation where the particle goes through the upper and lower slit, respectively. In this case, the probability to find the particle at position $x$ at time $t$ is given by
\be\label{e:im-both}
\mcP_\pm(x, t)\equiv \frac{1}{Z_{\rm one}}\mcP(0, x_\pm, x, x_\pm ), 
\ee
where 
\be
Z_{\rm one} = {Z_\pm}  \equiv \int \mcP(0, x_\pm, x, x_\pm )\mathrm{d} x ,
\ee
ensures that $\mcP_\pm$ is normalized. Due to the symmetry between slits, we have that $Z_\pm = Z_{\rm one} $ is the same for both $x_{\pm}$. 

Following Eqs.~\eqref{e:Pfactor-slits} and \eqref{e:im-both} we can write (cf. Eqs.~\eqref{e:im-p+S} and \eqref{e:im-p-S} in Appendix~\ref{s:quantum_diff})
\begin{widetext}
\BE\label{e:im-state}
\frac{1}{\sqrt{Z_{\rm one}}}F_0 (x_\pm , 0) F_1(x, x_\pm)  &=& \sqrt{\mcP_{\pm}(x, t)}e^{S_\pm(x,t)},\\
\frac{1}{\sqrt{Z_{\rm one}}}G_1(x_\pm , x) G_0(0, x_\pm ) &=& \sqrt{\mcP_{\pm}(x, t)}e^{-S_\pm(x,t)},\label{e:im-stateG}
\EE
\end{widetext}
which defines the imaginary-time phase $S_\pm$. 

If the two slits are open, instead, then the probability for the particle to be located at position $x$ at time $t$ is given by 
\begin{widetext}
\be\label{e:im-interf}
\begin{split}
\mcP_{\rm both}(x , t) &\equiv\frac{1}{Z_{\rm both}}\sum_{x_1, x_3\in\{x_+ , x_-\}} \mcP(0, x_1, x, x_3 )  = \\ 
&= C \left\{\frac{1}{2}\left[\mcP_+(x, t)  + \mcP_-(x, t)\right] + \sqrt{\mcP_+(x,t) P_-(x, t)}\cosh\left[\Delta S(x, t) \right]\right\} 
\end{split}
\ee
\end{widetext}
where $Z_{\rm both}$ is a normalization constant, $C = {2 Z_{\rm one}}/{Z_{\rm both}}$ and
\be
\Delta S(x, t) = S_+(x,t) - S_-(x,t).
\ee
The first two terms in the second line of Eq.~\eqref{e:im-interf} come from the elements of the sum with $x_1 = x_3 = x_\pm$ (see Eq.~\eqref{e:im-both}).  The last term in the second line of Eq.~\eqref{e:im-interf}, which is the imaginary-time interference term, comes from the elements of the sum with $x_1\neq x_3$, i.e., 
\begin{widetext}
\BE\label{e:im-interf-term}
F_0 (x_+ , 0) F_1(x, x_+) G_1(x_- , x) G_0(0, x_- ) &=& Z_{\rm one}\sqrt{\mcP_+(x, t)\mcP_-(x,t)}e^{\Delta S} ,\\
F_0 (x_- , 0) F_1(x, x_-) G_1(x_+ , x) G_0(0, x_+ ) &=& Z_{\rm one}\sqrt{\mcP_+(x, t)\mcP_-(x,t)}e^{-\Delta S}.
\EE
\end{widetext}
The right hand side of these equations is obtained by using Eqs.~\eqref{e:im-state} and \eqref{e:im-stateG}. 

Equation~\eqref{e:im-interf} is the imaginary-time version of Eq.~\eqref{e:q-interf}. Unlike the cosine term in Eq.~\eqref{e:q-interf}, whose integral can vanish due to its oscillatory behavior, the hyperbolic cosine term in Eq.~\eqref{e:im-interf} is always positive. So, the constant $C$ is required for $\mcP_{\rm both}$ to be properly normalized. 

Since the factors in Eqs.~\eqref{e:im-state} and \eqref{e:im-stateG} are products of factors like those in Eqs.~\eqref{em:F} in the main text, the phases $S_\pm (x, t)$ are proportional to the time step $\epsilon$. So, under a Wick rotation $\epsilon\to i\epsilon$ the right hand side of Eqs.~\eqref{e:im-state} and \eqref{e:im-stateG} turn into complex wave functions like that in Eq.~\eqref{e:wave+-} and the hyperbolic cosine turns into a cosine, as $\cos(i z) = \cosh(z)$ for any $z$. Thus combining the approach above with the SRC to implement the intrinsic perspective can lead to standard quantum interference.

This provides a fresh perspective to think about quantum interference. When Wigner's friend has information about which slit the particle goes through---e.g., when only one stlit is open---this has to be reflected in the physical correlates of the experiment ``internal'' to her. So, $x_3 = x_1$ (see Fig.~\ref{f:slits}). In this view, the imaginary-time version of quantum interference arises because, when Wigner's friend cannot access any information about which slit the particle goes through, the values of $x_1$ and $x_3$ do not have to coincide even though they refer to the same ``thing'' (i.e., the slits). This fact, along with the SRC described in the main text, can then be considered from this perspective as the source of standard quantum interference. This would be true no matter the reason for which Wigner's friend lacks `which-way' information. Furthermore, our approach considers the whole experimental setup, or context, from beginning to end. So, it could also potentially take account of variations of the two-slit experiments, such as delay choice or quantum erasure experiments. 

\subsubsection{Observations as dynamical patterns: matter comes and goes, but patterns persist}\label{s:patterns}

We would like to reflect now on the notion of observer and the process of observation. To do so, we would like to draw from the enactive approach to mind and life. In this view, a living organism is not a bunch of matter but rather a dynamical pattern supported on a flux of matter. While matter itself comes and go, the dynamical pattern persists.  In Thompson's words (emphasis ours):

\begin{myquotation}An organism is a material being, and its reality at any given moment coincides completely with its material constitution. Yet this identity cannot be based on the constancy of matter because its material composition is constantly renewed: `every five days you get a new stomach lining. You get a new liver every two months. Your skin replaces itself every six weeks. Every year, ninety-eight percent of the atoms in your body are replaced. This nonstop chemical replacement, metabolism, is a sure sign of life'... {\em Only at the level of form or pattern can we find constancy in flux...} 

An organism identity is not bound to its material constitution, for this constitution is constantly renewed; {\em its identity is accomplished dynamically at a formal level.} Yet with this freedom comes a correlative necessity: the organism has to change; stasis is impossible. The organism must eat and excrete; otherwise it dies. Without incessant metabolic exchange with the world there can be no emancipation of dynamic selfhood from mere material presence. (Ref.~\cite{thompson2010mind}, pp. 150-152)
\end{myquotation}

Furthermore, the enactive approach attempts to transcend dualisms---like that of action and perception or subject and object---by treating the poles of a duality as co-defining each other through the holistic process they are involved in. For instance, sensors and effectors are not considered as two separate structures completely independent of each other. Rather, they are understood as co-defined through the co-constitutive role they play in the adaptive engagements between agent and environment. In the words of di Paolo {\em et al.} (emphasis ours):

\begin{myquotation}[W]hen we speak of sensorimotor integration, or mastering the laws of [sensorimotor contingencies], {\em we are not starting from separate processes external to each other---the sensory and the effector processes---which we then bring together into an explanation of action and perception.} We start from already co-defined pairs (i.e. from moments of a same adaptive engagement between agent and environment). Distinguishable anatomic structures may become specialized during development and evolution, and so become reused and recombined into various different adaptive engagements [...] {\em We call this structures sensors and effectors, but this is a reification that hides the fact that what makes them so is the co-constitutive role they play in the adaptive enactments of an agent.} (Ref.~\cite{di2017sensorimotor}, p. 139)
\end{myquotation}

Similarly, rather than two polarities completely independent of each other, subject and world are considered as inseparable from each other. Referring to Merleau-Ponty~\cite{merleau1962phenomenology}, Thompson says (citations to Merleau-Ponty are omitted from this quote):

\begin{myquotation}Merleau-Ponty maintains that the relation between self and world is not that of subject to object but rather what he calls, following Heidegger, being-in-the-world. For a bodily subject it is not possible to specify what the subject is in abstraction from the world, nor is it possible to specify what the world is in abstraction from the subject: ``The world is inseparable from the subject, but from a subject  which is nothing but a project of the world, and the subject is inseparable fom the world, but from a world which the subject itself projects...'' 

Things in the world bring forth suitable intentional actions and motor projects from the subject (the subject is a project of the world), but things in the world have specific motor senses or affordances only in relation to the motor skills of the subject (the world is projected by the subject). This body-environment circuit of motor intentionality belongs to what Merleau-Ponty calls ``the intentional arc'' subtending the life of consciousness, which integrates sensibility and motility, perception and action. 
The intentional arc and being-in-the-world are neither purely first-personal (subjective) nor purely third-personal (objective), neither mental nor physical. They are existential structures prior to and more fundamental than these abstractions. (Ref.~\cite{thompson2010mind}, pp. 247-248)
\end{myquotation}

So, from an enactive perspective the observer and the observed could be seen as being co-defined through the process of observation. The process of observation itself could be seen as a circular dynamical pattern supported on a flux of matter that may come and go. It is not ``internal'' to the observer but emerges from the circular coupling between observer and world. It is analogous to a sensorimotor loop possibly enabled by experimental devices that scientists develop to couple to the environment in more fundamental ways. Thanks to these devices, such a dynamical pattern is not constrained by the cognitive limitations of a particular kind of observer. 

Furthermore, not only the observer and the observed are relational to each other, but the process of observation itself is relational. This was noticed early on by Varela {\em et al.}, who built on ideas from Eastern philosophy like the so-called ``Abhidharma'' (emphasis ours): 

\begin{myquotation}In the Abhidharma analysis of consciousness, each moment of experience takes the form of a particular consciousness that has a particular object to which it is tied by particular relations. For example, a moment of seeing consciousness is composed of a seer (the subject) who sees (the relation) a sight (the object)... 

[The Indian philosoher] Nagarjuna attacks the independent existence of all three terms---the subject, {\em the relation}, and the object. (Ref.~\cite{varela2017embodied}, p. 221)
\end{myquotation}

However, to our knowledge, there have not been attempts to explicitly formalize this additional level of relationalism in the enactive approach. The presence of Wigner as an external observer in this work explicitly acknowledges such a relational nature (see Fig.~\ref{f:circular}A in the main text). RQM also acknowledges that the process of observation is relative to an explicit external observer. However, it does not attempt to implement the intrinsic perspective. Here we allow Wigner and his friend to play alternative roles (as Alice and Bob) to implement observations from an intrinsic perspective (see Fig.~\ref{f:first}). 

By focusing on the process of observation itself rather than on the subject-object polarities, enactivism can help us address an apparent limitation of our approach. For simplicity, we have focused on modeling a single observer interacting with an experimental system. However, science is not limited to one scientist. For instance the initial state of an experimental system could be prepared by a scientist and the final state be measured by another. In this case, it would seem as if the experiment could be performed without necessarily closing the loop. However, for one of these scientists to realize that there is indeed a correlation between initial and final states, the other has to accurately communicate what she did or observed. This communication closes the loop. Similarly, a scientist could prepare the initial state of an experiment, write down clearly what he has done for other scientists to understand it, and then sadly die. Another scientist can read what the now deceased scientist wrote, continue the experiment, and observe the final state. This indirect communication again closes the loop. Matter may come and go, but the dynamical pattern persists.

\

\section{Markov processes and imaginary-time quantum dynamics}\label{s:MP_recasted}

Here we first discuss the principle of maximum dynamical entropy, or principle of maximum caliber. This is a general variational principle, similar to the free energy principle, from which a variety of models at, near, and far from equilibrium can be derived~\cite{presse2013principles}.  We have used this principle in the main text to derive the form of the stationary distribution over the dynamical trajectories characterizing a scientist interacting with an experimental system. We initially focus on chains for simplicity. 

Afterwards, we show that the belief propagation (BP) algorithm, which is exact on chains, can be written in a way formally analogous to imaginary-time or Euclidean quantum dynamics~\cite{Zambrini-1987}. More precisely, BP messages play the role of imaginary-time wave functions and the continuous-time limit of the BP iteration corresponds to the imaginary-time version of Schr\"odinger equation. However, in this case the  BP messages on the leaves of the chain, which initiate the BP iteration, can always be chosen constant. 

We then discuss the case of cycles which is more interesting. Indeed, the BP algorithm is not guaranteed to be exact anymore~\cite{Weiss-2000}. However, we can choose initial and final conditions for the probability marginals to effectively turn the cycle into a chain. This yield the formal analogue of the general imaginary-time quantum dynamics considered in Ref.~\cite{Zambrini-1987}. 

\subsection{Principle of maximum caliber and factor graphs}\label{s:MaxCal}
The principle of maximum entropy~\cite{jaynes2003probability} to derive some common equilibrium probability distributions in statistical physics can be extended to the so-called principle of maximum caliber to deal with non-equilibrium distributions on trajectories~\cite{presse2013principles}. In particular Markov chains and Markov processes can be derived from the principle of maximum caliber (see e.g. Sec. IX B in Ref.~\cite{presse2013principles}). We introduce this principle here with an example relevant for our discussion. 

Consider a probability distribution $\mathcal{P}_{\rm ch}(x_1,\dotsc ,x_n)$ on (discretized) paths $(x_1,\dotsc , x_n)$, where $x_\ell$ refers to the position at time $t=\ell\epsilon$. Here we will consider the case of an open chain. Assume that we only have information about the average energy on the (discretized) paths given by
\be\label{e:Eav}
\mathcal{H}_{\rm av}[\mathcal{P}_{\rm ch}] = \bra\frac{1}{T}\sum_{\ell=0}^{n-1}\mathcal{H}_\ell(x_{\ell+1}, x_\ell)\epsilon\ket_{\mcP_{\rm ch}} ,
\ee
where $T=n\epsilon$ is the total time duration of the path, and $\mathcal{H}_\ell$ is the energy function at time step $\ell$. Here
\be\label{e:Eav}
\bra f\ket_\mathcal{P} = \int \mathcal{P}(x_0,\dotsc ,x_n)f(x_0,\dotsc , x_n)\prod_{\ell = 0}^{n-1}\mathrm{d}x_\ell  ,
\ee
denotes the average value of a generic function $f$ of a path, with respect to a generic path probability distribution $\mcP$. For convenience, here we are using integrals instead of sums, as in the main text. However, our analysis is valid for discrete variables too by changing these integrals by sums, $\int\to\sum$. 

The principle of maximum caliber tells us that among all possible probability distributions we should choose the one that both maximizes the entropy
\be\label{e:entropy}
\mathcal{S}[\mathcal{P}_{\rm ch}] = -\bra\ln\mathcal{P}_{\rm ch}(x_0,\dotsc ,x_n)\ket_{\mcP_{\rm ch}},
\ee
and is consistent with the information we have, i.e. $\mathcal{H}_{\rm av}[\mathcal{P}_{\rm ch}] = E_{\rm av}$, where $E_{\rm av}$ is the fixed value of the average energy. Introducing a Lagrange multiplier $\lambda$ to enforce the constraint on the average energy, the constrained maximization of $\mathcal{S}[\mathcal{P}_{\rm ch}]$ becomes equivalent to the maximization of the Lagrangian $\mathcal{S}[\mcP_{\rm ch}] - \lambda \mathcal{H}_{\rm av}[\mathcal{P}_{\rm ch}]$. The solution to this problem is the distribution
\be\label{e:MaxCal}
\mathcal{P}_{\rm ch}(x_0,\dotsc , x_n) = \frac{1}{\mathcal{Z}}\exp\left[-\frac{\lambda}{T}\sum_{\ell=0}^{n-1}\mathcal{H}_\ell(x_{\ell+1} , x_{\ell})\epsilon\right],
\ee
where $\mathcal{Z}$ is the normalization factor. 

Notice that $\mathcal{P}_{\rm ch}$ in Eq.~\eqref{e:MaxCal} can be written as a product of factors
\be\label{e:P_prodF}
\mathcal{P}_{\rm ch}(x_0,\dotsc , x_n) = \frac{1}{Z}\prod_{\ell=0}^{n-1} F_\ell(x_{\ell+1} , x_{\ell}).
\ee
Without loss of generality, we can choose the factors as
\be\label{e:F}
F_\ell(x_{\ell + 1} , x_{\ell}) = \frac{1}{|\mathcal{A}|}\exp\left[-\frac{\lambda}{T} \mathcal{H}_\ell(x_{\ell+1} , x_{\ell})\epsilon\right],
\ee
with $|\mathcal{A}|= \sqrt{2\pi T\epsilon / m\lambda}$, so $Z = \mathcal{Z}/|\mathcal{A}|^{n}$ in Eq.~\eqref{e:P_prodF}.  

\

\subsection{Quantum-like formulation of stochastic processes via the cavity method}\label{s:QBP}
\begin{figure*}
\includegraphics[width=0.8\textwidth]{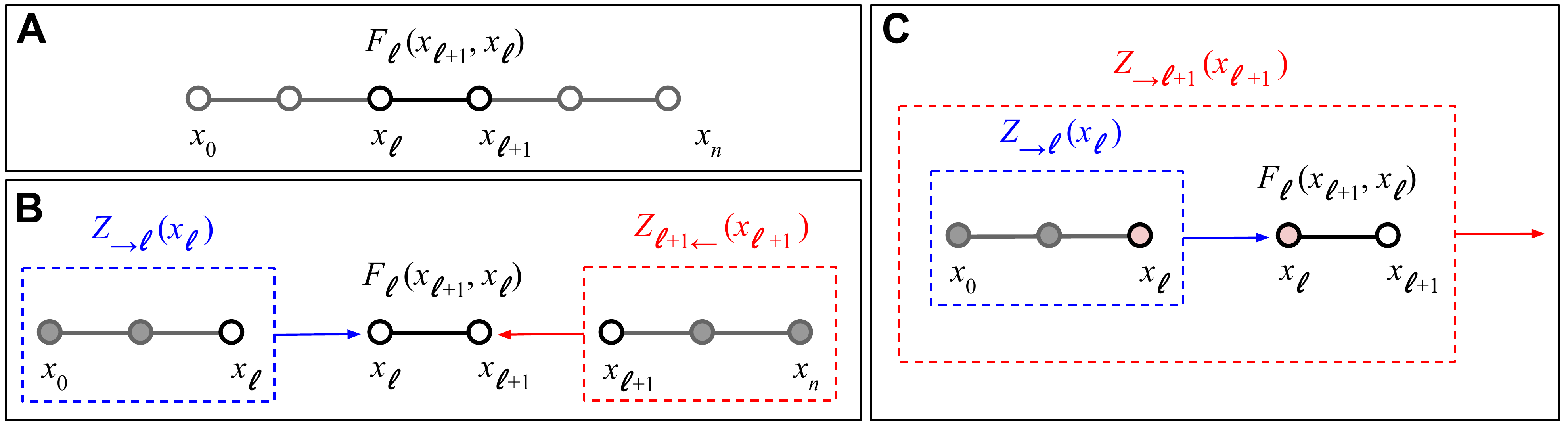}
\caption{ {\it Cavity method}. (A) Factor graph associated to a Markov chain (see Eq.~\eqref{e:P_prodF}). (B) Graphical expression for the pairwise marginal $\mathcal{P}_\ell(x_{\ell +1}, x_{\ell })$ (see Eq.~\eqref{e:P_BP}); the partial partition functions $Z_{\to\ell}(x_\ell)$ (blue; see Eq.~\eqref{e:Z->}) and $Z_{\ell+1\leftarrow}(x_{\ell+1})$ (red; see Eq.~\eqref{e:Z<-}) correspond to the sum over all variables on the cavity graphs inside the dashed rectangles, except for $x_\ell$ and $x_{\ell +1}$ which are clamped to be able to recover the whole graphical model by multiplying for $F_{\ell}(x_{\ell +1}, x_{\ell })$. (C) The partial partition function $Z_{\to\ell+1}(x_{\ell+1})$ (red) can be recursively computed by multiplying the partial function $Z_{\to\ell}(x_\ell)$ and the factor $F_\ell(x_{\ell +1}, x_{\ell})$ and tracing over $x_\ell$ (see Eq.~\eqref{e:BP->}). This is the content of the belief propagation algorithm~\cite{Mezard-book-2009} specified by Eqs.~\eqref{e:BP->} and \eqref{e:BP<-}.}
\label{f:cavity}
\end{figure*}

\subsubsection{Cavity messages as imaginary-time wave functions}\label{s:cavity}
Here we show how the belief propagation algorithm obtained via the cavity method~\cite{Mezard-book-2009} (ch. 14) can be formally written in terms of the imaginary-time Schr\"odinger equation and its conjugate. First, notice that by marginalizing the probability distribution defined in Eq.~\eqref{e:P_prodF} over all variables except $x_\ell$ and $x_{\ell+1}$ we obtain
\begin{widetext}
\BE
\mathcal{P}_\ell(x_{\ell +1} , x_{\ell }) &=& \frac{1}{Z} F_\ell(x_{\ell +1}, x_{\ell }) Z_{\to\ell}(x_\ell)Z_{\ell+1\leftarrow}(x_{\ell +1}),\label{e:P_BP}\\
p_\ell(x_\ell ) &=& \sum_{x_{\ell+1}}\mathcal{P}_\ell(x_{\ell +1} , x_{\ell }) = \frac{1}{Z} Z_{\to\ell}(x_\ell)Z_{\ell\leftarrow}(x_{\ell }),\label{e:p_BP}
\EE
\end{widetext}
where the partial partition functions $Z_{\to\ell}(x_\ell)$ and $Z_{\ell\leftarrow}(x_{\ell })$ of the original factor graph are given by the partition functions of the modified factor graphs that contain all factors $F_{\ell^\prime}$ to the left (i.e. $\ell^\prime < \ell$) and to the right (i.e. $\ell^\prime \geq \ell$) of variable $x_\ell$, respectively; i.e. (see Fig.~\ref{f:cavity}A,B; cf. Eq. (14.2) in Ref.~\cite{Mezard-book-2009}).
\BE
Z_{\to\ell}(x_\ell) &=& \int \prod_{\ell^\prime = 0}^{\ell-1}F_{\ell^\prime}(x_{\ell^\prime +1}, x_{\ell^\prime})\mathrm{d}x_{\ell^\prime},\label{e:Z->}\\
Z_{\ell\leftarrow}(x_{\ell }) &=& \int \prod_{\ell^\prime = \ell}^{n-1}F_{\ell^\prime}(x_{\ell^\prime + 1}, x_{\ell^\prime })\mathrm{d}x_{\ell^\prime +1}.\label{e:Z<-}
\EE
$Z_{\to\ell}(x_\ell)$ and $Z_{\ell\leftarrow}(x_{\ell })$ can be interpreted as information that arrives to variable $\ell$ from the left and from the right side of the graph, respectively. 

By separating factor $F_{\ell -1}$ and $F_\ell$ in Eqs.~\eqref{e:Z->} and \eqref{e:Z<-}, respectively, we can write these equations in a recursive way as (see Fig.~\ref{f:cavity}C; cf. Eq. (14.5) in Ref.~\cite{Mezard-book-2009})
\BE
Z_{\to\ell}(x_\ell) &=& \int F_{\ell -1}(x_{\ell }, x_{\ell -1})Z_{\to\ell-1}(x_{\ell-1})\mathrm{d}x_{\ell-1},\label{e:BP->}\\
Z_{\ell\leftarrow}(x_{\ell }) &=& \int Z_{\ell+1\leftarrow}(x_{\ell+1 }) F_{\ell}(x_{\ell +1}, x_{\ell }) \mathrm{d}x_{\ell+1}.\label{e:BP<-}
\EE
These recursive equations are usually referred to as the {\em belief propagation algorithm}. Since the partial partition functions are typically exponentially large, Eqs.~\eqref{e:BP->} and \eqref{e:BP<-} are commonly written in terms of normalized {\em cavity messages} ${\nu_{\to\ell}(x) = Z_{\to\ell}(x)/Z_{\to\ell}}$ and ${\nu_{\ell\leftarrow}(x)=Z_{\ell\leftarrow}(x)/Z_{\ell\leftarrow}}$, where $Z_{\to\ell}$ and $Z_{\ell\leftarrow}$ are the corresponding normalization constants. This choice of normalization has at least two advantages: (i) it allows us to interpret the messages as probability distributions and (ii) it keeps the information traveling from left to right separated from the information traveling from right to left. 

We will now show that a different choice of normalization, i.e. 
\be\label{e:mu_def}
\mu_{\to\ell}(x) = \frac{Z_{\to\ell}(x)}{\sqrt{Z}},\hspace{0.5cm} \mu_{\ell\leftarrow}(x) = \frac{Z_{\ell\leftarrow}(x)}{\sqrt{Z}},
\ee
which violates the features (i) and (ii) mentioned above, allows us to connect the BP equations, i.e. Eqs.~\eqref{e:BP->} and \eqref{e:BP<-}, with those of Euclidean quantum mechanics. Indeed, let us write
\BE
\mu_{\to\ell}(x) \mu_{\ell\leftarrow}(x) &=& p_\ell(x),\label{e:mu->*<-mu} \\
\frac{\mu_{\to\ell}(x)}{\mu_{\ell\leftarrow}(x)} &=& e^{2\phi_\ell(x)}, \label{e:mu->/<-mu}
\EE
where Eq.~\eqref{e:mu->*<-mu} comes from Eq.~\eqref{e:p_BP} and Eq.~\eqref{e:mu->/<-mu} is a definition of the ``effective field'' or ``phase'' $\phi_\ell$. Equations~\eqref{e:mu->*<-mu} and \eqref{e:mu->/<-mu} imply that we can parametrize the cavity messages in terms of $p_\ell$ and $\phi_\ell$ as
\BE
\mu_{\to\ell}(x)  &=& \sqrt{p_\ell(x)}e^{\phi_\ell(x)},\label{e:mu->} \\
\mu_{\ell\leftarrow}(x)  &=& \sqrt{p_\ell(x)}e^{-\phi_\ell(x)},\label{e:<-mu}
\EE
which are the analogue of a ``wave function'' in imaginary time.

\

\subsubsection{Belief propagation as imaginary-time quantum dynamics}\label{s:QBPdyn}
In terms of the messages $\mu_{\to\ell}$ and $\mu_{\ell\leftarrow}$ in Eq.~\eqref{e:mu_def}, the belief propagation equations \eqref{e:BP->} and \eqref{e:BP<-} become
\BE
\mu_{\to\ell}(x) &=& \int F_{\ell -1}(x, x^\prime)\mu_{\to\ell-1}(x^\prime)\mathrm{d}x^\prime,\label{e:muBP->}\\
\mu_{\ell\leftarrow}(x) &=& \int \mu_{\ell+1\leftarrow}(x^\prime)F_{\ell}(x^\prime , x)\mathrm{d}x^\prime,\label{e:muBP<-}
\EE
where we have done $x_\ell = x$, $x_{\ell - 1} = x^\prime$ in Eq.~\eqref{e:muBP->}, and $x_{\ell + 1} = x^\prime$ in Eq.~\eqref{e:muBP<-}. This contrasts with the standard formulation in terms of the $\nu$-messages described after Eq.~\eqref{e:BP<-}, where the messages must be renormalized at each iteration of the belief propagation equations (cf. Eq. (14.2) in Ref.~\cite{Mezard-book-2009}). Such iterative renormalization is avoided here because the normalization constant $\sqrt{Z}$ is the same for {\em all} quantum-like cavity messages. Equations~\eqref{e:muBP->} and \eqref{e:muBP<-} are formally analogous to Eq. (2.16) in Ref.~\cite{zambrini1986stochastic} and its adjoint, respectively, which describe imaginary-time quantum dynamics. 

For concreteness let us consider the energy function
\be\label{e:BP_Hl}
\mcH_\ell (x,x^\prime) = \frac{m}{2}\left(\frac{x - x^\prime}{\epsilon}\right)^2 + V(x).
\ee
Due to the Gaussian term in the corresponding factors $F_\ell$ in Eq.~\eqref{e:F}, the integrals in Eqs.~\eqref{e:muBP->} and \eqref{e:muBP<-} can be approximated to first order in $\epsilon$ (cf. Appendix~\ref{s:App_path}). Indeed, since $\epsilon\to 0$, the real Gaussian factor associated to the quadratic term in Eq.~\eqref{e:BP_Hl} is exponentially small except in the region where ${|x - x^\prime| = O(\sqrt{\hbar\epsilon/m})} $. This allow us to estimate the integral to first order in $\epsilon$ by expanding the $\mu$ terms in Eqs.~\eqref{e:muBP->} and \eqref{e:muBP<-} around $x$ up to second order in ${x - x^\prime}$. Consistent with this approximation to first order in $\epsilon$, we can also do $\exp{[- V(x)\epsilon / \hbar]} = 1 - V(x)\epsilon / \hbar + O(\epsilon^2) $ in factors $F_\ell$. In this way we get the equations 
\begin{widetext}
\BE
\mu_{\to\ell}(x) &=& \mu_{\ell-1}(x) -\frac{\lambda\epsilon}{T} V_\ell(x)\mu_{\to\ell}(x) + \frac{T\epsilon}{2m\lambda}\frac{\partial^2\mu_{\to\ell}(x)}{\partial x^2} + O(\epsilon^2),\\
\mu_{\ell\leftarrow}(x) &=& \mu_{\ell+1\leftarrow}(x) -\frac{\lambda\epsilon}{T} V_\ell(x)\mu_{\ell\leftarrow}(x) + \frac{T\epsilon}{2m\lambda}\frac{\partial^2\mu_{\ell\leftarrow}(x)}{\partial x^2} + O(\epsilon^2). 
\EE
\end{widetext}

Now, to get the continuous-time limit, let ${\mu_\to (x, \ell\epsilon) = \mu_{\to\ell}(x)}$ and ${\mu_\leftarrow (x, \ell\epsilon) = \mu_{\ell\leftarrow}(x)}$, and expand ${\mu_{\to}(x,t-\epsilon) = \mu_\to(x,t) - \epsilon\dot{\mu}_{\to}}(x,t)$ as well as ${\mu_{\leftarrow}(x,t+\epsilon) = \mu_{\leftarrow} + \epsilon\dot{\mu}_{\leftarrow}(x,t)}$, where $t = \ell\epsilon$ and the dot operator stands for time derivative. So, taking $\epsilon\to 0$ we obtain 
\begin{widetext}
\BE
-\frac{T}{\lambda} \frac{\partial\mu_{\to}(x, t)}{\partial t} &=& -\frac{T^2}{2m\lambda^2}\frac{\partial^2\mu_{\to}(x, t)}{\partial x^2} +V(x, t)\mu_{\to}(x, t)  ,\label{e:EQBP->} \\
\frac{T}{\lambda}\frac{\partial\mu_{\leftarrow}(x, t)}{\partial t} &=& -\frac{T^2}{2m\lambda^2}\frac{\partial^2\mu_{\leftarrow}(x, t)}{\partial x^2} +V(x,t)\mu_{\leftarrow}(x, t)  ,\label{e:EQBP<-}
\EE
\end{widetext}
which yields precisely the imaginary-time Schr\"odinger equation and its adjoint, with $\hobs = T/\lambda$ playing the role of Planck constant $\hbar$. Indeed, Eqs.~\eqref{e:EQBP->} and \eqref{e:EQBP<-} are formally analogous to Eqs. (2.1) and (2.17) in Ref.~\cite{Zambrini-1987}; the analogous of $\theta$ and $\theta^\ast$ therein are here $\mu_{\leftarrow}$ and $\mu_{\to}$, respectively (but see Appendix~\ref{s:EQM}).

\subsection{Euclidean quantum mechanics: From linear chains to cycles}\label{s:EQM}
\begin{figure}
\includegraphics[width=\columnwidth]{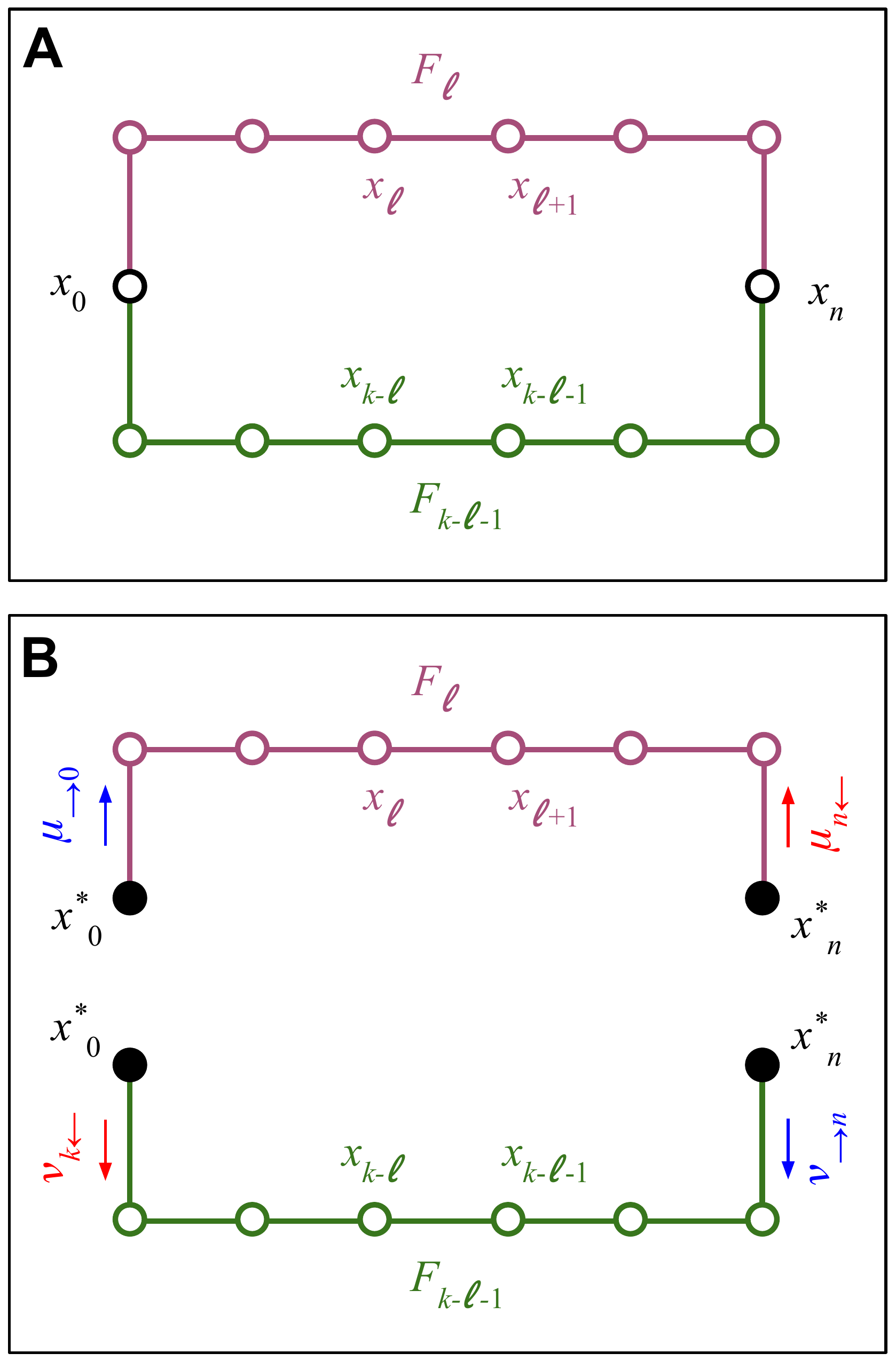}
\caption{ {\it Belief propagation on cycles}. (A) Factor graph with circular topology, in which belief propagation does not generally lead to the correct marginals \cite{Weiss-2000}. (B) By conditioning on $x_0$ and $x_n$ (black filled circles) we effectively turn the cycle into two chains---one ``external'' (purple) and another ``internal'' (green)---with $x_0$ and $x_n$ clamped to some given values $x_0^\ast$ and $x_n^\ast$. Such clamping can be implemented via initial and final marginals $p_0(x_0) = \delta(x_0 - x_0^\ast)$ and $p_n(x_n) = \delta(x_n - x_n^\ast)$. More generally, we can fix generic marginals $p_0$ and $p_n$ and search for solutions consistent with them. }
\label{f:cavity_cycle}
\end{figure}

Although Eqs.~\eqref{e:EQBP->} and \eqref{e:EQBP<-} look like imaginary-time quantum dynamics, things are not so interesting for chains. Indeed, the messages $\mu_{\to 0}$ and $\mu_{n\leftarrow}$, which correspond to the leaves of the chain and serve as the initial conditions for Eqs.~\eqref{e:muBP->} and \eqref{e:muBP<-}, can always be chosen constant. The reason is that there are no factors before $F_0$ nor after $F_n$. In contrast, imaginary-time quantum dynamics involve more interesting initial and final conditions. Indeed, besides Eqs.~\eqref{e:EQBP->} and \eqref{e:EQBP<-}, the imaginary-time quantum dynamics described in Ref.~\cite{Zambrini-1987} is characterized by general initial and final conditions (see Eq.~(2.14) therein). Moreover, belief propagation is not exact on cycles~\cite{Weiss-2000}. 

To have some initial insight, notice that the probability distribution for a chain can be written as (see Fig.~\ref{f:cavity}a)
\begin{equation}\label{em:linear}
        \mathcal{P}_{\rm ch}(x_0,\dots ,x_n) = \widetilde{p}_0(x_0)\prod_{\ell=0}^{n-1} \widetilde{\mcP}_\ell(x_{\ell+1}|x_{\ell}),
\end{equation}
where $\widetilde{p}_0(x_0)$ is the probability for the initial state to be $x_0$ and $\widetilde{P}_\ell(x_{\ell+1}|x_\ell)$ is the probability for the system to perform a transition from state $x_\ell$ to state $x_{\ell+1}$ at time step $\ell$. This implies that we can generate a path or ``history'' $(x_0,\dotsc , x_n)$ consistent with $\mcP_{\rm ch}$ by first sampling from $\widetilde{p}_0(x_0)$ and then subsequently sampling from the conditional distributions $\widetilde{\mcP}_\ell(x_{\ell + 1}|x_\ell)$, which leads to a Markov chain in states $x_\ell$. In analogy with physics, $\widetilde{p}_0$ could be said to be the initial state that is propagated forward in time according to the ``mechanism'' or ``law'' $\widetilde{\mcP}_\ell$. In principle, we can change the initial state $\widetilde{p}_0\to p_0$ without changing the ``mechanism'', so the probability of a path changes accordingly by replacing $\widetilde{p}_0$ by $p_0$ in Eq.~\eqref{em:linear}.

In contrast, for a cycle (see Fig.~\ref{f:cavity_cycle}A)
\begin{equation}\label{e:ring_EQM}
\widetilde{\mathcal{P}}(\widetilde\bx) = \frac{1}{Z}\prod_{\ell=0}^{k-1} F_\ell(x_{\ell +1},x_{\ell}),
\end{equation}
this is not necessarily true. Here $k = 2n$, $x_k = x_0$ and $\widetilde{\bx} = (x_0,\dotsc , x_n, \dotsc , x_{k-1})$. Consider the marginal $\mcP(\bx) = \sum_{x_{n+1},\dotsc , x_{k-1}}\widetilde{\mcP}(\widetilde{\bx})$, where $\bx = (x_0,\dotsc , x_n)$. In general, we have
\be\label{e:Bernstein} 
\mathcal{P}(x_0,\dotsc ,x_n) = {p}(x_0, x_n) \prod_{\ell = 0}^{n-2} \mathcal{P}_{\ell}(x_{\ell + 1}|x_\ell , x_n),
\ee
which has the structure of a Bernstein process (see, e.g., the integrand in Eq.~(2.7) in Ref.~\cite{Zambrini-1987}, where ${p}(x_0 , x_n) \to m(x, y)$ therein and $\mathcal{P}_{\ell}\to h$ therein). Equation~\eqref{e:Bernstein} is related to the fact that we can turn a cycle into two chains by conditioning on two variables, say $x_0=x_0^\ast$ and $x_n=x_n^\ast$ (see Fig.~\ref{f:cavity_cycle}b). This yields the factorization
\begin{equation}\label{em:histories}
\mathcal{P}(x_1,\dots ,x_{n-1}|x_0^\ast, x_n^\ast) = \prod_{\ell=0}^{n-2} P_\ell^{(x_0^\ast, x_n^\ast)}(x_{\ell+1}|x_{\ell}),
\end{equation}
where
\be 
\mcP_\ell^{(x_0^\ast, x_n^\ast)}(x_{\ell+1}|x_{\ell}) = \mcP_\ell(x_{\ell+1}|x_{\ell},x_0^\ast, x_n^\ast) = \mcP_\ell(x_{\ell+1}|x_{\ell}, x_n^\ast),
\ee	
which is not a Markov chain in variables $x_\ell$ alone. So, we cannot generate a path or history $(x_0,\dotsc , x_n)$ consistent with $\mcP$ by first sampling from a single variable marginal $\widetilde{p}_0(x_0)$ and a Markovian ``mechanism'' or ``law'' $\widetilde{\mcP}_{\ell}(x_{\ell+1}|x_\ell)$ on variables $x_\ell$. We certainly can use a Markov chain to sample from $\mcP$, but it would have to be run for a time long enough to build up the required correlations; it would not do the job in a single pass. So, while Markov processes could be said to naturally represent linear causality, so common in physics, Bernstein processes and its quantum-like features (see below) could be considered as natural examples of reciprocal causality, so common in cognitive science.

However, even though the system is not Markov with respect to a chain on variables $x_\ell$~\cite{pearl2009causality} (p. 16), we can still describe it by multiple ``Markov chains'' on $x_\ell$, i.e., one per each choice of $x_0^\ast$ and $x_n^\ast$ (see Eq.~\eqref{em:histories}). Colloquially, if we say that a ``world'' is characterized by a chain of cause-effect relationships, cycles could be said to induce a kind of ``many-world'' description of phenomena. Notice that the need to condition on initial and final variables parallels the need to fix initial and final conditions in physics, e.g., in the least action principle of classical mechanics or the path integral formulation of quantum mechanics. 
 
We now show that, even though the BP algorithm is not exact on cycles~\cite{Weiss-2000}, the dynamics can still be naturally formulated in terms of imaginary-time quantum dynamics. Assume that we know the initial and final single-variable marginals, $p_0$ and $p_n$. Conditioning on the initial and final variables turns the cycle into two chains (see Fig.~\ref{f:cavity_cycle} B): an ``external'' chain going from time step $0$ to time step $n$ (purple) and an ``internal'' chain going from time step $n$ to time step $k = 2n$ (green). So, BP can become exact again on each chain, as long as the messages are consistent with the initial and final conditions. We will see that if the ``external'' dynamics are given by the probability matrix $P_0^{\rm ext}$ then the ``internal'' dynamics are given by the transposed matrix, i.e., $P_\ell^{\rm int} = \left[P_\ell^{\rm ext}\right]^T$. For generic initial and final conditions, we have that $P_0^{\rm ext}$ and $P_n^{\rm ext}$ are not necessarily symmetric and can be given by ``pure states'', yet still satisfy an imaginary-time quantum dynamics. So, the results presented in the main text can in principle be extended to general probability matrices. 

Indeed, we can formulate the dynamics on the ``external'' chain in terms of messages $\mu_{\to\ell}$ and $\mu_{\ell\leftarrow}$ satisfying Eqs.~\eqref{e:muBP->} and \eqref{e:muBP<-} as well as the initial and final conditions (cf. Eq.~(2.14) in Ref.~\cite{Zambrini-1987})
\BE
p_0(x) &=& \mu_{\to 0}(x)\mu_{0\leftarrow}(x)\nonumber\\
&=& \mu_{\to 0}(x) \int \mu_{n\leftarrow }(x^\prime)\widetilde{F}_0(x^\prime , x)\mathrm{d} x^\prime,\label{e:mu00um}\\
p_n(x) &=& \mu_{\to n}(x)\mu_{n\leftarrow}(x)\nonumber\\
&=& \left[\int \widetilde{F}_0(x ,x^\prime)\mu_{\to 0 }(x^\prime)\mathrm{d} x^\prime\right]  \mu_{n\leftarrow}(x) ,\label{e:munnum}
\EE
where $\widetilde{F}_0 = F_{n-1}\cdots F_0$. In the continuous-time limit, this yields the formal analogue of the imaginary-time quantum dynamics presented in Ref.~\cite{Zambrini-1987}. 

We can also formulate the dynamics of the ``internal'' chain, which goes from time step $n$ to time step $k=2n$, in an analogous way. It is convenient to use $k - \ell$, with $\ell = 0,\dots , n-1$, as time step index. The dynamics of the messages $\nu_{\to k - \ell}$ and $\nu_{k -\ell\leftarrow}$ for the ``internal'' chain satisfy 
\BE
\nu_{\to k- \ell }(x) &=& \int {F}_{k-\ell-1} (x, x^\prime) \nu_{\to k-\ell -1 } (x^\prime) \mathrm{d} x^\prime , 
\label{e:|nu>eq0} \\
\nu_{k-\ell\leftarrow} (x) &=& \int\nu_{k-\ell +1 \leftarrow }(x^\prime) {F}_{k-\ell } (x^\prime ,x)\mathrm{d} x^\prime , 
\label{e:<nu|eq0}
\EE
which are analogous to Eqs.~\eqref{e:muBP<-} and \eqref{e:muBP->}. Furthermore, the $\nu$-messages must satisfy the boundary conditions
\BE
p_n(x) &=& \nu_{\to n}(x)\nu_{n\leftarrow}(x)\nonumber \\
&=& \nu_{\to n} (x) \int\nu_{k\leftarrow }(x^\prime)\widetilde{F}_n (x^\prime , x) \mathrm{d} x^\prime,\label{e:nu00un}\\
p_k(x) &=& \nu_{\to k}(x)\nu_{k\leftarrow}(x)\nonumber\\
&=& \left[\int \widetilde{F}_n (x, x^\prime) \nu_{\to n } (x^\prime) \mathrm{d} x^\prime\right] \nu_{k\leftarrow}(x) ,\label{e:nunnun}
\EE
which are analogous to Eqs.~\eqref{e:mu00um} and \eqref{e:munnum}. Here $\widetilde{F}_n = F_{k-1}\cdots F_n$ and $p_k = p_0$ as they refer to the same variable, $x_0$.

Now, taking into account that $F_{k-\ell-1}=F_\ell^T$ (see Eq.~\eqref{e:F=FT}), Eqs.~\eqref{e:|nu>eq0} and \eqref{e:<nu|eq0} become
\BE
\nu_{\to k- \ell }(x) &=& \int {F}_{\ell}^T (x, x^\prime) \nu_{\to k-\ell -1 } (x^\prime) \mathrm{d} x^\prime , 
\label{e:|nu>eq} \\
&=& \int \nu_{\to k-\ell -1 } (x^\prime) {F}_{\ell} ( x^\prime , x) \mathrm{d} x^\prime\nonumber \\
\nu_{k-\ell\leftarrow} (x) &=& \int\nu_{k-\ell +1 \leftarrow }(x^\prime) {F}_{\ell -1 }^T (x^\prime ,x)\mathrm{d} x^\prime , 
\label{e:<nu|eq}\\
&=& \int {F}_{\ell -1 } (x, x^\prime)\nu_{k-\ell +1 \leftarrow }(x^\prime)\mathrm{d} x^\prime . \nonumber 
\EE
Equations~\eqref{e:|nu>eq} and \eqref{e:<nu|eq} are equivalent to Eqs.~\eqref{e:muBP<-} and \eqref{e:muBP->}, respectively. We can see this by doing (notice that arrows point in opposite directions)
\BE
\nu_{\to k-\ell} &=& \mu_{\ell\leftarrow},\label{e:|munu>}\\
\nu_{k-\ell\leftarrow} &=& \mu_{\to \ell}.\label{e:<munu|}
\EE

So, every solution to the ``external'' chain dynamics---characterized by a probability matrix $P^{\rm ext}_\ell(x,x^\prime) = \mu_{\to \ell} (x)\mu_{\ell\leftarrow}(x^\prime)$---yields, through Eqs.~\eqref{e:|munu>} and \eqref{e:<munu|}, a solution to the ``internal'' chain dynamics---characterized by a probability matrix $P^{\rm int}_{k-\ell}(x,x^\prime)=\nu_{\to k-\ell}(x)\nu_{k-\ell\leftarrow}(x^\prime) = P^{\rm ext}_\ell(x^\prime, x)$. Using matrix notation, $P_{k_\ell}^{\rm int} = \left[P_\ell^{\rm ext}\right]^T$ is given by the transpose of $P_\ell^{\rm ext}$. In particular, $P_k^{\rm int} = \left[P_0^{\rm ext}\right]^T$ and $P_n^{\rm int} = \left[P_n^{\rm ext}\right]^T$. So, both solutions satisfy the corresponding boundary conditions since $p_0(x)=P^{\rm ext}_0(x,x) = P_k^{\rm int}(x,x) $ and $p_n(x)=P_n^{\rm ext}(x,x)=P_n^{\rm int}(x,x)$. 
In analogy with the case of a chain studied at the beginning of this subsection, $p_0$ and $p_n$ could be considered as fixed initial and final states connected via the ``mechanism'' or ``law'' $F_\ell$.  So, in this case we also obtain an imaginary-time quantum dynamics and its transposed for the ``external'' and ``internal'' chains, respectively. Furthermore, the initial state is also related by the transposed operation, which allows for the implementation of the SRC in the main text to yield a quantum dynamics (see Appendix~\ref{s:pair}).

There are proofs of existence and uniqueness of the positive solutions of Eqs.~\eqref{e:muBP->} and \eqref{e:muBP<-} with the boundary conditions in Eqs.~\eqref{e:mu00um} and \eqref{e:munnum}, for $p_0$ and $p_n$ without zeros, with various degrees of generality (see, e.g., Refs.~\cite{zambrini1986stochastic,Zambrini-1987} and references therein). Clearly, the same applies to Eqs~\eqref{e:|nu>eq0} and \eqref{e:<nu|eq0} with the boundary conditions in Eqs.~\eqref{e:nu00un} and \eqref{e:nunnun}. In particular, they hold for the Hamiltonian in Eq.~\eqref{em:H} in the main text.  However, after implementing the SRC we obtain a dynamics formally analogous to quantum dynamics, which is not anymore equivalent to the imaginary-time version. 

In our approach, the two dynamics only have to coincide on the initial state of the system. The imaginary-time quantum formalism is only an intermediate step regarding how things {\em would} look to an {\em imaginary} external scientist. The real-time quantum formalism is associated to how things {\em actually} look to a {\em genuine} scientist from her own intrinsic perspective, the only perspective she has as far as we know. So, we might not need to be concerned with the existence of solutions for the imaginary-time quantum dynamics, but rather with the existence of solutions for the real-time quantum formalism that results after the SRC is implemented---although this may be connected via a Wick rotation~\cite{Zambrini-1987}.

The joint marginal (see Eq.~\eqref{e:Bernstein}; cf. Eq.~\eqref{e:P_BP})
\be
\begin{split}
p(x_0, x_n) &= \mu_{n\leftarrow} (x_n) \widetilde{F}_0 (x_n , x_0) \mu_{\to 0} (x_0)\\
& = \nu_{k\leftarrow}(x_k) \widetilde{F}_n(x_k , x_n) \nu_{\to n}(x_n),
\end{split}
\ee
where $x_k = x_0$ and  the second equality comes from Eqs.~\eqref{e:|munu>} and \eqref{e:<munu|} and the fact that $\widetilde{F}_{n} = \widetilde{F}_{0}^T$. This joint marginal is associated to a Bernstein process that is also Markovian (cf. Eq.~(2.9) in Ref.~\cite{Zambrini-1987}; $\mu_{\to 0}$, $\mu_{n\leftarrow }$, and $\widetilde{F}_0$ here correspond, respectively, to $\theta^\ast$, $\theta$, and $h$ therein). Indeed, using Eqs.~\eqref{e:P_BP}, \eqref{e:p_BP} and \eqref{e:mu_def} we can write 
\BE
\mcP^+_\ell(x_{\ell+1}|x_{\ell})  &=& \frac{\mcP_\ell(x_{\ell+1}, x_\ell  )}{p_\ell (x_\ell)}\nonumber\\ 
&=& F_\ell (x_{\ell + 1} , x_\ell)\frac{\mu_{\ell + 1\leftarrow} (x_{\ell + 1})}{\mu_{\ell\leftarrow} (x_\ell)}, \\
\mcP^-_\ell(x_{\ell}|x_{\ell + 1})  &=& \frac{\mcP_\ell(x_{\ell+1}, x_\ell  )}{p_{\ell + 1} (x_{\ell + 1})}\nonumber\\
&=& F_\ell (x_{\ell + 1} , x_\ell)\frac{\mu_{\to\ell } (x_{\ell })}{\mu_{\to \ell + 1 } (x_{\ell +1 } )}, 
\EE
for the forward and backward transition probabilities of Markov process equivalent to the Bernstein process (cf. Eqs.~(2.12) and (2.11) in Ref.~\cite{Zambrini-1987};  $\mcP^+_\ell$ and $\mcP^-_\ell$ here correspond, respectively, to $q$ and $q^\ast$ therein.). So, in this case, we can generate paths or histories consistent with the whole distribution, Eq.~\eqref{e:Bernstein}, by first sampling from $p_0$ and propagating the sample forward with $\mcP^+_\ell$ as in a standard Markov chain---information about $p_n$ is already encoded in the BP messages. Similarly, we can first sample from $p_n$ and propagate the sample backwards with $\mcP^-_\ell$.

More generally, following the analogy with quantum mechanics, we could in principle search for ``mixed'' solutions where $p_0(x) = \sum_\alpha \lambda_\alpha p^\alpha_{0}(x)$ and $p_n(x) = \sum_\alpha \lambda_\alpha p_n^\alpha (x) $---the index $\alpha$ could be real, in which case the sum turns into an integral. Here, for each ``pure state'' $\alpha$, $p_0^\alpha$ and $p_n^\alpha$ are connected through BP messages, $\{\mu^\alpha_{\to\ell}, \mu^\alpha_{\ell\leftarrow}\}$ and $\{\nu^\alpha_{\to k-\ell}, \nu^\alpha_{k-\ell\leftarrow}\}$, via the ``external'' and ``internal'' chains, respectively. Furthermore, $\lambda_\alpha\geq 0$ denotes the probability (or probability density) associated to $\alpha$. This expands the set of possible solutions since these ``mixed'' solutions include the ``pure'' ones when $\lambda_\alpha \neq 0$ for only one value of $\alpha$. However, we are not aware of formal results on this more general setting. 

The approach discussed here can in principle extend the results of the main text to initial asymmetric probability matrices. However, these do not seem to lead to initial density matrices consistent with standard quantum theory. 

Consider a two-state system with initial pure probability matrix (cf. Eqs.~\eqref{e:mu->} and \eqref{e:<-mu})
\be\label{e:P0nophysical}
\begin{split}
P_0 &= \begin{pmatrix}
\sqrt{p} e^{\phi_1}\\
\sqrt{1-p} e^{\phi_2}\\
\end{pmatrix} 
\begin{pmatrix}
\sqrt{p} e^{-\phi_1} &
\sqrt{1-p} e^{-\phi_2}
\end{pmatrix} \\
&=
\begin{pmatrix}
p & \sqrt{p(1-p)} e^{\Delta\phi}\\
\sqrt{p(1-p)} e^{-\Delta\phi} & 1-p\\
\end{pmatrix},
\end{split} 
\ee
where $\Delta\phi = \phi_1 - \phi_2$. The corresponding initial density matrix $\rho = (P+P^T)/2 + (P-P^T)/2i$ is not necessarily a standard quantum pure state, unless $P_0 = P^T_0$ is symmetric, i.e., $\Delta\phi = 0$. 

Moreover, a generic mixed density matrix for a two-state system can be written as
\be
\rho = \sum_\alpha \lambda_\alpha \rho^\alpha,
\ee
where
\be
\rho^\alpha =
\begin{pmatrix}
p^\alpha & \sqrt{p^\alpha(1-p^\alpha)} e^{i\Delta\varphi^\alpha}\\
\sqrt{p^\alpha(1-p^\alpha)} e^{-i\Delta\varphi^\alpha} & 1-p^\alpha\\
\end{pmatrix}, 
\ee
is a pure density matrix, $\Delta\varphi$ is the phase difference, $0\leq \lambda_\alpha$ and $\sum_\alpha \lambda_\alpha = 1$. The magnitudes of the off-diagonal elements are given by
\be
\left|\sum_\alpha\lambda_\alpha \sqrt{p^\alpha(1-p^\alpha)} e^{\pm i\Delta\varphi^\alpha}\right| \leq \sum_\alpha\lambda_\alpha \sqrt{p^\alpha(1-p^\alpha)} \leq 1,
\ee
which are not larger than one because they are an average of quantities $\sqrt{p^\alpha(1-p^\alpha)}$ whose magnitudes are smaller than one. In contrast, the off-diagonal elements in Eq.~\eqref{e:P0nophysical} can be larger than one if $\Delta\phi\neq 0$. 
The analysis above suggests that only initial probability matrices that are symmetric can be associated to a standard density matrix. This is consistent with the presentation in the main text, since $P_0 = \widetilde{F}_0^T \widetilde{F}_0 = P_0^T$ therein. 
The question of whether the more general case can have a physical interpretation is left for the future.

\subsection{Standard Markov chains in terms of square roots of probabilities}\label{s:squareroot}

In the previous subsection we argued that asymmetric initial probability matrices cannot in general be mapped to an standard density matrix. So, a  consistent starting pure state in our approach must take the form $\rho_0(x,x^\prime) = \sqrt{p(x) p(x^\prime)}$. That is, it should be given by the square roots of the starting marginals $p(x)$. Here we discuss how such square roots of probabilities can naturally appear in standard Markov chains. Indeed, using the product rule of probability theory we can write the pairwise marginals in the three different ways
\be\label{e:MarkovBayes}
\begin{split}
\mathcal{P}_\ell(x_\ell, x_{\ell+1}) &=  \mathcal{P}^+_\ell(x_{\ell+1}| x_\ell) p_\ell(x_\ell)\\
&= \mathcal{P}^-_{\ell}(x_\ell|x_{\ell+1})p_{\ell+1}(x_{\ell+1})\\
&= \theta_{\ell+1}(x_{\ell+1}) K_\ell( x_{\ell+1} , x_\ell)\theta_\ell(x_\ell)
\end{split}
\ee
where $\mathcal{P}^+_\ell$ and $\mathcal{P}^-_\ell$ denote the forward and backward transition probabilities, respectively, and
\BE
\theta_\ell(x_\ell) &=& \sqrt{p_\ell(x_\ell)},\label{e:theta}\\
K_\ell(x_\ell, x_{\ell+1}) &=& \sqrt{\mathcal{P}^+_{\ell}(x_{\ell+1}|x_\ell)\mathcal{P}^-_{\ell}(x_{\ell}|x_{\ell+1})}.\label{e:K_P+*P-}
\EE
The less common, more symmetric alternative in the third line of Eq.~\eqref{e:MarkovBayes} is obtained by multiplying the first two lines in Eq.~\eqref{e:MarkovBayes} and taking the square root.

Equation~\eqref{e:K_P+*P-} can be written as
\be\label{e:KEQM+}
\begin{split}
K_\ell( x_{\ell+1}, x_\ell) =& \mathcal{P}_\ell^+(x_{\ell+1}|x_\ell)\sqrt{\frac{p_\ell(x_\ell)}{p_{\ell+1}(x_{\ell+1})}}\\
=& \mathcal{P}_\ell^+(x_{\ell+1}|x_\ell){\frac{\theta_\ell(x_\ell)}{\theta_{\ell+1}(x_{\ell+1})}},
\end{split}
\ee
by using Bayes rule to change $\mathcal{P}^-_\ell$ in Eq.~\eqref{e:K_P+*P-} for $\mathcal{P}^+_\ell$.
Using Eqs.~\eqref{e:theta}, \eqref{e:KEQM+}, and
\be\label{e:stochastic_update}
p_{\ell+1}(x_{\ell+1})= \int \mathcal{P}_\ell^+(x_{\ell+1}|x_\ell)p_\ell(x_\ell) \mathrm{d} x_\ell,
\ee
we can readily see that
\be\label{e:quantum-like}
\theta_{\ell+1}(x_{\ell+1})=\int K_\ell(x_{\ell+1}, x_\ell)\theta_\ell(x_\ell)\mathrm{d} x_\ell.
\ee

Equation~\eqref{e:K_P+*P-} can also be written as
\be\label{e:KEQM-}
\begin{split}
K_\ell( x_{\ell+1}, x_\ell) =& \mathcal{P}_\ell^-(x_\ell | x_{\ell+1})\sqrt{\frac{p_{\ell +1}(x_{\ell +1})}{p_{\ell}(x_{\ell})}}\\
=& \mathcal{P}_\ell^-(x_{\ell}|x_{\ell+1}){\frac{\theta_{\ell+1}(x_{\ell+1})}{\theta_{\ell}(x_{\ell})}},
\end{split}
\ee
by using Bayes rule to change $\mathcal{P}^+_\ell$ in Eq.~\eqref{e:K_P+*P-} for $\mathcal{P}^-_\ell$, instead. Using Eqs.~\eqref{e:theta}, \eqref{e:KEQM-}, and the reverse of \eqref{e:stochastic_update}, i.e.
\be\label{e:stochastic_update_back}
p_{\ell}(x_{\ell})= \int \mathcal{P}_\ell^-(x_{\ell}|x_{\ell +1})p_{\ell +1}(x_{\ell +1}) \mathrm{d} x_{\ell +1},
\ee
we can readily see that
\be\label{e:quantum-like*}
\theta_{\ell}(x_{\ell})=\int\theta_{\ell+1}(x_{ \ell+1}) K_\ell(x_{\ell +1}, x_{\ell})\mathrm{d} x_\ell.
\ee

Equations~\eqref{e:KEQM+} and \eqref{e:KEQM-} are similar to Eqs.~\eqref{e:muBP->} and \eqref{e:muBP<-} as well as to Eqs.~(2.12) and (2.11) in Ref.~\cite{Zambrini-1987}, respectively, where $K_\ell$ here plays the role of $h$ therein. There is a difference, though, in that here the wave-like functions $\theta_\ell$ do not have a phase; they are strictly equal to the square root of the corresponding probability.

\subsection{A potential alternative road to classical mechanics}\label{s:classical}

Here we show that the ``classical'' limit of imaginary-time quantum dynamics, i.e., $\hobs\to 0$, can follow a dynamics formally analogous to imaginary-time classical dynamics. We focus on energy functions of the form (cf. Eq.~\eqref{em:non-relativisticH} in the main text)
\be\label{e:nonH}
\mcH(x,x^\prime ) = m (x-x^\prime)^2/\epsilon^2 + V(x).
\ee
%


From an extrinsic perspective, the system composed of observer and experimental system is described by the probability distribution $\mcP(\bx)$ to observe a path $\bx = (x_0,\dotsc , x_n)$---see Fig.~\ref{f:circular} and Eqs.~\eqref{em:circular} and \eqref{em:F} in the main text. In the $\hobs\to 0$ limit, the system follows the most probable path, which is the one that minimizes the total energy function $\sum_{\ell }\mcH(x_{\ell+1}, x_\ell)$, given initial and final conditions, $x_0$ and $x_n$. Such optimal path satisfies therefore the equations
\be\label{e:opt_path}
\frac{\partial \mcH (x_{\ell+1}, x_\ell)}{\partial x_\ell} + 
\frac{\partial \mcH (x_{\ell}, x_{\ell -1})}{\partial x_\ell} = 0
\ee
for $0 < \ell < n-1$, except that the end points, $\ell = 0$ and $\ell = n-1$ are fixed. Using Eq.~\eqref{e:nonH} this yields
\be\label{e:NE_it}
m\left[\frac{x_{\ell +1} - 2x_\ell + x_{\ell -1}}{\epsilon^2}\right] = \frac{\partial V (x_\ell)}{\partial x_\ell} .
\ee
Since in the continuous-time limit, $\epsilon\to 0$, the term in square brackets in the left hand side of Eq.~\eqref{e:NE_it} is the acceleration, this is Newton equation in an inverted potential $-V$. Under a Wick rotation, $\epsilon\to i\epsilon$, Eq.~\eqref{e:NE_it} turns into the actual Newton equation in the potential $V$ with the correct sign (cf. Eqs.~\eqref{e:vN_Wick} and \eqref{e:Sch_Wick} in Appendix~\ref{s:quantum_nutshell}). 

We see that the need to condition on initial and final states, so common in physics---e.g., the least action principle formulation of classical mechanics and the path integral formulation of quantum mechanics---arises naturally here as a way to break the underlying circularity (see Appendix~\ref{s:EQM}). Such a circularity can also be broken by conditioning on two consecutive variables $x_0$ and $x_1$, which amounts at conditioning on initial position and velocity. This suggests that the reciprocal causality, associated to the embodied observer, might be responsible for the fact that the fundamental equations of physics are typically second-order differential equations in terms of a kind of physical variable, e.g., position only, instead of the more parsimonious, first-order equations.

Now, imaginary-time classical dynamics turns into real-time classical dynamics after a Wick rotation, acquiring a symplectic structure associated to ``energy'' function $\mcH_\ell$. Furthermore, the SRC in the main text effectively implements a Wick rotation that turns imaginary-time into real-time quantum dynamics. So, it is natural to expect that a suitably adapted version of the SRC coupling could turn imaginary-time into real-time classical dynamics too. The Wick rotation is common in special and general relativity too. So, while reciprocal causality might be the reason for the fundamental equations of physics to be second-order, the SRC, associated to the intrinsic perspective, might be the reason such equations acquire a symplectic structure associated to an energy function when these laws are written in terms of a {\em pair} of conjugated variables. 

\

\section{Self-reference and science}\label{s:self-reference}

Here we discuss how self-reference may play a key role in science in general, and physics in particular for the reader that may not be familiar with this concept that plays a central role in our approach. Our goal is to facilitate the review of the literature on self-reference and to provide some initial ideas on how this concept might help us build a more encompassing, manifestly self-referential (or reflexive) scientific framework~\cite{velmans2009understanding,bitbol2008consciousness,thompson2014waking,varela2017embodied} (see also Ref.~\cite{blackmore2018consciousness}, ch. 17,18). 

We first briefly describe some common strategies to deal with self-referential systems in biology, computer science, and logic.  
These strategies are usually based on a pair of objects that either mutually refer to each other or are initially treated as two generic objects that end up referring to the same system. In particular, we discuss Kleene's recursion theorem, which formalizes self-reference using the former type of strategy. In the main text we build on these ideas to implement the intrinsic perspective.

Afterwards, we argue that science may greatly benefit by following the example of computation that fully embraces self-reference, rather than trying to avoid it---as scientists have usually done perhaps due to the puzzling phenomena associated to self-reference. Indeed, computation---whose power is intrinsically tied to self-reference---is a concrete, well-defined mathematical example of the viability and potential advantages of embracing self-reference. A manifestly self-referential scientific framework would be in line with neurophenomenology, which combines refined first- and third-person methods for investigating consciousness (see Appendix~\ref{s:physical-phenomenal}). Finally, consistent with our approach, we discuss some potential relationships between quantum theory and some aspects of self-reference.

\subsection{Some aspects of self-reference}\label{s:aspects}
Here we discuss some aspects of self-reference with the aim to provide some intuition to the readers that may not be familiar with it. The main point we want to highlight is that the strategies to deal with self-reference often involve pairs of elements that, in a sense, play complementary roles by mutually referring to each other or by representing the same object in two different ways. In particular, we discuss Kleene's recursion theorem which could be considered as a mathematical formalization of these ideas. A similar strategy is used in the main text to build models of observers that can refer to themselves. Here we also highlight that self-reference is associated with both strength, e.g., universality, and weakness, e.g., incompleteness and undecidability. 

\subsubsection{A kind of ``complementarity''}\label{s:complementarity}

A common example of self-referential system is the DNA molecule, which can help produce a copy of itself. Key in achieving this is an architecture composed of two complementary subsystems, two strands, that enable the production of a copy of each other. In doing so, each strand plays dual roles, active and passive, when participating in the replication of and when being replicated by the other strand, respectively. A computational analog, though less symmetric, is the Python 3 program in Fig.~\ref{f:first}B in the main text which prints out a copy of its own code. Again, key in achieving this is an architecture composed of two strings. The top string passively refers to the string below, while the bottom string actively prints out the top string. This complementary architecture of Turing machines that can refer to themselves is formalized in Kleene's recursion theorem (see Appendix~\ref{s:recursion}). The Python 3 code in Fig.~\ref{f:first}B corresponds to the particular case of a Turing machine $\textsc{TM}$ that, for any input $w$, prints its own code $\textsc{``TM''}$, i.e. $\textsc{TM(}w\textsc{)} = \textsc{``TM''}$.

Generally, a self-referential system plays both an active role, as the ``subject'' that is doing the act of referring (to itself), and a passive role, as the ``object'' that is being referred to (by itself). In particular, the self-referential capabilities of computer programs sprout from the possibility to assign to each program (an active ``subject'') a string of characters (a passive ``object''), i.e., the code that generates it. Since programs output and take as input strings of characters, they can in particular output or take as input their own code. Indeed, it is possible to formulate computation in a manifestly self-referential way, e.g., via $\lambda$-Calculus (see Appendix~\ref{s:frameworks}).

\subsubsection{Universality}\label{s:universality}

In computation, self-reference is closely related to universality. In particular, without self-reference there cannot be universal Turing machines~\cite{moore2011nature} (ch. 7). One way in which a Turing machine can refer to itself relies on a Universal Turing machine
\be\label{e:UTM}
\textsc{U(``TM'',}w\textsc{)} = \textsc{TM(}w\textsc{)}.
\ee
This is a two-input Turing machine which builds the Turing machine \textsc{TM} whose code is given by the first input, \textsc{``TM'''}, runs it on the second input $w$, and outputs the result, \textsc{TM($w$)}. In this sense, the first and second inputs of \textsc{U} in Eq.~\eqref{e:UTM} play active and passive roles, respectively. Since $w$ can be any string of characters, it can be $w=\textsc{``TM''}$ which yields $\textsc{U(``TM'', ``TM'')} = \textsc{TM(``TM'')}$, i.e. a Turing machine \textsc{TM} running on (a description of) itself. 

\subsubsection{Undecidability and incompleteness}\label{s:incompleteness}

Self-reference can lead to puzzling phenomena like undecidability and incompleteness. Here we briefly discuss these two consequences of self-reference. In Appendix~\ref{s:planck-self} we briefly discuss some potential analogies of these phenomena in the case of (self-referential) observers, which may have implications for the ideas developed in the main text. 

The strategy of using functions of two variables playing effectively dual roles, such as \textsc{U} in Eq.~\eqref{e:UTM}, to study self-referential systems is rather common~\cite{yanofsky2003universal}. We now consider two additional examples associated to undecidability and incompleteness. 

First, assume there exists a special Turing machine~\cite{moore2011nature}
\be\label{e:Halts}
\textsc{Halts(``TM'',} w\textsc{)} = \begin{cases} 1,\: &\text{  if  {\sc TM} halts on } w,\\ 
0,\: &\text{  otherwise,} 
\end{cases}
\ee
that determines whether an arbitrary Turing machine \textsc{TM}, whose code is given by the first input \textsc{``TM''}, halts when run on the second input, $w$. It would then be possible to build another special Turing machine, \textsc{Catch22(``TM'')}, which loops forever if $\textsc{Halts (``TM'', ``TM'')} = 1$ and outputs $1$ otherwise. Then  \textsc{Halts(``Catch22'', ``Catch22'')} would be undefined since \textsc{Catch22(``Catch22'')} halts if and only if it does not. So, the halting problem is undecidable.

Now, consider a formal system that makes statements about natural numbers. Examples of statements are axioms and theorems. Proofs of theorems are also statements, made of sequences of statements, connecting axioms to theorems. Statements and numbers are the analog of Turing machines and the strings of characters on which they operate, respectively. As with Turing machines, it is also possible to assign to each statement $\mathcal{S}$ (an active ``subject'') a natural number ``$\mathcal{S}$'' (a passive ``object''), i.e., the so-called G\"odel number of the statement. This allows statements to refer to themselves. For instance, let~\cite{yanofsky2003universal}
\begin{widetext}
\be\label{e:Prov}
\mathrm{Prov}(x,y)\equiv \textrm{``}y\textrm{ is the G\"odel number of a proof of a statement whose G\"odel number is }x\textrm{.''} 
\ee
\end{widetext}
and consider the statement $\mathcal{S}(x)\equiv (\forall y)\neg\textrm{Prov}(x,y)$, which contains a numeric variable $x$. In words, $\mathcal{S}(x)$ states that there is no proof for the statement whose G\"odel number is $x$.  It is possible to find a natural number $x^\ast$ which equals the G\"odel number ``$\mathcal{S}(x^\ast)$'' of the corresponding statement $\mathcal{S}(x^\ast)$, i.e. ${x^\ast=\textrm{``}\mathcal{S}(x^\ast)\textrm{''}}$. The statement $\mathcal{S}(x^\ast)$ is referring to itself: it states that there is no proof for the statement whose G\"odel number is ``$\mathcal{S}(x^\ast)$'', i.e. that there is no proof for itself. So, the formal system is incomplete.
%

\subsubsection{Kleene's recursion theorem}\label{s:recursion}

\begin{figure*}
\includegraphics[width=0.6\textwidth]{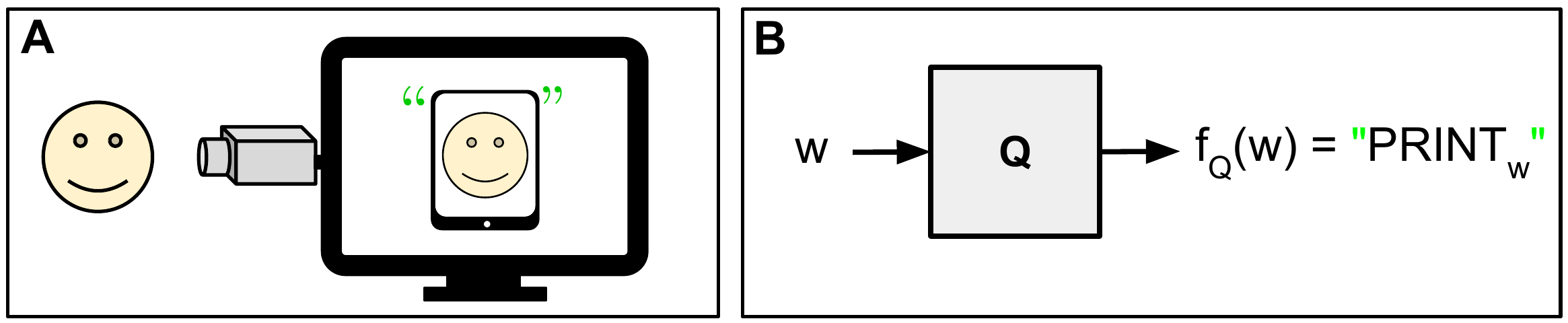}
\caption{ {\it Inferring the description of a printing machine.} (A) Cartoon example of a Turing machine (represented here by a computer) that, given some input (here a face), prints the {\it description} of a Turing machine (represented here by a Tablet within quotation marks) that prints the given input. (B) A more formal representation of the Turing machine in (A), here called $Q$. Ref.~\cite{sipser2006introduction} uses $Q$ to proof the recursion theorem. The input is a string $w$ of characters from a suitable alphabet, and the Turing machine that prints $w$ is called {\sc Print}$_w$. The Turing machine $Q$ basically {\it infers} $\textsc{Print}_w$, effectively implementing the function $f_Q$ that maps $w$ into ``$\textsc{Print}_w$''. Proving the existence of $Q$ is straightforward (see Ref.~\cite{sipser2006introduction}, ch. 6). }
\label{f:q}
\end{figure*}

\begin{figure*}
\includegraphics[width=0.6\textwidth]{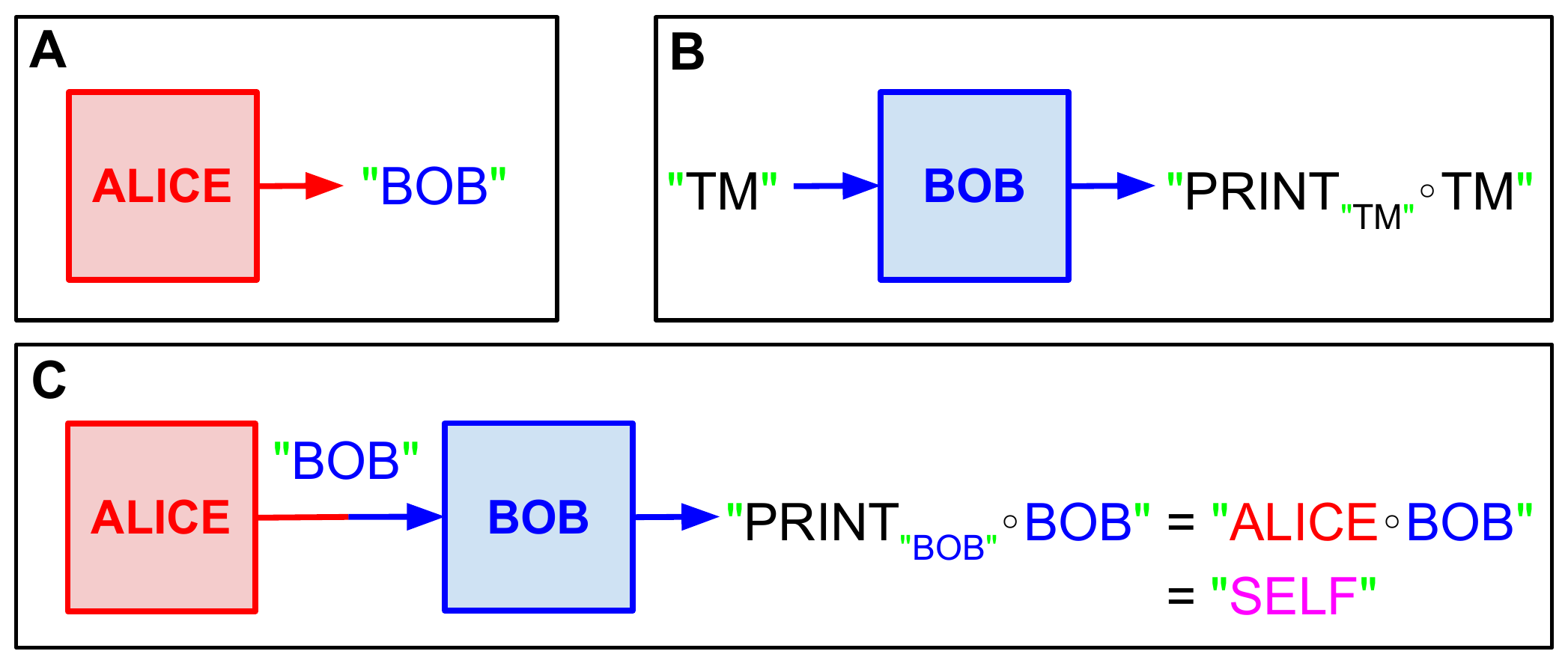}
\caption{ {\it Self-printing Turing machine.} (A) The Turing machine $\textsc{Alice} = \textsc{Print}_{\text{``\textsc{Bob}''}}$ prints a description of the Turing machine {\sc Bob}. But how is {\sc Bob} defined? (B) {\sc Bob } takes as input the description ``{\sc TM}'' of a generic Turing machine {\sc TM} and infers, via $Q$ (see Fig.~\ref{f:q}), the description of a Turing machine $\textsc{Print}_{\text{``\textsc{TM}''}}$ that prints ``{\sc TM}''. {\sc Bob} then composes the Turing machine $\textsc{Print}_{\text{``\textsc{TM}''}}$ with the Turing machine {\sc TM} and outputs the corresponding description, i.e. ``$\textsc{Print}_{\text{``\textsc{TM}''}}\circ \textsc{TM}$''; composition is represented here by the symbol $\circ$. (C) The Turing machine $\textsc{Self} = \textsc{Alice}\circ\textsc{Bob}$ that results from the composition of {\sc Alice} and {\sc Bob} outputs a description of itself, as it can be seen by doing {\sc TM} $=$ {\sc Bob} in (B) and using $\textsc{Alice} = \textsc{Print}_{\text{``\textsc{Bob}''}}$. }
\label{f:quine}
\end{figure*}

Here we briefly discuss the recursion theorem of computer science, which can be considered as a mathematical formalization of the concept of self-reference. Figure~\ref{f:first}B in the main text illustrates the core concept underlying the recursion theorem using the specific case of a self-printing program, or quine. A self-printing program is composed of two complementary sub-programs, say {\sc Alice} and {\sc Bob}, that, like the upper and lower strings in the Python 3 program in Fig.~\ref{f:first}B, essentially print each other. 

Interestingly, since Turing machines can be implemented using Recurrent Neural Networks (RNNs)~\cite{siegelmann1995computation,siegelmann1995computation,siegelmann1991turing,hyotyniemi1996turing}, we can expect that a RNN that can refer to itself should have a similar architecture. That is, a self-referential RNN, say $\mathcal{N}_\textsc{Self} = \mathcal{N}_\textsc{Alice}\circ\mathcal{N}_\textsc{Bob}$, should be composed of two subnetworks, $\mathcal{N}_\textsc{Alice}$ and $\mathcal{N}_\textsc{Bob}$, that mutually refer to (e.g., model) each other (see below). Since RNNs also serve as models of biological neural networks, this suggests that double-hemisphere architecture of the brain and the global architecture of the central nervous system, which is composed of the right and left neural subnetworks, may be a way for this neural system to implement or enhance self-reference. For instance, since neural networks are usually interpreted as encoding models of data, this architecture of the brain and nervous system may be a way to acquire or enhance self-modeling capabilities. It is natural to expect, though, that such capabilities are encoded at different scales to enhance the resilience of the system to failures and the like, as suggested by patients who have a single functional brain hemisphere or whose {\em corpus callosum} has been severed~\cite{corballis2018perceptual,pinto2017split}.

We now formalize the recursion theorem following Ref.~\cite{sipser2006introduction} (chapter 6; see also Ref.~\cite{recursion}). Let $\Sigma$ be an alphabet, i.e. a finite set of characters, and let $\Sigma^\ast$ denote the set of all possible strings of characters from alphabet $\Sigma$, here referred to as words.  A Turing machine is an abstract machine with no memory constraints which can manipulate the characters in an alphabet $\Sigma$ according to a pre-specified set of rules. In other words, a Turing machine is the implementation of an abstract mechanical process that transforms a given string of characters into another, effectively computing a given function $f:\Sigma^\ast\to\Sigma^\ast$. 

It is possible to associate to every Turing machine {\sc TM} a unique string of characters $\textsc{``TM''}\in\Sigma^\ast$, which is referred to as the description of the Turing machine; the quotation marks {\sc `` ''} can be considered here as an operator that transform Turing machines into strings in $\Sigma^\ast$.  A Turing machine {\sc TM} is the abstract version of a program, e.g. a search engine, that can run on a computer to perform a given task, e.g. search for websites related to a specific keyword. A description of a Turing machine {\sc ``TM''}, instead, is the abstract version of the code written in a specific programming language that is used to compile the corresponding program. This ability to associate a unique string of characters to a Turing machine is what allows a Turing machine to implement self-reference. Indeed, from this perspective we can think of a Turing machine {\sc TM} as a string of characters, i.e. its description {\sc ``TM''}, that can manipulate any string of characters $w\in\Sigma^\ast$, including its own description, i.e. $w=\textsc{``TM''}$. In this sense, it can manipulate (a description of) itself.

An example relevant for our discussion is the Turing machine {\sc Q} represented as a computer in Fig.~\ref{f:q}A. Given some input (e.g. the image of a face in Fig.~\ref{f:q}A), the Turing machine {\sc Q} prints the {\em description} of another Turing machine, represented by a tablet within quotation marks in Fig.~\ref{f:q}A, that prints the given input.  More formally, the Turing machine {\sc Q} implements a function $f_{\textsc{Q}}$, represented by a box in Fig.~\ref{f:q}B, that takes as input any word $w\in\Sigma^\ast$ and prints the description {\sc ``Print$_w$''} of another Turing machine {\sc Print}$_w$ that ignores its input and just prints $w$. The existence of such a Turing machine is proven in the

\

\noindent{\bf Lemma $6.1$ of Ref.~\cite{sipser2006introduction}:} There is a computable function $f_Q : \Sigma^\ast\to\Sigma^\ast$ such that, for any string $w$, $f_Q(w) = \text{``\textsc{Print}$_w$''}$ is the description of a Turing machine $\textsc{Print}_w$ that ignores its input, just print $w$ and halts. 

\

The Turing machine {\sc Q} is useful to build self-printing Turing machines, as illustrated in Fig.~\ref{f:quine}. As we said above, a self-printing Turing machine 
\be\label{e:SelfTM}
{\textsc{Self} =  \textsc{Alice}\circ\textsc{Bob}},
\ee
is composed of two Turing machines, {\sc Alice} and {\sc Bob}, that essentially print each other. Figure~\ref{f:quine}A shows the Turing machine 
\be\label{e:AliceTM}
\textsc{Alice} = \textsc{Print}_\textsc{``Bob''},
\ee
which ignores its input, prints a description {\sc ``Bob''} of the Turing machine {\sc Bob}, and halts. Now, if $\textsc{Bob} \stackrel{?}{=} \textsc{Print}_\textsc{``Alice''}$ were to similarly ignore its input, just print a description of {\sc Alice}, and halt, we would have a circular definition were the definition of {\sc Alice} depends on who {\sc Bob} is, and vice versa. 

To avoid such circular definition, {\sc Bob} essentially works backwards by inferring the description of {\sc Alice} from the output she produces, which is {\sc ``Bob''} (see Fig.~\ref{f:quine}B). This is precisely what the Turing machine {\sc Q} does: given an input $w=\textsc{``Bob''}$ it prints the description {\sc ``Print$_\textsc{``Bob''}$''} of a Turing machine $\textsc{Print}_\textsc{``Bob''}$ that ignores its input, prints $w=\textsc{``Bob''}$, and halts. 

So, when {\sc Q} takes the input {\sc ``Bob''} it outputs precisely {\sc ``Alice''}, since ${\textsc{Print}_\textsc{``Bob''} = \textsc{Alice}}$. 
But the full self-printing machine actually is ${\textsc{Self} =  \textsc{Alice}\circ\textsc{Bob}}$ (see Fig.~\ref{f:quine}C), so {\sc Bob} is designed such that (see Fig.~\ref{f:quine}B): (i) it takes as input the description {\sc ``TM''} of an arbitrary Turing machine {\sc TM} and infers via {\sc Q} the description of a Turing machine that prints {\sc ``TM''}; (ii) it then generates the composition $\textsc{Print}_{\textsc{``TM''}}\circ\textsc{TM}$ of the Turing machines associated to the description $``{\textsc{Print}_{\textsc{``TM''}}}\textsc{''}$, that it inferred via {\sc Q} from the given input {\sc ``TM''}, and to the description {\sc ``TM''} it receives as input; (iii) it finally prints the description $``{\textsc{Print}_{\textsc{``TM''}}\circ\textsc{TM}}\textsc{''}$ of such composition. This fully specifies {\sc Bob} in a way that is independent of who {\sc Alice} is, i.e.
\be\label{e:BobTM}
\textsc{Bob} = \prescript{}{\textsc{``TM''}}{\textsc{Print}}_{``{\textsc{Print}_{\textsc{``TM''}}\circ\textsc{TM}}\textsc{''}}.
\ee
Notice that the first {\sc Print} operator in the definition of {\sc Bob} in Eq.~\eqref{e:BobTM} has also a left subscript {\sc ``TM''}, which indicates its input; this contrasts with the definition of {\sc Alice} which does not have such a left subscript indicating that it always ignores its input.

So, we can now fully specify {\sc Alice} by replacing {\sc Bob} in Eq.~\eqref{e:AliceTM} by the left hand side of Eq.~\eqref{e:BobTM}. With both {\sc Alice} and {\sc Bob} fully specified, we can fully specify {\sc Self} in Eq.~\eqref{e:SelfTM} too. See Fig.~\ref{f:quine} and Ref.~\cite{sipser2006introduction} (ch. 6) for further details.

\begin{figure*}
\includegraphics[width=0.7\textwidth]{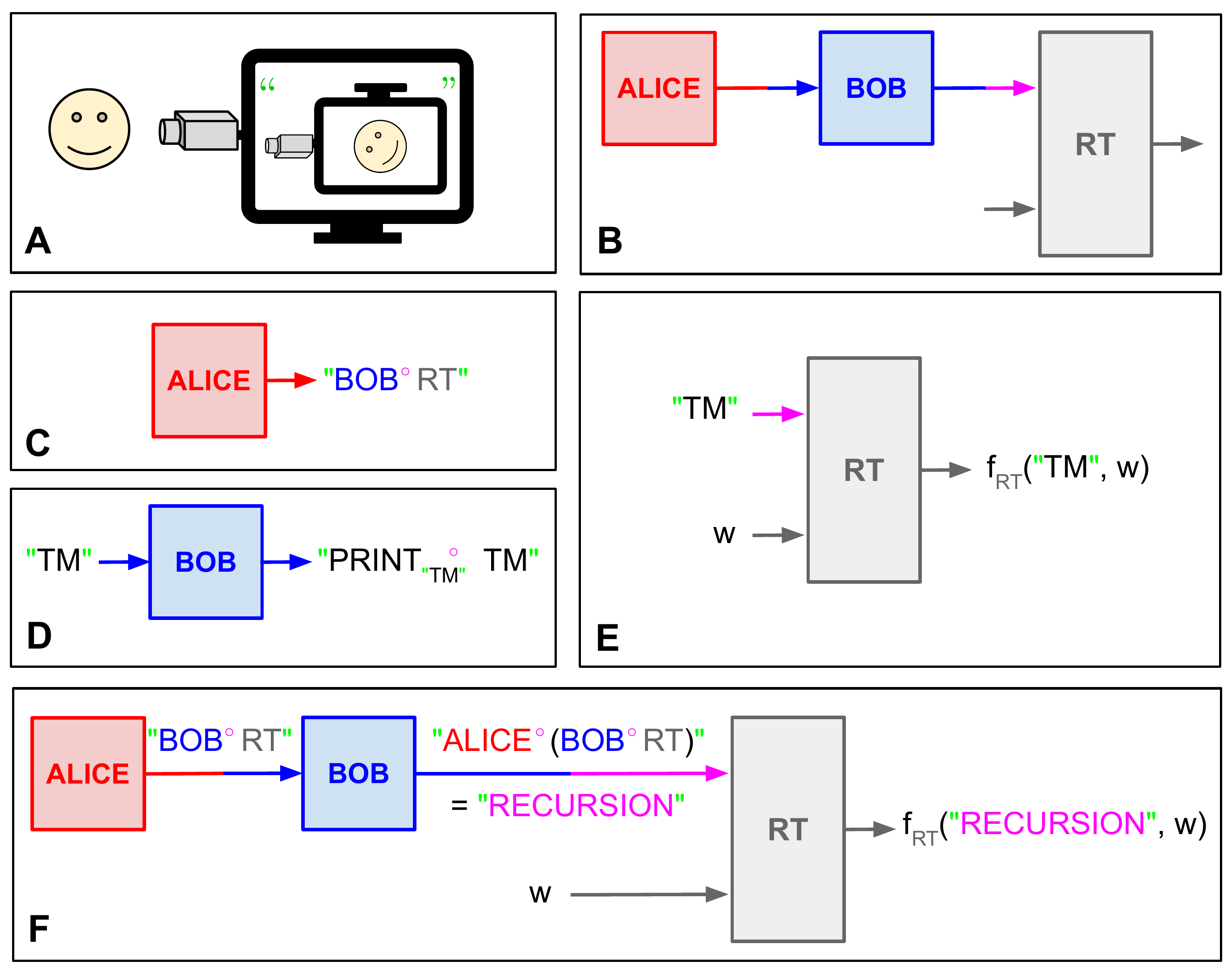}
\caption{ {\it Recursion theorem.} (A) The recursion theorem essentially states that a Turing machine (here a computer) can access a description of itself and manipulate it, along with the input (here a face), to produce a certain output (here a description of a rotated copy of itself printing a rotated face). (B) The architecture of a generic Turing machine, $\textsc{Recursion}$, that implements this idea is composed of three sub-machines: {\sc Alice} and {\sc Bob}, similar to those in Fig.~\ref{f:quine}, along with a 2-input Turing machine {\sc RT}. We can write this Turing machine as ${\textsc{Recursion} = \textsc{Alice}^\circ\textsc{Bob}^\circ\text{RT}}$, where the superindex $^\circ$ refers to composition along the upper input channel of {\sc RT}. But how exactly are {\sc Alice}, {\sc Bob}, and {\sc RT} defined? (C) $\textsc{Alice}= \textsc{Print}_{\text{``\textsc{Bob}$^\circ$\textsc{RT}''}}$ prints a description of the composition of {\sc Bob} and {\sc RT} along the upper input channel of {\sc RT}. (D) {\sc Bob} is defined in the same way as in Fig.~\ref{f:quine}B, only that now {\sc TM} is a 2-input Turing machine. (E) {\sc RT} takes as inputs the description ``{\sc TM}'' of a Turing machine {\sc TM} and a string $w$ and implements a general computable function $f_{\textsc{RT}}(\text{``\textsc{TM}'', w})$, that operates on both the description of {\sc TM} and the input $w$ to produce an output. (F) Proof of the recursion theorem by putting all pieces together and using the definitions of the Turing machines in (C-E). Since ${\textsc{Recursion} = \textsc{Alice}^\circ\textsc{Bob}^\circ\text{RT}}$ and $\textsc{Alice}= \textsc{Print}_{\text{``\textsc{Bob}$^\circ$\textsc{RT}''}}$, we can see that the whole Turing machine {\sc Recursion} implements a function $f_\textsc{RT}(\text{``\textsc{Recursion}''}, w)$ that can use its own description, ``{\sc Recursion}'', during the computation.}
\label{f:recursion}
\end{figure*}

Now, a Turing machine not only can print its own description, it can also use it as an input and perform general computational operations with it. Furthermore, a Turing machine (illustrated in Fig.~\ref{f:recursion}A by a big computer) can take a combined input composed of external data (e.g. the imagine of a face in Fig.~\ref{f:recursion}A) and its own description (illustrated in Fig.~\ref{f:recursion}A by a small computer printed within quotation marks on the screen of the big computer) and perform general computational operations with it. Figure~\ref{f:recursion}A illustrates this by a computer that takes as external input the image of a face and print a description of a rotated version of itself printing a rotated version of the face. 

The architecture of such general Turing machine, that we call here {\sc Recursion}, is composed of three Turing machines (see Fig.~\ref{f:recursion}b): {\sc Alice} and {\sc Bob}, which together generate the description of {\sc Recursion}, and another 2-argument Turing machine {\sc RT} that takes as inputs both an arbitrary word $w\in\Sigma^\ast$ provided from the outside and the description of {\sc Recursion} generated from the inside of {\sc Recursion} itself by the composition of {\sc Alice} and {\sc Bob}, i.e.
\be\label{e:RecursionRT}
\textsc{Recursion} = \textsc{Alice}\circ\textsc{Bob}^\circ\textsc{RT};
\ee
here the superscript $^\circ$ indicates the output of {\sc Bob} is passed to the upper input channel of {\sc RT}.  

The definition of $\textsc{Alice}$ is slightly modified to take into account the new Turing machine {\sc RT} (see Fig.~\ref{f:recursion}C), i.e.
\be\label{e:AliceRT}
\textsc{Alice} = \textsc{Print}_{\textsc{``Bob$^\circ$RT''}}.
\ee
The definition of {\sc Bob} instead remains the same as in Eq.~\eqref{e:BobTM} since it was defined in terms of a generic Turing machine {\sc TM} (see Fig.~\ref{f:recursion}D). The 2-argument Turing machine {\sc RT} implements the actual computations according to a given 2-argument function $f_{\textsc{RT}}$; it is similarly defined in terms of a generic Turing machine {\sc TM} whose description enters through the upper input channel, leaving its lower input channel free to receive external data $w\in\Sigma^\ast$ (see Fig.~\ref{f:recursion}E,F). This proves the (see Fig.~\ref{f:recursion} and chapter 6 in Ref.~\cite{sipser2006introduction} for further details)

\

\noindent{\bf Recursion theorem (Theorem $6.3$ in Ref.~\cite{sipser2006introduction}):} Let {\sc RT} be a Turing machine that computes a 2-argument function ${f_{\textsc{RT}} : \Sigma^\ast\times\Sigma^\ast\to\Sigma^\ast}$. Then there is another Turing machine {\sc Recursion} that computes a function ${f_{\textsc{Recursion}} : \Sigma^\ast\to\Sigma^\ast}$, where for every ${w\in\Sigma^\ast}$,

$$f_{\textsc{Recursion}}(w) = f_{\textsc{RT}}(\text{``\textsc{Recursion}''},w). $$

\subsection{Self-reference, relationalism, and non-duality}\label{s:nonduality}

Self-reference is both powerful and puzzling. On the one hand, without self-reference universal Turing machines cannot exist. On the other hand, self-reference can lead to puzzling phenomena such as incompleteness and undecidability. The latter feature has historically made scientists and mathematicians try to avoid self-reference. We start by briefly discussing why this may not be the most scientifically useful attitude and how self-reference has been actually fully embraced in some domains of mathematics in a rigorous and systematic way. Based on this initial discussion, we argue next that similar considerations may apply to the case of science in general, and that there may indeed be a close relationship between self-reference and relationalism, which has played a relevant role in scientific progress. Finally, we discuss the particular case of the potential relationalism between subject and object, or ``subject/object nonduality''. This opens up the possibility of enlarging science to incorporate ``self-referential experiments'', where scientists could rigorously and systematically both investigate themselves and agree on their personal experiences, e.g., via meditative techniques. In Appendix~\ref{s:planck-self}, we briefly discuss quantum indeterminacy and how it might be related to incompleteness and undecidability.

\subsubsection{Manifestly self-referential frameworks}\label{s:frameworks}

Historically, mathematics tried to avoid self-reference by holding a strict hierarchy with numbers at the bottom and functions above them---and functionals above functions, etc. In this hierarchy, functions can operate on numbers but not the other way around.  However, G\"odel's theorem and Turing's halting problem show that, at least in the case of arithmetics and computation, this hierarchical structure is rather artificial and cannot avoid self-reference (see Fig.~\ref{f:duality}A and Appendix~\ref{s:incompleteness}). Moreover, the power of axiomatic systems and Turing machines are intimately related to self-reference~\cite{moore2011nature} (ch. 7). 

So, why not rather embrace self-reference? There is at least one mathematical formalism that fully embraces self-reference. Church's $\lambda$-Calculus is a neat formulation of computation in what we here call a manifestly self-referential way---i.e., in a way that makes its self-referential nature easy to see. In $\lambda$-Calculus there are no hierarchies, only strings operating on strings~\cite{moore2011nature} (ch. 7.4). So, strings are both the elements of the domain and the transformations on that same domain (see Fig~\ref{f:duality}B). There are no {\em a priori} assumptions that strings cannot operate on themselves nor that some strings play special roles. 

To provide some intuition to readers that may not be familiar with $\lambda$-Calculus, let us consider the following toy example involving single-argument functions operating on numbers. To put numbers and functions on an equal footing, we can represent numbers with constant functions~\cite{moore2011nature} (ch. 7.4.2). For instance, the number two can be represented with the function $\textsc{Two()}$ which, for any value of its argument, always returns the value two. This example is limited in that the ultimate interpretation of functions is the standard interpretation, i.e., the way they operate on numbers. Again, its sole purpose is to provide some intuition. 

Now, the function $\textsc{Square}()$ returns the square of its argument. When a function takes another function as argument it returns the composition of the two functions. For instance, $\textsc{Square}(\textsc{Two}()) \equiv [\textsc{Square}\circ\textsc{Two}]()$, which is equivalent to the function $\textsc{Four}()$ since $\textsc{Two}()$ returns two for any argument and $\textsc{Square}()$ with argument two returns four. Now, in this representation, numbers can also operate on functions. For instance, we can have $\textsc{Two}(\textsc{Square}())=[\textsc{Two}\circ\textsc{Square}]()$ which equals $\textsc{Two}()$, since $\textsc{Square}()$ yields the square of any number it takes as argument, but $\textsc{Two}()$ will return the value two no matter what value $\textsc{Square}()$ passes to it. $\lambda$-Calculus is simpler in that it is based solely on strings of characters, but more powerful in that it naturally allows for recursion, which can lead to self-reference. 

\subsubsection{Self-reference and relationalism }\label{s:self-relationalism}
In science self-reference has been traditionally avoided by strictly relying on third-person methods. For instance, according to Rovelli (emphasis ours):

\begin{myquotation} If we want to describe [observer] $O$ itself quantum mechanically, we can, {\em but we have to pick a different system $O^\prime$ as the observer}, and describe the way $O$ interacts with $O^\prime$. In this description, the quantum properties of $O$ are taken into account, but not the ones of $O^\prime$, because this description describes the effects of the rest of the world on $O^\prime$. (Ref.~\cite{rovelli2007quantum}, Sec. 5.6.2)
\end{myquotation}

Granted, scientists are usually considered physical systems that can be investigated with the same kind of tools other physical systems are investigated. However, there has been a hierarchy wherein scientists are implicitly assumed to have the special status of being able to investigate other physical systems from outside, as if they were not physical systems themselves that therefore necessarily interact with the systems they investigate. Whether such an interaction can be neglected or not is another question that to our knowledge has not been systematically investigated as we try to do here. 

Furthermore, a strict adherence to third-person methods {\em a priori} prevents us from even exploring the possibility that physical systems, like scientists, could investigate themselves (see below). That we currently may not be aware of reliable methods to do so does not necessarily imply they cannot exist~\cite{varela2017embodied,thompson2014waking,wallace2007hidden,loy2012nonduality} (see Appendix~\ref{s:physical-phenomenal}). The existence of manifestly self-referential frameworks like Church's $\lambda$-Calculus suggests it should be possible to develop a similarly manifestly self-referential scientific framework, if we do not insist in pushing away the puzzling phenomena that can come along with the strengths associated to self-reference.

Moreover, self-reference is closely related to relationalism, which has been key in scientific progress. Unlike absolute notions, which tend to single out some things as special, e.g., the Earth in the Ptolemaic model of the solar system, time in Newtonian mechanics, spacetime in the special theory of relativity, etc., relational notions tend to be more democratic and put things on an equal footing. This is precisely what $\lambda$-Calculus does, for instance, by putting on equal footing not only all strings of a given domain but also the elements of such a domain with the transformations on it. In a sense, $\lambda$-Calculus is about {\em relationships} between strings.

As another example, consider a dictionary~\cite{harnad1990symbol,levary2012loops,vincent2017hidden} out of which we are supposed to learn the meaning of every word. Now, to understand the meaning of a word we need to look up in the dictionary the meaning of the words used to define it, and so on recursively. However, if we continue doing this sooner or later we will come full circle, unless there are some words whose meaning is grounded from outside the dictionary, e.g., from direct experience~\cite{harnad1990symbol}. But this would give an absolute, special status to some words. If we insist instead that every word in the dictionary can only be learned from inside it, i.e., that the dictionary is complete, we will never be able to learn the meaning of any word in an absolute sense. 
A complete dictionary is a collection of semantic loops~\cite{levary2012loops} which can only provide the {\em relationship} between words, not their meaning in any absolute or inherent way. 

In line with one of the main messages of Einstein's relativity, i.e., relationalism, some Eastern traditions maintain that the existence of all natural phenomena in the universe, like the meaning of words in a complete dictionary, is {\em relational} or {\em dependently arisen}~\cite{varela2017embodied,loy2012nonduality}. This does not imply that natural phenomena do not exist, in the same way that Einstein's relativity does not imply either that time does not exist. This rather implies that natural phenomena, like time, do not exists {\em in an absolute or inherent way}---this lack of absolute existence is sometimes referred to as {\em emptiness}~\cite{varela2017embodied,thompson2014waking,lama2005universe,loy2012nonduality} (see Appendix~\ref{s:physical-phenomenal}).

\subsubsection{Self-reference and non-duality}\label{s:self-non}

\begin{figure}
\includegraphics[width=0.8\columnwidth]{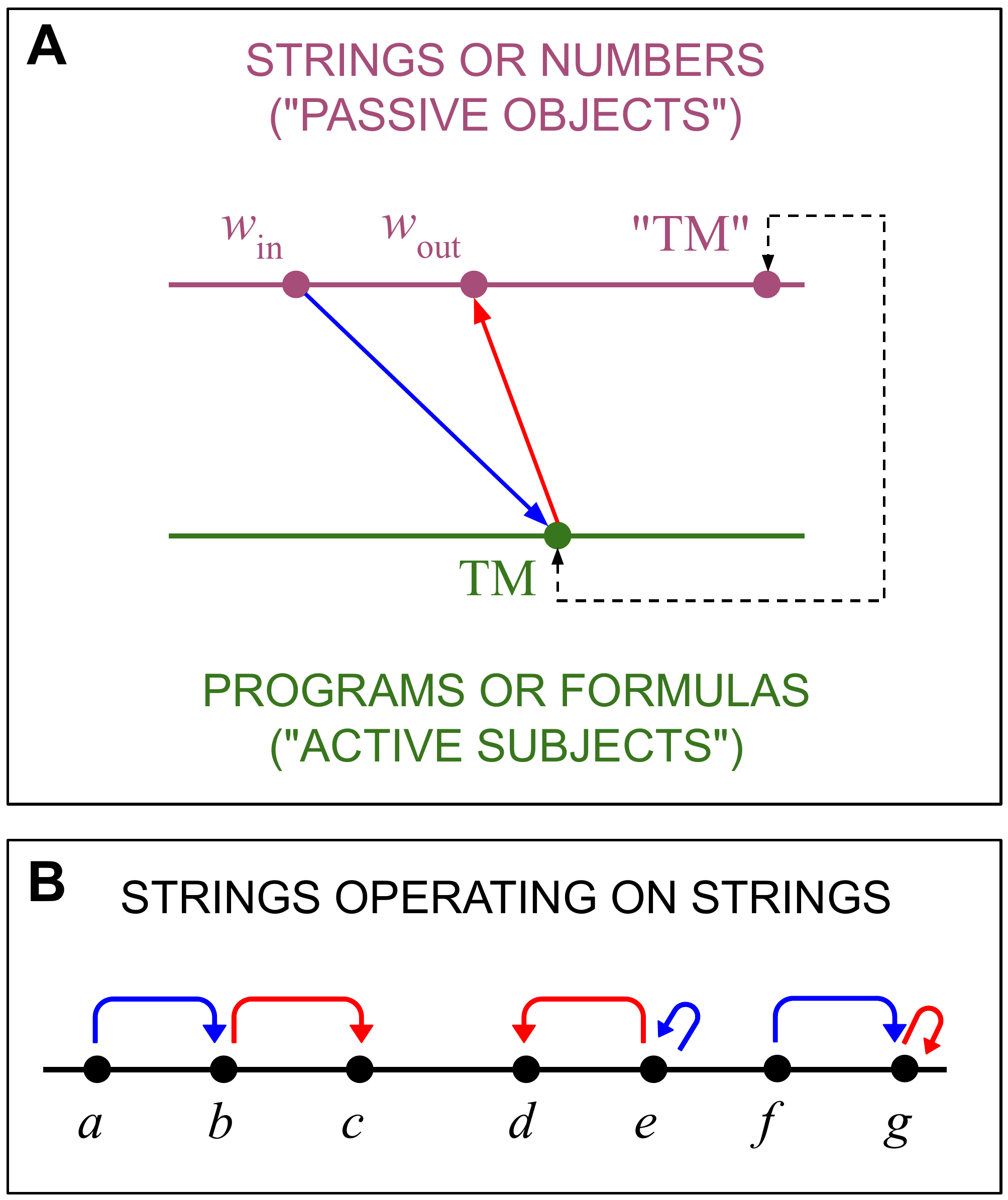}
\caption{{\em Self-reference and data-code duality:} (A) Turing machines manipulate strings while formulas in axiomatic systems make statements about numbers (see Appendix~\ref{s:aspects}). In this sense, Turing machines and formulas play active roles while strings and numbers play passive roles. Like subjects manipulating or talking about objects. However, both Turing machines and formulas can be mapped one-to-one to strings and numbers, respectively (dashed arrow). This property, usually referred to as ``data/code duality'', allows Turing machines and formulas to refer to themselves. (B) The existence of data/code duality suggests the hierarchy between Turing machines and strings, on the one hand, and formulas and numbers, on the other hand, is rather artificial. Indeed, Church's $\lambda$-Calculus is a neat formulation of computation in terms of strings only, which makes the self-referential abilities of computation clearly manifest. For instance, there are strings (like ``e'') that can take themselves as input and output another string, and others (like ``g'') that, for any string they take as input, output themselves. }\label{f:duality}
\end{figure}

As we already mentioned, the hierarchy between numbers and functions, or between strings (data) and Turing machines (code), do not avoid self-reference. $\lambda$-Calculus shows this distinction is rather artificial and provides an explicit example of how dual or complementary roles can be both present in self-referential systems (see Fig.~\ref{f:duality}). In other words, self-referential systems could be considered as single units that integrate dual roles. In common terminology this may be referred to as ``data/code duality'' to emphasize that the same system can play either role. In the language used by some Eastern traditions this might be considered a kind of {\em non-duality} (between data and code), emphasizing that data and code are not two separate entities. Now, within this context, data are data in the sense that, like passive objects, they are manipulated by codes. Similarly, codes are codes in the sense that, like active subjects, they manipulate data. In this sense, data and codes are relational to each other. 

The fully relational view embraced by some Eastern traditions implies that, like data and codes, the objects of observation and the observing subject are also relational to each other. This relational nature implies, in particular, that neither object nor subject exists in an absolute or inherent way (see Appendix~\ref{s:physical-phenomenal}). Granted, like the relative nature of time, this appears to contradict common sense. However, much like the relativity of time becomes apparent at unusually very high speeds, these Eastern traditions maintain that the relationalism between subject and object---or {\em subject/object non-duality},  as it is commonly referred to---becomes apparent at unusually very high mental concentrations~\cite{varela2017embodied,loy2012nonduality}. 

This is not necessarily as strange as it may sound. There is experimental evidence that the experience both of perceiving external objects and of being a subject that perceives them are associated to neural correlates within the same organism (cf. Fig.~\ref{f:lucid}). Although such experiences can be generated by external physical processes, according to mainstream consciousness neuroscience the organism's experience is intrinsically tied to the internal neural correlates such processes generate. For instance, a brain dead person would not experience such external processes even though they may be ``out there''. And the other way around, the experience of such external processes can be manipulated via direct interventions of the organism's neural system, e.g., via brain stimulation techniques, without changing the external processes themselves. 

If the experience of subject and object takes place within the same unit, i.e., the organism, there is the possibility that the organism could experience them as such, as not being two separate entities---see Fig.~\ref{f:lucid} and Appendix~\ref{s:thompson-consciousness} for an analogous situation that can happen in the case of a lucid dream. Under normal conditions the experience  of subject and object typically involve the integration of a wide variety of stimuli. However, according to some Eastern traditions, it is possible to reach unusually very high levels of mental concentration that can allow people to focus single-pointedly on essentially one single stimulus, e.g., the sensation of the breath going in and out through the nose. These can potentially remove most of the cognitive complexity of the perception process, leaving effectively one single object of observation that could in principle facilitate the experience of subject/object non-duality.

These kinds of self-referential experiments, where subjects investigate themselves, are qualitatively different from the kind of experiments traditionally performed in science~\cite{velmans2009understanding,blackmore2018consciousness}. However, it is in principle possible to compare the resulting subjective reports with traditional third-person investigations of the subjects' neural activity, for instance. Furthermore, these Eastern traditions have been exploring per centuries methods with the potential to rigorously train the mind to systematically experience these kinds of phenomena. They have also been exploring rigorous communication protocols that can potentially allow peers to reach inter-subjective agreements on their internal experiences, as done in mathematics---indeed, although it is not possible to show a mathematical theorem in the same way that we show a pointer in an experimental device, mathematicians can still agree on the proof of a theorem via the internal reasoning they carry out in their minds~\cite{wallace2007hidden}. So, there is previous work suggesting it may be possible to relax the {\em a priori} restriction that physical systems cannot refer to themselves and, in the spirit of $\lambda$-Calculus, build a manifestly self-referential scientific framework that combines third- and first-person methods, as it is being currently explored in the field of neurophenomenology (see Appendix~\ref{s:physical-phenomenal}).

\subsection{Incompleteness, undecidability, and quantum indeterminacy}\label{s:planck-self}

As we have discussed in Appendix~\ref{s:incompleteness}, self-reference can lead to puzzling phenomena like incompleteness and undecidability. If self-reference plays a key role in the derivation of the quantum formalism, as our approach suggests, it is natural to expect similar puzzling phenomena in quantum theory. Indeed, there have already been explorations of these ideas~\cite{dalla1977logical,brukner2009quantum, breuer1995impossibility,calude2004algorithmic,eisert2012quantum}. In particular, relationships between quantum randomness/uncertainty and incompleteness have been suggested~\cite{calude2004algorithmic,brukner2009quantum}. Here we provide a brief informal discussion of this possibility. 

Conceptually, the proofs of both undecidability and incompleteness are based on a binary relationship encoded in terms of a two-variable function (see Appendix~\ref{s:incompleteness}). The proof of undecidability is based on the function \textsc{Halts(``TM'', $w)$}, which encodes the relationship whether the Turing machine with code ``TM'' halts on input $w$ or not (see Eq.~\eqref{e:Halts}). Similarly, the proof of incompleteness is based on the function ${\rm Prov}(x, y)$, which encodes the relationship whether $y$ refers to a proof of the statement referred to by $x$ (see Eq.~\eqref{e:Prov}). A new object is then constructed in terms of these very relationships. For instance, the program {\sc Catch22}, introduced after Eq.~\eqref{e:Halts}, is defined in terms of the function {\sc Halts}, and the statement $\mathcal{S}$, introduced after Eq.~\eqref{e:Prov}, is defined in terms of the function Prov. Finally, it is shown that these new objects do not satisfy the very relationship in terms of which they are defined. More precisely, it is not possible to decide whether program {\sc Catch} would {\em halt} nor to find a {\em proof} for the statement $\mathcal{S}$.

Analogously, the key relationship in our work is {\em observation}. So, it is natural to expect that there may be some physical processes, defined in terms of this very relationship, which cannot be observed. Although we do not have a formal proof of this at the moment, there is indeed evidence that ``it is impossible for an observer  to distinguish all present states of a system in which he or she is contained, irrespective of whether this system is a classical or a quantum mechanical one and irrespective whether the time evolution is deterministic or stochastic"~\cite{breuer1995impossibility}. According to this result, observations of systems that contain the observer would always be inaccurate, there would always be an uncertainty associated. 

There is also a kind of unobservability in the nuerophenomenological framework discussed at length in Appendix~\ref{s:physical-phenomenal}. In this framework the most fundamental level of consciousness is considered to be non-dual awareness, which can be contentless and the very precondition for an observer to become aware of any content. Being the very precondition for any content to manifest in awareness, non-dual awareness cannot be objectified {\em as a content of awareness} itself. A toy model of this is an eye which is a precondition to see any visual object but it cannot directly see itself. Since the SRC implements the bare intrinsic perspective, it could be related to non-dual awareness (see Appendix~\ref{s:parallels}).

\section{On non-stoquastic Hamiltonians }\label{s:negative-general}

The analysis in the main text was restricted to factors $F_\ell$ with non-negative entries, which can be interpreted in probabilistic terms. Genuine quantum dynamics does not seem to have this restriction, though, since Hamiltonians can have negative, non-negative, and complex entries, which would lead to generic factors through the mapping $F_\ell = \id -\epsilon (H_{s, \ell}+H_{a, \ell})/\hbar$, where $H_{s, \ell}$ and $H_{a, \ell}$ are the real and imaginary parts of the Hamiltonian (see main text and Appendices~\ref{s:pair} and \ref{s:gaussian-matrix}). In Appendices~\ref{s:EM} and \ref{s:App_EM} we present a well known example of a quantum dynamics with a complex Hamiltonian that can be written in terms of real non-negative kernels with a probabilistic interpretation. Here we argue that our approach is not necessarily restricted to factors with non-negative entries either. This is relevant because it could be useful to compare our approach to the genuine (non-relativistic) quantum formalism (see below). 

We first discuss how Hamiltonians that directly lead to real kernels with negative off-diagonal entries can be obtained by truncating Hamiltonians on larger spaces associated to real kernels with non-negative entries. We will discuss a well-known example of an infinite-dimensional quantum system that can originally be described with factors with non-negative entries, which can therefore be interpreted in probabilistic terms. After a standard truncation of the original model to its first two energy levels, however, it turns into an effective two-dimensional system described by factors with negative off-diagonal entries. Thus it seems as if the original probabilistic interpretation is lost.  However, it is the former system which is actually implemented in the lab, where the experimenter has to make sure certain near resonance conditions are met for the effective two-dimensional model to be a good {\em approximation} of the actual system. 

We then discuss a simple example of a coin toss process, where negative numbers in probabilistic expressions can be given an operational meaning along this line. Afterwards, we briefly discuss how complex Hamiltonians could be understood as effective descriptions of more fundamental real Hamiltonians, at least in the examples we focus on here. 

In summary, here we discuss a few examples of so-called ``non-stoquastic'' Hamiltonians that, if {\em physically realizable}, turn out to be derivable as truncations or approximations of ``stoquastic'' Hamiltonians. Whether or not {\em physically realizable} (non-relativistic) non-stoquastic Hamiltonians can be derived in general from real non-negative kernels is an open question we leave for the future. This is interesting in that it may suggests deviations between our approach and the full quantum formalism. So, better understanding this may help test whether our approach indeed coincides with quantum theory and, if not, whether there may be some previously overlooked physical constraints that should be added to the quantum formalism. Indeed, the full quantum formalism presupposes that we can physically implement any unitary transformation. However, as the examples we briefly discuss here suggest, taking into account the actual physical implementation of such unitary operators may lead to some additional constraints. 

Importantly, this discussion is restricted to non-relativistic quantum mechanics. In relativistic quantum mechanics there might be some intrinsic sources of non-stoquasticity which may not be associated to non-negative real kernels. We leave the extension of these ideas to relativistic quantum mechanics for future work.  

\subsection{Hamiltonians with positive off-diagonal entries}\label{s:two_atom}

Here we discuss the example of an infinite-dimensional quantum system described by factors with non-negative entries which, after truncation to its first two energy levels, turns into an effective system described by factors with negative off-diagonal entries. The latter is known as a two-level atom interacting with a coherent radiation field~\cite{haken2005physics} (see Sec.~15.3 therein).

Indeed, consider the Hamiltonian of an atom modelled as an electron, described by the momentum operator $i\hbar\nabla_{\mathbf{r}}$, moving in a potential field $V(r)$ produced by the nucleus,
\be\label{e:H0_atom}
H_0 = -\frac{\hbar^2}{2m}\nabla_{\mathbf{r}}^2 + V(r) = \sum_{n=0}^\infty E_n \left.| n\ket\bra n\right|.
\ee
In the second equality we have expanded the Hamiltonian in terms of its eigenvalues $E_n$ and its eigenvectors $\left| n\ket$, where $n\in\mathbb{Z}$, $n\geq 0$. 

Now consider a perturbation 
\be\label{e:U(t)}
U(\mathbf{r},t) = e \mathbf{r}\cdot\mathbf{E}(t) = e\mathbf{r}\cdot\mathbf{E}_0\cos(\omega t),
\ee
so the perturbed Hamiltonian becomes ${H = H_0 + U}$. Notice that the full Hamiltonian operator, $H$, can be derived via a path integral with Lagrangian 
\be\label{e:L2}
L = \frac{m}{2}\dot{\mathbf{r}}^2 - V(r) - U(\mathbf{r},t).
\ee
Using the same ideas of Appendices~\ref{s:examples} and \ref{s:details}, we can also derive $H$ via a real non-negative kernel 
\be
\mathcal{K}_\epsilon(\mathbf{r}^\prime, \mathbf{r}) = e^{-\epsilon\mathcal{H}(\mathbf{r}^\prime, \mathbf{r})/\hbar}\geq 0,\label{e:K2}
\ee
where
\be
\mathcal{H}(\mathbf{r}^\prime, \mathbf{r}) = \frac{m}{2}\left(\frac{\mathbf{r}^\prime-\mathbf{r}}{\epsilon}\right)^2 + V\left(\frac{|\mathbf{r}^\prime + \mathbf{r}|}{2}\right) + U(t),\label{e:HK2}
\ee
is essentially the Wick rotation $\epsilon\to -i\epsilon$ of the Lagrangian $L$ in Eq.~\eqref{e:L2} (apart from an irrelevant global minus sign).

We will now see that, after a standard truncation of the full Hamiltonian $H=H_0+U$ (see Eqs.~\eqref{e:H0_atom} and \eqref{e:U(t)}) into an effective two-level system, we lose the equivalence with the positive kernel given by Eqs.~\eqref{e:K2} and \eqref{e:HK2}. Indeed, in the derivation of the Hamiltonian of a two-level atom it is usually assumed that the perturbation defined in Eq.~\eqref{e:U(t)} is near resonance with two relevant energy levels of the Hamiltonian $H_0$, say $E_0$ and $E_1$, i.e. {$|\omega - \omega_0|\lll\omega_0$}, where ${\hbar\omega_0 = E_1 - E_0}$. In this case, it is usually assumed that only the dynamics of these two energy levels matter. So, we can write
\be\label{e:H_R}
\begin{split}
H =& E_0\left| 0 \ket\bra 0\right| + E_1\left| 1 \ket\bra 1\right| + U_{01}\left(\left| 0 \ket\bra 1\right| + \left| 1 \ket\bra 0\right|\right)\\
&+ H_\mathcal{R},
\end{split}
\ee
where the first three terms in the right hand side of Eq.~\eqref{e:H_R} correspond to the transitions taking place within the subspace spanned by $\{\left|0\ket ,\left|1\ket\}$, and 
\be\label{e:Residual}
\begin{split}
H_\mathcal{R} =& \sum_{n=2}^\infty E_n \left.| n\ket\bra n\right|\\
&+ \sum_{m=0}^\infty\sum_{n> m, n\neq 0, 1}^\infty \left(U_{mn}\left| m \ket\bra n\right| + U_{nm}\left| n \ket\bra m\right|\right),
\end{split}
\ee
collects all the remaining transitions. Here we have written 
\be 
U_{mn} = U_{nm}^\ast =  \bra m\right| U(t) \left| n \ket,
\ee
for $m, n\in\mathbb{Z}$, $m,n\geq 0$. For the sake of illustration, we are restricting here to the case where $U_{01}=U^\ast_{01}$ can be chosen to be real and $U_{00} = U_{11} = 0$~\cite{haken2005physics} (see Sec.~15.3 therein); this explains the form of Eq.~\eqref{e:H_R}.

At this point it is argued that we can neglect $H_\mathcal{R}$ since the system is near resonance.  This yields the effective two-level Hamiltonian 
\be\label{e:H2eff}
H_{\rm eff} = \overline{E}\id^{(01)} -\frac{\hbar\omega_0}{2} \sigma_Z^{(01)} + D\cos(\omega t)\sigma_X^{01}, 
\ee
where $\overline{E}=(E_0+E_1)/2$, $D = \bra 0\right| \mathbf{r}\cdot\mathbf{E}_0\left| 1 \ket$, and
\BE
\id^{(01)} &=& \left| 0 \ket\bra 0\right| + \left| 1 \ket\bra 1\right|,\\
\sigma_X^{(01)} &=& \left| 0 \ket\bra 1\right| + \left| 1 \ket\bra 0\right|,\\
\sigma_Z^{(01)} &=& \left| 0 \ket\bra 0\right| - \left| 1 \ket\bra 1\right|.
\EE

If we now try to write this as a real kernel in the way we did in Appendices~\ref{s:pair}-\ref{s:details}
\be\label{e:F_eff_atom}
F_{\rm eff} = \id + \epsilon J_{\rm eff},
\ee
with $J_{\rm eff} = -H_{\rm eff}/\hbar$, we end up with off-diagonal negative entries due to the factor $\cos(\omega t)$ accompanying $\sigma_X$ in Eq.~\eqref{e:H2eff}. So, the full Hamiltonian $H=H_0 + U$ in Eq.~\eqref{e:H_R} can be represented in terms of the real positive kernel given by  Eqs.~\eqref{e:K2} and \eqref{e:HK2}, but the truncated effective Hamiltonian $H_{\rm eff}$ in Eq.~\eqref{e:H2eff} cannot. What happened? The full factor $F = \id-\epsilon H/\hbar$ associated to kernel $\mathcal{K}_\epsilon $ (see Eqs.~\eqref{e:K2} and \eqref{e:H_R}), which has only non-negative entries, can be written as 
\be\label{e:F_R_atom}
F = \mathcal{R} + F_{\rm eff} . 
\ee
So, even though the effective factor $F_{\rm eff}$ in Eq.~\eqref{e:F_eff_atom} can have negative off-diagonal entries, those would be ``corrected'' by the ``reference'' term $\mathcal{R}= -\epsilon H_\mathcal{R}/\hbar$ yielding only positive quantities with a clear probabilistic interpretation. In the next section we show a simple example of a similar situation.


We have discussed a physical system commonly studied in theoretical and experimental physics, which seemingly leads to negative numbers in probabilistic expressions. However, we showed that a description of the full system contains only positive numbers that can be clearly interpreted in probabilistic terms, and that negative numbers only appear after part of the system is neglected. We now discuss a simple example of a coin toss that has a similar structure to the example of the two-level atom presented above. Our aim is to provide some conceptual insight as to why negative numbers in probabilistic expressions do not need to be an issue more generally. 

Consider a biased coin which, when tossed, lands heads or tails with probability $1-p$ and $p$, respectively. So, the coin is described by the probability vector (cf. Eq.~\eqref{e:F_R_atom})
\be\label{e:coin_prob}
\mcP_{\rm coin} = \begin{pmatrix} 1-p \\ p\end{pmatrix} = \mathcal{R}_{\rm coin} + \mathcal{N}_{\rm coin}, 
\ee
where the first and second entries of the vectors involved refer to heads and tails, respectively, and 
\BE
\mathcal{R}_{\rm coin} &=& \frac{1}{2}\begin{pmatrix} 1 \\ 1\end{pmatrix} , \\
\mathcal{N}_{\rm coin} &=&  \frac{1-2 p}{2}\begin{pmatrix} 1 \\ -1\end{pmatrix}.
\EE
The rightmost side in Eq.~\eqref{e:coin_prob} can be formally understood as an abstract change of basis. We have done this to show that, even though we cannot directly interpret $\mcN_{\rm coin}$ in probabilistic terms because it contains negative numbers, we can still give it a probabilistic interpretation {\em relative} to the {\em reference} probability $\mcR_{\rm coin}$.

Indeed, we can generate a sample from $\mcP_{\rm coin}$ by first sampling from the flat reference probability distribution $\mcR_{\rm coin}$ and then {\em correcting} the sample generated according to $\mcN_{\rm coin}$. More precisely, $\mcN_{\rm coin}$ can be interpreted as encoding transitions from states associated to negative entries in $\mcN_{\rm coin}$ into states associated to positive entries. For concreteness, let $1-2p>0$. In this case, if the sample generated from $\mcR_{\rm coin}$ is in state tails, we have to flip it with probability $1-2p$. If it is in state heads, instead, we do nothing. So, the total probability for a sample to be heads is the probability for a sample to be heads in the first stage (i.e., $1/2$) plus the probability that it was tails in the first stage (i.e., $1/2$) and was flipped in the second stage (i.e., $1-2p$). Therefore, the probability for the sample to be in the heads state is $1-p = 1/2 + (1-2p)/2$ as required. Similarly, for tails. 

\subsection{Hamiltonians with complex entries}\label{s:non-stoquastic}

In Appendices~\ref{s:EM} and \ref{s:App_EM}  we present a well-known example of a Hamiltonian with complex entries which can nonetheless be interpreted in probabilistic terms. In Appendix~\ref{s:gaussian-complex} we also discuss the example of a quantum free particle which can originally be described in terms of real non-negative kernels. However, after a change of coordinates, the same free particle can be described in terms of a Hamiltonian operator with complex entries. Another example is a {\em real} Hamiltonian describing superconducting flux qubits which, under certain circumstances, can be approximated by an effective complex Hamiltonian~\cite{vinci2017non} (see Eqs. (1) and (11) therein). As a last example, consider the case of many interacting charged particles which can be described, in the non-relativistic limit, by the so-called Darwin Hamiltonian~\cite{essen1996darwin} (see Eq.~(62) therein). The Darwin Hamiltonian is real and so symmetric because it only contains terms quadratic in the canonical momenta, $\hat{p}_j = -i\hbar\nabla_{\bx_j}$. However, the case of a single charged particle in an electromagnetic field, which could in principle be generated by the remaining particles in the Darwin Hamiltonian, can be effectively described by a complex Hamiltonian (see Eq.~\eqref{e:Main_SchrodingerEM}). Relativistic correction might also yield complex Hamiltonians, but we are limiting here to non-relativistic systems.

Let us now briefly discuss the case of more general complex Hamiltonians, $H=H_s + H_a/ i$, which are associated to asymmetric dynamical matrices $J=J_s + J_a$ (see Appendices~\ref{s:diff-quantum}, \ref{s:pair}, \ref{s:EM} and \ref{s:App_EM}). The anti-symmetric part $J_a = -H_a/h$ is in principle associated to phenomena with an extrinsic directionality (see main text and Appendix~\ref{s:gaussian-quantum}) or that are irreversible---in the sense that the factor $F_\ell (x_{\ell + 1}, x_\ell)$ associated to the transition $x_\ell\to x_{\ell +1}$ differs from the factor $F_\ell(x_{\ell}, x_{\ell+1})$ associated to the reverse transition $x_{\ell+1}\to x_\ell$ only if the antisymmetric part of $F_\ell$ is non-zero (see Appendix~\ref{s:obj_constraints}). In the main text we have discussed how complex Hamiltonians can arise from the need to get rid of the part with extrinsic directionality before implementing the SRC because it only make sense for an observer external to the corresponding C-observer (cf. property (C3) in the main text). An alternative may be possible if we could understood irreversible phenomena as an effective description that arises when we do not take all dynamical variables into account, e.g. when we neglect the dynamics of the environment. In this case, when all phenomena are taken into account, there should not be irreversible phenomena. If so, complex Hamiltonians might be effective descriptions of more fundamental real (and so symmetric) Hamiltonians, as in the examples of superconducting flux qubits and the Darwin Hamiltonian above.

\section{Quantum observables and quantum measurements}\label{s:measurements}

In this work we have mostly focused on one observable: position. However, genuine quantum theory deals with different kinds of observables, e.g., momentum or energy (see Appendix~\ref{s:quantum_nutshell}). This naturally leads to consider other, sometimes counter-intuitive states different from position. For instance, it is common to speak of the momentum or energy eigenstates as if they physically exists as such, in the abstract. However, strictly speaking, this overlooks the fact that such abstract notions have to be implemented in the laboratory in terms of things we are familiar with, e.g., the position of a pointer in a measuring device. As pointed out by Feynman (emphasis ours):

\begin{myquotation} 
[A]ll measurements of quantum-mechanical systems could be made to reduce eventually to position and time measurements (e.g., the position of the needle on a meter or the time of flight of a particle). Because of this possibility {\em a theory formulated in terms of position measurements is complete enough in principle to describe all phenomena}. Nevertheless, it is convenient to try to answer directly a question involving, say, a measurement of momentum without insisting that the ultimate recording  of the equipment must be a position measurement and without having to analyze in detail that part of the apparatus which converts momentum to a recorded position. (Ref.~\cite{feynman2010quantum}, p. 96).
\end{myquotation}

So, our focus on position variables only should in principle allow for the description of all (non-relativistic) phenomena~\cite{feynman2010quantum} (p. 96).

Now, the view that a quantum measurement has to be implemented via a physical interaction of the system of interest with the measuring device is in line with our approach as it relies on the explicit modeling of all the parts of an experiment. Not only the system dynamics, as traditionally done, but also the preparation and the measurement steps, as well as the scientists involved. From this perspective, the abstract notion of quantum eigenstates different from position can pragmatically be considered as a shorthand notation for the kind of physical manipulations we have to do in a laboratory in order to measure such quantities. Again, strictly speaking, such physical manipulations can only be done in terms of things we are familiar with, like the position of a knob, a spatial pattern of lighting pixels in a display that we may interpret as digits, the angle between two mirrors, etc.
 
Here we first briefly discuss some aspects of quantum measurement theory using the measurement of momentum of a free particle as an illustration. In particular, we briefly discuss how the formalism of quantum measurements of momentum eigenstates arise from measurements of position of a pointer in a measuring device suitably interacting with the system. Afterwards, we discuss how the Hamiltonian operators required to implement the quantum measurement naturally arise from non-negative real kernels. This discussion is somehow informal as it is only for the sake of illustration. We leave a more formal discussion for future work. 

Now, in quantum measurement theory it is often assumed that the interaction between the measuring device and the observed system is almost instantaneous and the motion of the system during that time is ignored---this limit is usually referred to as impulsive measurement~\cite{peres2006quantum,aharonov2008quantum}. This usually requires the coupling constant between system and device to be large. However, as pointed out by Asher Peres: ``This drastic simplification is not always justified. Coupling constants occurring in nature are finite, and sometimes very small. It may therefore be necessary to couple the system and the measuring apparatus during a finite, possibly long time''~\cite{peres2006quantum} (ch. 12.6). In line with this, the coupling constants that we obtain from non-negative real kernels seem to be weak. It is not clear at this point whether this is necessarily so, or it is only an artifact of the preliminary approach we have taken here. We foresee two possible outcomes of a future more formal treatment, both of which are interesting in their own right. 

First, if the weakness of such coupling constants is indeed just an artifact, it may disappear in a more formal treatment of general measurements in our formalism. In this case our approach may be fully consistent with the standard formalism of (non-relativistic) quantum theory. Second, if the weakness of such coupling constants is not an artifact, our approach might be inconsistent with the full standard formalism of quantum theory. Such inconsistency may suggest some experiments to choose between the two. If the result of such experiments were to favor our formalism, we may learn about some physical constraints that we may have overlooked in the standard quantum formalism. 

\subsection{Some aspects of quantum measurement theory}\label{s:measAspects}

Here we briefly review some aspects of quantum measurement theory~\cite{peres2006quantum,aharonov2008quantum} (see chapters 12 and 7-9, respectively, therein). 
For concreteness and simplicity, we here consider the measurement of the momentum operator
\be\label{e:psys}
p = -i\hbar\frac{\partial}{\partial x}
\ee
of a one dimensional free particle of mass $m$, whose Hamiltonian operator is
\be\label{e:Hsys}
H_{\rm sys} = \frac{p^2}{2 m} = -\frac{\hbar^2}{2 m} \frac{\partial^2}{\partial x^2}.
\ee

One way to implement a measurement in the laboratory is via a suitable physical interaction of the system of interest with another system, i.e., a measuring device, which has a property we can directly see, e.g., the position of a pointer.  As measuring device, let us consider another one dimensional free particle of mass $M$, whose Hamiltonian operator is 
\BE\label{e:Hdev}
H_{\rm dev} = -\frac{\hbar^2}{2 M} \frac{\partial^2}{\partial X^2},
\EE
and whose position, $X$, plays the role of the pointer. The corresponding momentum operator 
\be\label{e:Pdev}
P = -i\hbar\frac{\partial}{\partial X}
\ee
is the generator of translations of the pointer state, i.e., of the position $X$ of the device particle. So, to generate translations of the pointer state according to the momentum $p$ of the system particle, which is what we want to measure, we can consider an interaction given by the Hamiltonian operator (cf. Eq.~(7.5) in Ref.~\cite{aharonov2008quantum})
\be\label{e:Hint}
H_{\rm int} = g p P .
\ee
So, the total Hamiltonian of the coupled system and device particles is
\be\label{e:Hmeas}
H_{\rm meas} = H_{\rm sys} + H_{\rm dev} + H_{\rm int}.
\ee

Assume the joint state of the system and device particles is initially uncorrelated, i.e., $\Psi_0(x, X) = \psi_{\rm sys}(x)\psi_{\rm dev}(X)$. Assume also that the initial state of the system particle is
\be\label{e:psiSys}
\psi_{\rm sys}(x) = \frac{1}{{\mathcal{A}}}\int c_k \frac{e^{i k x}}{\sqrt{2\pi}} \mathrm{d} k,
\ee
where $|\mathcal{A}|^2$ is a normalization constant and $e^{i k x}/\sqrt{2\pi}$ are the eigenstates of $p$ in Eq.~\eqref{e:psys}, which are also the eigenstates of $H_{\rm sys}$ in Eq.~\eqref{e:Hsys} as this is specified solely by $p$. 

Equation~\eqref{e:psiSys} and the well known result
\be\label{e:DiracDelta}
\int  \frac{e^{i x (k-k^\prime)}}{{2\pi}}\mathrm{d}x = \delta(k-k^\prime),
\ee
imply that
\be
\int |\psi_{\rm sys}(x)|^2 \mathrm{d} x = \frac{1}{|\mathcal{A}|^2}\int |c_k |^2  \mathrm{d} k = 1;
\ee
here we have used the fact that $\psi_{\rm sys}$ is normalized to get the rightmost result. 

Now, assume that the initial state of the device particle is 
\be\label{e:psiDev}
\psi_{\rm dev}(X) = \frac{e^{-X^2/4\sigma^2}}{\sqrt[4]{2\pi\sigma^2}}.
\ee
That is, the position of the device particle is initially centered around zero with an uncertainty characterized by $\sigma$.

Following Refs.~\cite{peres2006quantum,aharonov2008quantum} (see chapters 12 and 7-9, respectively, therein), let us neglect for simplicity the free particle Hamiltonians $H_{\rm sys}$ and $H_{\rm dev}$ in Eq.~\eqref{e:Hmeas}. So, the joint state of the system and device particles at time $\tau$ is given by 
\be\label{e:psi_tauMom}
\begin{split}
\Psi_\tau(x,X) &= e^{- i \tau g p P /\hbar}\Psi_0(x, X) \\
&=\int c_k e^{i k x}\left[e^{- \tau g \hbar k \frac{\partial}{\partial X}}\psi_{\rm dev}(X)\right]\mathrm{d} k\\
&=\int c_k e^{i k x}\frac{e^{-(X-\tau g \hbar k )^2/4\sigma^2}}{\sqrt[4]{2\pi\sigma^2}}\mathrm{d} k .
\end{split}
\ee
We have obtained this expression as follows. First, we have used Eq.~\eqref{e:psi'} in Appendix~\ref{s:quantum_nutshell} with $U= e^{-i H_{\rm int}\tau/\hbar}$, where $H_{\rm int}$ is given by Eq.~\eqref{e:Hint}. Second, we have replaced $\psi_{\rm sys}(x)$ and $\psi_{\rm dev}(X)$ in  $\Psi_0(x, X) = \psi_{\rm sys} (x)\psi_{\rm dev}(X)$ by the right hand side of Eqs.~\eqref{e:psiSys} and \eqref{e:psiDev}, respectively. Third, inside the integral defining $\psi_{\rm sys}(x)$, we have replaced the momentum operator $p=-i\hbar\partial / \partial x$ by its eigenvalues $\hbar k$ because $e^{i k x}/\sqrt{2\pi}$ are the corresponding eigenfunctions. Finally, we have used the fact that $e^{-\xi \frac{\partial}{\partial X}}\psi_{\rm dev}(X) = \psi_{\rm dev} (X - \xi)$, since the exponential of a derivative implements a Taylor expansion.  

Now, the probability to observe at time $\tau$ the system and device particle in positions $x$ and $X$, respectively, is $\mcP(x, X) = |\Psi_\tau(x,X)|^2$. However, we are {\em not} interested in directly observing the position of the system particle, $x$. Indeed, the whole point of using a measuring device is to {\em indirectly infer} the {\em momentum} of the system by observing {\em only} the pointer state. So, the probability we are interested in is actually the marginal 
\be
\begin{split}
\mcP_{\rm dev}(X) &= \int\mcP(x, X)\mathrm{d}x \\
&= \int\frac{|c_k|^2 }{|\mathcal{A}|^2} \frac{e^{-(X-\tau g \hbar k )^2/2\sigma^2}}{\sqrt{2\pi\sigma^2}}\mathrm{d}k,
\end{split}
\ee
where we have used Eq.~\eqref{e:DiracDelta}. 

Since the Gaussian distribution tends to a Dirac delta function when $\sigma^2\to 0$, for $\sigma$ small enough we have
\be
\mcP_{\rm dev}(X) \approx \int\frac{|c_k|^2}{|\mathcal{A}|^2} \delta(X-\tau g \hbar k )\mathrm{d}k.
\ee
This means that with probability $|c_k|^2/|\mathcal{A}|^2$ the position of the pointer is $X= g\tau\hbar k$, which is proportional to the momentum eigenstate $\hbar k$. So, the momentum of the particle $\hbar k = X/g\tau $ can indeed be measured from the position $X$ of the pointer, according to the measurement postulate of quantum theory (see Appendix~\ref{s:quantum_nutshell}; see, e.g., Refs.~\cite{schwabl2005quantum,nielsen2002quantum,aharonov2008quantum, peres2006quantum}).

\

\subsection{Implementation via real non-negative kernels}\label{s:measKernels}
Here we briefly discuss how the quantum measurements discussed in the previous subsection can be implemented via real non-negative kernels, following the framework described in Appendices~\ref{s:pair}-\ref{s:details}. 
Consider again the example of measuring the momentum of a one dimensional free particle of mass $m$ by coupling it to another one dimensional free particle of mass $M$, which plays the role of measuring device. As in the previous section, the position of the device particle plays the role of the pointer. 

As we shall see below, the total Hamiltonian-like function for this system is
\begin{widetext}
\be\label{e:mcHmeas}
\mcH_{\rm meas}(x, x^\prime ; X, X^\prime) = \mcH_{\rm sys}(x,x^\prime) + \mcH_{\rm dev}(X, X^\prime) + \mcH_{\rm int}(x, x^\prime ; X, X^\prime) 
\ee
\end{widetext}
where
\begin{widetext}
\BE
\mcH_{\rm sys}(x,x^\prime) &=& \frac{m}{2 (1-a^2)}\left(\frac{x-x^\prime}{\epsilon}\right)^2, \label{e:mcHsysMom} \\
\mcH_{\rm dev}(X, X^\prime) &=& \frac{M}{2 (1-a^2)} \left(\frac{X-X^\prime}{\epsilon}\right)^2,\\
\mcH_{\rm int}(x, x^\prime ; X, X^\prime) &=&  -\frac{a\sqrt{m M}}{1-a^2} \left( \frac{x-x^\prime}{\epsilon}\right)\left(\frac{X-X^\prime}{\epsilon}\right).\label{e:mcHintMom}
\EE
\end{widetext}
The factor $(1-a^2)^{-1}$, with $|a| < 1$, renormalizing the mass of the system and device particles is important to cancel out the effect of the interaction (see below). 

Using Eqs.~\eqref{e:mcHsysMom}-\eqref{e:mcHintMom}, Eq.~\eqref{e:mcHmeas} can be written as
\begin{widetext}
\be\label{e:NoetherMom}
\mcH_{\rm meas}(x,x^\prime ; X, X^\prime) = \frac{M}{2 } \left(\frac{\Delta X}{\epsilon}\right)^2 + \frac{m}{2(1-a^2)\epsilon^2}\left(\delta x - a\sqrt{\frac{M}{m}}\Delta X\right)^2,
\ee
\end{widetext}
where $\Delta X = X-X^\prime$ and $\delta x= x-x ^\prime$. So, the implementation of the momentum measurement could be interpreted as a perturbation of the system average {\em change} in position---from $\bra\delta x\ket = 0$ to $\bra\delta x\ket = a\sqrt{M/m}\Delta X$. There is also  a renormalization of the mass of the system---from $m$ to $m/(1-a^2)$. Furthermore, in the spirit of the theory of quantum measurements briefly outline in the previous subsection, the size of the perturbation depends on the {\em change} in the position of the device's pointer. 

We now show that the Hamiltonian function in Eq.~\eqref{e:mcHmeas} indeed leads to the corresponding Hamiltonian operator in Eq.~\eqref{e:Hmeas}. Using Eqs.~\eqref{e:mcHmeas}-\eqref{e:mcHintMom}, the corresponding transition kernel
\be
\mathcal{K}_{\rm meas}(x,x^\prime ; X, X^\prime) = \frac{e^{ -\epsilon\mcH_{\rm meas}(x, x^\prime ; X, X^\prime)/\hbar}}{Z},
\ee
where $Z$ is a suitable normalization factor, can be written as
\be\label{e:Kmeas}
\mathcal{K}_{\rm meas}(\delta x; \Delta X) = \frac{e^{-(\delta x , \Delta X) C^{-1}(\delta x , \Delta X)^T / 2}}{\sqrt{(2\pi)^2\det C}}.
\ee
Here $\delta x = x - x^\prime$ and $\Delta X = X - X^\prime$; furthermore,
\be
C^{-1} = 
\frac{1}{\hbar\epsilon (1-a^2)}
\begin{pmatrix}
m & -a\sqrt{m M} \\
-a\sqrt{m M} & M
\end{pmatrix}
\ee
is the inverse of the covariance matrix
\be\label{e:Cov}
C \equiv 
\begin{pmatrix}
\bra\delta x^2\ket & \bra\delta x\Delta X\ket \\
\bra\delta x\Delta X\ket & \bra\Delta X^2\ket
\end{pmatrix}
=
\hbar\epsilon
\begin{pmatrix}
\frac{1}{m} & \frac{a}{\sqrt{m M}} \\
\frac{a}{\sqrt{m M}} & \frac{1}{M}
\end{pmatrix} .
\ee

So, if the joint state of the system and device particles at time $t$ is $\Psi_t(x, X)$ the corresponding joint state at time $t+\epsilon$, with $0<\epsilon\ll 1$, is given by
\begin{widetext}
\be\label{e:PsiMeas}
\begin{split}
\Psi_{t + \epsilon}(x, X) &= \int \mathcal{K}_{\rm meas}(x,x^\prime ; X, X^\prime)\Psi_t(x^\prime , X^\prime)\mathrm{d} x^\prime \mathrm{d}X^\prime\\ 
&= \int\mathcal{K}_{\rm meas}(\delta x ; \Delta X) \Psi_t(x-\delta x , X-\Delta X)\mathrm{d} \delta x \mathrm{d} \Delta X\\
&= \Psi_t(x, X) + \frac{\hbar\epsilon}{2 m}\frac{\partial^2}{\partial x^2}\Psi_t(x, X) 
+ \frac{\hbar\epsilon}{2 M}\frac{\partial^2}{\partial X^2}\Psi_t(x, X) 
+ \frac{\hbar\epsilon a}{\sqrt{m M}}\frac{\partial^2}{\partial x\partial X}\Psi_t(x, X),
\end{split}
\ee
\end{widetext}
where we have written $x^\prime = x-\delta x$ and $X^\prime = X-\Delta X$. To get the third line in Eq.~\eqref{e:PsiMeas} we expanded $\Psi_t(x - \delta x, X- \Delta X)$ up to second order in $\delta x$ and $\Delta X$. Furthermore, we used the fact that the first-order moments of the Gaussian in Eq.~\eqref{e:Kmeas} are zero and the corresponding second-order moments are given by Eq.~\eqref{e:Cov} (cf. Eq.~\eqref{e:realGaussian} in Appendix~\ref{s:App_path}).

Now, we can build a Schr\"odinger equation from Eq.~\eqref{e:PsiMeas} by implementing the analogue of Eq.~\eqref{e:GaussianSchrodinger} in Appendix~\ref{s:App_path}, which effectively implements the SRC. This yields
\begin{widetext}
\be
\begin{split}
\epsilon\frac{\partial\Psi_t(x, X)}{\partial t} &= i\left[\int\mathcal{K}_{\rm meas}(\delta x ; \Delta X) \Psi_t(x-\delta x , X-\Delta X)\mathrm{d} \delta x \Delta X -\Psi_t(x , X) \right]\\
&= -i\frac{\epsilon}{\hbar} H_{\rm meas}\Psi_t(x, X),
\end{split}
\ee
\end{widetext}
where $H_{\rm meas}$ is given precisely by Eq.~\eqref{e:Hmeas} with $g = -a/\sqrt{m M}$. However, since $|a| < 1$ the coupling constant, $g = -a/\sqrt{m M}$, is somewhat weak, which contrasts with the impulsive limit commonly assumed in the theory of quantum measurements (see chapters 12 and 7-9 in Refs.~\cite{peres2006quantum,aharonov2008quantum}, respectively). Finally, multiplying both sides by $i\hbar/\epsilon$ we get the standard form of Schr\"odinger equation, i.e., $i\hbar\partial\Psi_t/\partial t = H_{\rm meas}\Psi_t$.

\

\section{Comparison to some related approaches to quantum theory}\label{s:comparison}
Here we briefly discuss two prominent approaches to quantum theory---QBism~\cite{debrota2018faqbism,fuchs2013quantum} and a recent derivation from information principles~\cite{d2017quantum}. We also briefly discuss a different approach~\cite{mueller2017law} to model the observer based on algorithmic information theory. Although still at a toy-model level, this recent algorithmic approach also points to some quantum-like phenomenology. We focus on how we currently understand these alternative approaches relate to ours. 

In particular, we describe some of the main similarities and differences with QBism in the next section. Afterwards, we argue that the ``purification principle'', which singles out quantum theory in the derivation from information principles~\cite{d2017quantum}, may be related to the self-referential coupling combined with the possibility to turn classical non-Markovian models into Markovian ones by enlarging the state space. Finally, we argue that the definition of ``first-person perspective'' in the algorithmic information approach~\ref{s:Muller}, which plays a central role in that theory, is not the genuine first-person perspective as experienced by the observer itself. It is rather a description of how such a ``first-person perspective'' should look from the third-person perspective of another, external observer---more like a non-relational version of Fig.~\ref{f:circular}A in the main text. 

As far as we know, though, these approaches stay mute about the potential origin of Planck's constant. In contrast, our approach suggests that its origin is rooted in the physics of the observer (see main text). 

\subsection{QBism}\label{s:qbism}

QBism is defined in Ref.~\cite{debrota2018faqbism} as ``an interpretation of quantum mechanics in which the ideas of agent and experience are fundamental. A `quantum measurement' is an act that an agent performs on the external world. A `quantum state' is an agent's encoding of her own personal expectations for what she might experience as a consequence of her actions. Moreover, each measurement outcome is a personal event, an experience specific to the agent who incites it. Subjective judgments thus comprise much of the quantum machinery, but the formalism of the theory establishes the standard to which agents should strive to hold their expectations, and that standard for the relations among beliefs is as objective as any other physical theory.''

Our approach shares with QBism the central role given to experience and the users of the quantum formalism, i.e., the scientists. Because this is the key fact in most of the resolutions proposed by QBism to the conceptual difficulties of quantum theory, such resolutions would be valid in our framework too. There are, however, some significant differences with QBism too. 

On a conceptual side, QBism treats scientists as rather abstract agents immersed in a publicly shared physical universe. The quantum formalism is seen as a normative criterion~\cite{debrota2018faqbism} (see Sec.~18 therein) setting ``the standard to which agents should strive to hold their expectations''. Such agents are betting, implicitly or explicitly, on their subsequent experiences, based on earlier ones, and the quantum formalism is a tool to help them place better bets~\cite{mermin2018making} (p. 8).  In contrast, we treat both scientists and the rest of the universe as physical systems with a dynamical role to play. As such, we explicitly model agents, using tools from cognitive science, as  physical systems learning from their interactions with the external world. So, we see quantum measurements as specific {\em physical} interactions between agents and the world, and  quantum states as {\em physically} encoding an agent's ``expectations for what she might experience as a consequence of her actions''.

On a technical side, QBism focuses on finite-dimensional systems and singles out, for any given dimension, a particular {\em reference measurement} as special---the so-called symmetric informationally complete measurements~\cite{fuchs2013quantum,debrota2018faqbism}. A proof of existence of such measurements for any finite dimension to our knowledge is still to be found. In contrast, we derive a quantum formalism directly by explicitly modeling scientist interacting with experimental systems, using standard probability theory as a starting point, with no restrictions on dimensionality. We hope our approach may suggest some alternative perspectives for the ongoing effort of the QBism community to derive the quantum formalism.

\subsection{An informational derivation of the quantum formalism}\label{s:informational}
The abstract formalism of quantum theory has been derived from sixth informational principles~\cite{d2017quantum}. Only one of these principles, the so-called {\em purification principle}, is highlighted by the authors as the distinctive axiom of quantum theory. This principle is interpreted as expressing a law of conservation of information, i.e., as the fact that every irreversible physical process can be regarded as arising from a reversible interaction of the system with an environment, which is eventually discarded~\cite{chiribella2010probabilistic,Chiribella-2011} (see also ch. 1.4 in Ref.~\cite{d2017quantum}). 

An irreversible transformation of a system $S$ in state $\rho$ can be written as $\rho^\prime = \sum_k E_k \rho E_k^\dag$, where $E_k$ are suitable operators (see, e.g., chapter 8.2 in Ref.~\cite{nielsen2002quantum}). This contrasts with the reversible transformations that we have focused on in this work, which can be written in terms of a single unitary operator $U$, i.e., $\rho^\prime = U\rho U^\dag$ with $U U^\dag = \id$ (see Appendix~\ref{s:quantum_nutshell}). Roughly, the purification principle essentially states that any irreversible transformation on a system $S$, characterized by a set of operators $\{E_k\}_k$, can always be implemented via a suitable reversible (unitary) transformation $U_{SE}$ acting on the system $S$ plus an environment $E$. More precisely, if $\rho^\prime = \sum_k E_k \rho E_k$ then we can always write $\rho^\prime = \tr_E \left[U_{SE}\left(\rho\otimes\sigma\right) U_{SE}^\dag\right]$, where the partial trace $\tr_E$ marginalizes over the environment, effectively discarding its degrees of freedom. Here $\sigma$ is the initial state of the environment and the tensor product $\rho\otimes\sigma$ specifies that the system and the environment are initially uncorrelated. See Ref.~\cite{d2017quantum} for a more precise description of the purification principle. 

In our approach the purification principle seems to be closely related to the SRC combined with the possibility to embed classical non-Markovian processes into classical Markovian ones with a larger state space. As a simple example of the latter, consider the non-Markovian process where the probability, $\mathcal{P}_{\rm nM}(x_{\ell +1}|x_\ell, x_{\ell -1})$, for the system to transition into state $x_{\ell +1}$ at time steps $\ell+1$ depends on the states, $x_\ell$ and $x_{\ell-1}$, at the previous two time steps. We can define a new variable $y_\ell = (x_\ell , x_{\ell-1})$, which takes values in a state space with twice the dimension of the original state space. In terms of these new variables, the process becomes Markovian with transition probability $\mathcal{P}_{\rm M}(y_{\ell +1}|y_\ell)$.

Indeed, our approach starts with classical Markovian-like processes, which are encoded by using either real probability matrices or BP messages (see Eq.~\eqref{em:P_l+1} in the main text and Appendix~\ref{s:EQM}). The role of the transition kernels is played by factors $F_\ell(x_{\ell+1},x_\ell )$, which only depend on the state of the system at two consecutive states (see Eq.~\eqref{em:F} in the main text). A unitary, time-reversible quantum formalism only arises {\em after} implementing the SRC between two Markov processes (see Fig.~\ref{f:first}C,F and Eqs.~\eqref{em:1PP_PAoB}-\eqref{em:free1PPb}). 

Now, if we have a factor that depends on more than two consecutive states, we would not get a time-reversible quantum formalism after implementing the SRC---at least not in the way we have implemented it here. However, we could in principle follow an approach similar to that described above. For instance, a factor like ${F}^{\rm nM}_\ell(x_{\ell + 1}, x_\ell, x_{\ell -1})$, which is analogous to the non-Markovian transition probability $\mcP_{nM}$ above, could be rewritten in terms of the same kind of variables, $y_\ell = (x_\ell , x_{\ell-1})$, as a Markovian-like factor ${F}^{\rm M}_{\ell}( y_{\ell +1}, y_\ell)$. Implementing the SRC in terms of the new variables would therefore lead to a time-reversible quantum formalism on such a larger state space. In this way our approach might lead to a derivation of the purification principle. 

\subsection{An attempt to model the observer via algorithmic information theory}\label{s:Muller}

A different, algorithmic-information theoretic approach introduced by M\"uller~\cite{mueller2017law} to investigate the observer also suggests the emergence of some quantum-like phenomena. However, as acknowledged by M\"uller~\cite{mueller2017law}, in its current version this approach does not attempt to derive actual known laws of motion. Our approach focuses instead on actual laws of nature, like Schr\"odinger or von Neumann equations. 

M\"uller characterizes the state of the observer with a binary string $s$, designate it as the observer's ``first-person perspective'', and assigns a conditional (algorithmic) probability $\mcP_{\rm algo}(s^\prime | s)$ for the state of the observer to transition from $s$ to $s^\prime$~\cite{mueller2017law} (see, e.g., Eqs.~(1), (3) and (4) therein). However, this looks to us much like a {\em non-relational} version of the {\em external} or third-person perspective described in Fig.~\ref{f:circular}A in the main text, where the observer is Wigner's friend but the external scientist (Wigner) who describes her is neglected (see Appendix~\ref{s:on_internal}).

Our approach suggests that this could indeed lead to some (imaginary-time) quantum-like features (see, e.g., Eqs.~\eqref{em:H} and \eqref{em:real_vN}), but it contrasts with the self-referential intrinsic perspective described in Fig.~\ref{f:first} in the main text, which has been both non-trivial to implement and key to obtain a (real-time) quantum dynamics. So, it is not clear to us that, in its current version, M\"uller's approach can lead to a fully fledged quantum-like formalism. It is a very promising approach, though, because its computational roots may facilitate the implementation of self-reference, and therefore the intrinsic perspective as defined here. In particular, it should be possible to incorporate $\mcP_{\rm algo}$ as part of the state of the observer using ideas similar to those we have used in this work. 

M\"uller~\cite{mueller2017law} mentions that his approach, in its current version, has a rather solipsistic flavor. At the moment it is not clear to us how to interpret this. However, our approach attempts to place both the phenomenal and the physical worlds on an equal but non-dual---i.e., relational---footing (see Appendix~\ref{s:physical-phenomenal}).


\end{document}